\pdfoutput=1

\documentclass[preprint2]{aastex}

\usepackage{graphicx}

\newcommand{\CII}{C~{\sc ii}}
\newcommand{\CIII}{C~{\sc iii}}
\newcommand{\CIV}{C~{\sc iv}}
\newcommand{\SiII}{Si~{\sc ii}}
\newcommand{\SiIII}{Si~{\sc iii}}
\newcommand{\SiIV}{Si~{\sc iv}}
\newcommand{\NIV}{N~{\sc iv}}
\newcommand{\NV}{N~{\sc v}}
\newcommand{\HeI}{He~{\sc i}}
\newcommand{\HeII}{He~{\sc ii}}
\newcommand{\HeIII}{He~{\sc iii}}
\newcommand{\HI}{H~{\sc i}}

\newcommand{\AlII}{Al~{\sc ii}}
\newcommand{\AlIII}{Al~{\sc iii}}
\newcommand{\FeII}{Fe~{\sc ii}}
\newcommand{\OI}{O~{\sc i}}
\newcommand{\OII}{O~{\sc ii}}
\newcommand{\OVI}{O~{\sc vi}}
\newcommand{\NiII}{Ni~{\sc ii}}
\newcommand{\MgII}{Mg~{\sc ii}}
\newcommand{\Cloudy}{C{\sc loudy}}

\slugcomment{Accepted by the Astrophysical Journal Supplement}

\shorttitle{QSO Metal Line Absorption Systems}
\shortauthors{Boksenberg \& Sargent}

\begin{document}

\title{PROPERTIES OF QSO METAL LINE ABSORPTION SYSTEMS AT HIGH REDSHIFTS: 
NATURE AND EVOLUTION OF THE ABSORBERS AND NEW EVIDENCE ON ESCAPE OF IONIZING 
RADIATION FROM GALAXIES\altaffilmark{*}}

\author{Alec Boksenberg}
\affil{University of Cambridge, Institute of Astronomy, Madingley Road, 
Cambridge, CB3 0HA, UK}
\email{boksy@ast.cam.ac.uk}

\and

\author{Wallace L.W. Sargent}
\affil{Division of Physics, Mathematics, and Astronomy, California 
Institute of Technology, 1200~E.~California Blvd., Pasadena, California 91125}

\altaffiltext{*}{The data presented herein were obtained at the W.M. Keck 
Observatory, which is operated as a scientific partnership among the 
California Institute of Technology, the University of California and the 
National Aeronautics and Space Administration. The Observatory was made 
possible by the generous financial support of the W.M. Keck Foundation.}

\begin{abstract}

Using Voigt-profile-fitting procedures on Keck HIRES spectra of nine 
QSOs we identify 1099 \CIV\ absorber components clumped in 201 systems 
outside the Lyman forest over $1.6 \lesssim z \lesssim 4.4$. With 
associated \SiIV, \CII, \SiII\ and \NV\ where available we investigate 
bulk statistical and ionization properties of the components and 
systems and find no significant change in redshift for \CIV\ and 
\SiIV\ while \CII, \SiII\ and \NV\ change substantially. The \CIV\ 
components exhibit strong clustering but no clustering is detected for 
systems on scales from 150 km s$^{-1}$ out to 50000 km s$^{-1}$. We 
conclude the clustering is due entirely to the peculiar velocities of 
gas present in the circumgalactic media of galaxies. Using specific 
combinations of ionic ratios we compare our observations with model 
ionization predictions for absorbers exposed to the metagalactic ionizing 
radiation background augmented by \emph{proximity} radiation from their 
associated galaxies and find the generally accepted means of radiative 
escape by transparent channels from the internal star-forming sites is 
spectrally not viable for our stronger absorbers. We develop an active 
scenario based on runaway stars with resulting changes in the efflux of 
radiation that naturally enable the needed spectral convergence and in 
turn provide empirical indicators of morphological evolution in the 
associated galaxies. Together with a coexisting population of relatively 
compact galaxies indicated by the weaker absorbers in our sample 
the collective escape of radiation is sufficient to maintain the IGM 
ionized over the full range $1.9 \lesssim z \lesssim 4.4$. 
\end{abstract}


\section{INTRODUCTION}

The spectra of exceptional quality delivered by the Keck~I High Resolution 
Spectrograph (HIRES; Vogt et al. 1994) have revealed individual metal 
absorption features related to the high-redshift Lyman forest for a large 
fraction of the stronger lines \citep{cow95,tyt95,soc96,wsl96}. By redshift 
{\it z} $\sim 3$ such absorbers are found to have a median carbon abundance 
approximately $10^{-2}$ of solar (with substantial scatter) and Si/C abundance 
similar to Galactic halo stars, although these values are based on rather 
uncertain ionization corrections \citep{soc96,rhs97}.\footnote{Damped Lyman 
$\alpha$ (DLA) absorbers suffer much less from this limitation (e.g., Pettini 
et al. 2008).} There is evidence for some metal enrichment also at 
considerably lower \HI\ column densities (Ellison et al. 2000; Schaye et al. 
2000a, 2003). While it is not yet fully clear how the pollution of the general 
forest material has come about, the observed metal absorbers provide a 
powerful \emph{in situ} probe of early stages in the growth of structure and 
the formation and evolution of galaxies and give an observational approach to 
determining the spectral character of the cosmological ionizing radiation 
background at those times.

Detailed hydrodynamical simulations of cosmological structure formation in the 
presence of a photoionizing background which yield direct quantities for 
comparison with observations have been developed by several groups 
\citep{cen94,zan95,her96,dat01,vie02} and much new work continues in this 
field. In these simulations it is straightforward to compute the neutral 
hydrogen absorption that would be produced in the light of a background QSO 
along an arbitrary line of sight through the simulation volume. It is 
impressive that such results can reproduce the evolving spectral appearance 
and statistical properties of cosmologically distributed \HI\ absorbers in 
considerable detail, spanning the range from the weakest detected to those 
showing damped Lyman~$\alpha$ profiles. An important insight gained from such 
simulations is that galaxies and \HI\ absorbers develop naturally together in 
the hierarchical formation of structure. High column density lines 
({\it N}(\HI) $\gtrsim 10^{17}$ cm$^{-2}$) arise from radiatively cooled gas 
associated with forming galaxies in collapsed, high density, compact regions. 
Lower column density absorption ({\it N}(\HI) $\lesssim 10^{15}$ cm$^{-2}$) 
occurs in the shallower dark matter potential wells, containing gas in various 
stages of gravitational infall and collapse; typically these are in the form 
of flattened or filamentary structures of moderate overdensity with Doppler 
parameters that are often set by peculiar motions or Hubble flow in addition 
to thermal broadening. Gravitational, pressure and ram-pressure confinement 
all play significant roles. Such a scenario in which metal absorption arises 
in gas assumed to be homogeneously enriched is discussed by Rauch et al. 
(1997). In their simulation the large velocity widths of some metal absorbers 
arise from interactions between associated protogalactic clumps or from 
alignments of groups of such objects along chance filaments in the line of 
sight. This simple model does not include stellar ``feedback'' by galaxies 
which could strongly modify the local gas distribution and kinematic state by 
producing outflows both enriching the nearby medium and opposing the general 
infalling motion. Increasing attention has since been given to accounting for 
stellar processes in galaxy formation simulations 
\citep{the02,cro02,sah02,maw03,nsh04} and recent much enhanced sophistication 
in the treatment of star formation and feedback has enabled the study of 
associated metal absorbers in ever more realistic simulations (e.g., Kawata 
\& Rauch 2007; Oppenheimer et al. 2009, 2012; Cen \& Chisari 2011; Smith et 
al. 2011; Shen et al. 2012, 2013) alongside considerable observational 
evidence that has been reported (e.g., Rauch et al. 1999, 2001; Bond et al. 
2001; Adelberger et al. 2003, 2005b; Aguirre et al. 2005; Simcoe et al. 2006; 
Steidel et al. 2010).
 
The strength and spectrum of the metagalactic ionizing radiation background at 
high redshift and the nature of the sources which ionized the intergalactic 
medium (IGM) have long been outstanding issues in cosmology. Although at 
relatively low to intermediate redshifts QSOs had been accepted as the main 
contributors to the metagalactic ionizing radiation, there was speculation on 
whether they dominate at the highest redshifts 
\citep{bec87,das87,sag87,bdo89,mam93,ham96}. Most of later studies demonstrate 
that QSOs fall short of producing enough flux to satisfy measurements at 
redshifts significantly beyond $z \sim 3$ where the space density of bright 
QSOs is sharply decreasing \citep{cec97,mhr99,sco00,bck01}. The presence of a 
strong population of high redshift Lyman-break galaxies \citep{ste96,ste99} 
and the apparent observations of significant flux beyond the \HI\ Lyman limit 
escaping from such galaxies \citep{spa01,sha06,iwa09} lend strong support to 
the idea that star-forming regions dominated the ionizing background at early 
times \citep{hae01,bol06}. The spectral shape of the ultraviolet background 
radiation should be reflected in the ionization pattern of QSO metal system 
absorbers \citep{cha86,bes86,sas89,var93,gas97,bsr01,ber02,aga07} and this can 
be used to identify the character of the ionizing sources if the spectral 
modifications due to propagation of the radiation through the IGM are properly 
accounted for (Haardt \& Madau 1996, 2012; Madau \& Haardt 2009). In turn, 
this can contribute to determinations of heavy element abundances in denser 
regions of the IGM as well as in outer circumgalactic regions (CGM).

In this paper we study the statistical properties and ionization states of a 
large sample of metal absorbers and trace their evolution in redshift, greatly 
extending our earlier work \citep{bok97,bsr01,bsr03}. In \S2 we describe 
the observations and initial data reduction and in \S3 outline our analysis of 
the absorbers and show that these exist consistently as multi-phase-ionization 
systems containing individually identifiable \emph{single-phase component 
regions} of widely-ranging ionization state. In \S4 we give full tables of 
results with supporting information, comments and displays and in \S5 define 
samples which we use in the following sections. In \S6 we derive statistical 
properties and the redshift evolution of absorber system and component 
quantities, also contrasting properties of absorbers physically close to the 
sightline QSOs. In \S7 we present clustering studies from which we deduce the 
nature of the contributing absorbers. In \S8 we present redshift distributions 
of ionic ratios and ionic ratio combinations. In investigative studies aiming 
to fit the latter observations we compare these with results of photoionization 
modeling using the \Cloudy\ code \citep{fer98,fer11} within which we apply 
exposure by the metagalactic ionizing radiation background from QSOs and 
galaxies at large \citep{ham12}, also considering possible effects of delayed 
reionization of intergalactic \HeII, collisional ionization and absorber
parameters, but find all these insufficient to match our data for the strong 
absorbers in our sample particularly at the higher redshifts in our range. We 
then explore the addition of direct radiation at the absorbers emitted from 
their local galaxies and discover that to meet the ionic patterns we observe 
requires considerable modification of the spectral distribution of the escaping 
ionizing radiation from that commonly assumed. In \S9 we draw strong new 
conclusions about the mechanism needed to bring this about and demonstrate 
empirically that an active process based on runaway stars can, in natural and 
robust fashion, enable quantitative matches to the observed ionic ratio 
combinations for the strong absorbers. Then, together with similarly derived 
indications for a coexisting population of galaxies associated with the 
numerous weaker absorbers in our sample we are able to conclude with a 
specific evolving picture of galaxy geometrical properties that can yield 
levels of escape fraction adequate for necessary contributions to the assumed 
metagalactic ionizing radiation background at least over $1.9 < z < 4.4$. We 
summarize our results in \S10. 

This paper follows from \citet{bsr03}, using the same data base with some 
adjustment to sample membership and addition of a few measured values obtained
from the original spectra. However the analytical content and conclusions are 
brought up to date and greatly extended and benefit from the remarkable 
intervening developments in the associated fields. 

Unless otherwise stated, at appropriate sections below we assume a 
($\Omega_{M}$, $\Omega_{\Lambda}$, $\Omega_{b}$, h) = (0.3, 0.7 0.045, 0.7) 
cosmology.

\section{OBSERVATIONS AND DATA REDUCTION}

The work presented in this paper is based on HIRES observations of a set 
of nine QSOs with redshifts $2.32 < z_{\rm em} < 4.56$, listed with associated 
information in Table 1. Most of the spectra are obtained using a slitwidth of 
$0\farcs86$ yielding a resolution $\sim 6.6$~km~s$^{-1}$ FWHM covered by 
roughly three pixels. The exception is for the gravitationally-lensed object 
Q1422$+$2309 \citep{pbw92} of which, to confine our observations to a single 
sightline consistent with the rest, we take the data from image component C 
obtained in excellent seeing ($\lesssim 0\farcs6$) using a narrower slitwidth, 
$0\farcs574$, yielding $\sim 4.4$~km~s$^{-1}$ FWHM, and with position angle 
set to minimize contamination from the closely-spaced neighbouring components 
A and B \citep{rsb99,rsb01}.

Two partially overlapping configurations for each wavelength region gives 
complete coverage of the free spectral range for the HIRES echelle format. The 
data are reduced as described in \citet{bas97}, with the individual exposures 
for each QSO wavelength-shifted to heliocentric, vacuum values and added 
together with weight according to the signal-to-noise ratio (S/N). Continuum 
levels are delineated by means of polynomial fits to regions apparently free 
of absorption features and these are used to produce continuum-normalized 
spectra ready for the analysis of the absorption systems. 

The spectra show complex and quite large changes in S/N along their length due 
to the variations in instrument efficiency, the sharply uneven signal over the 
range resulting from the overlapped setups, the intrinsic spectral variation 
of signal intensity from the QSOs themselves, and the incidence of atmospheric 
emission and absorption features. In order to account for these individual 
patterns of S/N in the subsequent analysis procedure (see \S3.2), matching 
statistical 1$\sigma$ error arrays are built up during the reduction stages 
for use in association with each completed spectrum file. A general 
indication of minimum signal quality is given in Table 1 by S/N values 
sampled at a few rest-wavelength positions common to each spectrum and 
avoiding the QSO broad emission lines (within which the S/N is higher).

The primary regions of the nine spectra that are used for obtaining the data 
in this paper are displayed in Figure 1. Atmospheric emission and absorption 
features become progressively more dense with increasing redshift. For clarity 
we have removed these from the displays and bridge the gaps with \emph{thin 
continuous lines}. The positions of the atmospheric emission features are 
identified by vertical ticks with lengths crudely graded to indicate 
intensity, and the complex pattern of absorption is shown by indicative full 
spectra set below the QSO spectra; both reflect the accumulated individual 
heliocentric shifts applied discretely over the observing runs to correct the 
QSO spectra. More on this and much additional information is given in 
footnotes to Tables 2--10 (see \S4.1).

\section{DETERMINATION OF ABSORPTION-LINE PARAMETERS}

\subsection{Selection Strategy}

Outside the Lyman forest the metal absorbers typically appear as well-defined 
clumps with velocity structure ranging up to a few hundred km s$^{-1}$ in 
width. In general there are wide expanses of apparently clear redshift space 
between such clumps. We classify these absorbing entities as {\it systems} and 
identify them by the presence of \CIV, the prevalent ion and the only one 
detected in the weakest systems. Stronger systems also contain lines of some, 
or occasionally most, of the species \SiIV, \CII, \SiII, \NV, \OI, \AlII, 
\AlIII, \FeII\ and \NiII, if available in the observed spectral range. Within 
each system we define a population of physical ``clouds'', termed 
{\it components}, each having a Gaussian velocity distribution of specific 
width and which collectively reproduce the system velocity structure in detail.
  
Because of blending with \HI\ absorption for metal lines in the Lyman forest, 
which becomes increasingly severe to higher redshifts, our data sample is built 
primarily on measurements made outside the forest; only in some exceptionally 
favourable cases are metal lines in the forest included. 

For the work presented in this paper we concentrate on the lines 
\CIV~$\lambda\lambda$1548.2049,1550.7785, 
\SiIV~$\lambda\lambda$1393.7602,1402.7730, 
\CII~$\lambda$1334.5323, \SiII~$\lambda$1260.4221 and 
\NV~$\lambda\lambda$1238.821,1242.804 (vacuum wavengths).\footnote{All 
rest-frame wavelengths and related atomic data used here are from 
\citet{mor03}.} If a line is not detected at the wavelength expected from the 
presence of other species in a given component of the same system, an upper 
limit is determined (see \S3.2). In systems with \SiII~$\lambda$1260.4221 
obscured in the forest we substitute $\lambda$1526.7070 and include 
$\lambda$1304.3702 if available, but these are relatively much weaker and 
detected only when \SiII\ is strong; when not detected, the upper limits 
obtained from these are often too high to be useful.

\subsection{Profile-Fitting Analysis}

For the analysis of our spectra we use the Voigt profile-fitting code VPFIT 
developed by Carswell and collaborators \citep{car02} and kindly made 
available to us. VPFIT is a $\chi^{2}$-minimization program capable of making 
detailed fits to the absorption profiles of several different transitions 
simultaneously. It estimates redshift, $z$, Doppler parameter, {\it b} 
($= \sqrt{2}\sigma$) which includes thermal and Gaussian turbulence broadening 
terms, and column density, $N$, all with their associated errors, for the 
individual components of the systems in the defined fitting regions. For the 
instrumental resolution included in this procedure we take 
${\it b}_{instr} = 2.83$ km~s$^{-1}$ for Q1422$+$239C and 
${\it b}_{instr} = 3.96$ km~s$^{-1}$ (6.61 km~s$^{-1}$ FWHM) for the rest 
(see \S2). Since the $\chi^{2}$-minimization technique operates on the reduced 
spectra there is a degree of correlation between neighbouring pixels arising 
from the rebinning of the data in the reduction process. To compensate for this 
slight smoothing we derive an error-correction file for each spectrum 
obtained by comparison of values at many wavelengths in the accumulated error 
array with directly measured values of the root-mean-square fluctuation in the 
final continuum. The error-correction factor is relatively small, typically in 
the range 1.1--1.3, and is applied from this file automatically within VPFIT.  
 
While the lower ionization species often are dominated by narrower components 
and the higher by broader, we find, consistent with the S/N, 
that \emph{regardless of width} a matching range of components \emph{having 
representation across the ionized species} invariably are present, albeit with 
different strengths. We conclude that the different components observed in any 
system trace individual physical cloud regions, and, depending on the specific
ionization states of these regions, the associated species contribute 
appropriately to the individual balance of component absorption strengths. In 
other words, the individual components identified in the metal systems quite 
closely represent \emph{single-phase-ionization absorbing regions co-existing 
in multi-phase system complexes}. We demonstrate this here with a simple 
example case. Following the analysis procedure we describe in detail in this 
section, and using a system at ${\it z} = 2.285$ in Q1626$+$6433, we show 
selectively in Figure 2 (taken from Figure 4 and Table 2) how the derived 
components are distributed in radial velocity among the low to high ionic 
species. While different components dominate in \SiIV, \CII, \SiII\ and \NV, 
all coexist as blended, strong constituents of the \CIV\ overall profile, 
well-separable in our Voigt profile-fitting analysis. Note in particular the 
two nearly coincident (in redshift) components 2 and 3 having markedly 
different widths: both are strong in \CIV\ but change greatly in dominance 
between the observed low and high ionization counterpart species. Among these 
\SiIV\ shows related intermediate behaviour. Being closely at the same 
redshift, simplistically these two identifiable regions could be separated 
along the line of sight or separated radially within a single structure. 
Either way, the physical components of this simple absorption system 
straightforwardly represent individual regions having quite different degrees 
of ionization unambiguously traceable through the species. Along the same 
lines, a considerably more complex example is discussed in \S4. We deal with 
component ionization structure more quantitatively in \S8 and \S9. 

The unifying assumptions we therefore adopt in our analysis are (i) that the 
component redshift structure seen in one ion corresponds exactly to that in 
any other ion of the same system, while allowing that in general the relative 
strengths of the corresponding components be different, and (ii) that the 
{\it b-}values of corresponding components in different ionic species of a 
given system each include the same temperature and same Gaussian turbulent 
motion associated by the expression \citep{rau96}
\begin{equation}
{b_{ion}^{2} = b_{turb}^{2} + \frac{2kT}{m_{ion}}},   
\end{equation}
\noindent where $m_{ion}$ denotes the atomic mass of the ion. This relation is 
formally handled within the VPFIT procedures.

For each QSO the first step in our procedure is to identify all \CIV\ 
absorption doublets evident outside the forest. In VPFIT, doublet or other 
members of the same ion automatically are fitted with the same parameters when 
the wavelength regions where they occur are specified. We begin our analysis 
by deriving initial redshifts of all necessary components that can be 
identified in the fitting of the profiles of the \CIV\ doublets alone. Next, 
in a given system, all other members of our defined set of ionic species 
potentially present in the available wavelength range, \emph{whether apparent 
or not}, are assigned the same initial set of component redshifts and linked 
to track together with \CIV\ in the subsequent fitting stages. Doublet species 
are accepted even when only one well-defined member is outside the Lyman 
forest. The component {\it b-}values for all ionic species of the same atom 
(\CIV\ with \CII\ and \SiIV\ with \SiII) are also linked to track together. 
Finally, to enable VPFIT to derive mutually consistent {\it b-}values among 
the different species, following the requirements discussed above, we assign 
initial trial component temperatures and within VPFIT physically link the 
{\it b-}values of all atomic species present. To set these initial 
temperatures we use, as an indicative aid, results given by \citet{rau96} 
from a formal decomposition into thermal and Gaussian non-thermal motions in 
a sample of related \CIV\ and \SiIV\ absorption components, shown in their 
Figure 3 as a plot of \CIV\ thermal against total {\it b-}values. Finally, 
VPFIT then is allowed to attempt simultaneous integral fits to the 
distributed common components making up the line profiles of the different 
species in a system by iteratively adjusting in physically consistent fashion 
the components' common temperatures and individual total {\it b-}values and 
column densities, while refining the common redshifts, in a first complete 
pass. Potential components not detected ``drop out'' of the analysis and 
subsequently are assigned upper limits in the manner described below. Equally, 
if a member of the set of associated species in a system has components which 
are mildly saturated but others have clear profiles, the procedure enables the 
saturated components to be identified in the analysis with parameters in 
common and be assigned consistent column densities with appropriate errors.

In this consolidated procedure involving all species in a system, most members 
are well separated from lines at other redshifts and matters are generally 
straightforward. When blending with any species at other redshifts does occur 
we find reliable values usually can be obtained by simultaneously fitting all 
members of the different systems present. In similar fashion, where necessary 
we accomodate all but the most severe interference from atmospheric emission 
and absorption within our profile fitting analysis. In the relatively few 
cases where some or all components of one doublet member are too badly 
contaminated, we make do solely with the equivalent components of the other 
member. Equally, if both members are partially contaminated we select the 
available component structure from free areas between the two. Component 
parameter values judged to be too uncertain due to blending or contamination 
are excluded from the subsequent scientific analysis. 

\citet{rau96} derive a mean temperature $3.8 \times 10^{4}$ K for their 
sample but note that their analysis is dominated by narrower components which 
have smaller measurement errors than the more uncertain broader features. 
They also suggest that a tail in the {\it b-}value distribution towards large 
values, apparently indicating temperatures beyond $10^{5}$ K, may well 
represent blends of components. Our analysis supports this view. Since \CIV\ 
is more common and generally stronger than the other species, we note that a 
blend of components representing a typical mix of relatively high ionization 
regions tends often to appear significantly broader in \CIV\ than \SiIV\ due 
to the relative weakness of \SiIV. If interpretated as a {\it single} feature 
such a blend indeed is likely to indicate an erroneously high temperature. 
Furthermore, we show in \S8.4 that collisional ionization at temperatures 
near $10^{5}$ K cannot be a significant contender for the absorbing regions 
we observe. 

For the relatively few \emph{broader} components present we find our fits to 
the observed profiles are generally consistent with a dominant turbulence 
contribution, there being relatively little difference in the {\it b-}values 
among the species, although with considerable uncertainty in the balance 
between the thermal and turbulence contributions in some shallow cases. In 
our analysis we set the temperature of all these broader components at the 
nominal mean value $3.8 \times 10^{4}$ K given by 
\citet{rau96}.\footnote{Although uncertain as we have already commented, such 
a value may well be consistent for relatively low density regions, as we 
indicate in \S8.5.} Nevertheless, we show now in the following that there is 
little dependence on temperature in our results involving the observed broader 
lines. 

To explore the sensitivity of this approximation we use a complex 
system at $z = 2.291$ in Q1626$+$6433 described in \S4.2 and shown in detail 
in Figures 4 and 6. This is an unusual system containing several broad 
components which overlap numerous narrow ones. It is an appropriate example 
because in the fitting process components are not treated in isolation but 
adjusted relative to one another to achieve the overall fit to the data; the 
degree to which a given component is influenced by others in this process then 
depends on their relative strengths over its range of overlap. We compare the 
column density results for all components in the complex system from two 
separate computational runs using the widely-spaced fixed temperatures 
$1\times10^{4}$ K and $1\times10^{5}$ K collectively assigned to the three 
broadest components, 2, 4 and 14 identified in Figure 6 (where we use the 
nominal temperature $3.8\times10^{4}$ K). These two temperatures exceed the 
bounds expected for clouds of low metallicity and density photoionized by the 
intergalactic ultraviolet background radiation (see \S8.5), and make a 
stringent test (note however, as already indicated above, in using this 
artificial range we do not imply that at the upper temperature the broad 
components are collisionally ionized). In the two cases the initial temperature 
assignments of the remainder of the components are made identical to those in 
the nominal case, following the procedure already described. In Figure 3 we 
show the resultant column densities for \CIV, \SiIV, \CII\ and \SiII\ obtained. 
The two sets of values are not significantly different, nor do they differ 
significantly from our results using the adopted nominal temperature for the 
three broadest components. It is apparent that the derived column densities 
for the individual components are very little affected by the adopted thermal 
properties of the \emph{broad} components in an absorption complex and gives 
us confidence that our necessarily approximate approach for these 
nevertheless gives a sound procedure for obtaining reliable column densities 
for all the components. We give the broad absorption features further 
attention in \S4.

The more constrained temperature bounds set by the widths of the 
\emph{narrower} components in effect make the derived column densities for 
these, also, relatively insensitive to temperature. Such insensitivity to 
profile width applies equally for the instrumental profile, which varies by 
$\sim \pm3.8\%$ in velocity width over the spectral range \citep{bas97}, and 
allows us to use a single, averaged, figure for each spectrum, as given 
earlier.

In some occasional instances a minor velocity shift relative to \CIV, 
typically of a small fraction of 1 km~s$^{-1}$, is made during the fitting 
process to improve the relative alignment of one or other of the profiles of 
widely separated species in the same system. Shift patterns are different 
among the QSOs and not all show detectable internal misalignments. The need 
for such slight shifts both between \CIV\ and \SiIV\ and, when available, 
between widely spread \SiII\ transitions, indicates that these are not due to 
any physical ionization phenomena. As indicated by \citet{bas97}, we 
attribute the observed small misalignments to slight departures from true of 
the polynomial fits to the wavelength scale. Notwithstanding, as we 
establish from ``before and after'' cases, even without correction there is 
no significant effect on the derived column densities or {\it b-}values. 

Component parameter errors as given by VPFIT are nominally $1\sigma$ values, 
but confusion between too-closely overlapping components with comparable 
parameters can give very large apparent errors due to uncertainty in value 
assignments between them.\footnote{Nevertheless, the combined column 
densities in such cases remains accurate and VPFIT then gives the correct 
error for the \emph{total} column density of a set of adjacent components in 
a complex.} Consequently, in making our profile fits we avoid ``overfitting'' 
and adopt the general rule to end with the minimum number of components that 
give a reasonable fit after achieving a reduced $\chi^{2}$ close to 1 per 
degree of freedom globally for the set of spectral regions linked in the 
analysis. 

Generally our initial procedure results in a set of profiles which correspond 
well with the data for all of the components detected in each species. We 
iterate the process by making a further two or three VPFIT runs while 
minimizing the necessary number of components. For some strong, 
well-separated, components, reliable {\it b-}values sometimes can be obtained 
independently for Si as well as C, although such cases are relatively few. 

At the conclusion of the fitting process for each system we obtain the 
associated errors on the component column densities alone by fixing the 
corresponding values of $z$ and {\it b-}value in a final iteration. In this 
operation we also derive upper limits for all potential components within the 
different species which have been too weak to survive the first pass of the 
analysis. We do this by re-introducing them with the appropriate fixed values 
for $z$ and {\it b-}value and a small assigned column density well below the 
level of significance. The associated error values which are obtained become 
the adopted 1$\sigma$ upper limits; all such values are on the linear portion 
of the curve of growth.

It is important to note that the wide range of ionization conditions 
generally found within a complex system means that determinations of ionic 
\emph{ratios} cannot accurately be made from ratios of \emph{total} system 
quantities: to be physically meaningful they must be determined individually 
from the components of the multi-phase systems, as we do later in this paper. 
Moreover, when significant system structure is identifiable in an absorption 
spectrum the intrinsic ability of the Voigt profile-fitting technique to 
segregate different absorbing regions in a complex system gives it 
significant advantage over the apparently more direct method of computing 
optical depths throughout a system \citep{sas91,son98} because the latter 
cannot account for overlapping blends of adjacent components, interlopers 
from other systems or differential temperature broadening between species of 
the same physical component.\footnote{However, for very weak or heavily 
contaminated signals related to low-density regions of intergalactic gas, 
not easily detectable directly, the optical depth technique coupled with 
multiple stacking of spectra introduced by \citet{cas98} becomes a powerful 
tool in such circumstances \citep{ell00,sch00,agu02,ara04}.}

\section{PRESENTATION OF RESULTS}

\subsection{The Data}

Following the prescription outlined in the previous section, for each system 
we obtain excellent simultaneous fits over all species with a single pattern 
of component redshifts and appropriately linked {\it b-}values. As an 
illustration of the method, we show in Figure 4 a set of our detailed VPFIT 
results for Q1626+6433. We include all available target transitions 
superimposed on the observations for all detected systems having more than 
just \CIV\ accessible outside the Lyman forest. The spectra of all nine QSOs 
(refer to the condensed displays in Figure 1) are treated equally and in 
Tables 2--10 we list the full body of values derived, here with 
``unaccompanied'' \CIV\ features included. Column 1 gives the absorption 
redshift {\it z} as vacuum, heliocentric values; column 2 gives 
{\it b-}values for C; column 3 gives linked {\it b-}values for Si as 
described in \S3.2, shown bracketed, and independent values when left as a 
free parameter, unbracketed; columns 4 to 8 give column densities; and 
column 9 identifies components by number within a system. For N, derived 
{\it b-}values are not listed but with its comparable atomic mass these 
intrinsically are close to those of C when representing the same component. 
Each system is sub-headed in the tables with the mean redshift of its 
constituent components. Where there is severe contamination, strong 
saturation or too much confusion from blending by lines at other redshifts to 
give useful quantities, no entry is given and explained in a footnote. The 
listed {\it b-}values for C almost always are the result of fits to \CIV\ 
which usually dominates in strength over \CII. In the rare cases when 
component regions have low ionization and \CIV\ is very weak, independent 
{\it b-}values are derived solely from \CII\ (and additionally \SiII\ when 
left as a free parameter). In all, for the well-separated, stronger 
components the formal $1\sigma$ error in {\it z} typically is 
$\lesssim$ 0.000005, and in {\it b-}value (for C, and Si when 
independently derived), $\lesssim$ 0.5 km~s$^{-1}$.

Very few indeed of the absorption features in our spectra remain unidentified 
and of these most are very weak and some may be spurious. \emph{No dense 
forest of weak metal lines} is seen even in the spectra having the highest 
S/N in our sample (see also: Ellison et al. 2000; Aracil et al. 2004). To 
allow ready comparison among the QSOs, in Figure 5 we show ``spike diagrams'' 
displaying the column densities of all \emph{detected} components identified 
in Tables 2--10; note that the vertical scales for \SiIV, \SiII\ and \NV\ are 
lower by 1 dex than for the others. Systems containing one or more components 
with significant Lyman~$\alpha$ damping wings (we take the column density 
$N$(\HI) $\gtrsim 10^{18.7}$) are marked {\it d}. The missing indeterminate 
values indicated in the tables are relatively few and the gaps have only a 
minor effect on the appearance of these diagrams. The coverage in redshift 
outside the Lyman forest for the main transitions of our sample occurs between 
two dotted vertical lines shown in each frame except for \SiII\ where we 
include such limits for all three transitions 
$\lambda\lambda\lambda$1260,1304,1527. Present in the figures are the rare 
values reliably obtained in the Lyman forest for the standard lines of \SiIV, 
\CII, \SiII\ and \NV, clarified in the table footnotes.

\subsection{Broad Absorption Features}

As introduced in \S3 a number of systems contain one or more broad components 
(up to a few $\times 10$ km~s$^{-1}$), generally of high ionization, 
self-consistently present in both members of the \CIV\ doublet, that overlap 
in velocity space with many of the more numerous narrower components. Often a 
broad feature protrudes (in velocity) at the edge of a system boundary and can 
be seen directly as having a structureless appearance. In some cases the 
presence of a significant broad feature immersed among the body of narrow 
components in a system is required to reproduce the profile of one or more of 
the observed species. While we could contrive to construct the broad features 
from a solid blend of numerous narrow components we do not believe this 
is justified by the data, and good, consistent, fits using single 
broad components are obtained in general. Nevertheless, the implicit model 
envisaged in the VPFIT profile constructions characterizes each assumed cloud 
in a system simply by thermal and Gaussian turbulence broadening, while 
significant velocity \emph{gradients} could also feature in the true overall 
absorption profile. Broad, high ionization components thus might represent 
regions of low volume density dominated by bulk motions \citep{rau96}. 
Limitations to the detection of weak, broad components are discussed in \S6.5.

A simple example with partially exposed high ionization broad components is 
the ${\it z} = 2.056$ system in Q1626$+$6433: the full observed range of 
species is displayed in Figure 4 and the constituent components of the 
\CIV~$\lambda$1548 profile are shown in the upper left panel of Figure 6. In 
the latter, component 1, with {\it b} $\sim$ 21 km~s$^{-1}$, is well enough 
separated from the others for its smooth outline to be clearly seen, and 
component 4, with {\it b} $\sim$ 55 km~s$^{-1}$, although shallower (and 
partially overlapping with component 1) reveals a smooth extended wing. The 
associated two panels on the upper right separately show combinations of the 
broad and narrow components. 

In the lower panels of the same figure we show details of the 
${\it z} = 2.291$ system in Q1626$+$6433, already used in \S3.2, as a more 
complex as well as more comprehensive example, with broad features which are 
more immersed in the system. We go first to the panels on the left. The 
narrower components, {\it b}(\CIV) $\lesssim$ 10 km~s$^{-1}$ (see Table 2), 
here mostly have quite low ionization, with \CII\ relatively strong; 
components 11 and 17 are the exceptions. Of the broader components, 2 and 4 
unusually also have significant strength in \CII, while the remainder have 
relatively high ionization with \CIV\ strong and \CII\ very weak or 
undetected. 

The associated three sets of panels on the right highlight different subsets 
of these components. While the contributions of the combined high ionization 
components dominate the overall profile of \CIV\ (middle set), it is 
particularly striking that the embedded combined subset of narrow, lower 
ionization components (top set) closely resembles the \SiIV\ overall profile 
and indeed the \CII\ profile. As stressed before, this emphasises the 
successful segregation of differently ionized regions. Of the broadest 
components (bottom set) the high ionization component 14, with 
{\it b} $\sim$ 32 km~s$^{-1}$, is the strongest by far in \CIV\ and the only 
significant component identified in the weak \NV\ profiles shown in Figure 4.
\citet{fox07a} discuss the origin of broad and narrow \CIV\ components in 
DLA and sub-DLA systems.

\section{DEFINITION OF SAMPLES}

From the data in Tables 2--10 we define two samples {\it sa} and {\it sb}. 
Sample {\it sa} is selected, as far as possible, to be statistically 
homogeneous; we use this for the investigations of properties in \S6. Sample 
{\it sb}, enlarged as defined below, is used in \S8 and \S9 to probe the 
ionization state of the gas through individual ionic \emph{ratios} which 
does not, within defined limits, require a statistically homogeneous 
population of absorbers. For clustering studies in \S7 we use both samples, 
as there indicated.

We find from VPFIT trials using selected higher redshift systems having 
several available Lyman series members, for which we assume the \HI\ profiles 
contain at least the component populations identified in the metals, that the 
absorbers in our chosen samples generally are optically thin in the Lyman 
continuum. However, higher members of the Lyman series are not uniformly 
available in the redshift range of our data. From Lyman~$\alpha$ alone, the 
strongly saturated nature of most of the observed system profiles, as can be 
seen from the examples in Figure 4, compounded by the large thermal 
broadening experienced by \HI\ relative to the metal species studied here, 
causes severe confusion among adjacent \HI\ components in a great many of the 
systems. Consequently we did not attempt to set formal $N$(\HI) thresholds, 
except in the particular cases of components showing discernible 
Lyman~$\alpha$ damping wings as described in the sample definitions below. 
In \S8.5 we show how \HI\ column density, as well as metallicity, would 
influence the derived ionic ratios.

The samples are defined as follows:

sample {\it sa}---for the statistical studies in \S6, includes all \CIV, 
\SiIV, \CII, \SiII\ and \NV\ lines which fall outside the Lyman forest but 
limiting \SiII\ only to the strong transition $\lambda$1260, while excluding 
all components in the nine systems mentioned in \S4.1 showing Lyman~$\alpha$ 
damping wings indicating $N$(\HI) $\gtrsim 10^{18.7}$ cm$^{-2}$ for \emph{any} 
of its components, namely: $z = 1.927$ in Q1626$+$6433, $z = 2.438$ in 
Q1442$+$2931, $z = 2.761$ in Q1107$+$4847, $z = 2.904$ in Q0636$+$6801, 
$z = 2.770$ and 2.827 in Q1425$+$6039, $z = 4.031$ in Q1645$+$5520, 
$z = 3.317$ in Q1055$+$4611, and $z = 4.080$ in Q2237$-$0607 (see Figure 5 and 
footnotes to Tables 2--10);\footnote{The high \HI\ column density may have 
been a driver for obtaining the spectra of some of these systems. They are 
excluded from sample {\it sa} with the aim that this becomes homogeneous.}
used also as the baseline sample in the clustering studies in \S7.

sample {\it sb}---includes all components in sample {\it sa}; if 
\SiII~$\lambda$1260 is in the forest, adds observed lines of 
$\lambda\lambda$1304,1527 outside the forest; adds strong, unambiguous lines 
in relatively clear regions of the forest for any of the species except \CIV; 
for the ionization studies in \S8 and \S9 excludes the systems at 
$z =$ 2.827, 3.317 and 4.080 each showing densely distributed \HI\ absorption 
collectively having $N$(\HI) $\gtrsim 2 \times 10^{20}$ cm$^{-2}$ (i.e., being 
in the full ``damped Lyman~$\alpha$'' class) while including the other six of 
the related systems which show milder damping wings centred on very few 
specific components, but excludes the latter components to limit 
self-shielding effects; used without these stated exclusions in the clustering 
studies in \S7 as there indicated.

Sample {\it sa} contains 887 \CIV\ components in 192 systems and sample 
{\it sb}, 1099 \CIV\ components in 201 systems. When specified in \S\S6--8 
we set appropriate column density thresholds to reduce bias due to 
variations in S/N across the redshift range.

Unless otherwise stated, to avoid significant QSO proximity ionization effects 
we select systems having velocity $\gtrsim$ 3000 km s$^{-1}$ from the nominal 
redshifts of the background QSOs \citep{par01,wil08}. We find no significant 
differences in our results or conclusions if we extend this limit to 5000 
km s$^{-1}$ (\S6.1 and \S6.2).

\section{STATISTICAL PROPERTIES} 

\subsection{Evolution of C~{\scriptsize IV}, Si~{\scriptsize IV}, 
C~{\scriptsize II}, Si~{\scriptsize II} and N~{\scriptsize V}}

The distribution in redshift of the \CIV\ component column densities, 
$N$(\CIV), for the full sample {\it sa} is shown in the top panel of Figure 
7(a). For each system the values appear as vertical distributions of data 
points. To avoid confusion in the crowded figures, errors, listed in Tables 
2--10, are not indicated in this and subsequent similar figures; mostly these 
are relatively small. The histogram set in the bottom of the panel shows the 
aggregated pattern of sightlines covered in redshift outside the Lyman forest
to the nine QSOs. The second and third panels similarly show component 
subsets of the simpler and the more complex systems manifest in the \CIV\ 
sample, taking systems respectively having numbers $n \leqslant 6$ and 
$n \geqslant 7$ identified components. By these we wish to explore whether 
our wide observed range fundamentally reflects more than one population. 
While the precise demarcation we use here is not unique, this nevertheless 
serves to contrast two indicative groups we choose to compare. We tried 
populations with combinations of other, similar, values, for example 
$n \leqslant 5$ and $n \geqslant 9$, with insignificant change in outcomes 
that follow. Finally, the column densities \emph{summed} over each system are 
displayed in the bottom panel.   

The panels in Figure 7(b) give displays for the \SiIV, \CII, \SiII\ 
($\lambda$1260) and \NV\ component column densities corresponding to the top
panel of Figure 7(a) (note the vertical scale shifts for \SiIV, \SiII\ and 
\NV\ relative to \CIV). As before, the aggregated coverage in redshift 
outside the Lyman forest is indicated by the inset histogram for each ion. 
Within the individual permitted redshift ranges all system components 
detected in \CIV\ are represented in the other species either as determined 
values or upper limits. Only \SiIV\ and \CII\ cover the complete redshift 
range, while \SiII\ and \NV\ are rather poorly sampled. By comparing with 
\CIV\ it is evident that \SiIV\ and \CII\ are detected preferentially in the 
\emph{complex} systems (perhaps partly because these are generally the 
stronger systems) but recall these relate to sample {\it sa}, not the 
extended sample {\it sb} with its additional strong systems and used in \S8 
and \S9. \SiII\ mimics \CII\ quite closely within the redshift intervals in 
common. For \NV, despite the meagre coverage, we note that at the lower 
redshifts this is detected in most of the few narrow windows available 
suggesting that \emph{in this range the ion is quite prevalent}, while at 
higher redshifts there are no detections.

Within the sampling limitations, the general indication of the displays in 
the two figures is that at least \CIV\ and \SiIV\ are distributed quite 
uniformly over the observed redshift range. The data for these and the rest 
of the species, taken as systems, are treated quantitatively in \S6.3.

\subsection{Absorbers Physically Close to the Sightline QSOs}  

It is interesting to compare the relative ionization states of absorbers in 
close physical proximity to the sightline QSOs, excluded from the samples as 
is defined in \S5. We show in Figure 8, in condensed displays corresponding 
to the top panel in Figure 7(a) and the panels in Figure 7(b), the 
distribution in redshift of all component species with system velocities 
$\lesssim$ 3000 km s$^{-1}$ from the nominal redshifts of their related QSOs. 
Over the whole observed range in redshift the character of these results 
conforms consistently with expectations for gas illuminated by relatively 
hard radiation: \CII\ and \SiII\ are weak to the extent that they are almost 
always undetected, and there is relatively strong \NV. This is in significant 
contrast with the presentations in Figures 7(a) and (b). As we show in \S8
and \S9, at the higher redshifts the radiative environment of the widely 
distributed strong absorbers not directly influenced by the sightline QSOs 
increasingly becomes dominated by much softer spectral contributions from 
associated galaxies, thus consistent with non-detection of \NV.
 
\subsection{Evolution of System Ionic Number Density and Column Density}

In Figure 9(a) we display redshift evolution plots for all the observed 
species in the systems of sample {\it sa}. As a baseline for all species we 
select systems having total column density 
$N_{syst}$(\CIV) $> 1 \times 10^{12}$~cm$^{-2}$, and additionally apply the 
individual thresholds indicated in the figure for \SiIV, \CII, \SiII\ and 
\NV. Imposing these thresholds gives close to homogeneous sampling for each 
ion over the observed redshift range.

Looking first at \CIV, the six panels give the total number of systems per 
unit redshift interval $d{\cal N}_{syst}/dz$ and the aggregated column 
density per unit redshift interval $dN_{tot}/dz$ as a function of redshift 
for the sets of \emph{all}, \emph{simple} and \emph{complex} systems in 
sample {\it sa}. The data are corrected for multiple redshift coverage and 
summed over the adopted bins covered by the horizontal bars. The indicated 
$\pm1\sigma$ uncertainties for $d{\cal N}_{syst}/dz$ are defined by the 
number of systems in each bin detected above the column density threshold.
For $dN_{tot}/dz$ the uncertainties similarly are based on the number of 
systems present in each bin but now with each system weighted by its total 
column density, and are dominated by systems with the highest column 
densities.

To examine the evolution of the number density of systems with redshift,  
${\cal N}(z)$ ($\equiv d{\cal N}_{syst}/dz$), in the comoving volume we use 
\citep{mis02}
\begin{equation} 
{\cal N}(z) = 
{\cal N}_0\frac{(1 + z)^{2+\epsilon}}{\sqrt{\Omega_{\rm M}(1 + z)^3 
+ \Omega_{\Lambda}}},
\end{equation}
\noindent where ${\cal N}_0$ is the local value of ${\cal N}(z)$. If the 
absorbers have constant comoving volume density and constant proper size 
then $\epsilon = 0$. Contained in the expression is the $\Lambda$CDM 
cosmology-corrected absorption pathlength interval which for a given 
redshift interval $dz$ is given by
\begin{equation} 
dX = \frac{(1 + z)^{2}}{\sqrt{\Omega_{\rm M}(1 + z)^3  
+ \Omega_{\Lambda}}}\phd dz\phantom{.........}\Omega = 1.
\end{equation}
In the left-hand column of \CIV\ panels ($d{\cal N}_{syst}/dz$) we show in 
dotted lines representations of unevolving populations relating to the three 
data sets and vertically scaled for easy comparison. It is evident that the 
data are consistent with \emph{no evolution in number density}. \citet{mis02} 
using their combined sample EM15 including data from \citet{sbs88} and 
\citet{ste90} obtain $\epsilon = -1.18$, quite strongly evolving in the sense 
of increasing number density with cosmic time as the two earlier studies had 
found (largely with the same data). However, ours is a very sensitive survey 
of relatively few QSOs but containing a large number of weak systems; the 
other surveys have much lower spectral resolution and yield only strong 
systems, with rest-frame equivalent width $W_0 > 0.15$ \AA\ (implying 
$N_{syst}$(\CIV) $\gtrsim 5 \times 10^{13}$~cm$^{-2}$), from nearly an order 
of magnitude more QSOs yet with a comparable number of systems in the combined 
sample. Nevertheless, the complex, therefore strong, systems in our sample
{\it sa}, although relatively few in number, show no evolution in number 
density. 

The same form as equation 2 can be used for the comoving total column density 
of the absorber population by substituting $N(z)$ ($\equiv dN_{tot}/dz$) and 
$N_0$. In the corresponding right-hand column of panels we again show in 
dotted lines representations of unevolving populations. Within the 
statistical accuracy these data are consistent, also, with \emph{no evolution 
in total column density}. A difference in evolution between weaker and 
stronger systems was noticed by \citet{ste90} in his high column density 
sample but here we find no evidence of this. 

The data for \SiIV\ are treated in similar fashion to that described above. 
The outcome follows a pattern largely resembling that for \CIV. Within the 
statistical accuracy we can again conclude that in number density and column 
density the data are consistent with an \emph{unevolving population}.

For \CII\ the data are rather limited and the resultant errors are large. 
Complex systems now strongly dominate the column density population (note the 
large scale-change in $dN_{tot}/dz$ for the subset of simple systems), 
consequently the complex and full samples follow closely similar behaviour. 
While there appears to be a trend of positive evolution with cosmic time both 
in number density and column density, the significance of these is not high. 

The data are even more limited for \SiII\ but indicate an evolutionary trend 
consistent with that shown by \CII. 

\NV\ is sampled yet more poorly in redshift but, at face value, from 
$z \sim 3$ the data indicate a steep rise in total column density with cosmic 
time while there is an apparent complete absence of detection at earlier 
times. 

The indicated lack of evolution in \CIV\ and \SiIV\ column density is perhaps
surprising: over the extensive redshift range observed, large changes might 
be expected {\it a priori} in response to the evolving baryon density and 
character of the ionizing sources, even for constant metallicity. However, we 
need to be cautious in interpreting the observations in Figure 9 in this way. 
As we see later, we are not here taking a freely global view of isolated 
absorbing regions at large in the cosmos but deal predominantly with 
\emph{regions closely associated with star-forming galaxies}. This may tend 
to unify the properties of the absorbing systems despite the large time-spans 
existing among them. In \S8 and \S9 we return to this in our full analysis of 
the ionization conditions exhibited by the individual \emph{components} of 
the systems in the extended sample {\it sb}.

\subsection{Evolution of C~{\scriptsize IV} and Si~{\scriptsize IV} Mass 
Density}

The comoving \CIV\ mass density is given by \citep{lan91}
\begin{equation} 
\Omega_{\scriptsize\textrm{\CIV}}(z) = 
\frac{H_0 \textrm{\phantom{.}}m_{\scriptsize\textrm{\CIV}}}
{c \textrm{\phantom{.}}\rho_{crit}}
\textrm{\phantom{.}} \frac{\sum N_{tot}(\textrm{\CIV},z)}{\Delta X(z)},
\end{equation}
\noindent where $\rho_{crit} = 1.89 \times 10^{-29}h^2$ g cm$^{-3}$ is the 
cosmological closure density, $m_{\scriptsize\textrm{\CIV}}$ is the mass of 
the ion, $H_0 = 100h$ km s$^{-1}$ Mpc$^{-1}$ (we use h = 0.7) and 
$\Delta X(z)$ is the total redshift path over which it is measured. We 
calculate $\Delta X(z)$ from equation (3) and 
$\sum N_{tot}(\textrm{\CIV},z)$ from the sample indicated in Figure 9(a). 
The four bins we show for \CIV\ in the figure have mean redshifts 
$\langle z \rangle = 1.96$, 2.65, 3.35, 4.10. For the full sample (top right 
panel of the \CIV\ set) we obtain the respective values (1$\sigma$ 
uncertainties) $\Omega_{\scriptsize\textrm{\CIV}} = (4.14\pm1.74$, 
$3.65\pm1.26$, $3.26\pm0.87$, $2.70\pm0.89)\times10^{-8}$. Although there 
an apparent mild evolutionary trend, our values are formally consistent 
with being invariant over the range $1.6 < z < 4.5$. Songaila (2001, 2005) 
also found a more or less constant trend in the same redshift range. The mean 
over the full redshift range of our values above is 
$\langle \Omega_{\scriptsize\textrm{\CIV}} \rangle = (3.44\pm1.24)\times10^{-8}$ 
at $\langle z \rangle = 3.05$. \citet{dod10} in the range $1.6 < z < 3.6$ 
find somewhat more of a rise towards lower redshift than ours, but taken 
together the two trends merge quite well; additionally, \citet{dod13} extend 
their measurements to $z$ = 6. \citet{shu14} obtain a value at very low redshift 
and with a large compendium of available literature results trace the evolving 
progression over $0 \lesssim z \lesssim 6$.

Using the form of equation (4) to do the same for \SiIV, the mean redshifts now 
are $\langle z \rangle = 2.09$, 2.65, 3.35, 4.10, and for the full sample (top 
right panel of the \SiIV\ set) we obtain the respective values 
$\Omega_{\scriptsize\textrm{\SiIV}} = (1.44\pm0.71$, $1.02\pm0.50$, 
$0.90\pm0.41$, $1.57\pm0.91)\times10^{-8}$. Similarly to \CIV, these are 
formally invariant over $1.9 < z < 4.5$. The resultant mean is
$\langle \Omega_{\scriptsize\textrm{\SiIV}} \rangle = (1.23\pm0.66)\times10^{-8}$
at $\langle z \rangle = 3.20$. There is good agreement with Songaila (2001, 
2005), and again \citet{shu14} obtain a value at very low redshift and 
show available literature results over $0 \lesssim z \lesssim 5.5$.

We show both sets of our data in Figure 9(b). As noted in the previous 
sub-section, on the understanding that the systems we detect are closely 
associated with galaxies, we do not probe the metal content of the widespread 
IGM.

\subsection{C~{\scriptsize IV} Component Column Density and Doppler Parameter 
Distributions}

The top panel in Figure 10 shows values presented in the {\it b}--log$N$ plane 
for all \CIV\ components in sample {\it sa}, extending over the range 
$1.6 < z < 4.4$. In the other two panels are the same values separated into the
ranges $1.6 < z < 3.1$ and $3.1 < z < 4.4$ (having means 
$\langle z \rangle 2.51$ and 3.58), dividing the data into roughly equal 
numbers of components and systems. There is no marked difference between the 
lower and higher redshift plots except perhaps for a mild extension to higher 
column densities at lower redshifts. 

We note that observational biases are present or potentially present in these 
data \citep{rau92}. Foremost, the unfilled triangular zone tending to the top 
left of the diagrams arises because at lower column densities the broader, and 
therefore shallower, components become relatively more difficult to detect 
above the noise. Thus, the broadest components are detected preferentially at 
the higher column densities. Relatedly, there is a possible tendency for 
extremely broad, very weak, components to be masked by more numerous narrower 
components which, in general, overlap them. 
 
A second possible bias is for the intrinsic narrowness of some lines to be 
hidden if approaching saturation. This would affect the {\it b-}values for 
the narrower, higher column density components at the bottom right of the 
diagrams. However, trials showed this is not a significant effect within the 
range of column densities in our data set. 

A potential third effect can be expected from unseparated blends of closely 
overlapping components which would then appear as single components with 
larger {\it b-}values and higher column densities. While recognising that 
defining components as entities cannot be an exact process, and that we have 
aimed to introduce the minimum number of components that justifiably  fit the 
spectral profiles, we judge that beyond our resolution limit such possible 
blending occurs relatively rarely in our high quality data.

The lower bound in {\it b-}value seen in the diagram for these data comes 
both at about the minimum value resolvable and near the level of thermal 
broadening for gas at $\sim 10^{4}$ K. Here we see a small rise in 
{\it b-}value with increasing $N$(\CIV), amounting to $\sim 1$ km s$^{-1}$ 
over a factor of more than 100 in column density, possibly reflecting the 
above latter two bias effects at some level. This is somewhat less than the 
effect noted by \citet{rau96} using a smaller data set.

Below we keep in mind particularly the first of the above bias effects. 
 
Continuing to Figure 11, the left-hand panels show the distributions of 
$N$(\CIV) and {\it b}(\CIV) for all components of sample {\it sa}. Again we 
compare values for the ranges $1.6 < z < 3.1$ and $3.1 < z < 4.4$ and also 
separate the simple and complex systems as defined in \S6.1. The vertical 
error bars are $\pm 1\sigma$ values derived from the number in each bin. It 
is clear, however, that the apparent shapes of these distributions are 
strongly influenced by the incomplete sampling particularly in the major 
``exclusion'' zone explained above, which severely distorts the true shapes 
of the distributions. This effect is manifested in $N$(\CIV) by the sharp 
fall towards smaller column densities from the apparent peak at 
log $N$(\CIV) $\sim 12.5$. In {\it b}(\CIV), the presence of a peak and the 
fall towards  smaller values are probably real but the sampling deficiency 
brings about too-rapid a fall in the distribution extending to larger values. 

Confining attention to the better-sampled regions, say 
log$N$(\CIV) $\gtrsim$ 12.5 and {\it b}(\CIV) $\lesssim$ 12 km~s$^{-1}$, here 
we see more quantitatively than in Figure 10 that in our components there is 
statistically \emph{no discernible bulk change from higher to lower redshifts 
in the distributions of $N$(\CIV) and {\it b}(\CIV)}, either in aggregate or 
segregated into simple and complex systems.

\subsection{C~{\scriptsize IV} System Column Density and Velocity Spread 
Distributions} 

In the right-hand panels of Figure 11 we give histograms similar to those in 
the left-hand panels but now for the total system \CIV\ column density, 
$N_{syst}$(\CIV), and the total spread in \emph{central radial velocity} over 
all components in a system, $\Delta v_{syst}$(\CIV) (i.e., not including the 
widths of the components). 

In the consideration of systems it is expected that there be a strong 
demarcation in properties between simple and complex cases and this is what 
is seen in the figure. The effect of incomplete sampling of low column 
density components is more complicated in these cases than for the component 
distributions just described. On the one hand, the presence or absence of 
weak components embedded in multi-component systems with relatively high 
aggregate column densities cannot significantly influence the detection of 
these systems; on the other, single-component systems clearly mirror the 
individual component sampling effects; and there are gradations between. 
Indeed, we note that for $N_{syst}$(\CIV) the appearance of the fall towards 
smaller column densities in the simple systems is substantially the same as 
for the components. For $\Delta v_{syst}$(\CIV) the effect intrinsically is 
less direct and the distribution in simple systems is strongly concentrated 
towards smaller velocity spreads.

But again, here in our data for the systems, there is no discernible change 
with redshift (at about the $1\sigma$ level) in the total or segregated 
distributions of $N_{syst}$(\CIV) and $\Delta v_{syst}$(\CIV).

\subsection{C~{\scriptsize IV} System Column Density, Velocity Spread and 
Number of Components}


Figure 12 shows some significant relationships between $N_{syst}$(\CIV), 
$\Delta v_{syst}$(\CIV), and the number of components detected in a system, 
$n_{syst}$(\CIV), again comparing values for $1.6 < z < 3.1$ and 
$3.1 < z < 4.4$. $N_{syst}$(\CIV) is strongly dependent 
both on $n_{syst}$(\CIV) and $\Delta v_{syst}$(\CIV) and, in turn, there is a 
strong proportionality between the latter two. Note that in these logarithmic 
plots, systems with only one detected component ({\it i.e.} nominally of zero 
velocity extent) are excluded in $\Delta v_{syst}$(\CIV). \citet{pab94} find 
similar relationships involving total equivalent width instead of column 
density. As before, we find no significant systematic change with redshift in 
any of the relationships.

\subsection{C~{\scriptsize IV} Component and System Differential Column 
Density Distributions} 

Figure 13 gives the \CIV\ differential column density distribution function 
$f(N,X)$ both for the individual components and the systems, each for the 
same two redshift subsets as before and here also shown for the full range 
(shown \emph{dotted}). The function $f(N,X)$ gives the number of absorbers 
per unit column density per unit absorption pathlength, $d^2{\cal N}/dNdX$ 
(accounting for the multiple redshift coverage from the different sightlines), 
where the $\Lambda$CDM cosmology-corrected absorption pathlength interval $X$ 
is given by equation 3. The data are summed over the bin size $10^{0.3}N$ as 
shown, with vertical error bars of $\pm 1\sigma$ values derived from the 
number of absorbers in each bin. 

First, for the components, the incompleteness at low column densities seen in 
Figure 10 causes the turnover below $N \sim 10^{12.5}$ cm$^{-2}$ but otherwise 
the distributions can be approximated as a power-law 
$f(N,X) \propto N^{-\beta}$ up to $N \sim 10^{14}$ cm$^{-2}$, after which the 
observations become increasingly more uncertain because in our sample very 
few individual components have higher column density. As found above with 
other quantities, there is little significant difference between the slopes 
for the two redshift subsets. We therefore use the combined set and above the 
completeness limit obtain $\beta = 2.36\pm0.10$ and show this in the figure. 
For a lower resolution sample with $\langle z \rangle = 2.65$ and extending 
to $N \sim 10^{14}$ cm$^{-2}$ \citet{pab94} obtain $\beta = 1.64\pm0.10$.

For the systems, in similar fashion, we obtain $\beta = 1.68\pm0.09$ for the 
combined set as shown. With sparser data \citet{ell00} obtain 
$\beta = 1.44\pm0.05$ near $z= 3.2$, Songaila (2001) from a large sample 
having $N_{syst} > 10^{13}$ cm$^{-2}$ over the range 
$1.5 \lesssim z \lesssim 4.5$ finds her data consistent with 
$\beta = 1.8\pm0.10$, \citet{dod10} obtain $\beta = 1.71\pm0.07$ over the 
range $1.6 < z < 3.6$  and \citet{dod13} in three redshift 
intervals extending over $2.5 < z < 6.2$ obtain an average value 
$\beta = 1.70\pm0.08$.

It is interesting that for \HI\ lines in the Lyman forest, values quite 
similar to the latter are found over column densities 
$N$(\HI) $\sim 10^{14.0-17.0}$ cm$^{-2}$, e.g., 
$\beta = 1.75\pm0.15$ for $z > 1.5$ \citep{kim02}, $\beta = 1.60\pm0.03$ 
for $z$ = 0.5--1.9 \citep{jan06}, $\beta = 1.650\pm0.017$ for $z$ = 2--3  
\citep{rud13}.

\section{C~{\scriptsize IV} COMPONENT AND SYSTEM CLUSTERING}

\subsection{Two-Point Velocity Correlation Function}

Many past studies of the two-point correlation function show that \CIV\ 
components cluster strongly on velocity scales $\lesssim 200$ km s$^{-1}$, 
with significant clustering out to a few 100 km~s$^{-1}$ more 
\citep{sar80,ysb82,sbs88,pab94,rau96,wsl96}. In some cases clustering extends 
to scales of order 1000--10000 km s$^{-1}$ and this can be traced to a few 
unusually complex groups of systems \citep{sas87,hhw89,dai96,qvy96,sca06}. 
Clustering similar to that observed in \CIV\ is measured in \MgII\ and other
ions \citep{ssb88,pab90,sas92,cc98,cvc03,sca06}. It is suggested that the 
clustering signal reflects either galaxy clustering, clustering of clouds 
within the same galactic halo, or a combination of the two. The issue is 
complicated by the wide disparity in velocity resolution and sample size 
among the different observations. The clustering seen in metals seems to 
contrast with that observed in \HI\ which in general shows no significant 
clustering signal \citep{sar80,rau92}, or comparatively weak clustering 
\citep{mcd00} and accentuated at larger column densities \citep{cri97}. 
However, as already raised in \S5, in \HI\ much of the small-scale structure 
is obscured by unresolved blending of overlapping velocity components due to 
the greater thermal broadening \citep{fso96} and often saturated absorption 
profiles.\footnote{In Figure 4 the \HI\ profiles we show are selected by 
detection of associated \CIV\ absorption and generally are stronger and more 
complex than the majority of the forest lines.} Below we return to this 
apparent contrast in clustering properties.

In Figure 14 we give velocity two-point correlation functions (TPCF) for 
the \CIV\ absorbers. In the standard manner these are normalized to the 
expected number of pairs per bin computed for a set of random distributions 
in redshift space matching the individual wavelength ranges and number of 
\CIV\ absorbers observed for each of the nine QSOs in the sample. The 
simulated randomly distributed absorbers are collated by pairwise velocity 
separation in 1000 realizations and compared with the distribution of 
velocity separations in the data to derive the correlation 

\begin{equation}
\xi(\Delta v) = 
\frac{N_{\rm data}(\Delta v)}{N_{\rm random}(\Delta v)} - 1, 
\end{equation}

\noindent where $\Delta v$ is the velocity in km s$^{-1}$ of one cloud as 
measured by an observer in the rest frame of the other.

In the top panel of the figure we show the result for the individual \CIV\
components of sample {\it sa} (here without a column density threshold limit) 
spanning the total range $1.6 < {\it z} < 4.4$, with velocity resolution 
15 km s$^{-1}$ for $\Delta v < 370$ km s$^{-1}$ and 20 km s$^{-1}$ for 
$\Delta v > 370$. The data points indicate the middle of the bins, with 
the first bin excluded because its width is comparable with the velocity 
width of the \CIV\ components. The $\pm1\sigma$ errors in the simulated 
random distribution are smaller than the data points shown. Following the 
usual pattern, the clustering signal in our data is strong at small velocity 
separations and declines steeply with increasing separation. The velocity 
correlation length, defined as the pair separation for which 
$\xi(\Delta v_0) = 1$, is $v_0 = 230$~km~s$^{-1}$, with significant signal 
extending only to $\sim 300$~km~s$^{-1}$.

The second panel gives the result for all components of sample {\it sb} (see
\S5 for the specific inclusions). As mentioned in \S5 this larger sample may 
be statistically less homogeneous than {\it sa}. The resultant profile is 
somewhat lumpier than for sample {\it sa}, now having $v_0 = 330$ km s$^{-1}$ 
and significant signal extending to $\sim 400$ km s$^{-1}$. The increase in 
$\Delta v$ relative to sample {\it sa} is consistent with the generally large 
velocity widths and richness of the added highly complex systems (see 
Tables 2--10). 

Structure similar to that for sample {\it sb} is found for \CIV\ mixed 
samples at intermediate and high redshifts in earlier studies (Petitjean \& 
Bergeron 1994; Songaila \& Cowie 1996; Womble at al. 1996; Rauch et al. 1996) 
and more recently by \citet{sca06}. 
Petitjean \& Bergeron fit the shape of the TPCF by using two Gaussians, 
obtaining a best fit with velocity dispersions $\sigma = 109$ and 
525 km s$^{-1}$ for their full sample and $\sigma = 95$ and 450 km s$^{-1}$ 
for a selected subset. With a higher resolution sample Rauch et al. (1996) 
find need for a three-component Gaussian fit with $\sigma = 22$, 136 and 
300 km s$^{-1}$. Following the same procedure for our samples {\it sa} and 
{\it sb}, we parameterize the TPCF as a multi-component Gaussian

\begin{equation} 
\xi(\Delta v) = 
A_{1}\phd exp \left(- \frac{\Delta v^{2}}{2\sigma_{1}^{2}}\right) + 
A_{2}\phd exp\left(- \frac{\Delta v^{2}}{2\sigma_{2}^{2}}\right) \ldots ,
\end{equation}

\noindent where $A_{n}$ is the amplitude of the $n^{th}$ component; the 
results are shown in the figure. First, for {\it sa}, we achieve a very good 
two-component fit with $\sigma_{1} = 35.5$ km s$^{-1}$, $A_{1} = 8.8$, 
$\sigma_{2} = 105$ km s$^{-1}$, $A_{2} = 11.6$. A one-component fit is ruled 
out and more than two components is unnecessary. For {\it sb} we obtain a 
very good three-component fit, with a narrow component identical in width to 
that in {\it sa}, $\sigma_{1} = 35.5$ km s$^{-1}$, but with $A_{1} = 5.0$, 
and the others with $\sigma_{2} = 80$ km s$^{-1}$, $A_{2} = 9.5$, 
$\sigma_{3} = 185$ km s$^{-1}$, $A_{3} = 6.5$.

The set of three remaining panels in the figure gives results for the 
data treated as {\it systems}; here the $\pm1\sigma$ errors in the random 
distributions (significantly larger than for the components because of the 
smaller numbers) are shown by \emph{bounding thin lines}. In the first panel 
of this set, the result obtained using the 201 system redshifts for sample 
{\it sb} with velocity resolution 500 km s$^{-1}$ and extending to 
$\Delta v = 12,000$ km s$^{-1}$ is shown on an expanded vertical scale. The 
TPCF is notably flat over the whole range of separations, demonstrating 
clearly that the \emph{systems} of \CIV\ absorption lines \emph{in our 
sample} are randomly distributed. Using the purer sample {\it sa} gives an 
almost identical result because the relatively few highly complex systems 
in sample {\it sb} contribute to the total each with the same weight as do 
the rest of the systems (we give the {\it sb} case to demonstrate that this 
result is not specific to sample {\it sa}). We find the same pattern for 
velocity resolutions below 500 km s$^{-1}$ as well as up to several 1000 
km s$^{-1}$ and for $\Delta v$ as large as 50,000 km s$^{-1}$. The 
large-scale example is in the second panel of the set and in the bottom 
panel is a subset of the sample in which we include only the 146 systems 
having velocity spread $\Delta v_{syst}$(\CIV) $< 150$ km s$^{-1}$ and 
use velocity resolution 150 km s$^{-1}$. In all cases the result is 
indistinguishable from a \emph{random distribution}. 

We point out here that our approach to identifying systems is based on 
the observed physical separation between \emph{groups of closely 
associated components} in our samples. These include some with quite 
complex structures having a wide spread in velocity and which well 
reflect the strong absorbers closely related to \emph{individual 
galaxies} as we discuss in \S7.3. \citet{sca06} take a different approach, 
defining a velocity linking length $v_{link}$ and grouping together 
into \emph{nominal} systems all components whose separation from their 
nearest neighbour is less than $v_{link}$, using values 25, 50 and 100 
km s$^{-1}$. In this way they find little difference between system 
and component correlation functions. Such $v_{link}$ lengths are 
considerably shorter than the overall widths of most of the complex 
groups we identify (see the vertical associations in Figure 7 and the 
plots in Figures 11 and 12), suggesting that in such cases the 
$v_{link}$ technique acts in effect as a smoothing process resulting in 
data of lower velocity resolution while nevertheless reflecting 
structure within the complex associations that shows up in the
\emph{component} correlation but for which in the \emph{system} 
correlation we here count each as one unit like any other. Furthermore, 
the majority of members in our system sample have velocity width 
$<$ 100 km s$^{-1}$ so might fortuitously act in the $v_{link}$ regime, 
yet we find no contributing signal to system clustering. In 
measurements similar to ours \citet{fat13} also find no evidence of 
system clustering (see \S7.4).

Briefly summarizing, for \emph{components}, although the detailed 
shape of the TPCF depends on the specific contents of the samples used, we
observe that broadly the results are similar. Our large, high quality sample 
{\it sa} gives better statistical definition than before and by exclusion of 
the few systems containing significant Lyman~$\alpha$ damping wings (present 
in sample {\it sa}) we regard this as representing the ``normal'' great 
majority of \CIV\ absorption systems. With this sample of components we 
obtain a more compact TPCF than in most other work. Specifically, the shape 
of our observed TPCF does not require Gaussian elements as broad as those 
found by \citet {pab94} or \citet{rau96}. Churchill et al. (2003) come to a 
similar conclusion for \MgII\ absorbing clouds. Now for \emph{systems}, most 
importantly in both of our samples we see no evidence of clustering \emph{on 
any scale} from $\Delta v$ as small as 150 km s$^{-1}$, where the general 
\emph{component} clustering signal is still strong, out to very large 
separation values (we return to this in \S7.4).

\subsection{Comparison with Galaxy Clustering}

It seems that for our component samples in the top two panels of Figure 
14 the TPCF has a common narrow core component, with broader components 
whose characteristics depend on the precise definition of the sample. While 
this might give a clue to the dynamical makeup of the absorber population 
we should bear in mind that this is a contrived description of the TPCF 
(in some cases we find an exponential fit is almost as good). 

For galaxies, the TPCF is a simple, fundamental statistic of the galaxy 
distribution. Estimations of \emph{real-space} galaxy clustering yield a 
TPCF very close to power-law form over a broad range of scales 
\citep{lov95,zeh02,haw03}, although on detailed examination this is seen 
to be largely fortuitous \citep{zeh04}. The parameters of the power-law 
depend on the characteristics of the sample galaxies and dominantly on 
their luminosity \citep{nor02}. On the other hand, galaxy clustering in 
\emph{redshift-space} is strongly affected by gravitationally induced 
distortions. On small scales, random peculiar velocities will cause 
clustering to be underestimated, while on large scales coherent infalling 
bulk flows will lead to an overestimate of the clustering amplitude 
\citep{kai87}. Consequently, the form of the redshift-space TPCF departs 
considerably from a simple power-law \citep{zeh02,haw03}. Large scale 
$\Lambda$CDM simulations show good correspondence with the observed galaxy 
clustering \citep{ben01,wei04}. In particular, these simulations predict 
remarkably little comoving clustering evolution from high redshifts to the 
present epoch (although the underlying dark matter clustering evolves 
strongly).

The two-component Gaussian fit to the \emph{component} TPCF for sample 
{\it sa}  given in the top panel of Figure 14 is reproduced in logarithmic 
form in Figure 15. The stepped horizontal lines in the figure give the 
\emph{system} TPCF $+1\sigma$ bin errors from the bottom set of three 
panels of Figure 14, taken as upper limits, in to 500 km s$^{-1}$ from the 
first panel then to 150 km s$^{-1}$ from the third. We convert the 
velocity scale for our full sample at $\langle z \rangle = 3.0$ to 
$h^{-1}$ comoving Mpc using Hubble constant $h$ in units of 
100 km s$^{-1}$ Mpc$^{-1}$ and show this at the top of the figure for 
comparison with the galaxy estimations. The two differently-scaled 
\emph{short-long dashed lines} are results from the large 2dF Galaxy 
Redshift Survey \citep{haw03} for the galaxy real-space TPCF as fitted by 
the power-law $(r/r_0)^{-\gamma_r}$ with $r_0 = 5.05 h^{-1}$ Mpc, 
$\gamma_r = 1.67$ and the corresponding redshift-space TPCF as fitted 
peacemeal by the power-law $(s/s_0)^{-\gamma_s}$ with $s_0 = 13 h^{-1}$ 
Mpc, $\gamma_s = 0.75$ at small scales, and $s_0 = 6.82 h^{-1}$ Mpc, 
$\gamma_s = 1.57$ around $s_0$. The \emph{dotted line} is the best-fit 
correlation function to the angular clustering at $z$ = 2.94 of 
Lyman-break galaxies, with power-law parameters $r_0 = 4.0 h^{-1}$ Mpc, 
$\gamma_r = 1.57$ \citep{adb05}.

Figure 15 shows that above $\Delta v \sim 150$ km s$^{-1}$ the 
\emph{one-dimensional} TPCF for the \CIV\ \emph{components} dips 
significantly below the TPCF found for galaxy clustering. In addition, the 
lack of significant absorber \emph{system} clustering observed above 
$\Delta v = 150$ km s$^{-1}$ gives a very substantial deficit in clustering 
amplitude relative to galaxies out to $\Delta v \sim 1000$~km~s$^{-1}$. 

\subsection{Absorbers and Galaxies}

To complete the picture we now consider observations associating absorbers 
with galaxies. Attempts to establish directly how metal absorption systems 
and galaxies are connected initially focused on searches for galaxies near 
the line of sight to QSOs with redshifts closely similar to the absorbers 
\citep{bas78,bab91,ste94,leb97,ste02}. Most work of this kind was done at 
$z \lesssim 1$ for relatively strong systems selected by the presence of 
\MgII\ and associated with gas having $N$(\HI) $\gtrsim 10^{17}$ cm$^{-2}$ 
(i.e., Lyman limit systems). Investigating the kinematical properties of 
several such cases with projected impact parameter 
$20 \lesssim d \lesssim 100$ proper kpc \citet{ste02} find the identified 
galaxies to be relatively normal spirals, with circular velocities 
$100 \leq v_c \leq 260$ km s$^{-1}$. While the absorber characteristics are 
consistent with rotation being dominant also for the absorbing gas, the total 
range of velocities, typically 200--300 km s$^{-1}$, and their placing to one 
side of the galaxy systemic redshift, is not consistent with simple disc 
rotation viewed along a single sightline.\footnote{The expectation that a 
sightline through a galaxy at large radius would show only a small 
differential rotational velocity, contrary to the observed velocity spread of 
the absorption systems, earlier had counted against the idea that the 
velocity structure in the systems is due to motions associated with single 
galaxies \citep{sbs88}.} \citet{ste02} suggest that models to explain their 
observations require either extremely thick rotating gas layers, rotation 
velocities that vary with height above the extrapolated galactic plane, or a 
combination of both, with rotational motion dominating over radial infall or 
outflow even for gas well out of the galactic plane. The kinematics of the 
absorbing gas observed in damped Lyman~$\alpha$ systems (DLA) similarly 
appears consistent with rotating thick disc geometries \citep{paw98} although 
not uniquely so \citep{hsm98,led98,mam99}, but there is difficulty reconciling 
the high ionization species with the low in the same model \citep{wap00}. 
Extending to higher redshifts, from a large sample of \MgII\ absorbers 
\citet{men11} find a strong correlation of associated [\OII] luminosity, an 
estimator of star formation rate, while the velocity structure of some of the 
strongest \MgII\ absorbers is suggestive of superwinds arising in actively 
star-forming galaxies \citep{bon01} and more generally there is evidence for 
the importance of large-scale galactic winds \citep{pet01}. \CIV\ absorption 
kinematically is strongly correlated with \MgII\ and usually extends more 
widely in velocity \citep{chu01}, and at high redshifts complex \CIV\ 
absorption systems are observed commonly to coexist with galaxies (e.g., 
Adelberger et al. 2003, 2005; Steidel et al. 2010; Martin et al. 2010). We 
address the latter more fully in \S8 and \S9.
 
\subsection{Conclusions on the Clustering Properties of Metal Absorbers}

In the light of observations linking \MgII\ and \CIV\ absorption systems with 
specific galaxies close in line of sight to the background QSOs, and the fact 
that these systems have extended kinematic structure of a few 100 km$^{-1}$ 
similar to the \CIV\ systems in our samples, it seems inescapable that our 
velocity correlation results for the absorption \emph{components} in sample 
{\it sa}, contrasting with the lack of \emph{system} clustering, are entirely 
due to the peculiar velocities of the gas present in the outer extensions or 
circumgalactic media (CGM) of individual galaxies, \emph{not to general 
galaxy-galaxy clustering}. This conclusion is not changed when we substitute 
the component sample {\it sb} with its bias to highly complex systems. The 
different broader components of the distributions we find in the 
multi-Gaussian fits to the shape of the TPCF in the cases we show in Figure 
14 presumably reflect the distribution of the more disturbed cases of outflow 
into the extended regions probed, while the narrow components of the 
distributions may indicate underlying, more quiescent, motion. \citet{cc98} 
concluded with a similar picture in their study of \MgII\ absorbers.

The explanation for the observed lack of \emph{system} clustering in our 
\CIV\ sample, while the systems are known to be associated with galaxies, 
then must be simply geometrical selection. In conventional clustering 
studies all related galaxies in the plane of the sky are included (with 
defined sample specifications) and the spatial variance is completely 
sampled in two or three dimensions on all scales. The situation for the 
one-dimensional sightline available to each background QSO, although highly 
extended in redshift, is quite different. In this case the galaxy population 
is probed so sparsely that it is a rare occurrence for the gaseous extent of 
more than one galaxy to be encountered in a given cluster or group.  

The lack of system clustering in our sample implies \emph{there is no obvious 
observational distinction between the clustering of metal systems and strong 
absorbers of the Lyman $\alpha$ forest}. Because any complex component 
structure in \HI\ is largely hidden by thermal broadening and saturation, the 
\HI\ absorption lines effectively are counted as \emph{systems} like the 
\CIV\ systems. However, the far more numerous weaker \HI\ lines probe to much 
lower densities in the intergalactic medium, so do not represent the same 
population as the observed metal systems which in general are more directly 
associated with galaxies.

Highly extended correlated structures would appear when a sightline 
fortuitously passes along a large-scale filamentary or sheetlike 
distribution of galaxies in a supercluster, bringing about a rich complex 
of absorber systems well extended in redshift, but the incidence of such 
occurrences is low. \citet{rau96} consider a less extreme version of such 
a model to explain the tail of the TPCF found in their smaller \CIV\ 
sample, interpreting this as a Hubble flow velocity extension. On the other 
hand, \citet{ssr02} find that for the components of a limited sample (12) 
of \OVI\ systems tracing the warm-hot intergalactic medium the velocity 
TPCF is similar to the spatial power-law form seen for galaxies, with a 
comoving correlation length $\sim 7h^{-1}$ Mpc. They conclude that for this 
population the signal is dominated by large-scale structure. We stress 
again, however, that in the present work we seek to characterize the 
``normal'' situation obtaining for the large bulk of relatively weaker metal 
absorption systems in the cooler phase identified by \CIV. In this respect 
our work is different from large-scale studies using samples only of strong 
\CIV\ systems. Quashnock et al.(1996) use 373 QSO sightlines having 
a total of 360 strong \CIV\ absorption systems covering the redshift range 
1.2--4.5; in contrast, our sample has an average of about 20 systems per 
sightline. With a velocity resolution of about 600 km s$^{-1}$ they find 
weak clustering, $\xi \sim 0.4$, on scales of superclusters, with 
significant signal contributed by groups of absorbers in only 7 of the 
sightlines. \citet{qav98} extend this study to smaller scales using 
velocity resolution 180 km s$^{-1}$ and a restricted sample of 260 strong 
\CIV\ systems drawn from 202 sightlines, finding significant clustering of 
power-law form on scales of clusters and superclusters. It is interesting 
that \citet{fat13} in a high velocity resolution investigation towards the 
QSO PKS 0237-233, a line of sight known to intersect highly untypical complex 
groups of strong absorbing systems and interpreted as revealing the presence 
of a supercluster of galaxies, in fact find no \emph{system} clustering at 
all and obtain displays very similar to those we show in Figure 14.

Our sample of numerous relatively weak \CIV\ systems in few sightlines 
does not probe well such large-scale structure. Nor, as we have seen, do 
our systems strongly sample galaxies within clusters. Our high velocity 
resolution data perform the complimentary role of finely probing the 
environment in the vicinities of \emph{individual} galaxies.

\section{IONIZATION BALANCE} 

\subsection{Evolution of Ionic Ratios Si~{\scriptsize IV}/C~{\scriptsize IV}, 
C~{\scriptsize II}/C~{\scriptsize IV}, Si~{\scriptsize II}/Si~{\scriptsize IV},
Si~{\scriptsize II}/C~{\scriptsize II}, N~{\scriptsize V}/C~{\scriptsize IV}}

In Figure 16(a) we give the distribution in redshift of the observed column 
density ratios \SiIV/\CIV\ for absorbers of sample {\it sb} (for this and the
related quantities below see \S5 for the specific inclusions used here) but 
restricted to those in which \SiIV\ is potentially accessible. The different 
symbols indicate specific sub-sets of the data listed in Tables 2--10 
relating to sample {\it sb} and are identified in the caption. Like in Figure 
7(a), here we show \emph{component} values for the full sample and separately 
the simple and the complex systems, with again the obvious vertical 
associations indicating members of the same system, and in a fourth panel 
show values summed over each \emph{system}. The upper limits indicate values 
in which \SiIV\ is undetected. In all these cases we now take 
$N$(\CIV) $> 1 \times 10^{12}$~cm$^{-2}$ as an acceptance threshold. The 
accompanying panels on the right give the corresponding median values 
obtained over the redshift spans defined by the horizontal bars, with 
indicated uncertainties obtained as $1\sigma$ bootstrap values. Unlike the 
single-species distributions in Figure 7(a), here we do not need to correct 
for multiple redshift coverage because each component or system ionic ratio 
value represents an independent measure of absorber ionization \emph{balance} 
(although incomplete, as we explain later). The upper limits shown in the 
data panels are included in the assessment of median values; the occasional 
high upper limits coming above the level of the derived median value here do 
not significantly influence the accuracy of the results. 

It is striking that within the small uncertainties the component values for 
the full sample {\it sb} of \SiIV/\CIV\ are shown to be \emph{constant over 
the whole observed range $1.9 < z < 4.4$}. For the simple systems there is a 
significant continuous fall in the median with cosmic time, by $\sim 0.4$ dex 
over the observed range, while the complex systems, containing the bulk of 
the fully detected components, remain consistently invariant. 

The system ionic ratios in the bottom of the figure are obtained after first 
separately summing the individual \SiIV\ and \CIV\ component values in each 
system. Once again, there is no detectable change in the median over 
$1.9 < z < 4.4$. Although we have earlier commented on the shortcomings of 
using total system values for ionic ratio determinations we include these 
results for comparison with work by \citet{soc96} and Songaila (1998, 2005)
who argue that their observed evolution of \SiIV/\CIV, which shows a 
substantial jump in the median at $z \sim 3$, points to a sudden hardening of 
the ionizing background that is consistent with an abrupt reduction in the 
opacity of the evolving IGM to \HeII\ ionizing photons as \HeII\ completely 
reionizes to \HeIII. We previously report the lack of such a jump 
\citep{bok97,bsr01,bsr03}, also concluded by \citet{kcd02} and \citet{agu04}, 
and reiterate this here (see also \citet{dod13} for extension to higher 
redshifts). Because of the \HeII\ high ionization threshold (54.4 eV) and 
small photoionization cross-section coupled with the rapid recombination rate 
of \HeIII, the double ionization of He is expected to lag far behind the 
ionization of \HI\ and \HeI\ and be fully completed only at $z \sim 3$ by 
QSOs approaching the peak of their activity (e.g., Madau et al. 1999; Sokasian
et al. 2002; McQuinn et al. 2009). Several workers find support more directly 
than above for high \HeII\ opacity in the intergalactic medium at $z \gtrsim 3$ 
\citep{rei97,hea00,kri01,sme02,shu10,syp13}. \citet{str00} and \citet{the02} 
show that \HeII\ reionization at $z \sim 3.5$ results in an increase by a 
factor 2 in the temperature of the intergalactic medium at the mean density 
and find evidence for such a jump from observations of Lyman~$\alpha$ forest 
lines. However, analyses by \citet{mcd01} and \citet{zht01} do not show a 
significant temperature change at these redshifts, although the temperatures 
they find are higher than expected for photoionized gas in ionization 
equilibrium with cosmic background radiation and can be explained by gradual 
additional heating due to \emph{ongoing} \HeII\ reionization, reflected also 
in more recent measurements \citep{bec11,bec13,par11}. Notwithstanding, use 
simply of the column density ratio \SiIV/\CIV\ to measure the influence of 
\HeII\ ionizing radiation on the absorbers has significant limitations and 
we do not make any claims concerning the evolution of the ambient ionizing 
spectrum based on Figure 16(a) alone. We consider \HeII\ reionization further 
in respect of our fuller interpretation of the ionization measurements 
coming in \S8.2.

Again following the pattern of Figure 7(a) and (b), displays corresponding 
to the panel at the top of Figure 16(a) are given in Figure 16(b) for the 
component column density ratios \CII/\CIV, \SiII/\SiIV, \SiII/\CII\ and 
\NV/\CIV, although here for the first and fourth of these there is somewhat 
more uncertainty in the median values arising from included upper limits. 
\CII/\CIV\ shows a mild progressive rise with cosmic time by $\sim 0.4$ dex 
over $3.7 < z < 1.9$, although here not far from the flatter distribution of 
median values found by \citet{son05}. \SiII/\SiIV\ 
shows a steeper trend in the same direction, rising by $\sim 0.7$ dex over 
this range, while \SiII/\CII\ appears quite flat, although in both cases the 
data are rather sparse. \NV/\CIV\ is relatively flat from $z \sim 3.2$ to 
lower redshifts, while to higher redshifts, within the limited data allowed 
by the available narrow windows, only upper limit values are obtained. As 
we earlier commented for Figure 7(b) using sample {\it sa}, it is similarly 
evident here that the ionic ratios in Figure 16(b) using the enhanced sample 
{\it sb} (see \S5) are detected preferentially in the strong, \emph{complex} 
systems identified among the \SiIV/\CIV\ displays in Figure 16(a). 

Although at this stage some collective inferences might be made about the 
radiation environment from the data in Figure 16(a) and (b), the fact remains 
that individual ionic ratios give only partial evidence of ionization state. 
Therefore we move on instead to use combinations of these ratios as more 
complete indicators from which to derive information on the evolution of the 
ionizing spectral energy distribution as experienced by the absorbers.  

\subsection{Evolution of Ionic Ratio Combinations and Comparison with Models 
of Absorbers Ionized by the Cosmic UV/X-Ray Background} 

We consider the combination of column density ratios \SiIV/\CIV~:~\CII/\CIV, 
\NV/\CIV~:~\CII/\CIV, \SiII/\SiIV~:~\CII/\CIV\ and  \SiII/\CII~:~\CII/\CIV. 
To compare our observations with model ionization predictions we simulate the 
conditions in our absorbers by means of the \Cloudy\ code version C10 
\citep{fer98,fer11}\footnote{We find the use in our application of the later 
version C13 makes very little change indeed to the results presented in this 
paper.} for one-sided illumination of a plane-parallel slab of gas. For the 
radiant exposure we begin with the metagalactic ionizing background 
synthesised by the cosmological radiative transfer code introduced by 
\citet{ham96} and recently substantially updated \citep{ham12}, which 
gives the evolving ionizing radiation from QSOs and star-forming galaxies 
propagating in a clumpy, promordial IGM. Here, for detailed reference useful 
in our further developments below, we give an outline of the model radiative 
inputs from \citet{ham12}.\footnote{\citet{ham12} use the intensity 
$J_{\nu}$ (erg s$^{-1}$ cm$^{-2}$ Hz$^{-1}$ sr$^{-1}$) but here to 
accomodate the requirement of one-sided illumination in our \Cloudy\ 
modeling we use $F_{\nu}$ = 4$\pi J_{\nu}$ and to avoid directional effects 
work basically with absorber regimes optically thin in the Lyman continuum.}  

The QSO emissivity contains: (i) an optical--ultraviolet component for 
radio-quiet QSOs with spectral energy distribution given by a power law of 
form $F_{\nu} \varpropto \nu^{-1.57}$ extending from 1300 \AA\ to shorter 
wavelengths \citep{tel02} covering the ionizing range of interest here, with 
comoving emissivity at 1 Ryd closely fitting the results of \citet{hop07}; 
and (ii) an X-ray component with contributions accounting in detail for the 
observed cosmological X-ray background. The underlying assumption for the 
ionization state of the IGM is photoinization equilibrium in a pure H+He gas. 
To account for the effective opacity of the IGM the new code uses a piecewise 
power-law parameterization of the distribution in redshift and column density 
of intergalactic absorbers that fits recent measurements of the Lyman 
continuum mean free path of 1 Ryd photons \citep{pro09} as well as the 
observed effective Lyman~$\alpha$ optical depth \citep{fau08,fan06}, and 
together covering the redshift range concerning the observations in this 
paper. Firstly, in the  panels of Figure 17(a), for reference we give 
spectral displays of the metagalactic radiation background if QSOs are the 
only contributing sources, (Q$_{HM}$), shown in \emph{dashed lines} at the 
specific redshifts $z_{HM}$ = 1.9, 2.6, 3.4, 4.4, bounding the sample 
redshift intervals we select for our data in the three panels of the 
figure.\footnote{We are grateful to have this specially provided 
\citep{haa14}.} We focus on the frequency range effective for 
photoionization of the relevant species in our absorbers, and identify the 
positions of the significant ionization thresholds. The spectra contain the 
sawtooth modulation produced by resonant absorption in the Lyman series of 
intergalactic \HeII\ \citep{mah09}, not included in earlier versions, which 
becomes significant at $z \gtrsim 3$. The analogous modulation produced by 
the \HI\ Lyman series becomes significant only at $z \gtrsim 6$. In our 
\Cloudy\ computations below we add the cosmic microwave background 
pertaining to each redshift to account for Compton cooling of the absorbers, 
which becomes mildly significant at low densities at our highest redshifts 
but otherwise has negligible effect. This background is not included in the 
\citet{ham12} radiative transfer computations. 

While QSOs play a dominant role as sources of ionizing radiation over a wide 
band of energies at {\it z} $\lesssim$ 3, to higher redshifts their strongly 
declining population coupled with the rapid build-up of star-forming galaxies 
brings the latter to be the dominant sources of hydrogen-ionizing radiation 
in the reionization of the IGM from early times. To compute the evolving 
Lyman continuum emissivity from galaxies at all epochs Haardt \& Madau start 
with an empirical determination of the star formation history of the Universe 
to obtain the redshift-dependent, dust-reddened, galaxy far-UV (1500 \AA) 
luminosity density. After correction for dust attenuation using a 
\citet{cal00} extinction law this is compared with results of evolving 
spectral population synthesis models using the GALAXEV library \citep{brc03} 
for a \citet{sal55} initial mass function (IMF) covering 0.1--100 $M_{\Sun}$ 
and a metal-enrichment law of decreasing metallicity with redshift according 
to the expression $Z(z) = Z_{\Sun}10^{-0.15z}$ \citep{kak07}, while adjusting 
the star formation rate density iteratively until the computed far-UV 
luminosity density as a function of redshift achieves a good match to related 
observational data. Stellar synthesis models then are used again to compute 
the evolving frequency-dependent UV emissivity (i) dust-reddened at all 
photon energies below 1 Ryd, and (ii) following a luminosity-weighted escape 
fraction of hydrogen-ionizing radiation $\langle f_{esc} \rangle$ that is 
constant over all energies between 1 and 4 Ryd. In this treatment 
$\langle f_{esc} \rangle$ is a free parameter that represents the proportion 
of ionizing radiation leaking into the IGM from star-forming galaxies and 
here assumed as emanating not from sources in a semiopaque medium but 
from \emph{a small fraction of essentially unobscured sources} 
\citep{gne08}.\footnote{It is important to note that this is an 
\emph{absolute} escape fraction, being the ratio of escaping Lyman continuum 
flux to the intrinsic flux emitted in the same spectral region by the stars 
in the galaxy. Observationally this ionizing escape fraction is usually 
assessed relative to the readily-accessible flux at rest-frame 1500 \AA\ (or 
a nearby wavelength) in the non-ionizing far-ultraviolet, corrected for 
attenuation by dust, then used to obtain the intrinsic stellar Lyman 
continuum flux from spectral synthesis models with which finally to compare 
the level of observed Lyman continuum radiation (here also with correction 
for the average IGM opacity) and so obtain the escape fraction 
\citep{spa01,sha06}. If there is insufficient observational evidence to 
determine the dust attenuation a \emph{relative} escape fraction is taken 
simply as the ratio of escaping Lyman continuum radiation to the escaping 
radiation at 1500 \AA\ \citep{spa01}.} The form of the escape fraction, 
given as 
\begin{equation}
\langle f_{esc} \rangle = 1.8 \times 10^{-4}(1 + z)^{3.4} 
\end{equation}
\noindent (refer also to: Inoue et al. 2006; Boutsia et al. 2011), is a 
steeply rising function of redshift \emph{dictated} by the need to reproduce 
the hydrogen-ionization rates obtained from flux-decrement measurements over 
$z =$ 2--5 and to compensate for the decline in the star formation rate 
density to higher redshifts sufficient to enable reionization of the IGM at 
early enough epochs (and consistent with {\it WMAP} results), and is made 
zero at energies above 4 Ryd. 

We use this expression throughout the rest of the paper in our empirical 
investigations of the escape \emph{process} and therefore it is important to
point out that in this ``minimal cosmic reionization model'' the measurements 
included by \citet{ham12}, and to which there is a good fit by equation 7 
\emph{over the stated range in redshift}, are \emph{independent} of the 
actual mode of escape so apply equally well to the work we describe later. 
While Haardt \& Madau make clear that the escape fraction cannot be certain 
on the large scale in redshift we stress that our range of consideration 
$1.9 < z < 4.4$ shown in Figure 17 and others later is coincident with the 
region based on solid determinations of hydrogen-ionization 
rates. Indeed, \citet{kol14} recently find a large mismatch between the 
result from similar observations at very low redshift by the Hubble Space 
Telescope Cosmic Origins Spectrograph and the prediction of equation (7), 
but at the same time stress the excellent agreement obtained for $z$ = 2--5, 
which covers our range of interest here. Recent flux-decrement measurements at 
high redshifts \citep{bab13} indicate a relative increase in 
hydrogen-ionization rate over those implied in equation (7), starting from 
$z \sim 3.5$ and rising to a factor $2\pm1$ at $z$ = 4.4. While recognizing 
this, for our indicative purposes in this paper we can adequately stay with 
equation (7) both for the metagalactic background values here and in our 
additional work that follows in this paper.

In Figure 17(a) we add, in \emph{continuous lines}, spectral displays for 
the full QSO+galaxy (QG$_{HM}$) metagalactic radiation background. The 
respective values of the flux $F_{\nu}$ (erg s$^{-1}$ cm$^{-2}$ Hz$^{-1}$) 
at the \HI\ Lyman limit (1 Ryd, i.e., 912 \AA) at the redshifts 
$z_{HM}$ 1.9, 2.6, 3.4, 4.4 are numerically: $4.02 \times 10^{-21}$, 
$4.02 \times 10^{-21}$, $3.02 \times 10^{-21}$, $2.10 \times 10^{-21}$.
Contrasting with the Q$_{HM}$ model, the QG$_{HM}$ spectral 
intensity is highly raised at energies below the \HI\ ionization edge where 
the galactic radiation makes a strong contribution,\footnote{Note that in 
the context of this paper spectral radiation from either of these sources at 
energies $< 1$ Ryd has insignificant effect on the ionic ratios we measure 
for the absorbers.} while beyond this edge there is a large depression of 
the galaxy spectrum reflecting the imposed small escape fraction and 
extending to the \HeII\ ionization edge at which the galaxy contribution is 
truncated. From the \HeII\ edge to higher energies the further depression to 
below the Q$_{HM}$ intensity levels, a strong function of redshift, is due 
to the increased \HeII\ continuum opacity through the changed ionization 
balance of the IGM resulting from the large increase in the H-ionizing 
emissivity from the early galaxy population but here unaccompanied by a 
similar increase at the \HeII\ edge.

To typify the general state of the absorbing regions (e.g., Simcoe et al. 
2002, 2006) we assume a low metallicity gas optically thin in the \HI\ Lyman 
continuum and adopt the nominal properties $Z = 0.003$ solar 
(i.e., $[Z] = -2.5$)\footnote{In the usual fashion we express the logarithmic 
abundance of element X relative to element Y compared with the solar values 
as $[X/Y] =$ log(X/Y) $-$ log(X/Y)$_{\odot}$.} and neutral hydrogen column 
density $N$(\HI) $= 10^{15}$~cm$^{-2}$. We take these values for illustration 
and here do not imply that precisely these are demanded by the observations: 
in \S8.5 we explore the sensitivity of the model results to these parameters 
over a wide range of values. In general we use solar relative metal 
abundances \citep{gre10} as a baseline. For Si we also include a higher value 
to reflect the observed range for Galactic metal-poor stars which, in a 
common pattern with other $\alpha$-elements, extends from solar values to a 
substantial overabundance. For C, on the other hand, measurements for 
metal-poor stars generally give approximately the solar relative abundance, 
although with considerable scatter \citep{rnb96,mcw97}. A similar picture is 
gained from observations of cosmologically-distributed metal-poor DLA systems 
\citep{pet08,pet12}. However, overabundances comparable to metal-poor stars 
also occur at near-solar metallicities in Galactic stars (e.g., Pettini 
\& Cooke 2012 and references therein). Based on all the above, we take 
[Si/C] $= 0.0, 0.4$ as representative ``soft'' limits and assume these values 
at all redshifts covered by our data. The other special case is N, which has 
quite a complex evolutionary nucleosynthetic origin \citep{hek00,hap07,pls08}. 
In particular, the relative abundance of N in metal-poor DLA systems shows a 
downward scatter extending by about 1 dex within the primary and secondary 
production boundaries \citep{pet08,pet12}. We take [N/C] $= 0.0, -1.4$ 
as representative limits at all redshifts we cover.\footnote{However, we need 
to be cautious about the general interpretation of relative abundance values 
[Si/C] and [N/C] for the absorbers we observe. Given their apparent origin as 
ejecta from galaxies \citep{ste10} we should alternatively think of these as 
``beginning life'' within galaxies at relatively high metallicities (e.g., 
Erb et al. 2006; Mannucci et al. 2009) and then becoming diluted in their 
subsequent transit into the immediate IGM (and making up part of the enriched 
circumgalactic medium \citep{ade05,ste10}: consequently, while the observed 
absorbers then indeed are of low metallicity, their relative metal abundances 
may not reflect a simple metal-poor environment.}

In Figure 17(b), first looking just at the data points, we show our specific 
\emph{combinations} of the observed ionic column density ratios displayed in 
the top panel of Figure 16(a) and the panels of Figure 16(b) derived from the 
individual components in sample {\it sb}. Here in the first vertical set of 
panels we show \SiIV/\CIV~:~\CII/\CIV\ distributed over the three redshift 
intervals used in Figure 17(a). Again, the coded symbols indicate specific 
observational sub-sets of the data listed in Tables 2--10 and identified in 
the caption to Figure 16(a). In the three other vertical sets of panels in 
the figure, in similar fashion, we include \NV/\CIV, \SiII/\SiIV\ and 
\SiII/\CII\ as ratios with \CII/\CIV. It is particularly striking that the 
data here take on a more coherent appearance: values from components in a 
given system distributed vertically in Figures 16(a) and (b) now string out 
within a broad two-dimensional track, accomodating the different ionization 
states present. It is clear that such two-dimensional analyses help to make 
useful interpretations of the ionization properties of the absorbers. 

To gain some initial insight into the cosmological drivers which influence 
the trend of our observations we begin by comparing our results with 
\Cloudy\ models confining the radiative input to the simple case of a 
metagalactic ionizing radiation background involving the Q$_{HM}$ model 
alone, shown again in \emph{dashed lines}. The two curves in each of the 
panels for \SiIV/\CIV~:~\CII/\CIV\ and \SiII/\CII~:~\CII/\CIV\ give our 
individual model results for the adopted bounding values combining 
[Si/C] $= 0.0, 0.4$ with the respective lower and higher redshifts in the 
range. In use here of our \Cloudy\ model curves we make small vertical 
adjustments specific to each of the sets involving \SiIV/\CIV, \SiII/\SiIV\ 
and \SiII/\CII, in magnitude averaging $\sim$ 0.2 dex, to align with the 
relatively well-defined $1.9 < z < 2.6$ data (in practice we base this on 
the QG$_{HM}$ model, given below, but closely similar values apply here to 
the Q$_{HM}$ model), then propagate the same adjustments to the other two 
redshift intervals. This ``calibrates'' the assessment of the evolutionary 
trends from the lower to higher redshifts in our data. From this point, all 
the graphical results presented below use these same adjustments. We assume 
these small differences reflect the accuracy of values of atomic data and 
functional parameters employed in the code: earlier \Cloudy\ releases 
require different (still small) adjustments with the same data. We do not 
attempt this for \NV/\CIV\ because the range of the data and the possible 
range of the models do not give sufficient grounds for discerning similar 
small adjustments.

Still with just Q$_{HM}$, for \SiIV/\CIV~:~\CII/\CIV, in the 
lowest redshift interval ($1.9 < z < 2.6$) the two bounding curves fit our 
data quite well over much of the range. Progressing similarly to the higher 
redshifts, relative to the model results there is first an upward scatter 
in the data ($2.6 < z < 3.4$) then a collective rise by $\sim 0.5$ dex 
($3.4 < z < 4.4$). Additionally, we note that at the higher redshifts there 
are relatively more components having \CIV\ and \SiIV\ but only upper 
limits for \CII\ (the open symbols with arrows pointing out to low \CII/\CIV. 
Most of the components with only upper limits for both \SiIV\ and \CII\ have 
$N$(\CIV) $\lesssim 5 \times 10^{12}$~cm$^{-2}$. We return to these factors 
in more detail in \S9.

Next, for \NV/\CIV~:~\CII/\CIV, while our data here are rather sparse, a 
progressive evolutionary trend over the three redshift intervals is again 
indicated. With photoionization to \NV\ requiring energies beyond 4 Ryd the 
presence at the highest redshifts of only upper limits for \NV\ indicates 
a substantially softer spectral energy distribution for the ionizing flux 
than given by the initial model we show here, this pointing to the 
inclusion of radiative contributions from galaxies as incorporated in the 
\citet{ham12} full radiative transfer computations we address below.

For \SiII/\SiIV~:~\CII/\CIV\ (here there is no relative abundance 
dependence), again in the first redshift interval the Q$_{HM}$ model fits 
the data quite well, in the middle interval there is evidence for the 
beginning of a significant downward trend in the data, followed at the 
highest redshifts by a bulk shift down by $\sim 0.6$ dex from the model.
Note here that these measurements require positive detection of \SiIV\ as
well as the baseline presence of \CIV\ and this strongly restricts the 
range of upper limits relative to those appearing for 
\SiIV/\CIV~:~\CII/\CIV.
 
Finally, for \SiII/\CII~:~\CII/\CIV, the Q$_{HM}$ model generally gives 
good fits to the data in the first two redshift intervals, while in the 
highest interval the data show a downward trend not well followed at the 
$z_{HM} =$ 3.4 boundary in the model. Here the upper limits are restricted 
just to \SiII\ by the need for positive detection of \CII\ as well as \CIV. 

In summary of the above, while a radiation environment due only to QSO 
sources provides the basis for a reasonable match to the data in our lowest 
redshift interval, to higher redshifts the evident significant evolution in 
the data is not consistently reproduced, in particular for \emph{the strong 
absorbers}, by this simple model.

We now include background radiation from both QSOs and galaxies (QG$_{HM}$) 
with their individual evolutionary behaviour, using the full \citet{ham12} 
development of the metagalactic radiation environment. The \Cloudy-generated 
results based on these complete spectra are given, in \emph{thick continuous 
lines}, alongside the Q$_{HM}$ model in Figure 17(b). 

For \SiIV/\CIV~:~\CII/\CIV, in the first redshift interval the dominance of 
QSOs makes the QG$_{HM}$ and Q$_{HM}$ model boundaries almost coincident; 
for the most part, both fit the data quite well. In the middle interval the 
two models remain close but the data show a significant departure from these
models at the $z_{HM} =$ 3.4 boundary. In the highest redshift interval, 
because of the positively evolving galaxy component and diminishing 
contribution from QSOs the evolutionary development of the QG$_{HM}$ model 
becomes quite extensive and at the $z_{HM} =$ 4.4 boundary shows a 
substantial uplift relative to the Q$_{HM}$ model, increasing strongly to the 
left along the \CII/\CIV\ axis, and here overall this model approaches a far 
more accomodating fit to the data than does the Q$_{HM}$ model. 

The picture here for \NV/\CIV~:~\CII/\CIV\ in the first and middle redshift 
intervals again is similar for the two models and they remain consistent 
with the data. In the highest redshift interval the QG$_{HM}$ model at the 
$z_{HM} =$ 4.4 boundary moves substantially down from the Q$_{HM}$ model in 
consequence of the highly depressed continuum beyond the \HeII\ edge seen in 
Figure 17(a), and is consistent with the observational upper limits. 

Unlike the other sets, for \SiII/\SiIV~:~\CII/\CIV\ the QG$_{HM}$ and 
Q$_{HM}$ model curves are almost identical in all the redshift intervals, 
but therefore still showing no tendency to follow the large downward 
progression of the data with redshift as already commented. 

For \SiII/\CII~:~\CII/\CIV\ in the first two redshift intervals the two 
models are little different and both give good fits to the data. However, 
in the highest interval the $z_{HM} =$ 4.4 boundary for the QG$_{HM}$ model 
now lifts substantially away from both the Q$_{HM}$ model and the data, while 
the relative downward trend in the data relative to the $z_{HM} =$ 3.4 
boundary remains.

Again summarizing, while the QG$_{HM}$ metagalactic spectral energy 
distribution makes progress in meeting the observational data in the cases 
involving \SiIV/\CIV\ and \NV/\CIV, we cannot achieve simultaneously 
consistent fits to our full set of observed ionic ratio combinations over 
the complete redshift range \emph{for the strong absorbers} with this model 
alone. 

\subsection{Effect of Delayed Reionization of Intergalactic 
He~{\scriptsize II}?}

Here we ask whether substantial further suppression of the metagalactic 
radiation intensity in the \HeII\ continuum can improve the fit to our 
high-redshift data. While in \citet{ham12} the discrete physical nature of 
absorbers is taken into account in the definition of ``effective opacity'' 
of the IGM, the discrete nature of the emitting sources is not. The 
radiative transfer formalism implicitly assumes a homogeneous \emph{volume 
averaged} background intensity and yields correspondingly homogeneous 
ionization properties of the IGM. This is a good assumption after \HeII\
reionization and the computed background intensity is appropriate for 
exposure of isolated absorbers. However, before full reionization of \HeII\ 
the IGM is highly inhomogeneous (e.g., Bolton et al. 2006; Shull et al.
2010), with expanding \HeIII\ regions surrounding QSOs\footnote{These 
``bubbles'' have ``fuzzy'' boundaries due to the penetrating nature of the 
associated X-ray emission.} embedded in a medium limited to \HeII\ where at 
large there is little or no radiation beyond 54.4 ev. The ionization 
thresholds  relating to the appearance and loss of \SiIV\ and \CIV\ 
(\SiIII: 33.5 eV; \SiIV: 45.1 eV; \CIII: 47.9 eV; \CIV: 64.5 eV) straddle 
the \HeII\ ionization edge and those of \NIV\ and \NV\ (respectively: 
77.5 eV; 97.9 eV) lie well beyond, making the ionic balance of all these 
susceptible to such radiative inhomogeneities. 

Alluding to this patchy nature of the reionization process \citet{mah09} 
pictured representative cases for the delayed reionization of \HeII\ by 
artificially increasing \HeII/\HI\ in the IGM, with strong effect on the 
depth of absorption in the \HeII\ Lyman series and continuum of the 
background radiation. As we have already shown in \S8.1 a significant 
demarcation around $z = 3$ is not indicated by the behaviour of \SiIV/\CIV\ 
median values as a function of redshift. However, now we are considering 
the more complete behaviour of several different combinations of ionic 
ratios containing one or both of \SiIV\ and \CIV. To get a gross idea of 
the possible effect on the radiation background of delayed reionization 
of \HeII\ we notionally follow \citet{mah09} beyond $z = 3$ (i.e., for our
boundary values at $z_{HM} =$ 3.4 and 4.4) but here for each value of 
$N$(\HI) in the original model fix \HeII/\HI\ and \HeI/\HI\ at the value 
they have at $z \sim 6$.\footnote{Again, we are grateful for the provision 
of these additional results of the radiative transfer code \citep{haa14}.} 
The spectral result is included, in \emph{dotted lines}, in Figure 17(a). 
The \HeII\ continuum now is much deeper at $z_{HM} =$ 3.4 than in the 
original case and also deeper at $z_{HM} =$ 4.4 (but naturally to a lesser 
extent at this redshift because of the already existing incipient trend in 
this direction), while radiation over the rest of the spectra including at 
the \HeI\ ionization edge is not significantly affected. Then, 
interestingly, for $z_{HM} =$ 3.4 and 4.4 the corresponding ionic ratio 
outcomes included in Figure 17(b) show significantly better concordence 
with the data for \SiIV/\CIV~:~\CII/\CIV, giving a necessary rise in the 
higher ionization region (i.e., to the left of the diagram), and remains 
consistent for \NV/\CIV~:~\CII/\CIV, (albeit within the assumed range of 
N/C relative abundance given in \S8.2). However the outcome for 
\SiII/\CII~:~\CII/\CIV\ at $z_{HM} =$ 4.4 relative to the data is 
considerably poorer than in the original and, most significantly, there is 
virtually no change for \SiII/\SiIV~:~\CII/\CIV. Taken all together, these 
results show that simple modifications in the \HeII\ opacity do not explain 
our high redshift data for the strong absorbers. To account for the 
expected cosmic variance arising from variously merging \HeIII\ regions in 
the approach to full reionization we also try a combination of both the 
original and delayed reionization cases. Unsurprisingly this produces a 
result between the two and gets us no further.

In the following two sub-sections we continue with the QG$_{HM}$ model to 
explore both the assumed validity of photoionization equilibrium as the 
dominant ionization process for the observed absorbers and the effects of 
changes in absorber parameters.

\subsection{Photoionization Equilibrium or Collisional Ionization?}

In pure collisional ionization equilibrium the \CIV\ ionization fraction 
strongly peaks at $10^5$~K \citep{sad93} and there is a peak at about the same 
temperature in the more complicated case of a non-equilibrium radiatively 
cooling gas \citep{sam76,tri02,gna07}. Such a temperature corresponds 
to {\it b}~$\simeq 10$~km~s$^{-1}$ and is already ruled out for our narrower 
components (see the {\it b-}value distribution given in Figure 11(a)). In the 
same circumstances \SiIV\ peaks at a slightly lower temperature: 
$\sim 0.8 \times 10^5$~K. In contrast, for photoionization equilibrium, 
\Cloudy\ modelling gives a mean temperature $\lesssim 0.5 \times 10^5$ K 
within the absorber gaseous column in the \CII/\CIV\ range of our data, as we 
show in the next sub-section.

To indicate the \emph{additional} effects of collisional ionization, in 
Figure 18 (with downward extended axes) we show in \emph{coded dashed lines} 
idealized model results from \Cloudy\ runs with our fiducial absorber 
parameters at the two fixed temperatures (1.0, 1.2) $\times 10^{5}$~K in the 
presence of the model QG$_{HM}$ ionizing background radiation and cosmic 
microwave background as before, contrasting the two intervals with bounding 
redshifts $z_{HM} =$ 1.9, 2.6 and 3.4, 4.4. In \emph{continuous lines} we 
include for comparison displays of our four ionic ratio combinations obtained 
in pure photoionization equilibrium (PIE) with the model QG$_{HM}$ as shown 
in Figure 17(b). The fixed temperature collisional ionization curves 
terminate in the diagrams where the ionic ratios become independent of total 
hydrogen volume density in the model. In the absence of ionizing radiation 
these models at all densities give single values at these termination points. 

It is clear from the figure that at temperatures where collisional ionization 
would be dominant, these models, unlike the photoionization models at their
equilibrium temperatures, bear little structural resemblance to our 
observations in Figure 17(b). We can conclude that the great majority of the 
absorbers in our samples are in or close to a state of photoionization 
equilibrium. It is important to note that in this there is \emph{no 
significant distinction between the broader and narrower components} in the 
sample, which fit similarly well. This confirms that the broader components 
represent turbulent or bulk velocity-extended structures rather than regions 
at much higher temperature dominantly thermally broadened (see \citet{fox07b}
for observations relating to DLA and sub-DLA systems).

\subsection{Effect of Changes in Absorber Parameters?}

In Figure 19 we demonstrate how the outputs of the four ionic ratio 
combinations we have used in Figure 17(b) for the QG$_{HM}$ model are 
affected by changes in metallicity and \HI\ column density and here again 
contrast the \Cloudy\ outputs for the low and high redshift intervals. The 
absorber fiducial case of [$Z$] $= -2.5$ and $N$(\HI) $= 10^{15}$ cm$^{-2}$ 
is shown in \emph{continuous lines}; the rest are coded as indicated. 
Corresponding displays for total hydrogen volume density, $n$(H), and mean 
temperature within the absorber gaseous column, $\langle T_e \rangle$, are 
included in the bottom two sets of panels. 

In all cases shown, over the whole included \CII/\CIV\ range there is 
relatively little effect on the computed results for the values 
[$Z$] $\lesssim -1.5$ and $N$(\HI) $\lesssim 10^{16.0}$ cm$^{-2}$, while 
\SiII/\SiIV~:~\CII/\CIV\ is almost invariant for all the listed parameter 
values. Comparison with the data in the corresponding panels of Figure 17(b) 
indicates that most of the strong, \emph{detected}, components in our sample 
(i.e., those without upper limit values in their makeup) conform with these 
constraints. Even with a general extension to [$Z$]~$\lesssim -1.0$ and 
$N$(\HI) $\lesssim 10^{17.0}$ cm$^{-2}$, significant departures from the 
fiducial case occur only for log~$N$(\CII)/$N$(\CIV) $\lesssim -0.5$. We note 
that to some extent such departures notionally can be included by our 
observational \CII/\CIV\ upper limits in \SiIV/\CIV~:~\CII/\CIV.

Over the whole of the indicated parameter variations there is also 
comparatively little change in the $n$(H)~:~\CII/\CIV\ relationship. For the 
\emph{given radiation fields} in the QG$_{HM}$ model \CII/\CIV\ is thus a 
good indicator of the gas volume density of the absorbers. Using this, we see 
by comparison with at least the top two sets of panels in Figure 17(b) that 
the detected absorber components are there indicated to range in density over 
$n$(H)~$\sim 10^{-3.5}$--$10^{-2.5}$ cm$^{-3}$. 

Concerning $\langle T_e \rangle$, for log $N$(\CII)/$N$(\CIV) $\lesssim -0.5$, 
raising [$Z$] leads to a reduction in temperature (due to increased cooling). 
For the absorber components with Si and C {\it b-}values listed independently 
in Tables 2--10 (i.e., with {\it b}(Si) shown \emph{unbracketted}: see \S3 
and \S4) we obtain, following \citet{rau96}, an average absorber temperature 
of $(2.0\pm0.2)\times10^{4}$ K over the range 
$-1.0 \lesssim$ log~$N$(\CII)/$N$(\CIV)~$\lesssim 1.0$, broadly consistent 
with Figure 19. For $z < 3.1$ and $z > 3.1$ the respective values are 
$(1.9\pm0.3)\times10^{4}$ K and $(2.1\pm0.4)\times10^{4}$ K, showing no 
significant change with redshift. These values are lower than the mean 
$3.8\times10^{4}$ K found by \citet{rau96}, but as we have explained in \S3, 
on the one hand strictly we include here only the narrower components 
({\it b} $\lesssim 10$ km~s$^{-1}$), and on the other, in their analysis, 
differential blending effects in broader components may give a tendency to 
estimate apparently higher temperatures. 

An important conclusion from these results is that the systematic 
evolutionary effects exhibited in the ionic ratios are not caused simply by 
\emph{changes in absorber properties}. We must therefore look to evolution 
in the \emph{ionizing radiation environment} to explain the observations, as 
we develop next. 

\subsection{Template Galaxies Local to the Absorbers?}

On the now general recognition discussed in \S7 that absorbers such as in our 
sample are regions located in the outer extensions of galaxies (e.g., Steidel
et al. 2010), we seek to account for our observations by augmenting the 
metagalactic background radiation with direct radiation from the star-forming 
regions within the local, \emph{parent}, galaxy. 

It should be noted that in computing the mean intensity of the 
\emph{metagalactic} background radiation, all cosmologically distributed 
sources must be included from the outset. The results for additional such 
sources cannot just be added individually since the background radiation 
intensity determines the ionization balance of the clumpy intergalactic 
medium which in turn determines the character of the background radiation. 
However, in including the radiative effects of star-forming galaxies in the 
\emph{local} environment of the absorbers, such contributions \emph{can} be 
treated incrementally because we do not imply any addition of sources 
influencing the general background: in effect we are simply defining the 
\emph{location of the absorbers} relative to the galaxies in the already 
existing population. Thus, a sufficiently isolated absorber will be exposed 
only to the metagalactic radiation while an absorber close to a galaxy will 
in addition experience a direct ``proximity effect'' within the ionizing 
spectral range 1--4 Ryd. For significant impact on our strong absorbers, 
from what we have found in \S8.2, over much of this range the radiation 
from the local galaxy will need to \emph{dominate} over the general 
background radiation.

Stars in galaxies at high redshift are born in dense protogalactic clumps 
with sufficient resident neutral hydrogen within the full galactic volumes to 
absorb most of the emitted ionizing radiation. That the IGM must be 
reionized by the action of such populations within $12 \lesssim z \lesssim 6$ 
and remain ionized thereafter gives demanding requirements for escape of a 
significant fraction of the stellar ionizing radiation from such highly 
attenuating environments (e.g., Bolton \& Haehnelt 2007; Pawlik et al. 2009; 
Bouwens et al. 2012a; Haardt \& Madau 2012; and see \S8.2). While only very 
small escape fractions are observed in galaxies at $z \lesssim 1$ (e.g., 
Leitherer et al. 1995; Siana et al. 2010) the needed increase with redshift 
is observationally indicated in some degree (e.g., Steidel et al. 2001; 
Giallongo et al. 2002; Shapley et al. 2006; Inoue et al. 2006; Iwata et al. 
2009; Vanzella et al. 2010; Boutsia et al. 2011) albeit in an apparently 
patchy manner among galaxies at least as seen from our observational 
viewpoint. Simple model estimations of escape fraction 
\citep{das94,hal97,mhr99,das00,ras00,cao02,fuj03,fas11} as well as 
sophisticated simulations (e.g., Razoumov \& Sommer-Larsen 2006, 2007, 2010; 
Gnedin et al. 2008; Wise \& Cen 2009; Yajima et al. 2011; 
Smith et al. 2011; Kimm \& Cen 2014) generally predict a wide range of 
values within the large uncertainties of the formulations for star formation, 
dust attenuation and supernova feedback, as discussed by many of the latter 
authors. \citet{cao02} and \citet{fuj03} propose specific effects of repeated 
supernova explosions to create galactic outflows which make ``channels'' 
enabling enhanced escape of ionizing radiation. \citet{yaj11} in their 
simulations find that the average escape fraction for galaxies at $z =$ 3--6 
steeply decreases as the halo mass increases because those with massive halos 
($M_{h} \sim 10^{11} M_{\Sun}$) have a densely clumpy gaseous structure 
enveloping the star-forming regions while with lower-mass halos 
($M_{h} \sim 10^{9} M_{\Sun}$) conical regions of highly ionized gas often 
develop which allow the ionizing photons to escape, although with a large 
scatter in escape fraction. In common among such escape eventualities, high 
escape fractions require high ``porosity'' achieved only by very strong 
supernova feedback. \citet{ade03} stress the role of supernovae-driven 
superwinds in their observational study of the local gaseous environments of 
Lyman-break galaxies and the presence of \CIV\ absorbers.

In relation to this we consider the possible influence on the balance of 
our ionic ratios of spectrally-dependent extinction by dust occurring in 
the local galaxies. In all quantitative studies of high redhift galaxies 
determination of reddening by dust leading to empirical extinction 
corrections throughout the (non-ionizing) far-ultraviolet to $\sim 1200$ 
\AA\ is common practice (e.g., Calzetti et al. 1994, 2000; Seibert et al. 
2002; Reddy et al. 2008, 2010; Bouwens et al. 2009, 2011, 2012b) but very 
little is known about the dust composition, grain size distribution and 
degree of sublimation (some amount of dust is likely to be destroyed by 
ionizing radiation) that would enable accurate extension into the \HI\ 
Lyman continuum (e.g., Draine \& Lee 1984; Martin \& Rouleau 1990; Cruise 
1993; Weingartner \& Draine 2001; Binette et al. 2005). Broadly, the 
relevant extinction probably peaks near 800 \AA\ and diminishes thereafter 
(e.g., Pei 1992; Gnedin et al. 2008). However, in the prevailing 
environments we study here the \emph{need} for accurate knowledge of 
extinction by dust over 1--4 Ryd is rendered irrelevant if the ionizing 
radiation escapes only through channels sufficiently clear of absorpion 
that there will be simply a \emph{geometrical} attenuation not dependent 
on wavelength. With the spectral shape then preserved we may use a global 
escape fraction in this region, indeed following \citet{ham12} in their 
definition of $\langle f_{esc} \rangle$ for the metagalactic contribution 
(\S8.2). It is interesting that even in the presence of additional 
``translucent'' escape routes not ``totally'' cleared of significantly 
absorbing gas and dust the large effective absorption cross section of 
neutral hydrogen in this spectral region will overide the effect of dust 
if the gas is not excessively ionized (the dust content relates to 
metallicity, therefore to total hydrogen, while the comparative weight of 
optical depths relates also to the residual neutral hydrogen). 
\citet{gne08} support such a picture in their full model simulations, 
finding the escape fraction at energies above 1 Ryd to be almost 
independent of dust absorption because the contribution from translucent 
points having included dust is small (see also Razoumov 
\& Sommer-Larsen 2007). Thus hydrogen can be the dominant source of 
opacity in our range of interest above 1 Ryd while dust is the prime source 
below this. An obscuring hydrogen column density $N$(\HI) greater than only 
a few $\times 10^{18}$ cm$^{-2}$ is sufficient to absorb essentially all 
the stellar ionizing radiation leaving the galaxy, while the constraint from 
$\gamma$-ray burst observations at $z \geq 2$ shows the value is generally 
much higher than this \citep{che07}, so we shall assume in general that in 
this model any escape of ionizing radiation will come about through 
effective geometrical porosity alone. 

Accordingly, for a local galaxy component we use an intrinsic spectral energy 
distribution similar to those applied as the distributed template galaxies in 
the \citet{ham12} QG$_{HM}$ model, but now not modified by long passage 
through the IGM. For this we employ Starburst99 (Leitherer et al. 1999, 2010; 
we use the August 2010 release v6.0.2 throughout), which has ingredients 
close to the Bruzuel \& Charlot (2003) GALAXEV models \citep{val05}. We model 
a galaxy of age 100 Myr undergoing continuous star formation, with 
\citet{sal55} IMF over the mass range 0.1--100 $M_{\Sun}$, using Geneva tracks 
with ``high'' mass-loss rates (these are optimized for young massive stars, 
of particular relevance here), and Pauldrach/Hillier atmospheres. For 
representative galaxy metallicities we are guided by the relation given by 
Haardt \& Madau (2012), shown in \S 8.2, and take the closest options in 
Starburst99 of 0.4 solar at $z$ = 1.9 and 2.6 and 0.2 solar at $z$ = 3.4 and 
4.4 (the implied differential effects in the modeling amount to $\sim 10\%$, 
small compared with the possible range in the galaxy mass-metallicity 
relation (e.g., Erb et al. 2006). We obtain low-resolution spectra 
($\sim 20$ \AA) appropriate for input to the \Cloudy\ code. 

To represent the ultraviolet luminosities of typical bright galaxies over 
our redshift range we take observed values for the characteristic absolute 
AB magnitude \citep{oag83}, $M^{*}_{AB}$, as follows: $-20.7$ ($z \sim 2$) 
and $-21.0$ ($z \sim 3$) at rest-frame 1700 \AA\ from \citet{ras09} and 
$-21.0$ ($z \sim 4$) and $-20.6$ ($z \sim 5$) at rest-frame 1600 \AA\ from 
\citet{bou11}. Interpolating over these we obtain at our boundary redshifts 
the respective values: $-20.6$ ($z \sim 1.9$), $-20.9$ ($z \sim 2.6$) and 
$-21.0$ ($z \sim 3.4$) at rest-frame 1700 \AA\ and $-20.9$ ($z \sim 4.4$) at 
rest-frame 1600 \AA. Over the same redshift range at these characteristic 
magnitudes dust extinction is relatively unchanging 
\citep{red08,bou09}:\footnote{There is a strong redshift dependence beyond 
this, with extinction practically non-existent by $z \sim 7$ \citep{bob12}.} 
we use a representative extinction factor $\sim 6$ from \citet{bou09} to 
obtain the corresponding intrinsic (i.e., dust-corrected) values 
$M^{*}_{AB_{int}}$.\footnote{Additionally, at any redshift, due to 
correlation between star-formation rate and dust extinction, galaxies of 
lower luminosity in general show reduced extinction 
\citep{wah96,red06,ras09,bou09}. \citet{ham12} in their derivation of the 
dust-reddened contribution of the \emph{population} of galaxies to the 
metagalactic radiation background integrate well down the faint-end slope 
of the luminosity function and consequently apply the appropriate 
\emph{luminosity-weighted extinction}. In the application we develop here 
we focus on the additional exposure experienced by an absorbing system in 
the close locality of a typical \emph{individual} relatively bright galaxy 
(see below), therefore use a level of extinction consistent with the 
luminosity of the given galaxy we consider.} Then, in line with our model 
assumptions discussed above, we use the values $M^{*}_{AB_{int}}$ at our 
adopted boundary redshifts to scale our Starburst99 model respectively at 
1700 and 1600 \AA\ to obtain the corresponding intrinsic spectral energy 
distribution extending through the \HI\ Lyman continuum (the model intrinsic 
flux ratios at the \HI\ Lyman limit relative to these wavelengths are 
respectively 0.39 and 0.36). Finally, to set appropriate 1--4 Ryd escape 
fractions at this stage, for consistency we apply the absolute values 
$\langle f_{esc} \rangle$ given by the \citet{ham12} formulation in 
\S 8.2: 0.0067 ($z = 1.9$); 0.014 ($z = 2.6$); 0.028 ($z = 3.4$); 
0.056 ($z = 4.4$).\footnote{in \S9 we develop the issue of escape fraction 
on the basis of quantitative results obtained from our further modeling.} 
Translating the above, we obtain the following values for the 
\emph{escaped} intrinsic 1 Ryd flux at 10 physical pc, 
$F^{*}_{\nu_{esc}}$ (912~\AA) erg s$^{-1}$ cm$^{-2}$ Hz$^{-1}$: 
$9.9 \times 10^{-14}$ ($z = 1.9$); $2.7 \times 10^{-13}$ ($z = 2.6$);
$5.9 \times 10^{-13}$ ($z = 3.4$); $1.0 \times 10^{-12}$ ($z = 4.4$).

Next, to set typical distances of prospective QSO absorbers from their parent 
galaxies we relate our relatively strong, complex absorbers in particular 
(i.e., those containing positively detected \CIV, \SiIV, \CII\ and \SiII) to 
the strong absorbers containing the same species \citet{ste10} find from their 
direct observations of outflowing gas in the circumgalactic media (CGM) of 
relatively luminous star-forming (Lyman-break) galaxies at redshifts 
$2 \lesssim z \lesssim 3$ and which in character we extend here notionally to 
$z$ = 4.4.\footnote{\citet{ste10} use close angular pairs of galaxies of 
different redshifts, with the local environment of the nearer probed along the 
sightline to the further acting as source. This complementary technique vastly 
increases the sample size at these scales relative to the established method 
using QSOs.} They find strong \emph{in-filling} circumgalactic absorption with 
an effective edge at a galactocentric radius $\sim 80$ physical kpc beyond 
which no metal absorption is detected. Similar results are obtained by 
\citet{ade05} at $1.8 \lesssim z \lesssim 3.3$ for \CIV\ alone. Linking with 
this, we take such galaxies as typical objects in our computations.

Correspondingly, for the modeling in this section we use values 80 and 20 
physical kpc as indicative galactocentric distances and add the resultant local 
spectral energy contributions at the absorbers, obtained from the flux values we 
have defined above, to the respective \citet{ham12} metagalactic ionizing 
background spectra at the $z_{HM}$ boundaries.\footnote{We continue with the 
designations $z_{HM}$ in the figures because in all cases including those with 
additional local sources we ``ride'' on the \citet{ham12} metagalactic ionizing 
background.} We then include the absorber values [Si/C] $=$ 0.0, 0.4 as before 
to give the combined bounding values to proceed with the \Cloudy\ modeling.

In Figure 20(a), following the style of Figure 17(a), we give the 
corresponding spectral energy distributions in \emph{continuous lines}, with 
the original Haardt \& Madau QG$_{\it{HM}}$ cases in \emph{dashed lines} 
(here we no longer show the Q$_{\it{HM}}$ model): both cases coincide above 
4 Ryd because of the cutoff there in the galaxy spectrum. For reference, in 
\emph{thick dotted lines} we also show the corresponding galactic spectra 
without application of an escape fraction. Moving to Figure 20(b), following 
the style of Figure 17(b), it is obvious that the shape of the galaxy full 
spectrum with attenuation invariant over 1--4 Ryd, as the current escape 
fraction provides, only achieves an evolving pattern that is, in the trend in 
redshift down the panels, manifestly a significantly \emph{poorer} fit to the 
collective data for . In particular, at the highest redshifts: for 
\SiIV/\CIV~:~\CII/\CIV\ there is a significant drop in the right of the 
diagram; for \SiII/\SiIV~:~\CII/\CIV\ there is an overall rise still further 
from the data; and for \SiII/\CII~:~\CII/\CIV\ there is an increased large 
rise away from the data in the left half of the diagram. Again we stress that
these conclusions apply in particular to the \emph{strong absorbers} having
positive detections for all species within each set of ionic ratios we 
display (we consider the partial upper limit cases later). These model 
results are not critically sensitive to the distance of the absorber: for 
these three sets we find no significant changes when we vary the received 
intensity at the absorber by placement at galactocentric radii anywhere from 
closer in to positions widely spaced between the two adopted values. 
Evidently this is because the galaxy spectrum strongly dominates over the 
ultraviolet background spectrum in the whole range 1--4 Ryd and no effective 
change in spectral \emph{shape} occurs within wide exposure limits. There is 
less to comment about \NV/\CIV~:~\CII/\CIV\ other than the lack of galactic 
radiation in the \HeII\ ionization continuum reduces the computed values 
further in the direction below the observed upper limits, similarly to that 
shown in Figure 17(b), and which here for $z_{HM} =$ 4.4 takes it below the 
bottom of the diagram.

To explain the poor fits pervading Figure 20(b), as possible cause, the 
synthesized galaxy models that are assumed inevitably must be a relatively 
loose approximation in the description of the emergent ionizing flux at 
energies above 1 Ryd because of the uncertainties in modeling the properties 
of the hot star population, both in definition of evolutionary tracks and 
in treatment of atmospheres 
\citep{sas97,cro00,kew01,pau01,snc02,vas07,sas08,eas09,lev12}, and while 
synthesis models are available which give somewhat more emphasis to the hot 
star population (e.g., Fioc \& Rocca-Volmerange 1997; Leitherer et al. 1999; 
Smith et al. 2002) the lack of feasible corroborating observations at the 
short wavelengths of interest here compounds the uncertainty. Moreover, the 
star-formation rate is likely to be varying \citep{brc03,lei99,kol99} and 
because of the short lives of massive stars such variation is greatly 
amplified in the ionizing far ultraviolet and, to maintain a high average 
luminosity, the duty cycle needs to be appropriately high: thus, while a 
constant star formation rate may be a good approximation at the longer 
wavelengths this might be too simplistic for what is required here. 

However, the \emph{required} spectral change is greater than implied by 
such possible modeling deficiencies. We find in practice that we can 
readily adjust the stellar IMF to give changes in the resultant 
population-synthesised galaxy spectra that enable the \Cloudy-generated 
outcomes indeed to become reasonable fits to the data. Essentially this 
simply  means truncating the IMF at a maximum near 20 $M_{\Sun}$ to remove 
the contributions from all the hottest stars. But this will not be a 
physically sound solution. Galaxy spectra observed at high redshifts in the 
rest-frame far-ultraviolet below 1 Ryd (where the parent galaxy is 
transparent, albeit with attenuation by dust) are seen to be dominated by 
emission from O and B stars. In particular, the common presence of the 
P Cygni profiles of the \NV\ and \CIV\ resonance lines demonstrates the 
existence of short-lived very hot stars near the top of the stellar mass 
range. These galaxy spectra are best matched by regions undergoing 
continuous star formation with stellar populations having a Salpeter IMF 
showing no evidence for deficiency of the most massive stars (e.g., Pettini 
et al. 2000; Shapley et al. 2003; Steidel et al. 2003; Erb et al. 2010). 
The inclusion of this full range of stellar masses must therefore be a 
\emph{condition} of any star-forming model we adopt. 

To test whether the solution is simply to introduce some partial attenuation 
by \HI\ we include an absorbing region having $N$(\HI) $= 10^{18}$ cm$^{-2}$ 
between the emitted galaxy spectrum and the CGM absorbers at $z_{HM} =$ 4.4 
and in Figure 20(a) show the spectral result in \emph{fine dotted line} and 
reflecting the shape of the \HI\ photoionization cross-section which peaks 
at 1 Ryd and steeply falls to a small value by 4 Ryd). However, overall, the 
corresponding \Cloudy\ results in Figure 20(b) become yet further removed 
from coming to a match with the data compared.

As an alternative to continuous star-formation it might be argued that the 
answer lies in the functioning of \emph{instantaneous} starbursts having the 
necessary full range of initial stellar masses. With delay of a few Myr after 
onset, when the hottest stars have exploded but those hot enough to produce 
substantial ionizing radiation still remain, the collective spectral content 
could provide the needed match. This is indeed so. But here the effective 
age-band to yield the observed ionic characteristics of the absorbers is 
itself only a few Myr: for a given isolated pairing of a specific 
star-forming region and a nearby absorber observed along a QSO sightline 
this means an unlikely fine-tuning. Regardless of this, reliance on such a 
one-to-one relation is not plausible because the galaxy as a whole has to 
appear spectrally complete in the observable far-ultraviolet at energies 
below 1 Ryd and this means essentially continuous star-formation when 
integrated over all regions in the galaxy, which then brings the situation 
at the absorber back to square one. 

We conclude that the highly significant mismatch between the collection of 
observed ionic ratio combinations of the strong absorbers and the model 
predictions relating to the region above 1 Ryd reveals that purely 
geometrical leakage of the ionizing radiation through absorptively-clear 
channels is \emph{physically} an incomplete representation.

Changing direction, we show next that a dynamically-dependent stellar 
scenario well enables a robust match to our data requirements while still in 
the full spectral domain of the population synthesis model with continuous 
star formation that we use above.

\section{SPECTRALLY MODIFIED IONIZING RADIATION FROM TEMPLATE GALAXIES BY
ESCAPE OF RUNAWAY STARS}

It is well known that a significant fraction of massive stars in the Galaxy 
are moving with higher than usual velocities ($\gtrsim$ 30 km$^{-1}$). Two 
processes are proposed for the origin of these ``runaway stars'': dynamical 
ejection from dense young stellar systems soon after their birth (e.g., 
Poveda et al. 1967; Gies \& Bolton 1986) and delayed ejection by explosion 
of a companion star (e.g., Blaauw 1961; Stone 1991). Which scenario dominates 
remains uncertain (e.g., Poveda et al. 1967; Hoogerwerf et al. 2001; Tetzlaff 
et al. 2011). From a catalogue of young (up to $\sim$ 50 Myr) runaway 
Hipparcos stars within 3 kpc of the Sun \citet{tet11} indicate for these a 
runaway frequency $\sim$ 30\% and the same kinematic properties no matter 
whether of low or high mass within the observed range.\footnote{Following 
\citet{tet11} we do not include differential weighting to address possible 
tendencies for mass-related rates of ejection of runaway stars (e.g., Anosova 
1986; Gies \& Bolton 1986; Stone 1991). Nevertheless, as becomes clear in the 
development below, such aspects can have little effect on our results because 
of the overiding influence of the mass-selective runaway process in the 
application here.} They present a peculiar velocity distribution peaking at 
$\sim 35$ km s$^{-1}$ and tailing to $\sim 100$ km s$^{-1}$, which we show
in the top panel of Figure 22; however, they do not apportionate between the 
two scenarios. As was first noted by \citet{das94} and later advanced by 
\citet{cak12} using a simplified analytic treatment, the high speeds of 
runaway stars enable a useful proportion of the wide range of massive stars, 
i.e., those radiating strongly in the ionizing spectral region, to travel 
out of their dense environments of origin and through the surrounding highly 
absorbing gaseous envelope before they explode at the end of their several 
Myr and upwards lifetimes. Thus, the supernova feedback facilitated ``passive'' 
escape of internal ionizing radiation into the CGM and IGM through transparent 
channels, addressed in \S8.6, is transposed into the initial \emph{kinematic} 
escape of the ionizing \emph{sources} themselves. This process becomes more 
efficient at high redshifts when galaxies are much smaller than locally in 
the Universe \citep{bou04,per08,oes10,ono13}. \citet{cak12} show that the 
radiation from runaway stars that have migrated to the low-density outer 
regions of high-redshift galaxies can be the dominant contributors to the 
reionization of the Universe. They also find that runaways ejected through 
dynamical encounters are far more effective than by the supernova explosion 
of a close companion because the delay of the latter event means very 
significantly shorter runaway-active lifetimes for the more massive of the 
ejectees. However, \citet{kac14} find from detailed high-resolution 
cosmological radiation hydrodynamics simulations that inclusion of runaway OB 
stars increases only mildly the mean escape fraction of \HI\ ionizing 
radiation over that from their new supernova feedback model which has higher 
yield than the non-runaway counterpart of \citet{cak12}. Both studies deal 
with models aimed at the period of reionization for $z \gtrsim 7$ while this 
paper relates to our observed range $z =$ 1.9--4.4 with \emph{ongoing} 
ionization and involving somewhat more developed galactic structures.

In our new modeling below we develop the runaway concept considerably 
further by formally adding the resulting \emph{spectral} changes that are 
induced in the final efflux of radiation. In this we consider both the 
dynamical ejection and supernova ejection mechanisms. Depending on the placing 
of a given star-forming region within the dense gaseous envelope of a galaxy 
and the velocity distribution of runaway stars, the most massive runaway stars, 
having the shortest main sequence lifetimes, may not escape at all from the 
effective boundary of galactic 
absorption before their demise, while to lower masses there will be an age 
threshold for escape after formation when longer-living stars can spend a good 
fraction of their main sequence lifetimes in adequately clear space. In effect, 
for given physical parameters, this introduces specific limits on the maximum 
\emph{active} stellar mass, only from below which can ionizing radiation be 
received. Then progressively to lower masses the corresponding radiative 
contributions both increase in time-span and are moderated by attendant 
intrinsic diminution in emitted flux coupled with substantial spectral 
softening: we show a range of indicative examples in the top panel of Figure 
21 (explained in more detail in the following sub-section). In aggregate this 
results in a large modification of the emergent ionizing spectrum due to 
runaway stars to one much attenuated at the higher energies approaching 4 Ryd 
while remaining considerably less affected at 1 Ryd where the \HI\ ionizing 
radiation is most effective and is generally accounted quantitatively (the 
\HI\ photoionization cross-section drops steeply from 1 Ryd to higher 
energies). As already indicated in \S8.6, we find such changes are the key 
requirement for coming to convergence with our observations. 

We stress that in this modeling we do not compromise our underlying 
requirement for enabling the full intrinsic spectral characterization of the 
synthesized high-redshift galaxy spectra in the non-ionizing far-ultraviolet 
range. All stars in a continuous formation process extending over the full 
mass range are present here identically as in the model in \S8.6 and the 
galaxy would appear the same spectrally in the observationally transparent 
range (taking due account of internal attenuation by dust). The difference 
occurs in the practically unobservable spectral region at energies above 1 
Ryd, where our kinematic modification enters in a systematic manner to 
overcome the gaseous obscuration for stars that escape from it but leaves 
blocked the contribution of stars that do not make it out before terminating. 
In contrast, the porosity model needs to provide sightlines along channels 
transparent to ionizing radiation extending from stars in their 
original siting, so revealing the full spectral content in the ionizing 
region; but as we show in \S 8.6 the indications from the ionic ratios we 
observe for our strong absorbers count against such clear paths at least in 
these cases. Moreover, as we also show, if at any point along the channel 
sightline there remains significant gaseous attenuation, the shape of the 
\HI\ photoionization cross-section gives a resultant spectral distribution 
depressed at the lower energies in the range 1--4 Ryd which, as we see from 
the bottom panel of Figure 20(a) and the top panel of Figure 21, is modified 
in the ``wrong'' direction. 

Using simplified galactic structures we now proceed in an \emph{empirical} 
approach to explore the resultant implications leading from the runaway 
scenario to produce the spectral outcomes that through our \Cloudy\ modeling 
give a good match to our evolving observational results and in particular for 
the strong absorbers. We include related indications of the effect of the 
gravitational potential due to galaxy and halo mass on the motion of the 
runaways.

\subsection{Method}

\subsubsection{Spectral Modeling}

We model the runaway-modified 1--4 Ryd spectral emission from galaxies in 
continuous star-formation by use of Starburst99 with the same overall 
parameters as applied in \S 8.6 (including both applicable metallicities 0.4 
and 0.2 $Z_{\Sun}$ to accomodate our redshift range) but since the radiative 
product of the runaway process is dependent on stellar mass as well as galaxy 
geometry we apply the following set of procedures while still encompassing the 
full mass range 0.1--100 $M_{\Sun}$.

(i) For both of the two metallicities we segment the full mass range into 
narrow component slices (we use 2 $M_{\Sun}$) within the overlying Salpeter 
IMF, then with Starburst99 compute with parameters as in \S8.6 two sets of 
fifty individually-derived component spectra pertaining to the specific mass 
intervals of the slices, and in this way collectively cover the whole mass 
range as before. In the top panel of Figure 21 we show a selection of these 
individual component spectra which indicate the significant changes with mass 
that underly our approach. In particular we point to the strongly developing 
\HeI\ ionization edge from $\sim$ 25 $M_{\Sun}$ to lower masses.

(ii) Next, for each of the component mass slices from (i) we obtain the 
mass-dependent fraction of main-sequence lifetime remaining after each of an 
empirical set of trial \emph{potential} escape intervals. In dynamical 
ejection we take the time from formation of an individual star to its 
departure from the effective boundary of galactic absorption. In supernova 
ejection we include an appropriate precurser interval, taking 4 Myr as a 
reasonable trial indicator (with increasing interval there is rapidly 
diminishing residual ionizing radiation). For main sequence lifetimes we use 
the formulation given by \citet{hur00}, again at metallicities 0.4 and 
0.2 $Z_{\Sun}$ (however, within this interval the change in model lifetime 
with metallicity is slight). In the second panel of Figure 21 we show this 
function of stellar mass (at the average 0.3 $Z_{\Sun}$) and in the third 
give illustrative examples of tracks of fractional active lifetimes for a 
range of trial escape intervals. The relatively shallow trend of main 
sequence lifetimes extending over the higher mass range, with the resultant 
confinement of escapees to lower masses after intervals exceeding 
$\sim$ 4 Myr coupled with coincident development of the \HeI\ ionization 
edge, provides the progressive spectral consequences heralded by the 
composite ionic data we observe in the strong absorbers over our redshift 
range. 

(iii) Continuing from (ii), from the view of the \emph{circumgalactic 
absorbers}, at whatever point a star appears from the boundary of the 
absorptive confines of its galaxy it will illuminate the whole available 
hemisphere and therefore be widely accessible to a given absorber without 
critical placing other than being on that side of the galaxy. For putative 
simple galaxy structures we assume star formation occurs largely in a compact 
stellar core in spheroidals and closely confined to the central plane in 
disks. We indicate such geometrical constructs as idealized cartoons in the 
bottom panel of Figure 21. In spheroidal models, escaping stars of a given 
velocity individually will reach the absorptive boundary of the galaxy after 
the same time interval from initiation. As these appear, a relatively distant 
absorber will receive ionizing radiation from sites only at the boundary of 
the hemisphere in its direct view. However, depending on remaining main 
sequence lifetime, as stars continue to move outward in their paths, those 
exiting from parts of the surface hidden to the absorber will rise above the 
boundary horizon and can add significantly to the illumination of the 
absorber. We illustrate this after an interval $P_{\rm{v,t}}$ defining stars 
of given velocity and time after initiation: in this example those in the 
large portion of the spherical surface shown in \emph{dashed line} illuminate 
the absorber while those in the small region shown in \emph{continuous line} 
are obscured by the galaxy. Such obscuration relates only to the radius of 
the absorptive boundary of the galaxy (we show this region shaded by vertical 
\emph{dotted lines}), so the larger is $P_{\rm{v,t}}$ the rapidly greater is 
the illuminating fraction (i.e., the fractional surface area of the 
\emph{dashed line} cap relative to the surface area of the underlying whole 
sphere). Conversely, for disks the body of the galaxy continues to obscure 
the stars departing from the far side. Furthermore, the internal stellar 
trajectories are more variated, with stars originating from a specific region 
in the plane reaching the absorptive boundary after transit times resulting 
from a wide range of slant distances. We show one such transit with the same 
interval $P_{\rm{v,t}}$ and corresponding spherical surface similarly coded 
and here it can be seen that the resulting illuminating fraction is much 
smaller. Thus in similar circumstances stars in disk-like galaxies proceed 
out of the confining gas with intrinsically lower geometrical efficiency than 
in spheroidals. Nevertheless, disks naturally present thinner gas layers to 
be overcome than spheroidals and this enables more massive, therefore more 
luminous, stars to escape within their lifetimes. Together, these two forms 
give due scope for the evolution in spectral outcomes our observations 
indicate.

(iv) For both the spheroidal and disk structures we include the fractional 
main sequence lifetimes within each of the set of trial escape intervals in 
(ii) together with the corresponding illuminating fractions in (iii) to scale 
the associated component spectra in (i) for each defined escape interval then 
sum all individual products to obtain the total spectral outcome for each 
interval. We find that a set of trial escape intervals covering the range 
2--15 Myr for dynamical ejection and a time-shifted set covering 6--15 Myr 
for supernova ejection are sufficient for application in the further 
procedures here described.

(v) Now to relate these trial escape \emph{intervals} to a set of trial escape 
\emph{distances} within a galaxy from the site of star-formation to the 
effective boundary of absorption, we use the velocity distribution of the 
population of runaway stars \citep{tet11} given in the top panel of Figure 22 
and for each of the trial escape intervals scale the integral spectral outcome 
obtained in (iv) by the relative weighting of the specific stellar velocity 
linking each escape interval with each trial escape distance defined. Then for 
each of such \emph{distance-based} composite sets we sum all the related 
velocity-based ($\gtrsim$ 30 km$^{-1}$) spectral products to obtain the 
corresponding total spectral contribution, and scale each relative to the 
whole stellar population by 0.3 to account for the fraction of runaway stars 
\citep{tet11}. For each model case we define the escape fraction relevant at 
the absorber, which here we term $f_{esc-abs}$, by the ratio at 1 Ryd of the 
received computed flux to the possible intrinsic flux from the same 
star-forming site in the galaxy but without attenuation or other modification, 
obtained from the summation of the basic component spectra in (i). 

(vi) Recognising the far greater effectiveness of dynamical ejection over  
supernova ejection (Conroy \& Kratter (2012); and as we find in our work 
below), but that the balance between these scenarios is uncertain, in the 
computations that follow we take each case separately for quantitative 
comparison.

(vii) To complete the spectra by extending to the observable non-ionizing 
region at energies below 1 Ryd we use the same unmodified, 
dust-extinction-corrected spectra as in \S8.6. Finally, at this stage, across 
the redshift range we add, as explained in \S8.6, the \citet{ham12} 
metagalactic flux QG$_{HM}$ to the local galaxy spectral contributions before 
\Cloudy\ modeling. With these we make iterative trials of boundary escape 
distances for both the galaxy structures, comparing the generated \Cloudy\ 
values of ionic ratio combinations with the values from our observations for 
each adopted redshift interval. We find evolving escape distances in the range 
100--600 pc most useful in associating the resultant spectral outcomes with 
our data, as we explain more fully in \S9.2.

\subsubsection{Gravitational Modeling}

Additionally, we assess the possible dynamical consequences to the runaway 
stars from the gravitational potentials relating to their parent galaxies. 

While pure dark matter simulations produce a consistent picture of the 
formation and evolution of large-scale structure in the $\Lambda$CDM cosmology 
(e.g., Navarro et al. 1997), far more detailed simulations are needed to 
describe the combination of dark matter and bayonic matter distributions 
within the scale of single galaxies. \citet{zem11} in a structural study of 
galaxies at $z$ = 2--4 using cosmological simulations which include 
modeling of the interstellar medium and star formation show the matter 
distribution as individual radial profiles for the star, gas and dark matter 
components. Within the overall baryon to dark matter universal mass fraction 
for the system (reached in aggregate at the outskirts of the halo) there is 
large radial variation among these components, with a strong concentration to 
a dominant baryonic core and a halo which is very markedly depressed generally 
within a few kpc of the centre. We reflect such mass distributions in our 
simple analytical density models below. While we always include such halos in 
our models, within our redshift range we concern ourselves only with escape 
beyond the gaseous boundary of a galaxy not from its dark-matter halo, and 
find that our runaway stars are influenced predominantly by the baryonic mass 
distributions with halo mass having relatively minor effect on the 
trajectories.

To represent our spheroidal galaxy model we use Plummer's (1911) simple 
approximation for the radial mass density distribution function given by

\begin{equation}
\rho(r) = \frac{3b^{2}M} {4{\pi}}\textrm{\phantom{.}} \frac{1} 
{[r^{2} + b^2]^{5/2}}\textrm{\phantom{.}}, 
\end{equation}

\noindent where $b$ is the radial softening length. For disk-like 
structures, we use the \citet{min75} three-dimensional model which in 
cylindrical coordinates (r,z) is given by 


${\phantom{.}}$

\noindent$\rho(r,z) = $
\begin{equation}
\frac{b^{2}M} {4\pi}\textrm{\phantom{.}} \frac{ar^{2} + 
[(a + 3(z^{2} + b^{2})^{1/2}][a + (z^{2} + b^{2})^{1/2}]^{2}} 
{[r^{2} + (a + (z^{2} + b^{2})^{1/2})^{2}]^{5/2}\textrm{\phantom{.}}(z^{2} + 
b^{2})^{3/2}}\textrm{\phantom{.}},
\end{equation}

\noindent where $a$ and $b$ are axial softening lengths, and reducing to the 
Plummer model when $a = z = 0$. In both cases we
approximate their dark matter halos by a spheroidal form \citep{zem11} using 
Plummer distributions with radial softening length 2--3 kpc according to halo 
mass, and make the baryon content 0.17 of halo mass \citep{kom11}. The paths 
of test particles in the relevant gravitational potentials are computed by 
integration using the fourth-order Hermite method \citep{maa92}. In the 
middle and bottom panels of Figure 22 we show trajectories in a selection of 
indicative cases for particles originating with velocities 30, 50 and 
70 km s$^{-1}$ moving radially from the core of spheroids and vertically from 
the plane of disks. We make this restriction in the latter case to avoid 
crowding in the figure; however in our spectral modeling above we include 
ejection fully into spherical space from any point in the plane of the disk, 
as we indicate in the cartoon in Figure 21. Depending on its mass a given star 
will terminate in its trajectory at the end of its main sequence lifetime.
If initiated dynamically shortly after its formation, in Figure 22 this can 
be referred to the absissa: along this axis in the bottom panel (but applying
equally to the middle panel) we mark some examples of such end points with 
indicated masses. For delayed supernova ejection the masses at the marked 
end points are correspondingly lower.

The figure shows that for spheroidal forms the stellar trajectories at 
velocities $\gtrsim 30$ km s$^{-1}$ are relatively little influenced at halo 
mass $10^{8}$ and $10^{9} M_{\Sun}$ while stars remain strongly contained 
from $\sim 10^{10} M_{\Sun}$. However for disk-like forms at halo mass 
$\sim 10^{10} M_{\Sun}$ the trajectories still allow substantial periods 
outside the galaxy, and if they have sufficient remaining lifetime can return, 
pass through the obscuring confines of the galaxy and out again on the other 
side. While from the point of view of the absorber such oscillating paths 
include the prospect of reappearing on the near side after passage back again 
through the galaxy, this loses some advantage by significant expenditure of 
lifetime. For contribution to the metagalactic radiation background (see 
\S9.2.2) stars are available from wherever they emerge from the galaxy. Such
re-entrant transits also apply to the spheroidal case but the different 
geometry scales make this ineffective for the relatively short-lived stars 
applicable here.

Having indicated the range of feasibility for escape of runaway stars from 
the simple galactic structures considered here we go no further in engaging 
such outcomes in detail in our spectral modeling in \S9.2 following. In a 
simplified trial, however, we find that in suppressing the low end of the 
velocity distribution, at which escape is more difficult (Figure 22), only 
minor spectral changes are induced. This is due largely to the relatively 
small intrinsic radiative contribution by stars escaping with low velocity 
which selectively are biased to lower mass.

At much higher redshifts \citet{cak12} consider a halo mass range 
$10^{7}-10^{10} M_{\Sun}$ and find that their runaway stars effectually peak 
in radiative output relative to the non-runaway population at 
$\sim 3 \times 10^{8} M_{\Sun}$ (the effect of the potential due to the 
galaxy and halo mass on the motion of the runaways is considered negligible).

\subsection{Results}

\subsubsection{Ionizing Radiation at the Local Absorbers}

Following the pattern of Figure 20(a), we show in Figure 23(a) an 
empirically derived set of our new spectra obtained from the computations 
defined in \S9.1.1 and containing physical parameters identified below. For 
reference we include in \emph{dashed lines} the displays of the 
metagalactic ionizing radiation background alone (QG$_{HM}$) and in 
\emph{fine dotted lines} these combined with full intrinsic galaxy spectra 
selected from Figure 20(a) with the adopted nominal escape fractions 
referred at 1 Ryd and received at the galactic distance 20 physical kpc 
(QG$_{HM}$+G$_{FULL}$). However, the new spectral results we now present 
contain the \emph{derived} runaway-generated escape fractions pertaining at 
the local absorbers, $f_{esc-run}$, not these nominal values. Therefore to 
assist visual comparison with the intrinsic spectral shapes from Figure 
20(a) we  normalize the runaway spectra at 1 Ryd by making corresponding 
adjustments of the galactic distance for these. For the dynamical ejection 
case shown in \emph{continuous lines} we do this directly while maintaining 
the outer limit at 80 physical kpc. For the supernova ejection counterpart, 
in \emph{bold dotted lines}, we use the same derived distances as a simple 
means of relative comparison of both ejection outcomes. All such resultant 
distances (identified in the figure) come within the observed range for the 
circumgalactic absorbers defined by \citet{ste10}.

Evidently there is a clear evolutionary trend with increasing redshift, 
showing development of a deep \HeI\ ionization discontinuity together with 
diminishing spectral intensity to higher energies characteristic of stars 
having mass from around 25 $M_{\Sun}$ to lower values as indicated in 
Figure 21, contrasting with the corresponding QG$_{HM}$+G$_{FULL}$ cases. 
The \HeI\ discontinuity has a marked effect on the lower ionization species 
of C and Si and for the strong absorbers at our highest redshifts is the 
dominant feature providing the necessary changes to bring together the 
predicted and observed ionic ratios.\footnote{The ionization potentials of 
\CII\ and \HeI, respectively 24.376 eV and 24.581 eV, nearly coincide, with 
the consequence of enhancing the ionization of \SiII\ relative to \CII\ and 
\SiIV\ thus preferentially reducing \SiII/\SiIV\ and \SiII/\CII.} Within 
this trend, at $z_{HM}$ = 1.9 and 2.6 we do find the need for significant 
contributions from the more massive stars, but with the concomitant 
requirement for much smaller escape fractions for galactic ionizing 
radiation than at the higher redshifts as formalized in equation (7). 
Nevertheless, the full intrinsic spectrum still has too much emphasis at 
the higher energies and the modified runaway contribution gives a 
significantly better fit than seen in the panels of Figure 20(b). Any 
``toleration'' for the presence of the hotter stars disappears by 
$z_{HM}$ = 3.4 and continues to our observational boundary at 
$z_{HM} =$ 4.4. We discuss the underlying physical consequences of these 
spectra more fully in association with the related ionization results 
which follow below. 

But first there is need for some important clarifying remarks. Foremost, 
as we have already highlighted in \S8.6 and indicate again above, we take 
it that our data containing solid measurements of all the species \CIV, 
\CII, \SiIV\ and \SiII\ relate directly to the observational regime in 
\citet{ste10}, namely \emph{strong absorbers within circumgalactic media 
associated with relatively luminous galaxies}. We have in addition 
numerous partial and full upper limits which we must expect represent 
weak absorbers present in a continuum of corresponding environments 
presumably including regions around more compact galaxies, especially 
at our higher redshifts, as well as extending beyond the CGM nominal 
perimeter at any redshift. We address this in \S9.2.2 in consideration of 
the collective contributions to \emph{metagalactic} radiation. Second, 
while the \emph{location} of a strong absorber relative to its parent 
galaxy is arbitrary for the QSO sightlines that we probe, and can be 
regarded as a natural degree of freedom, the ionic results we obtain 
here give information on the \emph{specific spectral shape} of the 
ionizing radiation received from the \emph{specific galaxy} and our data 
enable us to identify this shape which in turn gives evidence on the 
geometry of the galaxy. Here we deduce indicative ``absorptive boundary'' 
cases which spectrally achieve a good matching \emph{envelope} to our 
observational distributions within each displayed redshift interval. 
While these necessarily demonstrate feasibility, they represent only a 
part of the population and generally not the most \emph{emissive} 
examples within the given redshift interval. 

Now to Figure 23(b). Following the layout of the panels in Figure 20(b), 
we show the resulting \Cloudy\ predictions for the applied spectra 
presented in Figure 23(a) (these spectra represent the \emph{conclusions} 
of the iterative procedure explained in \S9.1.1 which yields the results 
shown here) and computed for the same fiducial absorber parameters 
defined earlier. As usual, to avoid overcrowding in the displays we 
combine the physical absorber distances with the values [Si/C] = 0.0, 0.4 
to give extreme bounding values for the modeling displayed within each 
redshift interval, but it should be visualized that both these relative 
abundance values (and values between) associate with \emph{each} 
individual redshift. Taking all redshift intervals together, these new 
models give good fits to the data, contrasting strongly with the set of 
non-runaway nominal full spectrum cases in Figure 20(b). Information on 
this set of results now follows.

In the top panels ($1.9 < z < 2.6$) we show runaway results for a disk 
model with dense \HI\ boundary of semi-height 200 pc (D200) for both 
redshifts. As experienced by an individual circumgalactic absorber, the 
value of $f_{esc-run}$ for \emph{dynamical} ejection that we derive for 
this (assuming this is the sole contributor) is 0.020, larger than the 
respective values 0.014 and 0.0067 at $z_{HM}$ = 2.6 and 1.9 given by 
equation (7). As indicated in Figure 23(a), to normalize the 1 Ryd 
spectral flux at $z_{HM}$ = 2.6 to that in Figure 20(a), here we make the 
corresponding absorber distance 24 kpc while at $z_{HM}$ = 1.9 leaving 80 
kpc as the CGM effective perimeter. Good fits are obtained in all four 
ionic ratio combinations in this redshift interval. For the associated 
\emph{supernova} ejected case at $z_{HM}$ = 1.9 and 2.6 (again assuming 
this is the sole contributor) we obtain $f_{esc-run}$ = 0.0015, about 7 
percent of the dynamically ejected value and giving results in the first 
and fourth panels which by themselves are poor fits to the data but can be 
accomodated if these are a contribution together with dynamical ejection. 
Notwithstanding its existence we conclude that supernova ejection is a 
negligible contributor to the emitted ionizing radiation in this redshift 
interval, in agreement with the findings by \citet{cak12}. However, if 
such stars are part of the population of directly observed runaway stars 
\citep{tet11} this needs to be accounted within the nominal 30\% total. 
On arbitrary supposition that both mechanisms are represented in equal 
numbers the total becomes $f_{esc-run}$ = 0.011 (i.e., (0.020+0.0015)/2). 
A spheroidal model of boundary radius 200 pc (S200) fits equally well to 
the data, as we show in the figure for the example of dynamical ejection 
at $z_{HM}$ = 2.6. For this model we obtain $f_{esc-run}$ = 0.096 and 
corresponding absorber distance 53 kpc for solely dynamical ejection and 
$f_{esc-run}$ = 0.052 if both mechanisms are there in equal numbers. 
While such a compact structure, by itself, may not be as plausible as a 
thin disk model, it may well represent component regions within extended 
lumpy structures such as are observed in this redshift range (e.g., 
Papovich at al. 2005; Law et al. 2007, 2012).

In the middle panels ($2.6 < z < 3.4$) we use the same D200 model at 
$z_{HM}$ = 2.6 and 80 kpc, and at $z_{HM}$ = 3.4 show results for a 
spheroidal model of boundary radius 400 pc (S400). For dynamical ejection 
good fits to the data are obtained across the four panels in this redshift 
interval; if this is the sole contributor S400 gives 
$f_{esc-run}$ = 0.020, comparable with the value 0.028 from equation (7), 
and to normalize the flux at the absorber to that in Figure 20, the 
corresponding absorber distance is 17 kpc. For supernova ejection, if 
alone, we obtain $f_{esc-run}$ = 0.0042, about 20 percent of the 
dynamical ejection value and giving \Cloudy\ results which again can be 
accomodated within the data (indeed quite helpfully in the fourth panel) 
but, again, only as a minor contributor together with dynamical ejection. 
As before, if both mechanisms exist in equal numbers, 
$f_{esc-run}$ = 0.012. A spheroidal model at $z_{HM}$ = 3.4 with boundary 
radius 300 pc (S300), gives results still fitting the data quite well for 
dynamical ejection, and if the sole contributor we obtain 
$f_{esc-run}$ = 0.035 with corresponding absorber distance 23 kpc. If 
both mechanisms are represented in equal numbers, $f_{esc-run}$ = 0.021. 
The spheroidal model S200 at $z_{HM}$ = 3.4 gives much poorer fits to the 
strong absorbers across all panels here.

In the bottom panels ($3.4 < z < 4.4$) we use the same S400 spheroidal 
model for $z_{HM}$ = 3.4 now at boundary radius 80 kpc and for 
$z_{HM}$ = 4.4 have a model with boundary radius 600 pc (S600) giving 
$f_{esc-run}$ = 0.0068 for solely dynamical ejection, much lower than the 
value 0.056 from equation (7). The corresponding normalizing absorber 
distance is 7 kpc. However, good fits are obtained throughout the four 
panels. The resultant particular outcomes in this defining redshift range 
are: \SiIV/\CIV~:~\CII/\CIV\ now lifts to accomodate the data and 
extensive upper limits; \SiII/\SiIV~:~\CII/\CIV\ shows the necessary 
substantial downward shift; \SiII/\CII~:~\CII/\CIV\ fits closely on the 
data; and \NV/\CIV~:~\CII/\CIV\ remains consistent with the data. In 
addition, at the closer galactic distance now applying at $z_{HM}$ = 4.4 
we indicate the effect of raising the metallicity of the model absorbers 
from the fiducial level [$Z$] $= -2.5$ to $-1.3$, and is useful 
especially for \SiII/\SiIV~:~\CII/\CIV. As touched on in \S8.2, this can 
illustrate a relatively early stage in the presumed progressive dilution 
of the outflowing galactic material as it merges with the rest of the CGM 
in its transit. Back to the nominal [$Z$] $= -2.5$, for the associated 
supernova ejected case we obtain $f_{esc-run}$ = 0.0013, again about 20 
percent of the dynamically ejected value, therefore still a minor 
contributor, but here giving results at $z_{HM}$ = 4.4 in the panels 
which well fit within the data. Clearly this is because dynamical 
ejection here already has inherent long transit times due to the 
relatively long physical pathlengths required for the fits to the data at 
our high redshifts. Thus for sufficiently long transit times relative to 
the supernova delay the two processes tend to converge in their outcomes 
(but then with little residual radiative output). Again, if both 
mechanisms exist in equal numbers, $f_{esc-run}$ = 0.0041. In these bottom 
panels we also include from the middle panels the S400 $z_{HM}$ = 3.4 
result solely for supernova ejection at absorber distance 17 kpc and show 
this at the lower relative abundance value [Si/C] = 0.0, not 0.4 as in 
the middle panels. It is interesting that for \SiII/\CII~:~\CII/\CIV\ 
this encompasses the data points outlying the lowest dynamical ejection 
curve and enables a better overall fit to the data. Fits closely similar 
to these for S600 are obtained with a model of boundary radius 500 pc 
(S500), giving $f_{esc-run}$ = 0.0099 and corresponding normalizing 
absorber distance 8.5 kpc if solely dynamical ejection and 
$f_{esc-run}$ = 0.0059 if both mechanisms contribute equally in 
numbers. 

Here in summary, for the \emph{strong} absorbers the above trials 
demonstrate the general overwhelming filtering effect on the ultimate 
stellar spectral contributions by: (a) the runaway process being 
differentially constrained by the short lifetimes of a wide swathe of 
massive stars; coupled with (b) the escaping, cooler, stars 
progressively having the necessary spectral features but limited by 
diminishing ionizing radiation (Figure 21). At $z_{HM}$ = 4.4 these 
together confine the band of effective stellar contributions to masses 
$\sim$ 7--25 $M_{\Sun}$. However, at our \emph{lower redshifts} 
runaway stars are included to the highest stellar masses 
(100 $M_{\Sun}$) although with radiative contributions still moderated 
in correspondence with loss in active lifetime depending on the 
character of the escape paths taken. We stress that these are 
empirical conclusions founded on our observational evidence of the 
specific, evolving, radiative environments experienced by the strong 
absorbers in our sample and having apparent properties similar to the 
CGM absorbers studied by \citet{ste10}. 

Now concerning the \emph{general metagalactic} ionizing radiation, 
which we address in the next sub-section: at the lower redshifts 
the runaway process is well able to deliver the output necessary to 
maintain the ionization of the IGM at the observed levels; however at
our higher redshifts we recognize that based on the strong absorbers 
the residual populations are unable to do this. Consequently, in the 
following we give attention to the weaker, apparently more highly 
ionized, absorbers we detect in our sample alongside the strong 
absorbers.

\subsubsection{Contribution to the Metagalactic Ionizing Radiation 
Background}

Above we obtain values for $f_{esc-run}$ on the supposition that both 
the dominant dynamically ejected and minor supernova ejected runaway 
mechanisms are included in equal numbers within the $30\%$ total from 
\citet{tet11}, although the actual partition between these currently is
uncertain. On this basis, and focusing on our spectral observations of 
\emph{strong} absorbers containing positively detected \CIV, \CII, 
\SiIV\ and \SiII, we obtain the values $f_{esc-run}$ = 0.011 for 
the disk model D200 at $z_{HM}$ = 1.9 and 2.6, close to the respective 
values 0.0067 and 0.014 obtained from observed \HI\ photoionization 
rates in the IGM implied by equation (7). If instead we take the 
spheroidal model S200, $f_{esc-run}$ = 0.052 and if this case better 
represents components of the observed lumpy structures within galaxies 
at these redshifts there is considerable radiative margin to accomodate 
possible reductions in total outcome due to the complex geometry. At 
$z_{HM}$ = 3.4 the spheroidal models S400 and S300 give respective 
values $f_{esc-run}$ = 0.012 and 0.021, with the latter being quite 
close to the value 0.028 from equation (7); at $z_{HM}$ = 3.0 there 
is correspondence. However, at $z_{HM}$ = 4.4 the spheroidal models 
S600 and S500 yield respectively $f_{esc-run}$ = 0.0041 and 0.0059, 
only about $10\%$ of the value 0.056 from equation (7). 

We conclude that the fractions of ionizing radiation given in equation 
(7) can be emitted in the approximate range $1.9 < z < 3.0$ from 
galaxies by means of runaway stars produced within galactic structures 
having respective escape boundary distances of order 200 to 300 pc. At 
redshifts beyond this there is rapid structual change bringing 
increasing boundary distances with emitted ionizing radiation falling 
far short of the level required to maintain the observed concurrent 
IGM \HI\ photoionization rates.

Nevertheless, it remains that the specific spectral characteristics we 
obtain in \S9.2.1 are strongly required by our empirically derived 
observations and that in general \emph{the spectrally modified outcomes 
of the stellar runaway modeling in concept are intrinsically robust and 
not reliant on special fine-tuning}. In an attempt to raise the emitted 
ionizing radiation at $z_{HM}$ = 4.4 for a given individual galaxy 
through a hybrid situation by adding radiation possibly emanating from 
stars escaping from star-forming regions separated from the boundary by 
only $\sim$ 200 pc, \emph{mixing} with that from stars traveling from 
the deeper seated sites already quoted, we find this model quickly is 
dominated by the hotter stars and so brings the \Cloudy\ predictions 
strongly out of step with the observations, as we have already 
demonstrated in Figure 20(b). We must therefore conclude that there 
coexists an extended population of relatively low luminosity compact 
galaxies with typical effective boundary distance $\lesssim$ 200 pc, 
thus individually emitting ionizing radiation with efficiency 
sufficiently high to accomodate the observed \HI\ ionization rates. 
Reionization models based on observed UV luminosity functions at high 
redshifts point to a similar need in order to complete reionization by 
$z$ = 6 \citep{ouc09,kaf12,boa12,rob13}. Can we identify such a 
population within our existing data at least within $3.0 < z < 4.4$?

In \S8.2 we note that at our higher redshifts there is a relatively 
large number of components having strong \CIV\ and detected \SiIV\ but 
only upper limits in \CII. Focusing on the interval $3.4 < z < 4.4$, 
in the upper panels of Figure 24 we show separately the set of 
\SiIV/\CIV~:~\CII/\CIV\ observations from Figure 23(b) having one or 
both of \SiIV\ and \CII\  as upper limits (notice there is no case 
with \emph{detected} \CII\ and upper limit \SiIV) and the set of 
strong absorbers with detected \CIV, \CII, \SiIV\ and \SiII. We add 
corresponding \Cloudy\ model results for the fiducial absorbers 
exposed to metagalactic ionizing radiation at the bounding redshifts 
$z_{HM}$ = 3.4, 4.4 and local radiation received at the indicated 
distances from dynamically ejected runaway stars escaping a spheroidal 
galactic structure having its dense gaseous boundary at radius 200 pc 
from the centre of star formation (S200), already used at 
$z_{HM}$ = 1.9 and 2.6 (with appropriate metallicity) in Figure 23(b), 
and in each case placed at the appropriate redshift-related 
normalizing distance linking with the nominal distances in Figure 20(b) 
for comparison, as earlier explained. The object is to contrast the 
distributions of the absorbers relative to the S200 \Cloudy\ results 
here with the other models shown in Figure 23(b). It is evident that 
the upper limit cases mostly identify quite well with the S200 model 
while the full detections in the other panel are better contained by 
the models represented in Figure 23(b). In the lower panels we add 
the Figure 23(b) related counterpart cases for \SiII/\SiIV~:~\CII/\CIV\ 
and \SiII/\CII~:~\CII/\CIV, now integrally including all available 
upper limits (as noted earlier, in addition to the baseline presence of 
\CIV, the former requires positive detection of \SiIV\ because it is 
the base for \SiII/\SiIV\ and the latter, of \CII\ as the base for 
\SiII/\CII). For the former, it can be seen that the values containing 
\CII\ upper limits fit more closely with the S200 model than do the 
values for the strong absorbers which certainly require models such as 
in Figure 23(b). For the latter, the more constrained set of values 
here relate well only with the models in Figure 23(b). 

With computations as before, for a spheroidal galaxy model of boundary 
radius 200 pc but now at metallicity 0.2 solar appropriate at our higher 
redshifts, we obtain $f_{esc-run}$ = 0.056 if dynamically
ejected and supernova ejected runaway mechanisms both are there in equal 
numbers, which is the same value as given by equation (7) (if entirely 
dynamically ejected we obtain 0.104). A spheroidal model of boundary 
radius 100 pc gives $f_{esc-run}$ = 0.17 if both mechanisms are 
included in equal numbers, well above the metagalactic requirement here, 
and indicative of the potential for extending to higher redshifts. For 
example, using equation (7) this value is reached at 
$z \sim 7$.

Guided by the panels in Figure 24 it is possible to see a similar and
evolving picture in the panels for the intervals $1.9 < z < 2.6$ and 
$2.6 < z < 3.4$ in Figure 23(b). In the top panels there is convergence 
on a transit geometry based on 200 pc and here the strong absorbers are 
associated with galaxies themselves emitting ionizing radiation at  
adequate levels to feed the metagalactic background. Within the middle 
panels this becomes marginal at $z \sim 3$ but now there is a developed 
buildup of the weaker absorbers we highlight in Figure 24 and which are 
well able to make up the deficit.  

Within the stellar runaway scenario, therefore, we can conclude that 
our strong absorbers, relating well to apparently similar absorbers in 
the CGM of galaxies directly observed by \citet{ste10}, represent 
regions around a relatively luminous subset of individual galaxies whose 
emitted ionizing radiation is strongly modified by a process which can be 
explained by the natural transit of runaway stars and galaxy parameters 
that substantially evolve over our observed range in redshift. We find 
such galaxies are able to maintain the ionization of the general IGM over 
redshifts $1.9 < z < 3.0$ but not beyond. We have in addition clear 
spectral indications for a large population of weaker and consistently 
more highly ionized absorbing regions also developing over our full 
observed range in redshift and apparently representing the CGM of a 
parallel set of galaxies with relatively compact absorbing regions 
emitting less modified spectra, so having greater radiative efficiency, 
which collectively dominate the production of metagalactic ionizing 
radiation at our higher redshifts extending from $z \sim 3$ and on. In 
particular, these meet the observed IGM \HI\ photoionization requirements 
not achieved by individual galaxies such as those located with our strong 
absorbers at $z$ beyond 3. 

The stellar runaway model brings possible spectral implications for the 
galaxy contribution to the \citet{ham12} QG$_{HM}$ isotropic metagalactic 
ionizing radiation background used in all our related computations. 
However, as is now indicated above, the spectral character of 
$f_{esc-run}$ from the D200/S200 models for $1.9 < z < 2.6$, coupled with 
inclusion of a dominant contribution from galaxies of similar 
$f_{esc-run}$ for $3.4 < z < 4.4$ and likewise extending into 
$2.6 < z < 3.4$, the equivalent runaway-generated radiation background 
over 1--4 Ryd will be spectrally close to that of QG$_{HM}$. As is clear 
in Figure 23(b), this is reinforced by the indications at the 80 kpc 
perimeter (where the local galaxy contribution is weak in all cases) that 
the derived \Cloudy\ results also for these runaway models in 
general align well with the results for the QG$_{HM}$ models alone.

\subsection{Discussion}
                     
The evident evolution in escape parameters that we find necessary to 
explain the ionization conditions in the strong absorbers we observe 
is a significant empirical indicator of morphological evolution in 
the population of luminous galaxies with cosmic time over our range 
$4.4 < z < 1.9$, requiring the distance between dense star-forming 
regions and the effective outer boundary of the gaseous volume that
prevents the escape of ionizing radiation to become progressively 
shorter. Observations of relatively luminous galaxies taken as 
simple structures show the \emph{luminous size} to be evolving in 
cosmic time with average half-light radius growing from about 
$\sim$1000 pc at $z$ = 5 to $\sim$2500 pc at $z$ = 2, with values at 
lower luminosity being several 100 pc smaller \citep{oes10,ono13}. For 
our presumed relatively luminous galaxies we find respective physical 
stellar escape distances to the outer absorptive boundary over our 
redshift range in cosmic time to be 600--200 pc. Can these distances 
be aligned with the apparent luminous dimensions? 

Galaxies within the lower redshifts of our range are straightforward to 
accomodate. A disk structure of observed radius and semi-height 200 pc 
or less can give a reasonable match to the emissive escape requirements 
through the runaway process. In this same redshift range, as earlier 
mentioned, galaxies also frequently exhibit complex structures with 
multiple sites of dense star formation \citep{pap05,law07,law12}, which 
can be regarded nominally as small spheroids with high emissive escape 
levels and is well accomodated by the runaway process. 

Now for the higher redshifts, taking spheroidal structures (or at the 
least, very "puffy" compact disks) as an approach to obtaining the 
longer escape paths which we find necessary for the strong absorbers,
additional aspects come to the fore. On the one hand, in general the 
presence of runaway stars manifestly will increase the \emph{apparent} 
luminous radial profile of galaxies. On the other, at the higher 
redshifts where the metagalactic ionizing radiation is stronger, there 
will be significant fully-ionized inverse Str\"{o}mgren regions (e.g., 
Dove \& Shull 1994) which effectively reduce inwards the distance from 
the edge of the \emph{absorptive} \HI\ boundary to the site of 
star-formation. Together these factors in their effect could bring to 
physical convergence the measured apparent luminous dimension of 
galaxies and the stellar radial escape dimensions fitting our 
conclusions for relatively luminous galaxies which require up to 
500--600 pc, shown in Figure 23. Such factors apply equally for the 
inferred coexisting population of galaxies having relatively lower 
luminosity, therefore intrinsically more compact, requiring typical 
escape distance 200 pc or less, shown in Figure 24.

In support of the runaway scenario it is worth mentioning the work 
of \citet{gne08} who model the \emph{radiative} escape of ionizing 
radiation from galaxies over $3 < z < 9$ using three-dimensional 
radiative transfer simulations with star formation and feedback 
(including via supernova explosions). Overall they find rather small 
values and for disk galaxies at $z \sim 3$ show that at lower galactic 
masses young stars are embedded deep in the neutral disk because the 
density and pressure required for star formation is reached only in a 
very thin (100--200 pc) region near the midplane, while at higher 
masses the disk is denser and star formation also occurs in regions 
closer to the outer edge. Interestingly for our own conclusions, 
they find the absolute escape fraction is negligible in the former, 
while only mildly positive in the latter due mostly to a small 
fraction of stars near the edge of the \HI\ boundary \emph{becoming 
exposed} through their ``normal'' movement (i.e., with low velocities, 
not as runaways) and equally so exposed by gross relative oscillation 
of the stars and the gaseous galactic disk. However, the necessary 
exposure of stars beyond the \HI\ boundary can be achieved by strongly 
\emph{aiding the movement} of stars as comes naturally in the runaway 
process we apply here.

Conversely, as an example already recounted in \S8.6, \citet{yaj11} 
in their simulations find that at $z$ = 3--6 the escape fraction 
increases as the halo mass decreases due to supernova-induced conical 
regions of highly ionized gas which allow ionizing radiation to escape 
more easily than in high mass cases. For our strong absorber population 
we give empirical evidence in \S8.6 and \S9.2 that the progressive 
spectral outcomes from such a porosity approach over our range 
$1.9 < z < 4.4$ do not represent well the escape of ionizing radiation 
from luminous galaxies having, as shown by \citet{ste10}, absorbers 
within their CGM similar to the strong absorbers in our sample. 

However we cannot thus exclude a porosity mechanism for our weaker 
absorbers, highlighted in Figure 24. While we show the runaway process 
can account both for the ionization state of these and the spectrally 
implied radiative contribution to keep the Universe ionized up to high 
redshifts, this may also apply to the porosity case for low mass, high 
redshift galaxies. Both mechanisms acting together in such circumstances 
then is to be expected.

\section{SUMMARY}

From high resolution spectral observations of nine QSOs (Table 1, Figure 1) we 
compile a large sample of metal-line systems identified as \CIV\ absorbers 
outside the Lyman forest in the redshift range $1.6\lesssim z \lesssim 4.4$ and 
include \SiIV, \CII, \SiII\ and \NV\ in these where available. By Voigt 
profile-fitting procedures using VPFIT \citep{car02} we can closely represent 
these multi-phase systems as complexes of co-existing 
``single-phase-ionization'' component regions (Figures 2, 4, 5, 6). We obtain 
column densities or upper limits for individual component ions of each species, 
with Doppler parameters for C and in selected cases for Si (Tables 2--10). 
Both for components and systems we study statistical distributions, clustering 
properties and ionization properties, and trace their evolution in redshift. 
Using combinations of ionic ratios as probes of the radiation environment we 
study the escape of ionizing radiation from galaxies, bearing on the ongoing 
cosmic ionization of the IGM, and introduce and develop a new process which we 
find necessary to enable the models successfully to meet the observations.

In all, we arrive at the following principal conclusions:

1. The \CIV\ \emph{component} column density and Doppler parameter number 
distributions (Figures 7(a) 10, 11), \emph{system} column density and velocity 
spread number distributions (Figures 7(a), 11), and differential column density 
distributions of components and systems (Figure 13), show no significant 
evolution over the observed redshift range.

2. The \CIV\ \emph{system} total number per unit redshift and total column 
density per unit redshift also show no significant evolution over the observed 
redshift range (Figure 9(a)). We find a mean \CIV\ comoving mass density 
$\langle \Omega_{\scriptsize\textrm{\CIV}} \rangle = (3.44\pm1.24) \times10^{-8}$ 
(1$\sigma$ uncertainty limits) (Figure 9(b)). \SiIV\ presents a similar 
evolutionary picture (Figures 7(b), 9(a)), with a mean comoving mass density 
$\langle \Omega_{\scriptsize\textrm{\SiIV}} \rangle = (1.23\pm0.66) \times10^{-8}$
(Figure 9(b)). Both are in broad agreement with Songaila (2001; 2005). \CII, 
\SiII\ and \NV\ change substantially with redshift, heralding changes in 
ionization state (Figures 7(b), 9).

3. We confirm for consistency that absorbers physically close to the sightline 
QSOs ($\lesssim$ 3000 km s$^{-1}$ from the QSO redshifts) have characteristics
of gas illuminated by relatively hard radiation (Figure 8).

4. The \CIV\ \emph{components} exhibit strong clustering out to velocity 
separations $\lesssim 300$ km s$^{-1}$ for our uniform statistical sample and 
to $\lesssim 400$ km s$^{-1}$ with the addition of some systems with densely 
complex components but for both there is no clustering signal detected for 
\emph{systems} on any scale from 150 km s$^{-1}$ out to 50000 km s$^{-1}$ 
(Figure 14). Neither of these (one-dimensional) distributions shows 
similarities with galaxy clustering (Figure 15). We argue that the results 
from our data are entirely due to the peculiar velocities of gas present 
in the outer extensions of galaxies (probed in the confined sightlines to the 
background QSOs, and which in general do not encounter more than one galaxy in 
a given cluster or group. 

5. We find no significant change in the component or system median column 
density ratio \SiIV/\CIV\ with redshift and in particular no large change near 
$z = 3$ (Figure 16(a)), contrary to previous observations coupled with claims 
that this can indicate the onset of complete reionization of \HeII. The 
component ratio \SiII/\CII\ is similarly unchanging while \CII/\CIV, 
\SiII/\SiIV\ and \NV/\CIV\ vary (continuously) with redshift (Figure 16(b)). 
However, individually these are only partial indicators of ionization state.

6. Using the four combinations of ionic ratios \SiIV/\CIV~:~\CII/\CIV, 
\NV/\CIV~:~\CII/\CIV, \SiII/\SiIV~:~\CII/\CIV\ and \SiII/\CII~:~\CII/\CIV\ we 
compare our observations of the \emph{strong} absorbers with model ionization 
predictions using the \Cloudy\ code (version C10) for absorbers exposed 
to the empirically founded evolving metagalactic ionizing radiation background 
\citep{ham12} due to the cosmologically distributed contributions of QSOs and 
galaxies (Figure 17(a)). While good fits are obtained at our lowest redshifts 
($z \lesssim 2.6$) where QSOs dominate the metagalactic ionizing radiation, it 
is apparent that simultaneously consistent fits to our ionic ratio 
combinations over the full redshift range, and in particular at our highest 
redshifts, cannot be achieved with this model alone (Figures 17(b), 23(b)).

7. With simple modification to the cosmic radiation model \citep{ham12,haa14}, 
we investigate the possible effect of delayed reionization of intergalactic 
\HeII, not formally included in its geometrical complexity in the original 
model, and find that this too cannot explain our observations (Figures 
17(a), (b)).

8. We model the effect of collisional ionization of the absorbers in the 
presence of the metagalctic ionization radiation background (Figure 18) and 
conclude that the great majority of the absorbers in our sample are in or 
close to a state of photoionization equilibrium, confirming the procedures 
used in our radiation modelling. Also we investigate the effects of changes 
of metallicity and \HI\ column density in our model absorbers (Figure 19) 
and find that the observed systematic evolutionary effects exhibited in the 
ionic ratios are not caused simply by changes in absorber properties. 

9. Our data support the presence in the absorbers of a range in relative 
abundance [Si/C] $\sim$ 0.0--0.4, consistent with $\alpha$-element 
enhancement in galactic metal-poor stars (Figures 17(b), 23(b)).

10. Our conclusions in points 6--8 sharply narrow the cause of the evolving 
ionization state in the absorbers we observe to substantial \emph{spectral 
changes} in their \emph{local} ionizing radiation environment, not included 
in previous work. In the remaining summary we outline the path leading to a 
new radiative process which enables the empirical convergence over our full 
range in redshift of our ionization modeling with our observations of the 
absorbers in our sample. Focusing first on our \emph{strong} component 
absorbers with positive detections of \CIV, \SiIV, \CII\ and \SiII, we seek 
only to match the \emph{spectral} outcomes of our modeling with the 
spectrally-induced signatures within our observed ionic ratio combinations 
to arrive at \emph{empirically} implied structurally-related geometrical 
properties of the \emph{associated} local galaxies. Following this we 
include our \emph{weaker} component absorbers in consideration of the 
production of the necessary ionizing radiation \emph{levels} to maintain the 
evolving ionization rate of the general IGM observed throughout our range 
in redshift.

11. On the recognition that absorbers such as the strong components in 
our sample represent regions located within a galactic CGM \citep{ste10}, 
and with our clustering observations more generally also consistent with 
this (point 4), we augment the metagalactic background radiation with 
\emph{proximity} radiation escaping from star-forming regions in the 
locally sited galaxy. A general resort to explain the escape of a fraction 
of stellar ionizing radiation from inside such highly attenuating 
environments is the formation of supernova-generated transparent channels 
through which some of the radiation can pass, largely spectrally unmodified. 
For a local galaxy component we employ Starburst99 \citep{lei99,lei10} to 
model a galaxy with \citet{sal55} IMF undergoing continuous star formation, 
producing an intrinsic spectral energy distribution similar to those 
applied in \citet{ham12}, then relate this to observed luminosities in the 
non-ionizing far-ultraviolet and apply the evolving escape fraction used 
by Haardt \& Madau over 1--4 Ryd in context of the metagalactic 
background. We place our absorbers at the typical distances 20 and 80 
physical kpc from the local galaxies \citep{ste10} and add the received 
radiative contributions to the evolving metagalactic ionizing background 
spectrum before the \Cloudy\ modeling. While reasonable fits to our full 
data set are obtained at our lower redshifts, in extension to higher 
redshifts there is increasing divergence from our data for the strong 
absorbers and taken collectively we find the fit to our data now becomes 
very significantly \emph{poorer} than before (Figure 20). We conclude that 
this \emph{geometrical} model describing the escape of Lyman continuum 
radiation, by itself, is \emph{not a viable representation} for our strong 
absorbers.

12. We therefore move to a more kinematically-dependent scenario in which the 
passive escape of ionizing radiation into the IGM is transformed into the 
dynamic escape of the ionizing \emph{sources} themselves. It is established 
that a significant fraction of all stars in the Galaxy are moving with higher 
than usual velocities, ejected as ``runaway stars'' from dense young stellar 
systems not long after their birth. Depending on the placing of a given 
star-forming region in the galaxy, the most massive stars, having the 
shortest main sequence lifetimes, may not escape at all from the boundary of 
galactic absorption, while longer-living lower mass stars can spend a 
fraction of their main sequence lifetimes in relatively clear space. We 
develop this considerably further by exploiting in detail the resulting 
\emph{spectral} changes in the final efflux of radiation that are induced. 

13. We model the runaway-modified galactic spectral energy distribution in the
Lyman continuum by use of Starburst99 with the same overall parameters as 
applied in point 12, but since the radiative product of the runaway process is 
dependent on stellar mass we now segment the full mass range into narrow slices 
within the underlying Salpeter IMF and compute a set of individually-derived 
population synthesis spectra which collectively cover the same mass range as 
before. Then we associate each component spectrum with the fraction of main 
sequence lifetime remaining after escape for a set of escape intervals 
stemming from a defined stellar velocity distribution and trial galactic escape
boundary distances. In this we consider star-forming sites in both spheroidal 
and disk-like galactic structures and include geometrical factors to assess the 
unshadowed radiative aspect at the absorbers for stars when escaping, and 
obtain the aggregated resultant spectral outcomes including both dynamically 
ejected and delayed supernova ejected runaways. As in point 12 we combine these 
with the metagalactic background radiation at each redshift before \Cloudy\ 
modeling. 

14. By iterative trials making comparison with our observations of the strong 
absorbers in our sample we find different geometries are indicated for each 
redshift interval in a pattern evolving with redshift, with a balance to 
contributions from higher stellar masses increasing with cosmic time. The 
spectra at the higher redshifts now feature a deep \HeI\ ionization 
discontinuity and diminishing strength to higher energies, characteristic of 
stars having mass around 20 $M_{\Sun}$ and lower (Figures 21, 23(a)). The 
\HeI\ discontinuity is the dominant feature providing the necessary changes to 
bring together the predicted and observed ionic ratio combinations for our 
strong absorbers at our higher redshifts. At our lower redshifts, runaway stars 
present to the highest stellar masses (100 $M_{\Sun}$) are needed to match 
the data for the strong absorbers although with radiative contribution still 
moderated in correspondence with loss in active lifetime depending on the 
character of the transit escape paths. Taking all redshift intervals together 
there is now a good match to these data over our full range in 
redshift (Figure 23(b)). The evolution in escape parameters is a significant 
indicator of morphological evolution in the population of the associated 
galaxies, requiring the positions of dense star-forming regions and the 
effective boundaries of the gaseous volume to become progressively closer 
with cosmic time.

15. Within the stellar runaway scenario we can conclude that our \emph{strong} 
absorbers, relating well to apparently similar absorbers in the CGM of galaxies 
directly observed by \citet{ste10}, represent regions around a relatively 
luminous subset of individual galaxies whose emitted ionizing radiation is 
strongly modified by a process which can be explained by the natural transit of 
runaway stars with parameters that substantially evolve over our observed range 
in redshift. We stress that these are empirical conclusions founded on our 
observational evidence of the specific, evolving, radiative environments 
experienced by the strong absorbers in our sample.

16. Now coming to direct consideration of the \emph{general metagalactic} 
background ionizing radiation, we find that galaxies corresponding to those 
associated with the strong absorbers in the conclusions above have radiative 
escape fractions at energies higher than 1 Ryd able to maintain the ionization 
of the general IGM over redshifts $1.9 < z < 3.0$ but not beyond. However, 
alongside the strong absorbers, we have in our data additional spectral 
indications for a large population of weaker and consistently more highly 
ionized absorbing regions (detected with upper limits in \CII) 
apparently representing the CGM of a parallel set of more compact galaxies 
having much greater radiative efficiency (Figure 24), rapidly developing in 
proportion to the strong absorbers over our observed redshift range 
(Figure 23(b)). In particular, through the runaway process these individually 
can reach radiative escape fractions at 1 Ryd that exceed the recognized level
needed to maintain the IGM ionized at $z$ = 4.4. This population also can 
contribute significantly to the production of metagalactic ionizing radiation 
over our lower redshifts then dominate at $z \sim 3$ and beyond, collectively 
meeting the observed IGM \HI\ photoionization requirements over our full range 
in redshift.

\acknowledgements

Foremost I want to acknowledge Wallace Sargent who passed away in October 2012.
He has been my best friend from the time we first worked together at Palomar 
exactly 41 years ago. A true gentleman, he excelled not only in science but 
also widely in many cultures. He inspired and helped me in countless ways and
I am forever grateful to him. And we had great fun together. He is the founding 
worker of all in this paper.
A.B.

We are grateful to the anonymous referees, whose advice improved this paper
significantly. We are greatly indebted to Michael Rauch for generous provision 
of some of his QSO spectra which added much to our results, Tom Barlow for the 
initial data reduction, Robert Carswell for being so unstinting with his time 
for schooling, advice and assistance for use of VPFIT, Francesco Haardt with 
Piero Madau for their great generosity in providing many new trial 
computations for the metagalactic background radiation and for very many 
greatly helpful discussions, Roderick Johnstone for enormous help, assistance 
and advice in the use of \Cloudy, Max Pettini for spending so much time on 
many clarifying discussions and for invaluable help and encouragement, Robin 
Catchpole for continuous patient discussion, advice and encouragement and 
shared humour over many years, Michael Shull for a great many extremely 
illuminating, inspiring and course-setting discussions and his trenchant 
critical reading of the manuscript, Sverre Aarseth for fundamental advice and 
extreme generosity and good will in providing the means for gravitational 
modeling and personally conducting the complex procedures, George Becker for 
long-standing serial advice and guidance, Gary Ferland for direct advice about 
\Cloudy, Claus Leitherer for help and advice on use of Starburst99, Robert 
Lupton for provision of a new facility within Supermongo, Roberto Abraham and 
Robert Simcoe for setting up analysis software and other assistance, 
Christopher Tout for much advice and assistance on stellar matters and Floor 
van Leeuven for analytical help with Hipparcos runaway star material. We 
benefited from informative discussions with Cathie Clarke, Crystal Martin, 
David Valls-Gabaud, Samantha Rix, Paul Crowther, Martin Haehnelt, Andy Fabian, 
Robert Kennicutt, Maciej Hermanowicz, Stephen Smartt, Michael Irwin, George 
Efstathiou, Gerry Gilmore and Michael Murphy, among many others. We also 
thank the Keck Observatory staff for their assistance with the observations. 
Finally, we extend special thanks to those of Hawaiian ancestry on whose 
sacred mountain we are privileged to be guests. Without their generous 
hospitality, the observations presented herein would not have been possible. 
A.B. gratefully acknowledges support from The Leverhulme Trust, the UK 
Particle Physics and Astronomy Research Council and the ongoing wide-ranging 
support of the Institute of Astronomy. W.L.W.S was supported by NSF 
grants AST-9900733 and AST-0206067.



\clearpage
\begin{deluxetable}{lccccccc}
\tabletypesize{\scriptsize}
\tablecaption{Log of Observations}
\tablewidth{0pt}
\tablecolumns{9}
\tablehead{
\colhead{Object} & \colhead{$z_{em}$\tablenotemark{a}} & \colhead{V(R)\tablenotemark{b}} & {Dates\tablenotemark{c}} & 
\colhead{$\lambda$-Range\tablenotemark{d}} & \colhead{Exposure\tablenotemark{e}} &
\colhead{FWHM\tablenotemark{f}} & \colhead{S/N\tablenotemark{g}} \\
{} & {} & \colhead{mag} & {} & \colhead{\AA} & \colhead{s}  & \colhead{km~s$^{-1}$} & {}  
}

\startdata
Q1626$+$6433 & 2.32 & 15.8 & 1995 05 19 --- 1996 07 22 & 3580 --- 6179\tablenotemark{h} & 24063 & 6.6 & \phm{0}88 --- 128 --- 137 \\
Q1442$+$2931 & 2.66 & 16.0 & 1994 06 12 --- 1995 06 21 & 3624 --- 6150\tablenotemark{h} & 22500 & 6.6 & 113 --- 107 --- 121 \\
Q1107$+$4847 & 2.97 & 16.6 & 1994 12 26 --- 1995 02 22 & 3643 --- 6663 & 26000 & 6.6 & \phm{0}90 --- \phm{0}94 --- \phm{0}81 \\
Q0636$+$6801 & 3.18 & 16.4 & 1993 11 13 --- 1995 02 22 & 3643 --- 6663 & 36200 & 6.6 & 112 --- 107 --- 100 \\
Q1425$+$6039 & 3.18 & 16.2 & 1995 05 18 --- 1995 05 19 & 3736 --- 6540 & 37200 & 6.6 & 168 --- 140 --- 113 \\
Q1422$+$2309C & 3.63 & 17.3 & 1998 02 01 --- 1999 04 14 & 3569 --- 7347 & 51200 & 4.4 & \phm{0}92 --- \phm{0}90 --- \phm{0}61 \\ 
Q1645$+$5520 & 4.10 & 17.0 & 1997 05 09 --- 1997 05 30 & 4557 --- 8123 & 72200 & 6.6 & 102 --- 119 --- \phm{0}81 \\
Q1055$+$4611 & 4.15 & 16.9 & 1995 04 07 --- 1996 05 24 & 4587 --- 8009 & 27037 & 6.6 & \phm{0}71 --- \phm{0}47 --- \phm{0}31 \\
Q2237$-$0607 & 4.56 & 18.3 & 1994 10 10 --- 1995 08 21 & 4933 --- 8800 & 54000 & 6.6 & \phm{0}62 --- \phm{0}42 --- \phm{0}28 \\
\enddata

\tablenotetext{a}{Redshifts largely obtained from the literature, supplemented by the positions of the onset of the Lyman forest and the peak 
of the Lyman $\alpha$ emission line in the HIRES spectrum.}

\tablenotetext{b}{Q2237$-$0607 has R mag; the rest have V mag.}

\tablenotetext{c}{Total period over which the observations were made.}

\tablenotetext{d}{Total available wavelength range, extending from within the Lyman forest to beyond the C~{\tiny IV} emission line.}

\tablenotetext{e}{Total time spent on the collection of exposures.}

\tablenotetext{f}{These are approximate values. Along the echelle orders the velocity resolution differs from the mean by $\sim \pm3.8\%$.}

\tablenotetext{g}{S/N over three pixels (roughly equal to a resolution element for all but Q1422$+$2309C) outside the Lyman forest 
sampled near the three wavelengths $\lambda\lambda\lambda$1270,1380,1500 in the rest frame of the QSO: by avoiding the main emission 
lines these tend to conservative estimates.}

\tablenotetext{h}{In subsequent observations used here coverage of the Lyman forest is extended to $\sim$ 3200 \AA.}

\end{deluxetable}

\clearpage
\begin{deluxetable}{crrcccccc}
\tabletypesize{\scriptsize}
\tablecaption{Heavy Element Absorption Lines: Q1626+6433  $z_{em} = 2.32^*$}
\tablewidth{0pt}
\tablecolumns{9}
\tablehead{
\colhead{$z$} & \colhead{$b$(C)}  & \colhead{$b$(Si)} &
\colhead{$N$(\CIV)} & \colhead{$N$(\SiIV)} &
\colhead{$N$(\CII)} & \colhead{$N$(\SiII)} &
\colhead{$N$(\NV)} &  \colhead{n} \\
{} & \colhead{km~s$^{-1}$} & \colhead{km~s$^{-1}$} & \colhead{10$^{12}$cm$^{-2}$} &
\colhead{10$^{11}$cm$^{-2}$} & \colhead{10$^{12}$cm$^{-2}$} & \colhead{10$^{11}$cm$^{-2}$} & \colhead{10$^{12}$cm$^{-2}$} & {} \\
\colhead{1} & \colhead{2} & \colhead{3} & \colhead{4} & \colhead{5} & \colhead{6} & \colhead{7} & \colhead{8} & \colhead{9} 
}

\startdata
\sidehead{1.779}
 1.778611 &    11.5 &   \nodata &    1.91 $\pm$    0.18 &    \nodata &    \nodata &    \nodata &    \nodata &  1  \\ 
 1.778811 &     4.4 &   \nodata &    0.55 $\pm$    0.13 &    \nodata &    \nodata &    \nodata &    \nodata &  2  \\ 
 1.778988 &     9.9 &   \nodata &    3.33 $\pm$    0.20 &    \nodata &    \nodata &    \nodata &    \nodata &  3  \\ 
 1.779124 &     7.0 &   \nodata &    7.40 $\pm$    0.20 &    \nodata &    \nodata &    \nodata &    \nodata &  4  \\ 
 1.779316 &    10.8 &   \nodata &   17.8 $\pm$    0.3\phm{0} &    \nodata &    \nodata &    \nodata &    \nodata &  5  \\ 
 1.779536 &    10.5 &   \nodata &    4.99 $\pm$    0.20 &    \nodata &    \nodata &    \nodata &    \nodata &  6  \\ 
 1.779771 &     4.7 &   \nodata &    0.42 $\pm$    0.13 &    \nodata &    \nodata &    \nodata &    \nodata &  7  \\ 
 1.780041 &     6.5 &   \nodata &    0.68 $\pm$    0.14 &    \nodata &    \nodata &    \nodata &    \nodata &  8  \\ 
 1.780331 &     9.3 &   \nodata &    7.07 $\pm$    0.19 &    \nodata &    \nodata &    \nodata &    \nodata &  9  \\ 
\sidehead{1.817}
 1.816490 &    16.2 &   \nodata &   13.4 $\pm$    0.5\phm{0} &    \nodata &    \nodata &    \nodata &    \nodata &  1  \\ 
 1.816527 &     7.0 &   \nodata &   23.7 $\pm$    0.5\phm{0} &    \nodata &    \nodata &    \nodata &    \nodata &  2  \\ 
 1.816715 &     6.2 &   \nodata &   10.7 $\pm$    0.2\phm{0} &    \nodata &    \nodata &    \nodata &    \nodata &  3  \\ 
 1.816823 &     7.5 &   \nodata &   23.9 $\pm$    0.3\phm{0} &    \nodata &    \nodata &    \nodata &    \nodata &  4  \\ 
 1.816977 &    11.2 &   \nodata &    1.34 $\pm$    0.20 &    \nodata &    \nodata &    \nodata &    \nodata &  5  \\ 
\sidehead{1.843}
 1.842835 &    11.6 &   \nodata &    2.52 $\pm$    0.23 &    \nodata &    \nodata &    \nodata &    \nodata &  1  \\ 
 1.842971 &    20.4 &   \nodata &    2.44 $\pm$    0.29 &    \nodata &    \nodata &    \nodata &    \nodata &  2  \\ 
\sidehead{1.847}
 1.846881 &    12.4 &   \nodata &    0.72 $\pm$    0.18 &    \nodata &    \nodata &    \nodata &    \nodata &  1  \\ 
 1.847040 &     4.3 &   \nodata &    1.39 $\pm$    0.13 &    \nodata &    \nodata &    \nodata &    \nodata &  2  \\ 
 1.847159 &     8.9 &   \nodata &    1.97 $\pm$    0.16 &    \nodata &    \nodata &    \nodata &    \nodata &  3  \\ 
\sidehead{1.880\tablenotemark{a}}
 1.879914 &     5.6 &   \nodata &    1.39 $\pm$    0.16 &    \nodata &    \nodata &    \nodata &    \nodata &  1  \\ 
 1.879991 &     4.5 &   \nodata &    2.70 $\pm$    0.16 &    \nodata &    \nodata &    \nodata &    \nodata &  2  \\ 
 1.880131 &     5.4 &   \nodata &    1.58 $\pm$    0.14 &    \nodata &    \nodata &    \nodata &    \nodata &  3  \\ 
 1.880287 &     7.4 &   \nodata &    1.22 $\pm$    0.17 &    \nodata &    \nodata &    \nodata &    \nodata &  4  \\ 
\sidehead{1.927\tablenotemark{b}}
 1.925731 &    13.9 &    (12.8) &   13.3 $\pm$    0.5\phm{0} &   35.8 $\pm$    3.6\phm{0} &    3.10 $\pm$    1.77 &    \nodata &    \nodata &  1  \\ 
 1.925743 &     4.7 &     (4.1) &    7.28 $\pm$    0.35 &   31.8 $\pm$    2.3\phm{0} &    4.39 $\pm$    1.14 &    \nodata &    \nodata &  2  \\ 
 1.925874 &     7.4 &     (5.9) &   13.1 $\pm$    0.3\phm{0} &   51.9 $\pm$    1.7\phm{0} &    8.53 $\pm$    0.89 &    \nodata &    \nodata &  3  \\ 
 1.926006 &     9.8 &     (6.9) &   17.9 $\pm$    0.3\phm{0} &   31.5 $\pm$    1.4\phm{0} &    \nodata &    \nodata &    \nodata &  4  \\ 
 1.926185 &     8.7 &     (7.1) &   18.5 $\pm$    0.3\phm{0} &   53.2 $\pm$    1.5\phm{0} &    \nodata &    \nodata &    \nodata &  5  \\ 
 1.926349 &     9.2 &     (6.6) &   38.7 $\pm$    0.9\phm{0} &   52.0 $\pm$    1.8\phm{0} &    \nodata &    \nodata &    \nodata &  6  \\ 
 1.926478 &    10.6 &     (9.1) &   16.0 $\pm$    0.6\phm{0} &   58.4 $\pm$    2.1\phm{0} &    \nodata &   15.0 $\pm$    4.5\phm{0} &    \nodata &  7  \\ 
 1.926632 &     6.2 &     (4.8) &    4.13 $\pm$    0.61 &   24.6 $\pm$    2.9\phm{0} &    \nodata &  141 $\pm$    8\phm{00} &    \nodata &  8  \\ 
 1.926703 &    21.2 &    (20.5) &   28.2 $\pm$    1.9\phm{0} &  175 $\pm$    8\phm{00} &    \nodata &   64.4 $\pm$   20.3 &    \nodata &  9  \\ 
 1.926733 &     6.4 &     (5.1) &   12.7 $\pm$    0.6\phm{0} &   41.4 $\pm$    2.7\phm{0} &    \nodata &   37.8 $\pm$    5.8\phm{0} &    \nodata & 10  \\ 
 1.926791 &     6.3 &     (4.9) &    2.85 $\pm$    0.62 &    7.03 $\pm$    2.63 &  [145 $\pm$   67]\phm{0} &  185 $\pm$    8\phm{00} &    \nodata & 11  \\ 
 1.927016 &    10.9 &     (9.5) &   28.2 $\pm$    0.5\phm{0} &  153 $\pm$    2\phm{00} &  188 $\pm$   21\phm{0} &  532 $\pm$    8\phm{00} &    \nodata & 12  \\ 
 1.927203 &     9.1 &     (6.7) &   52.6 $\pm$    1.6\phm{0} &  163 $\pm$    3\phm{00} &   91.9 $\pm$   10.3 &   88.1 $\pm$    3.5\phm{0} &    \nodata & 13  \\ 
 1.927299 &     7.5 &     (5.8) &   34.8 $\pm$    1.8\phm{0} &  167 $\pm$    4\phm{00} &   43.6 $\pm$    5.1\phm{0} &   62.3 $\pm$    3.4\phm{0} &    \nodata & 14  \\ *
 1.927498 &    17.0 &    (16.1) &   74.7 $\pm$    1.7\phm{0} &  248 $\pm$    5\phm{00} &   20.8 $\pm$    7.1\phm{0} &   53.3 $\pm$    7.3\phm{0} &    \nodata & 15  \\ *
 1.927552 &     7.2 &     (6.0) &    2.30 $\pm$    0.78 &   76.8 $\pm$    4.3\phm{0} &  [264 $\pm$   40]\phm{0} &  287 $\pm$    7\phm{00} &    \nodata & 16  \\ *
 1.927659 &     8.6 &     (7.6) &    2.22 $\pm$    0.34 &   13.2 $\pm$    1.6\phm{0} &   21.6 $\pm$    2.0\phm{0} &   38.2 $\pm$    3.3\phm{0} &    \nodata & 17  \\ 
 1.927852 &    14.5 &    (13.4) &   23.6 $\pm$    0.4\phm{0} &   39.0 $\pm$    1.2\phm{0} &    8.16 $\pm$    1.16 &    \nodata &    \nodata & 18  \\ 
 1.928104 &    18.1 &    (17.3) &    8.92 $\pm$    0.24 &   13.0 $\pm$    1.3\phm{0} &    \nodata &    \nodata &    \nodata & 19  \\ 
\sidehead{2.048\tablenotemark{c}}
 2.047870 &    13.4 &    (12.2) &   10.7 $\pm$    0.2\phm{0} &  $<$    0.90 &  $<$    0.43 &    \nodata &    \nodata &  1  \\ *
 2.048096 &     6.9 &     (5.8) &    2.36 $\pm$    0.16 &  $<$    0.68 &  $<$    0.33 &    \nodata &    \nodata &  2  \\ *
 2.048277 &     3.1 &     (2.1) &    0.60 $\pm$    0.12 &  $<$    0.57 &  $<$    0.26 &    \nodata &    \nodata &  3  \\ *
\sidehead{2.056\tablenotemark{d}}
 2.054804 &    20.6 &    (19.8) &    8.29 $\pm$    0.25 &  $<$    2.08 &    \nodata &    \nodata &    \nodata &  1  \\ 
 2.055106 &     8.2 &     7.4\phm{)} &   17.1 $\pm$    0.3\phm{0} &   19.8 $\pm$    1.1\phm{0} &    2.10 $\pm$    0.42 &    \nodata &    \nodata &  2  \\ 
 2.055356 &    10.0 &     8.8\phm{)} &   57.5 $\pm$    0.6\phm{0} &   51.1 $\pm$    1.4\phm{0} &    4.15 $\pm$    1.22 &    \nodata &    \nodata &  3  \\ 
 2.055439 &    54.6 &    (54.3) &   19.3 $\pm$    0.5\phm{0} &    9.00 $\pm$    1.16 &    \nodata &    \nodata &    \nodata &  4  \\ 
 2.055492 &     9.4 &     (7.6) &   11.9 $\pm$    0.8\phm{0} &  $<$    4.43 &    \nodata &    \nodata &    \nodata &  5  \\ 
 2.056236 &     4.1 &     (2.9) &    0.45 $\pm$    0.11 &  $<$    0.89 &    \nodata &    \nodata &    \nodata &  6  \\ 
\sidehead{2.069}
 2.069272 &    11.4 &    (10.0) &    1.86 $\pm$    0.13 &    1.71 $\pm$    0.83 &  $<$    0.42 &    \nodata &    \nodata &  1  \\ 
\sidehead{2.099\tablenotemark{e}}
 2.097631 &     8.3 &     (6.7) &    2.38 $\pm$    0.15 &  $<$    0.74 &  $<$    0.37 &    \nodata &    \nodata &  1  \\ 
 2.097875 &     5.9 &     (5.2) &   10.7 $\pm$    0.2\phm{0} &    5.01 $\pm$    1.15 &    0.64 $\pm$    0.36 &    \nodata &    \nodata &  2  \\ 
 2.097977 &     3.7 &     (2.4) &    1.14 $\pm$    0.17 &    3.02 $\pm$    0.93 &  $<$    0.35 &    \nodata &    \nodata &  3  \\ 
 2.098043 &    21.9 &    (21.2) &   15.4 $\pm$    0.3\phm{0} &    3.30 $\pm$    2.44 &  $<$    0.78 &    \nodata &    \nodata &  4  \\ 
 2.098433 &     3.7 &     (2.4) &    0.42 $\pm$    0.11 &  $<$    0.46 &  $<$    0.27 &    \nodata &    \nodata &  5  \\ 
 2.098740 &     7.6 &     (6.5) &    2.33 $\pm$    0.14 &    4.02 $\pm$    0.70 &  $<$    0.35 &    \nodata &    \nodata &  6  \\ 
 2.098945 &     4.3 &     (3.6) &    9.93 $\pm$    0.20 &    3.16 $\pm$    0.54 &  $<$    0.34 &    \nodata &    \nodata &  7  \\ 
 2.099050 &     9.2 &     (7.3) &   13.6 $\pm$    0.2\phm{0} &    1.81 $\pm$    0.67 &  $<$    0.42 &    \nodata &    \nodata &  8  \\ 
 2.099317 &     3.7 &     (2.4) &    1.03 $\pm$    0.14 &  $<$    1.07 &  $<$    0.36 &    \nodata &    \nodata &  9  \\ 
 2.099406 &    19.0 &    (18.1) &    7.65 $\pm$    0.27 &  $<$    1.13 &  $<$    0.66 &    \nodata &    \nodata & 10  \\ 
 2.099903 &    17.2 &    (16.4) &    0.54 $\pm$    0.19 &  $<$    0.92 &  $<$    0.61 &    \nodata &    \nodata & 11  \\ 
\sidehead{2.110\tablenotemark{f}}
 2.108103 &    17.3 &    (16.3) &    0.55 $\pm$    0.16 &  $<$    0.89 &    \nodata &    \nodata &    \nodata &  1  \\ 
 2.109149 &     5.6 &     4.4\phm{)} &   17.4 $\pm$    0.3\phm{0} &   81.5 $\pm$    1.6\phm{0} &   21.1 $\pm$    1.0\phm{0} &   16.7 $\pm$    2.7\phm{0} &    \nodata &  2  \\ 
 2.109168 &    26.4 &    (25.4) &    3.89 $\pm$    0.35 &  $<$    1.86 &    \nodata &    \nodata &    \nodata &  3  \\ 
 2.109378 &     4.7 &     4.4\phm{)} &    3.59 $\pm$    0.13 &   16.9 $\pm$    0.7\phm{0} &    \nodata &    \nodata &    \nodata &  4  \\ 
 2.109494 &     5.1 &     3.4\phm{)} &    1.37 $\pm$    0.12 &    9.28 $\pm$    0.60 &    7.84 $\pm$    0.46 &   13.9 $\pm$    1.9\phm{0} &    \nodata &  5  \\ 
 2.109604 &     9.1 &     (4.5) &    1.95 $\pm$    0.15 &    4.29 $\pm$    0.57 &    2.61 $\pm$    0.48 &    \nodata &    \nodata &  6  \\ 
 2.109749 &     6.8 &     (5.4) &    1.34 $\pm$    0.13 &    7.31 $\pm$    0.64 &    2.58 $\pm$    0.38 &    4.63 $\pm$    2.08 &    \nodata &  7  \\ 
 2.109938 &    18.2 &    (17.4) &    3.67 $\pm$    0.21 &    2.89 $\pm$    1.06 &  $<$    0.64 &    \nodata &    \nodata &  8  \\ 
 2.110120 &     5.8 &     (4.2) &    5.43 $\pm$    0.15 &   20.9 $\pm$    0.8\phm{0} &   22.5 $\pm$    0.8\phm{0} &   19.7 $\pm$    1.2\phm{0} &    \nodata &  9  \\ 
 2.110213 &     4.6 &     (2.9) &    4.63 $\pm$    0.16 &   15.8 $\pm$    0.7\phm{0} &   98.2 $\pm$    6.5\phm{0} &  286 $\pm$    7\phm{00} &    \nodata & 10  \\ 
 2.110289 &     3.6 &     (2.8) &    2.45 $\pm$    0.13 &    7.22 $\pm$    0.58 &    2.39 $\pm$    0.64 &    2.42 $\pm$    0.89 &    \nodata & 11  \\ 
 2.110398 &     6.1 &     (4.5) &    4.23 $\pm$    0.14 &    9.20 $\pm$    0.60 &    3.80 $\pm$    0.39 &    \nodata &    \nodata & 12  \\ 
 2.110506 &     6.1 &     (4.2) &    0.97 $\pm$    0.12 &  $<$    0.78 &    4.34 $\pm$    0.36 &    7.95 $\pm$    0.88 &    \nodata & 13  \\ 
 2.110688 &     6.7 &     (5.1) &    1.12 $\pm$    0.18 &    3.58 $\pm$    0.67 &    6.16 $\pm$    0.56 &    5.85 $\pm$    0.88 &    \nodata & 14  \\ 
 2.110734 &     3.1 &     (2.0) &    0.75 $\pm$    0.15 &    7.07 $\pm$    0.76 &  $<$    0.45 &    \nodata &    \nodata & 15  \\ *
 2.110907 &    11.1 &     (9.7) &    2.15 $\pm$    0.26 &  $<$    2.08 &  $<$    1.04 &    \nodata &    \nodata & 16  \\ *
 2.111442 &     7.4 &     (5.8) &    1.20 $\pm$    0.11 &  $<$    0.63 &  $<$    0.34 &    \nodata &    \nodata & 17  \\ 
\sidehead{2.208}
 2.208260 &     9.1 &     (7.3) &    0.62 $\pm$    0.12 &  $<$    0.68 &  $<$    0.41 &    \nodata &    \nodata &  1  \\ 
 2.208462 &     8.6 &     (6.7) &    4.84 $\pm$    0.13 &  $<$    0.67 &  $<$    0.40 &    \nodata &    \nodata &  2  \\ 
\sidehead{2.244\tablenotemark{g}}
 2.242553 &     4.0 &     3.0\phm{)} &    5.76 $\pm$    0.12 &    5.74 $\pm$    0.46 &    2.99 $\pm$    0.29 &    1.86 $\pm$    0.34 &    \nodata &  1  \\ 
 2.242713 &    13.7 &    (12.6) &   15.6 $\pm$    0.2\phm{0} &    2.57 $\pm$    0.69 &    0.60 $\pm$    0.42 &  $<$    0.59 &    \nodata &  2  \\ 
 2.243085 &     5.3 &     (4.0) &    0.60 $\pm$    0.17 &    2.24 $\pm$    0.74 &    1.20 $\pm$    0.45 &    0.74 $\pm$    0.55 &    \nodata &  3  \\ 
 2.243097 &    15.8 &    (14.8) &    8.30 $\pm$    0.26 &  $<$    1.21 &  $<$    0.73 &  $<$    0.89 &    \nodata &  4  \\ 
 2.243599 &    15.7 &    (14.7) &    0.93 $\pm$    0.14 &  $<$    1.52 &  $<$    0.42 &  $<$    1.31 &    \nodata &  5  \\ 
 2.243834 &     5.0 &     (3.9) &    0.95 $\pm$    0.09 &  $<$    0.73 &    \nodata &  $<$    0.39 &    \nodata &  6  \\ 
 2.244101 &     7.8 &     7.5\phm{)} &   13.5 $\pm$    0.2\phm{0} &    7.76 $\pm$    0.58 &    2.36 $\pm$    0.33 &    1.58 $\pm$    0.45 &    1.07 $\pm$    0.42 &  7  \\ 
 2.244483 &     8.4 &     (6.5) &   27.6 $\pm$    0.4\phm{0} &    3.46 $\pm$    0.97 &  $<$    0.61 &    \nodata &    6.57 $\pm$    0.75 &  8  \\ 
 2.244609 &    32.7 &    (32.2) &   14.8 $\pm$    0.9\phm{0} &  $<$    3.66 &  $<$    1.76 &    \nodata &   24.3 $\pm$    2.4\phm{0} &  9  \\ 
 2.244682 &     7.9 &     (5.7) &   [96.2 $\pm$    1.2]\phm{0} &   21.6 $\pm$    1.0\phm{0} &    0.69 $\pm$    0.56 &    \nodata &    8.71 $\pm$    0.80 & 10  \\ 
 2.244786 &     4.6 &     3.3\phm{)} &  \nodata &    5.76 $\pm$    0.62 &  $<$    0.38 &    \nodata &    2.16 $\pm$    0.38 & 11  \\ 
 2.244943 &    15.0 &    (14.0) &  [282 $\pm$    7]\phm{00} &   40.5 $\pm$    1.3\phm{0} &    0.81 $\pm$    0.66 &    \nodata &   24.0 $\pm$    0.8\phm{0} & 12  \\ 
 2.245051 &     3.3 &     (2.7) &   14.1 $\pm$    2.9\phm{0} &    6.52 $\pm$    0.62 &    2.54 $\pm$    0.34 &  $<$    0.51 &    \nodata & 13  \\ 
 2.245173 &     2.7 &     1.8\phm{)} &  \nodata &   10.3 $\pm$    0.7\phm{0} &  $<$    0.39 &  $<$    0.33 &    \nodata & 14  \\ 
 2.245253 &     8.3 &     5.1\phm{)} &   70.4 $\pm$    0.6\phm{0} &   11.4 $\pm$    0.8\phm{0} &  $<$    1.87 &  $<$    0.52 &    9.16 $\pm$    0.45 & 15  \\ 
 2.245413 &    19.9 &    (19.5) &    0.89 $\pm$    0.20 &  $<$    1.01 &  $<$    6.24 &    \nodata &    \nodata & 16  \\ 
\sidehead{2.285\tablenotemark{h}}
 2.284657 &    11.2 &     (9.8) &    0.62 $\pm$    0.14 &  $<$    0.51 &  $<$    0.34 &  $<$    0.43 &  0.58 $\pm$    0.32 &  1  \\ 
 2.285080 &     4.5 &     3.2\phm{)} &   17.8 $\pm$    0.7\phm{0} &   36.4 $\pm$    1.3\phm{0} &    5.62 $\pm$    0.70 &    2.65 $\pm$    0.70 &  $<$    0.62 &  2  \\ 
 2.285087 &     9.9 &     (8.7) &   20.5 $\pm$    0.5\phm{0} &   12.9 $\pm$    1.2\phm{0} &  $<$    0.87 &  $<$    0.95 &    4.80 $\pm$    0.83 &  3  \\ 
 2.285347 &     6.1 &     (5.0) &   14.7 $\pm$    0.2\phm{0} &    1.08 $\pm$    0.46 &    0.39 $\pm$    0.32 &  $<$    0.36 &    3.82 $\pm$    0.31 &  4  \\ 
 2.285440 &     5.3 &     2.9\phm{)} &   11.2 $\pm$    0.2\phm{0} &    7.54 $\pm$    0.47 &    3.22 $\pm$    0.33 &    1.82 $\pm$    0.34 &  $<$    0.27 &  5  \\ 
 2.285592 &     7.8 &     (6.4) &    1.39 $\pm$    0.12 &  $<$    0.48 &  $<$    0.31 &  $<$    0.36 &    1.64 $\pm$    0.29 &  6  \\ 
\sidehead{2.291\tablenotemark{i}}
 2.289840 &     3.9 &     (3.3) &  $<$    0.10 &    1.34 $\pm$    0.46 &   10.8 $\pm$    0.4\phm{0} &   12.5 $\pm$    0.6\phm{0} &  $<$    0.31 &  1  \\ 
 2.290023 &    21.4 &    (20.7) &    6.94 $\pm$    0.24 &   29.9 $\pm$    1.1\phm{0} &   25.8 $\pm$    0.8\phm{0} &   19.9 $\pm$    0.9\phm{0} &  $<$    0.69 &  2  \\ 
 2.290181 &     5.1 &     (5.2) &    2.11 $\pm$    0.16 &   12.1 $\pm$    0.8\phm{0} &   12.9 $\pm$    0.6\phm{0} &   14.3 $\pm$    0.7\phm{0} &  $<$    0.40 &  3  \\ 
 2.290256 &    16.8 &    (15.9) &    7.73 $\pm$    0.28 &   23.7 $\pm$    1.3\phm{0} &    6.03 $\pm$    0.82 &  $<$    1.26 &  $<$    0.81 &  4  \\ 
 2.290508 &    11.6 &    (10.3) &    3.17 $\pm$    0.28 &  $<$    1.01 &  $<$    0.66 &  $<$    0.72 &  $<$    0.83 &  5  \\ 
 2.290583 &     6.8 &     (5.9) &    1.80 $\pm$    0.21 &    7.69 $\pm$    0.78 &    2.05 $\pm$    0.49 &    \nodata &  $<$    0.52 &  6  \\ 
 2.290746 &     3.1 &     (2.0) &    0.15 $\pm$    0.14 &    1.16 $\pm$    0.54 &    3.86 $\pm$    0.40 &    3.25 $\pm$    0.48 &    \nodata &  7  \\ 
 2.290866 &    15.3 &    (14.3) &    2.69 $\pm$    0.46 &  $<$    1.53 &  $<$    1.11 &  $<$    1.17 &    \nodata &  8  \\ 
 2.290927 &     6.3 &     5.4\phm{)} &    0.46 $\pm$    0.27 &    4.99 $\pm$    0.93 &   16.7 $\pm$    0.7\phm{0} &   17.8 $\pm$    0.9\phm{0} &    \nodata &  9  \\ 
 2.291421 &     3.6 &     2.3\phm{)} &    5.80 $\pm$    0.13 &   13.0 $\pm$    0.5\phm{0} &    4.60 $\pm$    0.28 &    5.50 $\pm$    0.40 &    \nodata & 10  \\ 
 2.291573 &    10.0 &     (8.4) &   19.4 $\pm$    0.3\phm{0} &    5.32 $\pm$    0.69 &    0.68 $\pm$    0.43 &  $<$    0.61 &  $<$    0.92 & 11  \\ 
 2.291724 &     5.5 &     (4.9) &   22.3 $\pm$    0.4\phm{0} &   43.2 $\pm$    0.9\phm{0} &   12.9 $\pm$    0.5\phm{0} &    7.90 $\pm$    0.64 &  $<$    0.61 & 12  \\ 
 2.291836 &     3.5 &     (3.1) &   15.2 $\pm$    0.5\phm{0} &   31.2 $\pm$    0.9\phm{0} &   27.3 $\pm$    0.9\phm{0} &   25.4 $\pm$    1.2\phm{0} &  $<$    0.54 & 13  \\ 
 2.291866 &    32.1 &    (31.6) &   68.8 $\pm$    0.9\phm{0} &   17.4 $\pm$    2.7\phm{0} &  $<$    1.99 &    \nodata &   13.6 $\pm$    2.3\phm{0} & 14  \\ 
 2.291906 &     7.4 &     (6.4) &   16.7 $\pm$    0.5\phm{0} &   23.7 $\pm$    1.2\phm{0} &   10.1 $\pm$    0.7\phm{0} &    6.65 $\pm$    1.16 &  $<$    0.91 & 15  \\ * 
 2.292207 &     5.0 &     3.4\phm{)} &    1.67 $\pm$    0.14 &    5.75 $\pm$    0.50 &    9.38 $\pm$    0.36 &   10.7 $\pm$    1.2\phm{0} &  $<$    0.47 & 16  \\ * 
 2.292370 &     6.8 &     (5.6) &    1.46 $\pm$    0.12 &  $<$    0.46 &  $<$    0.27 &    \nodata &  $<$    0.60 & 17  \\ 
\sidehead{2.321\tablenotemark{j}}
 2.320312 &     5.8 &     (5.1) &    4.78 $\pm$    0.14 &    8.39 $\pm$    0.50 &  $<$    0.34 &  $<$    0.44 &  $<$    0.52 &  1  \\ 
 2.320398 &     6.2 &     (5.4) &    5.37 $\pm$    0.16 &    3.01 $\pm$    0.51 &  $<$    0.36 &  $<$    0.39 &  $<$    0.26 &  2  \\ 
 2.320591 &    16.8 &    (15.9) &   10.2 $\pm$    0.2\phm{0} &    1.73 $\pm$    0.67 &    \nodata &  $<$    0.52 &  $<$    0.37 &  3  \\ 
 2.321017 &    19.4 &    (18.7) &    1.18 $\pm$    0.16 &  $<$    0.70 &    \nodata &  $<$    0.53 &  $<$    0.72 &  4  \\ 
\sidehead{2.323\tablenotemark{k}}
 2.322443 &    15.0 &    (14.0) &    5.12 $\pm$    0.15 &  $<$    1.15 &    \nodata &  $<$    0.53 &    0.86 $\pm$    0.31 &  1  \\ 
 2.322665 &     7.9 &     (6.5) &    4.43 $\pm$    0.12 &  $<$    0.63 &    \nodata &  $<$    0.38 &  $<$    0.23 &  2  \\ 
 2.322896 &     6.9 &     (6.3) &    0.55 $\pm$    0.10 &  $<$    0.42 &    \nodata &  $<$    0.50 &  $<$    0.22 &  3  \\ 
 2.323238 &    19.0 &    (18.2) &    6.49 $\pm$    0.21 &  $<$    1.93 &    \nodata &  $<$    0.65 &  $<$    0.40 &  4  \\ 
 2.323344 &     5.4 &     (4.7) &    3.34 $\pm$    0.14 &    1.92 $\pm$    0.39 &  $<$    0.50 &  $<$    0.37 &  $<$    0.30 &  5  \\ 
 2.323611 &    18.2 &    (17.4) &    8.13 $\pm$    0.17 &  $<$    1.21 &  $<$    0.86 &  $<$    0.56 &    0.67 $\pm$    0.34 &  6  \\ 
 2.324029 &     9.9 &     (8.2) &    2.78 $\pm$    0.12 &  $<$    0.46 &  $<$    0.31 &  $<$    0.43 &  $<$    0.32 &  7  \\ 
\enddata

\tablenotetext{*}{See Figures 1, 2, 4 and 6.}

\tablenotetext{a}{Component 4 of \CIV~$\lambda$1551 coincides with \SiII~$\lambda$1527 at $z = 1.927$ and is not used but 
$\lambda$1548 is clear.}

\tablenotetext{b}{This complex system has very strong \CII\ and \SiII\ and weak \CIV\ in components 11 and 16: from the extent 
of the \HI\ Lyman $\alpha$ wings associated with these the limit on the combined column densities is log~$N$(\HI) $\lesssim 18.7$. 
\SiIV~$\lambda$1394 contains weak \NV~$\lambda$1239 at $z = 2.291$ and \CII\ at $z = 2.056$ but $\lambda$1403 is clear. \SiII\ values 
come from $\lambda$1527; $\lambda$1260, in the forest, gives a consistent fit to the data (shown in Figure 4). Although \CII\ is in the 
forest useful values are obtained for all components except 4--10, for which comparison with \SiII\ suggests have substantial underlying 
\HI\ (see Figure 4), and component 19 which is weak and may be significantly contaminated; furthermore, the strong components 11 and 16 
are mildly saturated and the values, given in square brackets, are highly approximate. An anomalous, weak, narrow feature appearing at 
$z = 1.927950$ in \CII\ and another at $z = 1.926104$ in \SiII, not detected in the other species, are indicated in Figure 4 but not 
included in the table.}

\tablenotetext{c}{\CII\ is outside the forest.}

\tablenotetext{d}{\SiIV~$\lambda$1394 is blended with very weak \NiII~$\lambda$1455 at $z = 1.927$ but $\lambda$1403 is clear. \CII\ 
is largely blended with \SiIV~$\lambda$1394 at $z = 1.927$ and broad \NV~$\lambda$1239 at $z = 2.291$ but components 2 and 3 have 
good values; component 6, which seems to appear in the VPFIT analysis (see Figure 4), is too uncertain to use. \SiII\ is 
available only in $\lambda$1527 and too weak to yield useful measures or upper limits here.}

\tablenotetext{e}{\SiIV~$\lambda$1394 is blended with strong \AlIII~$\lambda$1855 at $z = 1.328$ and values come mostly from 
$\lambda$1403.}

\tablenotetext{f}{In this complex system component 10 is very strong in \CII\ and \SiII\ and relatively weak in \CIV: mild damping 
wings centred on this position appear to be present in the \HI\ Lyman~$\alpha$ profile and indicates log~$N$(\HI) $\sim 18.3$. 
\SiIV~$\lambda$1394 is peripherally blended with \AlIII~$\lambda$1863 at $z = 1.328$, and \CIV~$\lambda$1548 in the $z = 1.817$ system 
encroaches slightly on $\lambda$1403, but together good values are obtained for all components. \CII\ contains \SiII~$\lambda$1260 
at $z = 2.291$ but in the main separation of these is straightforward (see Figure 4). \SiII~$\lambda$1260 is in the forest where it is 
partially clear (although here blended with very weak \SiII~$\lambda$1193 at $z = 2.285$) and together with the strong $\lambda$1527 
(see Figure 4) self-consistent values are obtained for the listed components.}

\tablenotetext{g}{This complex system has relatively high ionization. \CIV~$\lambda$1548 contains weak \SiII~$\lambda$1527 at 
$z = 2.291$ but $\lambda$1551 is clear; components 10--12 and 14 are strongly saturated in $\lambda$1548 and close to saturation or 
too uncertain in $\lambda$1551 and the two determined values (in square brackets), despite the apparent small errors, must be considered 
as highly approximate. \SiII~$\lambda$1260 is outside the forest but is generally weak and partially mixed with \NV~$\lambda$1243 at 
$z = 2.291$, making some component values too uncertain to use. \NV\ is in the forest but relatively strong and some useful values are 
obtained using both members of the doublet.}

\tablenotetext{h}{\CIV~$\lambda$1551 is blended with weaker \CIV~$\lambda$1548 at $z = 2.291$ but $\lambda$1548 is clear. \NV\ is 
outside the forest.}                 

\tablenotetext{i}{In this is a complex system component 1 is narrow and relatively strong in \CII\ and \SiII\ while undetected in \CIV\ 
but the \HI\ Lyman~$\alpha$ profile gives only log~$N$(\HI) $\lesssim 15.8$. Component 13 also is narrow and in \CII\ and \SiII\ is 
the strongest in the system; while \CIV\ is also quite strong, mild \HI\ Lyman~$\alpha$ damping wings centred on this component yields 
log~$N$(\HI) $\sim 17.7$. \CIV~$\lambda$1548 includes very weak \NiII~$\lambda$1742 at $z = 1.927$ and strong \CIV~$\lambda$1551 at 
$z = 2.285$ but $\lambda$1551 is clear. Many of the smaller values for $b$(C) are from \CII. \SiII~$\lambda$1260 is partially blended 
with \CII\ at $z = 2.110$ making values for components 14 and 17 too uncertain to include; additionally, component 6 appears 
significantly contaminated and is not used. \NV~$\lambda$1239 is blended with \CII\ at $z = 2.056$ and strong \SiIV~$\lambda$1394 at 
$z = 1.927$ and $\lambda$1243 is somewhat contaminated with weak \SiII~$\lambda$1260 at $z = 2.244$: only the more certain values are 
listed.} 

\tablenotetext{j}{This is close to $z_{em}$. In components 3 and 4 \CII\ seems contaminated, possibly with weak absorption related to 
nearby \AlIII~$\lambda$1855 at $z = 1.390$, and values are not included.}

\tablenotetext{k}{This is close to $z_{em}$ and may be associated with the $z = 2.321$ system. \CII\ is heavily blended with 
\MgII~$\lambda$2796 at $z = 0.586$, masking components 1--4.}

\end{deluxetable}

\clearpage
\begin{deluxetable}{crrcccccc}
\tabletypesize{\scriptsize}
\tablecaption{Heavy Element Absorption Lines: Q1442+2931 $z_{em}=2.66^*$}
\tablewidth{0pt}
\tablecolumns{9}
\tablehead{
\colhead{$z$} & \colhead{$b$(C)}  & \colhead{$b$(Si)} &
\colhead{$N$(\CIV)} & \colhead{$N$(\SiIV)} &
\colhead{$N$(\CII)} & \colhead{$N$(\SiII)} &
\colhead{$N$(\NV)} &  \colhead{n} \\
{} & \colhead{km~s$^{-1}$} & \colhead{km~s$^{-1}$} & \colhead{10$^{12}$cm$^{-2}$} &
\colhead{10$^{11}$cm$^{-2}$} & \colhead{10$^{12}$cm$^{-2}$} & \colhead{10$^{11}$cm$^{-2}$} & \colhead{10$^{12}$cm$^{-2}$} & {} \\
\colhead{1} & \colhead{2} & \colhead{3} & \colhead{4} & \colhead{5} & \colhead{6} & \colhead{7} & \colhead{8} & \colhead{9} 
}

\startdata
\sidehead{1.910}
 1.910407 &     7.9 &   \nodata &    2.12 $\pm$    0.12 &    \nodata &    \nodata &    \nodata &    \nodata &  1  \\ 
 1.910582 &     3.6 &   \nodata &    0.53 $\pm$    0.09 &    \nodata &    \nodata &    \nodata &    \nodata &  2  \\ 
\sidehead{1.974}
 1.973299 &    17.2 &   \nodata &    1.81 $\pm$    0.27 &    \nodata &    \nodata &    \nodata &    \nodata &  1  \\ 
 1.973547 &    17.6 &   \nodata &    2.53 $\pm$    0.28 &    \nodata &    \nodata &    \nodata &    \nodata &  2  \\ 
 1.974406 &    28.6 &   \nodata &    1.82 $\pm$    0.29 &    \nodata &    \nodata &    \nodata &    \nodata &  3  \\ 
 1.974843 &    11.5 &   \nodata &   10.9 $\pm$    0.2\phm{0} &    \nodata &    \nodata &    \nodata &    \nodata &  4  \\ 
 1.975128 &    11.0 &   \nodata &    3.11 $\pm$    0.17 &    \nodata &    \nodata &    \nodata &    \nodata &  5  \\ 
\sidehead{2.090\tablenotemark{a}}
 2.089562 &    15.4 &   \nodata &    2.90 $\pm$    0.28 &    \nodata &    \nodata &    \nodata &    \nodata &  1  \\ 
 2.089756 &     4.7 &   \nodata &    0.88 $\pm$    0.18 &    \nodata &    \nodata &    \nodata &    \nodata &  2  \\ 
 2.090285 &     5.4 &   \nodata &    0.63 $\pm$    0.18 &    \nodata &    \nodata &    \nodata &    \nodata &  3  \\ 
 2.090856 &     7.6 &   \nodata &    1.26 $\pm$    0.21 &    \nodata &    \nodata &    \nodata &    \nodata &  4  \\ 
 2.091048 &     8.6 &   \nodata &   11.0 $\pm$    0.3\phm{0} &    \nodata &    \nodata &    \nodata &    \nodata &  5  \\ 
\sidehead{2.157}
 2.156817 &    11.4 &   \nodata &    5.43 $\pm$    0.23 &    \nodata &    \nodata &    \nodata &    \nodata &  1  \\ 
 2.156911 &     7.3 &   \nodata &    2.44 $\pm$    0.18 &    \nodata &    \nodata &    \nodata &    \nodata &  2  \\ 
\sidehead{2.330}
 2.328970 &    39.3 &    (39.0) &    5.42 $\pm$    0.42 &    5.68 $\pm$    1.76 &    \nodata &    \nodata &    \nodata &  1  \\ 
 2.329049 &     9.2 &     (7.6) &    1.60 $\pm$    0.21 &  $<$    0.82 &    \nodata &    \nodata &    \nodata &  2  \\ 
 2.329390 &     3.8 &     (3.0) &    0.69 $\pm$    0.12 &    0.71 $\pm$    0.50 &    \nodata &    \nodata &    \nodata &  3  \\ 
 2.329668 &    15.4 &    (14.4) &    1.15 $\pm$    0.20 &    3.46 $\pm$    0.84 &    \nodata &    \nodata &    \nodata &  4  \\ 
 2.329941 &    10.0 &     (8.3) &   13.4 $\pm$    0.2\phm{0} &    4.47 $\pm$    0.72 &    \nodata &    \nodata &    \nodata &  5  \\ 
 2.330100 &     8.9 &     (7.2) &   11.4 $\pm$    0.3\phm{0} &    3.13 $\pm$    0.68 &    \nodata &    \nodata &    \nodata &  6  \\ 
 2.330248 &     7.3 &     5.7\phm{)} &   10.1 $\pm$    0.3\phm{0} &   13.5 $\pm$    0.8\phm{0} &    \nodata &    \nodata &    \nodata &  7  \\ 
 2.330420 &    25.8 &    (25.2) &   24.5 $\pm$    0.6\phm{0} &    9.78 $\pm$    1.83 &    \nodata &    \nodata &    \nodata &  8  \\ 
 2.330533 &     6.6 &     5.2\phm{)} &   14.3 $\pm$    0.3\phm{0} &    4.14 $\pm$    0.81 &    \nodata &    \nodata &    \nodata &  9  \\ 
\sidehead{2.410\tablenotemark{b}}
 2.409441 &     9.7 &     (8.0) &    5.37 $\pm$    0.26 &  $<$    0.67 &  $<$    0.23 &    \nodata &    \nodata &  1  \\ 
 2.410556 &    27.7 &    (27.2) &    4.99 $\pm$    0.39 &  $<$    1.38 &  $<$    0.45 &    \nodata &    \nodata &  2  \\ 
 2.410997 &    11.1 &     (9.6) &    3.21 $\pm$    0.26 &  $<$    1.04 &  $<$    0.31 &    \nodata &    \nodata &  3  \\ 
\sidehead{2.438\tablenotemark{c}}
 2.436407 &    (11) &    (10.0) &  $<$    0.45 &    4.97 $\pm$    1.96 &  $<$    1.14 &    4.11 $\pm$    1.47 &    \nodata &  1  \\ 
 2.436487 &    10.0 &     (8.3) &  $<$    0.69 &    4.72 $\pm$    2.30 &   33.1 $\pm$    1.9\phm{0} &   55.8 $\pm$    2.9\phm{0} &    \nodata &  2  \\ 
 2.436570 &     5.8 &     (4.7) &  $<$    0.29 &    5.53 $\pm$    1.31 &   50.2 $\pm$    2.1\phm{0} &   99.4 $\pm$    3.6\phm{0} &    \nodata &  3  \\ 
 2.436678 &     6.4 &     (5.2) &    0.31 $\pm$    0.19 &    3.75 $\pm$    0.94 &   21.7 $\pm$    0.9\phm{0} &   45.1 $\pm$    1.7\phm{0} &    \nodata &  4  \\ 
 2.436858 &    14.0 &    (12.9) &    0.62 $\pm$    0.26 &    6.47 $\pm$    1.34 &    6.78 $\pm$    0.57 &   12.2 $\pm$    1.8\phm{0} &    \nodata &  5  \\ 
 2.437026 &     7.5 &     (6.1) &  $<$    0.28 &  $<$    1.30 &   20.3 $\pm$    0.6\phm{0} &   32.6 $\pm$    1.3\phm{0} &    \nodata &  6  \\ 
 2.437178 &    14.3 &    (13.2) &    0.55 $\pm$    0.28 &  $<$    1.91 &    3.25 $\pm$    0.57 &    \nodata &    \nodata &  7  \\ 
 2.437357 &     5.7 &     (4.5) &    0.34 $\pm$    0.17 &    1.23 $\pm$    0.75 &   10.5 $\pm$    0.4\phm{0} &   16.2 $\pm$    1.1\phm{0} &    \nodata &  8  \\ 
 2.437588 &     7.4 &     (6.4) &    1.52 $\pm$    0.24 &   12.8 $\pm$    1.2\phm{0} &   13.5 $\pm$    0.5\phm{0} &   18.5 $\pm$    1.3\phm{0} &    \nodata &  9  \\ 
 2.437716 &     9.9 &     (8.3) &   16.9 $\pm$    0.4\phm{0} &   94.6 $\pm$    2.8\phm{0} &   16.0 $\pm$    0.6\phm{0} &   13.4 $\pm$    1.8\phm{0} &    \nodata & 10  \\ 
 2.437922 &    10.7 &     (9.2) &   38.6 $\pm$    0.5\phm{0} &  136 $\pm$    4\phm{00} &   27.4 $\pm$    0.7\phm{0} &   19.5 $\pm$    2.4\phm{0} &    \nodata & 11  \\*
 2.438059 &     5.2 &     (4.1) &    3.45 $\pm$    0.50 &   40.7 $\pm$    2.7\phm{0} &   12.2 $\pm$    1.2\phm{0} &    \nodata &    \nodata & 12  \\* 
 2.438110 &     5.1 &     (4.1) &    8.94 $\pm$    0.47 &   31.2 $\pm$    2.0\phm{0} &   30.2 $\pm$    1.7\phm{0} &   62.7 $\pm$    3.3\phm{0} &    \nodata & 13  \\* 
 2.438221 &     4.6 &     (3.6) &    5.51 $\pm$    0.28 &   48.0 $\pm$    1.8\phm{0} &   67.4 $\pm$    3.0\phm{0} &  119 $\pm$    4\phm{00} &    \nodata & 14  \\* 
 2.438373 &    11.7 &    (10.4) &   20.5 $\pm$    0.5\phm{0} &   36.9 $\pm$    1.4\phm{0} &   12.9 $\pm$    0.6\phm{0} &   13.9 $\pm$    2.8\phm{0} &    \nodata & 15  \\* 
 2.438535 &     4.8 &     (3.9) &    4.03 $\pm$    0.64 &   61.9 $\pm$    2.8\phm{0} &    6.26 $\pm$    0.62 &   10.4 $\pm$    1.0\phm{0} &    \nodata & 16  \\* 
 2.438675 &    19.8 &    (19.1) &   84.9 $\pm$    2.8\phm{0} &   64.0 $\pm$    3.6\phm{0} &   18.1 $\pm$    1.6\phm{0} &   17.7 $\pm$    2.3\phm{0} &    \nodata & 17  \\* 
 2.438728 &     3.3 &     (2.2) &    1.16 $\pm$    0.85 &    5.68 $\pm$    1.38 &    6.03 $\pm$    0.68 &   10.0 $\pm$    1.2\phm{0} &    \nodata & 18  \\ 
 2.438833 &     4.6 &     (3.6) &    \nodata &  151 $\pm$   13\phm{0} &   25.5 $\pm$    1.0\phm{0} &   44.2 $\pm$    2.0\phm{0} &    \nodata & 19  \\ 
 2.438906 &     7.2 &     (5.6) &    \nodata &  100 $\pm$    7\phm{00} &    7.43 $\pm$    1.05 &    7.92 $\pm$    1.13 &    \nodata & 20  \\ 
 2.439079 &    13.6 &    (12.5) &    \nodata &  215 $\pm$    8\phm{00} &    2.25 $\pm$    1.45 &    \nodata &    \nodata & 21  \\ 
 2.439098 &     3.3 &     (2.2) &    \nodata &  $<$    3.24 &    5.09 $\pm$    0.69 &    8.62 $\pm$    1.08 &    \nodata & 22  \\ 
 2.439173 &     4.3 &     (3.2) &    \nodata &   29.4 $\pm$    2.9\phm{0} &    2.68 $\pm$    0.48 &    \nodata &    \nodata & 23  \\ 
 2.439283 &     4.3 &     (3.2) &   74.3 $\pm$   19.5 &  134 $\pm$   13\phm{0} &   10.0 $\pm$    1.5\phm{0} &   23.1 $\pm$    2.3\phm{0} &    \nodata & 24  \\ 
 2.439372 &    11.3 &     (9.9) &  $<$   18.00 &  $<$   23.90 &   12.5 $\pm$    5.6\phm{0} &   10.7 $\pm$    5.3\phm{0} &    \nodata & 25  \\ 
 2.439420 &     9.7 &     (8.0) &   92.1 $\pm$   14.2 &  201 $\pm$   16\phm{0} &   14.5 $\pm$    4.9\phm{0} &   12.2 $\pm$    6.3\phm{0} &    \nodata & 26  \\ 
 2.439482 &    16.5 &    (15.6) &   49.3 $\pm$    3.6\phm{0} &   16.0 $\pm$    5.7\phm{0} &   17.4 $\pm$    2.5\phm{0} &   16.3 $\pm$    8.2\phm{0} &    \nodata & 27  \\ 
 2.439550 &     5.0 &     (4.0) &    8.64 $\pm$    1.36 &   17.3 $\pm$    2.4\phm{0} &   17.3 $\pm$    1.1\phm{0} &   38.6 $\pm$    3.1\phm{0} &    \nodata & 28  \\ 
 2.439823 &     9.7 &     (8.1) &   37.8 $\pm$    0.8\phm{0} &   32.9 $\pm$    1.3\phm{0} &  $<$    0.29 &    \nodata &    \nodata & 29  \\ 
 2.439998 &    12.8 &    (11.5) &   44.0 $\pm$    1.0\phm{0} &   13.5 $\pm$    1.6\phm{0} &    1.09 $\pm$    0.47 &    \nodata &    \nodata & 30  \\ 
 2.440086 &     5.2 &     (4.1) &   10.5 $\pm$    0.6\phm{0} &   15.4 $\pm$    1.2\phm{0} &  $<$    0.34 &    \nodata &    \nodata & 31  \\ 
 2.440252 &    14.1 &    (13.0) &    4.91 $\pm$    0.39 &  $<$    1.47 &  $<$    0.40 &    \nodata &    \nodata & 32  \\ 
\sidehead{2.468\tablenotemark{d}}
 2.465518 &     9.8 &     (8.2) &    0.82 $\pm$    0.21 &  $<$    0.94 &  $<$    0.30 &    \nodata &    \nodata &  1  \\ 
 2.466079 &    13.7 &    (12.5) &    2.64 $\pm$    0.27 &    1.58 $\pm$    0.87 &  $<$    0.34 &    \nodata &    \nodata &  2  \\ 
 2.467037 &     8.4 &     (7.1) &    3.54 $\pm$    0.27 &   12.8 $\pm$    0.7\phm{0} &    \nodata &    \nodata &    \nodata &  3  \\ 
 2.467190 &     9.5 &     (7.7) &    3.80 $\pm$    0.34 &    5.06 $\pm$    0.70 &    \nodata &    \nodata &    \nodata &  4  \\ 
 2.467409 &    12.0 &    (10.7) &   12.0 $\pm$    0.4\phm{0} &   36.8 $\pm$    0.9\phm{0} &    5.37 $\pm$    0.36 &    \nodata &    \nodata &  5  \\ 
 2.467755 &    15.2 &    (14.2) &    7.90 $\pm$    0.40 &    6.71 $\pm$    0.84 &    0.78 $\pm$    0.36 &    \nodata &    \nodata &  6  \\ 
 2.468117 &    18.5 &    (17.7) &    3.68 $\pm$    0.42 &    2.15 $\pm$    0.91 &  $<$    0.37 &    \nodata &    \nodata &  7  \\ 
 2.468760 &    10.0 &     (8.4) &    1.03 $\pm$    0.31 &    1.64 $\pm$    0.74 &    0.47 $\pm$    0.30 &    \nodata &    \nodata &  8  \\ 
 2.469769 &    13.8 &    (12.7) &    0.59 $\pm$    0.27 &  $<$    1.14 &  $<$    0.45 &    \nodata &    \nodata &  9  \\ 
\sidehead{2.474\tablenotemark{e}}
 2.473032 &     6.3 &     (5.1) &    0.85 $\pm$    0.18 &    2.35 $\pm$    0.55 &  $<$    0.28 &    \nodata &    \nodata &  1  \\ 
 2.473651 &    11.6 &    (10.2) &   39.3 $\pm$    0.5\phm{0} &   17.3 $\pm$    0.8\phm{0} &    1.06 $\pm$    0.38 &    \nodata &    5.39 $\pm$    0.39 &  2  \\ 
 2.473833 &     8.9 &     (7.3) &   23.3 $\pm$    0.4\phm{0} &   13.9 $\pm$    0.7\phm{0} &    0.88 $\pm$    0.35 &    \nodata &    0.86 $\pm$    0.32 &  3  \\ 
 2.474193 &     5.2 &     (3.8) &   36.2 $\pm$    0.5\phm{0} &   58.8 $\pm$    1.1\phm{0} &   11.7 $\pm$    0.3\phm{0} &    6.91 $\pm$    1.82 &    1.17 $\pm$    0.26 &  4  \\ 
 2.474560 &     6.3 &     (5.1) &    4.30 $\pm$    0.30 &    1.32 $\pm$    0.84 &  $<$    0.48 &    \nodata &    \nodata &  5  \\ 
 2.474597 &    20.0 &    (19.2) &   11.8 $\pm$    0.4\phm{0} &    8.59 $\pm$    1.46 &  $<$    1.17 &    \nodata &    \nodata &  6  \\ 
\sidehead{2.484\tablenotemark{f}}
 2.483641 &     5.3 &     (4.8) &    5.58 $\pm$    0.16 &    2.18 $\pm$    1.07 &  $<$    0.26 &    \nodata &    \nodata &  1  \\ 
\sidehead{2.502\tablenotemark{g}}
 2.501575 &     3.8 &     (3.0) &    0.58 $\pm$    0.14 &  $<$    0.74 &    \nodata &    \nodata &    \nodata &  1  \\ 
 2.501685 &    59.3 &    (59.1) &    4.15 $\pm$    0.46 &  $<$    1.92 &    \nodata &    \nodata &    \nodata &  2  \\* 
 2.502266 &    10.7 &     (9.2) &    8.62 $\pm$    0.22 &  $<$    0.69 &    \nodata &    \nodata &    \nodata &  3  \\* 
 2.502459 &     4.4 &     (3.7) &    0.74 $\pm$    0.14 &  $<$    0.49 &  $<$    0.30 &    \nodata &    \nodata &  4  \\ 
\sidehead{2.555\tablenotemark{h}}
 2.554824 &     8.9 &     (7.1) &    2.24 $\pm$    0.14 &  $<$    2.53 &    \nodata &    \nodata &    \nodata &  1  \\ 
 2.555470 &    24.8 &    (24.2) &    4.31 $\pm$    0.22 &  $<$    1.60 &    \nodata &    \nodata &    \nodata &  2  \\ 
\sidehead{2.616\tablenotemark{i}}
 2.614806 &     5.7 &     (4.8) &    1.30 $\pm$    0.12 &  $<$    0.49 &    \nodata &  $<$    0.25 &  $<$    0.37 &  1  \\ 
 2.615306 &     6.8 &     4.8\phm{)} &   16.4 $\pm$    0.3\phm{0} &   15.0 $\pm$    0.6\phm{0} &    \nodata &  $<$    0.26 &  $<$    0.42 &  2  \\ 
 2.615599 &    10.4 &    (10.1) &    1.12 $\pm$    0.15 &  $<$    0.66 &    \nodata &  $<$    0.33 &  $<$    0.34 &  3  \\ 
 2.615859 &    10.1 &     (8.5) &    3.16 $\pm$    0.16 &  $<$    0.69 &    \nodata &  $<$    0.31 &    0.47 $\pm$    0.18 &  4  \\ 
 2.616065 &     6.9 &     (5.8) &    7.44 $\pm$    0.18 &  $<$    0.71 &    \nodata &  $<$    0.32 &  $<$    0.24 &  5  \\ 
 2.616292 &     9.9 &     (8.3) &    4.23 $\pm$    0.16 &  $<$    0.63 &    \nodata &  $<$    0.34 &  $<$    0.20 &  6  \\ 
 2.616516 &     8.4 &     (6.9) &    2.74 $\pm$    0.15 &    1.35 $\pm$    0.57 &  $<$    0.49 &  $<$    0.31 &  $<$    0.18 &  7  \\ 
 2.616812 &     3.7 &     (2.4) &    0.42 $\pm$    0.10 &  $<$    0.43 &  $<$    0.25 &  $<$    0.29 &  $<$    0.13 &  8  \\ 
 2.617053 &     8.8 &     (7.3) &    9.57 $\pm$    0.20 &  $<$    0.58 &  $<$    0.70 &  $<$    0.29 &    0.87 $\pm$    0.17 &  9  \\ 
 2.617345 &     7.5 &     (6.5) &    0.45 $\pm$    0.13 &  $<$    0.59 &  $<$    0.86 &  $<$    0.28 &  $<$    0.21 & 10  \\ 
 2.617572 &     4.9 &     (4.1) &    5.28 $\pm$    0.18 &  $<$    0.50 &  $<$    0.28 &  $<$    0.25 &    1.08 $\pm$    0.19 & 11  \\ 
 2.617702 &     4.2 &     (3.2) &    1.43 $\pm$    0.20 &  $<$    0.57 &  $<$    0.32 &  $<$    0.28 &    0.76 $\pm$    0.22 & 12  \\ 
 2.617803 &    19.0 &    (18.2) &   23.0 $\pm$    0.4\phm{0} &    \nodata &  $<$    0.62 &  $<$    0.60 &    1.85 $\pm$    0.41 & 13  \\ 
 2.618136 &     4.2 &     3.1\phm{)} &    7.83 $\pm$    0.20 &    7.96 $\pm$    0.53 &    0.80 $\pm$    0.26 &    0.81 $\pm$    0.23 &  $<$    0.41 & 14  \\ 
\sidehead{2.623\tablenotemark{j}}
 2.622674 &     3.7 &     (2.4) &    0.57 $\pm$    0.12 &  $<$    0.47 &    \nodata &  $<$    0.45 &  $<$    0.17 &  1  \\ 
 2.623208 &    11.9 &    (10.6) &    4.99 $\pm$    0.20 &  $<$    0.79 &  $<$    0.51 &  $<$    0.33 &  $<$    0.35 &  2  \\ 
 2.623486 &    11.5 &    (10.1) &    4.15 $\pm$    0.21 &    3.23 $\pm$    0.78 &    0.91 $\pm$    0.38 &    0.73 $\pm$    0.33 &    0.39 $\pm$    0.22 &  3  \\ 
\enddata

\tablenotetext{*}{See Figure 1.}

\tablenotetext{a}{\CIV~$\lambda$1551 is immersed in \SiIV~$\lambda$1394 at $z = 2.438$ but $\lambda$1548 is clear.}

\tablenotetext{b}{\CIV~$\lambda$1551 in component 1 is contaminated but $\lambda$1548 is clear. \SiIV~$\lambda$1403 is masked by 
\CIV~$\lambda$1548 at $z = 2.090$ and the upper limit values come from $\lambda$1394. \CII\ is outside the forest.}

\tablenotetext{c}{This is highly complex system with components 3 and 14 particularly strong in \CII\ and \SiII\ and relatively weak in 
\CIV: in these, respective values log~$N$(\HI) $\sim19.7$ and $\sim19.5$ give a good fit to the \HI\ Lyman $\alpha$ damping wings. 
Overall \CIV\ is also very strong and in components 19--23 $\lambda\lambda$1548,1551 both are too saturated to be useful; here \CII\ 
provides the values for $b$(C). \SiIV~$\lambda$1394 contains \CIV~$\lambda$1551 at $z = 2.090$ but $\lambda$1403 is clear and good values
are obtained throughout. \SiII~$\lambda$1260, in the forest but only mildly contaminated, is used to supplement $\lambda\lambda$1304,1527 
and all are collectively self-consistent; errors are from $\lambda\lambda$1304,1527.}

\tablenotetext{d}{\CIV~$\lambda$1551 is partially blended with \CIV~$\lambda$1548 at $z = 2.474$ but $\lambda$1548 is clear. In \CII\
apparent contamination of components 3 and 4 makes these values too uncertain to include.} 

\tablenotetext{e}{\CIV~$\lambda$1548 is partially blended with \CIV~$\lambda$1551 at $z = 2.468$ but $\lambda$1551 is clear. The \SiII\ 
value is from $\lambda$1527. \NV\ is in the forest but for components 2--4 useful values come from $\lambda$1239, in a relatively clear 
region, which show velocity structure corresponding to these components in \CIV.}

\tablenotetext{f}{\SiIV~$\lambda$1394 is contaminated and the upper limit comes from $\lambda$1403.}

\tablenotetext{g}{\CII\ in components 1--3 is masked by \MgII~$\lambda$2804 at $z = 0.667$.} 

\tablenotetext{h}{\CII\ is very weak and too noisy to give useful upper limits. \SiII~$\lambda$1260 is outside the forest but masked by 
\NV~$\lambda$1239 at $z = 2.616$.}

\tablenotetext{i}{\CIV~$\lambda$1551 is blended with weaker \CIV~$\lambda$1548 at $z = 2.623$ but $\lambda$1548 is clear. \CII\ in 
components 1--6 is masked by \SiIV~$\lambda$1403 at $z = 2.438$. \SiII~$\lambda$1260 is outside the forest. \NV\ is outside the forest; 
in components 1--3 $\lambda$1239 is blended with \OI~$\lambda$1302 at $z = 2.438$, and in component 14 with \SiII~$\lambda$1304 in the 
same system and only $\lambda$1243 is used for these upper limits.}

\tablenotetext{j}{\CIV~$\lambda$1548 is blended with \CIV~$\lambda$1551 at $z = 2.616$ but $\lambda$1551 is clear. In \CII\ the close 
proximity to \SiIV~$\lambda$1393 at $z = 2.468$ makes component 1 too uncertain to use.}

\end{deluxetable}

\clearpage
\begin{deluxetable}{crrcccccc}
\tabletypesize{\scriptsize}
\tablecaption{Heavy Element Absorption Lines: Q1107+4847 $z_{em}=2.97^*$}
\tablewidth{0pt}
\tablecolumns{9}
\tablehead{
\colhead{$z$} & \colhead{$b$(C)}  & \colhead{$b$(Si)} &
\colhead{$N$(\CIV)} & \colhead{$N$(\SiIV)} &
\colhead{$N$(\CII)} & \colhead{$N$(\SiII)} &
\colhead{$N$(\NV)} & \colhead{n} \\
{} & \colhead{km~s$^{-1}$} & \colhead{km~s$^{-1}$} & \colhead{10$^{12}$cm$^{-2}$} &
\colhead{10$^{11}$cm$^{-2}$} & \colhead{10$^{12}$cm$^{-2}$} & \colhead{10$^{11}$cm$^{-2}$} & \colhead{10$^{12}$cm$^{-2}$} & {} \\
\colhead{1} & \colhead{2} & \colhead{3} & \colhead{4} & \colhead{5} & \colhead{6} & \colhead{7} & \colhead{8} & \colhead{9} 
}

\startdata
\sidehead{2.117}
 2.117248 &    17.1 &   \nodata &    2.24 $\pm$    0.11 &    \nodata &    \nodata &    \nodata &    \nodata &  1  \\ 
\sidehead{2.142}
 2.141490 &     6.4 &   \nodata &    0.54 $\pm$    0.10 &    \nodata &    \nodata &    \nodata &    \nodata &  1  \\ 
\sidehead{2.143}
 2.143093 &     4.6 &   \nodata &    2.23 $\pm$    0.12 &    \nodata &    \nodata &    \nodata &    \nodata &  1  \\ 
 2.143184 &     6.3 &   \nodata &   14.2 $\pm$    0.2\phm{0} &    \nodata &    \nodata &    \nodata &    \nodata &  2  \\ 
 2.143335 &     7.8 &   \nodata &    2.33 $\pm$    0.12 &    \nodata &    \nodata &    \nodata &    \nodata &  3  \\ 
\sidehead{2.153}
 2.153335 &     4.5 &   \nodata &    0.80 $\pm$    0.08 &    \nodata &    \nodata &    \nodata &    \nodata &  1  \\ 
 2.153493 &     9.5 &   \nodata &    2.78 $\pm$    0.11 &    \nodata &    \nodata &    \nodata &    \nodata &  2  \\ 
\sidehead{2.662\tablenotemark{a}}
 2.661835 &     9.2 &     (7.4) &    3.36 $\pm$    0.15 &    1.04 $\pm$    0.72 &  $<$    0.23 &    \nodata &    \nodata &  1  \\ 
 2.662468 &     7.4 &     (5.9) &    1.43 $\pm$    0.17 &  $<$    0.62 &  $<$    0.17 &    \nodata &    \nodata &  2  \\ 
\sidehead{2.685}
 2.684123 &     3.7 &     (2.4) &    0.53 $\pm$    0.19 &    1.46 $\pm$    0.71 &  $<$    0.22 &    \nodata &    \nodata &  1  \\ 
 2.684549 &     5.5 &     (4.5) &    2.78 $\pm$    0.80 &    9.10 $\pm$    1.54 &  $<$    0.35 &    \nodata &    \nodata &  2  \\ 
 2.684715 &    10.9 &     (9.4) &    1.42 $\pm$    0.87 &  $<$    2.32 &  $<$    0.63 &    \nodata &    \nodata &  3  \\ 
\sidehead{2.696}
 2.695956 &    43.6 &    (43.2) &    5.04 $\pm$    0.57 &  $<$    2.00 &  $<$    0.71 &    \nodata &    \nodata &  1  \\ 
 2.696317 &    23.6 &    (23.0) &    2.71 $\pm$    0.43 &  $<$    1.52 &  $<$    0.54 &    \nodata &    \nodata &  2  \\ 
 2.696644 &     7.8 &     (6.3) &    5.02 $\pm$    0.20 &    1.95 $\pm$    0.67 &  $<$    0.76 &    \nodata &    \nodata &  3  \\ 
 2.696901 &     4.9 &     (4.2) &    0.92 $\pm$    0.14 &    1.26 $\pm$    0.58 &  $<$    0.21 &    \nodata &    \nodata &  4  \\ 
\sidehead{2.724\tablenotemark{b}}
 2.722186 &     6.0 &     (4.9) &    9.77 $\pm$    0.19 &    1.53 $\pm$    0.64 &  $<$    0.24 &    \nodata &    \nodata &  1  \\ 
 2.723566 &     9.8 &     (8.2) &    0.88 $\pm$    0.17 &    0.91 $\pm$    0.70 &    \nodata &    \nodata &    \nodata &  2  \\ 
 2.723861 &     8.2 &     (6.8) &    4.15 $\pm$    0.17 &   11.3 $\pm$    0.7\phm{0} &    0.44 $\pm$    0.27 &    \nodata &    \nodata &  3  \\ 
 2.724090 &     5.2 &     (4.3) &    3.06 $\pm$    0.27 &   29.4 $\pm$    1.5\phm{0} &    1.01 $\pm$    0.37 &    0.74 $\pm$    0.41 &    \nodata &  4  \\ 
 2.724189 &    14.9 &    (13.9) &   14.2 $\pm$    0.5\phm{0} &   55.2 $\pm$    2.2\phm{0} &    4.30 $\pm$    0.68 &    4.19 $\pm$    0.73 &    \nodata &  5  \\ 
 2.724466 &    16.6 &    (15.6) &    9.61 $\pm$    0.54 &   11.2 $\pm$    2.3\phm{0} &  $<$    0.87 &    \nodata &    \nodata &  6  \\ 
 2.724542 &     6.1 &     (5.0) &    4.34 $\pm$    0.33 &    8.01 $\pm$    1.34 &  $<$    0.58 &    \nodata &    \nodata &  7  \\ 
 2.724852 &    13.2 &    (12.0) &    1.33 $\pm$    0.21 &    7.72 $\pm$    1.04 &    1.76 $\pm$    0.36 &    \nodata &    \nodata &  8  \\ 
 2.725055 &     6.0 &     (4.9) &    0.50 $\pm$    0.15 &    6.75 $\pm$    0.77 &    1.01 $\pm$    0.26 &    \nodata &    \nodata &  9  \\ 
\sidehead{2.761\tablenotemark{c}}
 2.758802 &    12.2 &    (10.9) &    1.05 $\pm$    0.22 &  $<$    0.89 &  $<$    0.52 &    \nodata &    \nodata &  1  \\ 
 2.759076 &     7.3 &     (6.0) &   52.9 $\pm$    1.2\phm{0} &   23.2 $\pm$    1.0\phm{0} &    0.88 $\pm$    0.27 &    \nodata &    4.64 $\pm$    0.44 &  2  \\ 
 2.759297 &     9.8 &     (8.1) &  425 $\pm$   32\phm{0} &  193 $\pm$    4\phm{00} &   10.7 $\pm$    0.4\phm{0} &    6.02 $\pm$    2.32 &   11.9 $\pm$    0.6\phm{0} &  3  \\ 
 2.759444 &     3.6 &     (2.7) &    7.36 $\pm$    3.53 &   83.1 $\pm$    4.6\phm{0} &    7.55 $\pm$    0.34 &   10.7 $\pm$    1.8\phm{0} &    \nodata &  4  \\ 
 2.759587 &    19.6 &    (18.8) &    6.30 $\pm$    0.39 &  $<$    1.38 &  $<$    0.39 &    \nodata &    \nodata &  5  \\ 
 2.759974 &     4.8 &     (4.0) &    3.95 $\pm$    0.18 &  $<$    0.63 &  $<$    0.22 &    \nodata &    \nodata &  6  \\ 
 2.760197 &    10.4 &     (8.9) &    3.87 $\pm$    0.29 &  $<$    1.48 &  $<$    0.39 &    \nodata &    \nodata &  7  \\ 
 2.760287 &     4.4 &     (3.7) &   14.4 $\pm$    0.4\phm{0} &  $<$    0.78 &  $<$    0.29 &    \nodata &    5.54 $\pm$    0.51 &  8  \\ 
 2.760492 &    10.3 &     (8.8) &    7.97 $\pm$    0.26 &  $<$    1.18 &  $<$    0.48 &    \nodata &    \nodata &  9  \\ 
 2.760643 &     4.5 &     (3.9) &    7.41 $\pm$    0.27 &    9.75 $\pm$    0.77 &  $<$    0.23 &    \nodata &    \nodata & 10  \\* 
 2.760785 &     7.9 &     (6.3) &   20.1 $\pm$    0.3\phm{0} &   13.2 $\pm$    0.8\phm{0} &    0.93 $\pm$    0.28 &    \nodata &    \nodata & 11  \\* 
 2.760985 &     6.8 &     (5.7) &   18.5 $\pm$    0.5\phm{0} &   13.0 $\pm$    0.8\phm{0} &    0.57 $\pm$    0.27 &    \nodata &    \nodata & 12  \\* 
 2.761139 &     8.4 &     (6.6) &   64.2 $\pm$    1.2\phm{0} &   27.5 $\pm$    1.0\phm{0} &    2.29 $\pm$    0.33 &    \nodata &    \nodata & 13  \\ 
 2.761307 &     7.8 &     (6.6) &   10.1 $\pm$    0.6\phm{0} &   16.2 $\pm$    1.1\phm{0} &    0.82 $\pm$    0.36 &    \nodata &    \nodata & 14  \\ 
 2.761430 &     7.2 &     (5.4) &   43.6 $\pm$    0.9\phm{0} &   69.5 $\pm$    1.7\phm{0} &    9.65 $\pm$    0.39 &    9.42 $\pm$    2.08 &    \nodata & 15  \\ 
 2.761591 &    10.6 &     (9.0) &    7.66 $\pm$    0.29 &   14.3 $\pm$    1.0\phm{0} &  $<$    0.37 &    \nodata &    \nodata & 16  \\ 
 2.761831 &     7.5 &     (6.2) &    3.68 $\pm$    0.20 &   17.8 $\pm$    0.9\phm{0} &    7.40 $\pm$    0.31 &    6.90 $\pm$    2.01 &    \nodata & 17  \\ 
 2.761996 &     7.0 &     (5.7) &    5.53 $\pm$    0.23 &   20.7 $\pm$    0.9\phm{0} &    4.51 $\pm$    0.32 &    3.33 $\pm$    1.95 &    \nodata & 18  \\ 
 2.762142 &     6.7 &     (5.4\phm{)} &    5.66 $\pm$    0.31 &   14.1 $\pm$    0.9\phm{0} &    5.57 $\pm$    0.44 &    9.86 $\pm$    2.31 &    \nodata & 19  \\ 
 2.762315 &     9.7 &     (8.1) &   20.8 $\pm$    1.0\phm{0} &   46.0 $\pm$    2.1\phm{0} &   70.4 $\pm$    1.4\phm{0} &  106 $\pm$    5\phm{00} &    \nodata & 20  \\ 
 2.762392 &     4.9 &     (4.1) &  224 $\pm$   17\phm{0} &  234 $\pm$   15\phm{0} &  $<$    1.17 &  $<$    3.77 &   \nodata & 21  \\ 
 2.762516 &     3.3 &     (2.5) &    6.97 $\pm$    0.45 &    9.51 $\pm$    1.03 &    7.57 $\pm$    0.39 &    \nodata &    \nodata & 22  \\ 
 2.762624 &    15.0 &    (14.0) &    7.58 $\pm$    0.33 &   17.8 $\pm$    1.3\phm{0} &    7.92 $\pm$    0.49 &    \nodata &    \nodata & 23  \\ 
 2.762901 &     4.8 &     (4.4) &    3.56 $\pm$    0.24 &   18.9 $\pm$    1.2\phm{0} &    2.69 $\pm$    0.32 &    4.65 $\pm$    2.83 &    \nodata & 24  \\ 
 2.762995 &     4.6 &     (3.8) &    3.26 $\pm$    0.28 &   20.6 $\pm$    1.3\phm{0} &   12.6 $\pm$    0.5\phm{0} &   28.9 $\pm$    3.2\phm{0} &    \nodata & 25  \\ 
 2.763076 &     4.8 &     (4.0) &    9.34 $\pm$    0.31 &   21.3 $\pm$    1.1\phm{0} &  $<$    0.35 &    \nodata &    \nodata & 26  \\ 
 2.763246 &    11.4 &    (10.0) &    3.67 $\pm$    0.31 &    7.32 $\pm$    1.20 &    3.72 $\pm$    0.48 &    \nodata &    \nodata & 27  \\ 
 2.763334 &     3.3 &     (2.4) &    1.04 $\pm$    0.20 &    8.65 $\pm$    0.91 &    8.02 $\pm$    0.39 &   19.1 $\pm$    2.9\phm{0} &    \nodata & 28  \\ 
 2.763478 &     6.5 &     (5.2) &    1.91 $\pm$    0.18 &    6.64 $\pm$    0.74 &    9.11 $\pm$    0.32 &    9.25 $\pm$    2.52 &    \nodata & 29  \\ 
 2.763659 &    10.6 &    (10.4) &    0.70 $\pm$    0.20 &    3.22 $\pm$    0.90 &    3.97 $\pm$    0.33 &    \nodata &    \nodata & 30  \\ 
 2.763976 &     8.2 &     (6.7) &  $<$    0.27 &    1.80 $\pm$    0.72 &    9.51 $\pm$    0.31 &   15.0 $\pm$    2.4\phm{0} &    \nodata & 31  \\ 
\sidehead{2.870\tablenotemark{d}}
 2.869957 &    11.3 &     (9.9) &    3.37 $\pm$    0.26 &  $<$    2.25 &    \nodata &  $<$    0.48 &    \nodata &  1  \\ 
 2.870287 &    10.9 &     (9.4) &   12.4 $\pm$    0.3\phm{0} &   31.7 $\pm$    2.3\phm{0} &  $<$    0.38 &  $<$    0.60 &    \nodata &  2  \\ 
 2.870562 &    13.2 &    (12.0) &    5.29 $\pm$    0.31 &  $<$    2.54 &  $<$    0.41 &  $<$    0.28 &    \nodata &  3  \\ 
\sidehead{2.909\tablenotemark{e}}
 2.908768 &    10.1 &   \nodata &    2.17 $\pm$    0.16 &    \nodata &    \nodata &    \nodata &  $<$    0.18 &  1  \\ 
\sidehead{2.956\tablenotemark{f}}
 2.956154 &    11.4 &    (10.0) &    1.14 $\pm$    0.15 &  $<$    0.72 &    \nodata &  $<$    0.53 &  $<$    0.28 &  1  \\ 
\enddata

\tablenotetext{*}{See Figure 1.}

\tablenotetext{a}{\SiIV~$\lambda$1403 in component 1 appears contaminated and is not used but $\lambda$1394 is clear. \CII\ is outside
the forest.}

\tablenotetext{b}{\SiIV~$\lambda$1403 in components 1--4 is blended with \MgII~$\lambda$2804 at $z = 0.863$ but $\lambda$1394 is clear. 
\CII\ in component 2 is too noisy to be useful. \SiII~$\lambda$1260 is in the forest but is used together with weak $\lambda$1527 and 
consistent values are obtained for components 4 and 5.}

\tablenotetext{c}{This is a highly complex system. Component 20 has strong \CII\ and \SiII\ but \CIV\ is not unusually weak in comparison; 
for this log~$N$(\HI) $\sim19.1$ gives a good fit to the outer parts of the associated \HI\ Lyman $\alpha$ damping wings. The adjoining 
component 21 is very strong in \CIV\ and \SiIV\ but undetectable in the low ionization species. \SiII~$\lambda$1260 is in the forest but 
this and $\lambda$1193 are used to supplement $\lambda$1527 and available portions of $\lambda$1304: the values obtained for the stronger 
components are listed with errors from $\lambda\lambda$1304,1527. \NV\ is in the forest but good values for components 2, 3 and 8 come 
from both members of the doublet.}

\tablenotetext{d}{\SiIV~$\lambda$1394 is not used because the feature at this position is too strong relative to $\lambda$1403. 
\SiII~$\lambda$1260 is outside the forest.}

\tablenotetext{e}{\SiIV~$\lambda$1393 coincides with \AlIII~$\lambda$1855 at $z = 1.938$ and $\lambda$1403 is too noisy to give a reliable 
value. \CII\ and \SiII\ both are too contaminated to be useful. \NV\ is outside the forest.}

\tablenotetext{f}{This is close to $z_{em}$. \CII, which here would yield only an upper limit, is masked by \SiIV~$\lambda$1403 at 
$z = 2.761$.}

\end{deluxetable}

\clearpage
\begin{deluxetable}{crrcccccc}
\tabletypesize{\scriptsize}
\tablecaption{Heavy Element Absorption Lines: Q0636+6801 $z_{em}=3.18^*$}
\tablewidth{0pt}
\tablecolumns{9}
\tablehead{
\colhead{$z$} & \colhead{$b$(C)}  & \colhead{$b$(Si)} &
\colhead{$N$(\CIV)} & \colhead{$N$(\SiIV)} &
\colhead{$N$(\CII)} & \colhead{$N$(\SiII)} &
\colhead{$N$(\NV)} &  \colhead{n} \\
{} & \colhead{km~s$^{-1}$} & \colhead{km~s$^{-1}$} & \colhead{10$^{12}$cm$^{-2}$} &
\colhead{10$^{11}$cm$^{-2}$} & \colhead{10$^{12}$cm$^{-2}$} & \colhead{10$^{11}$cm$^{-2}$} & \colhead{10$^{12}$cm$^{-2}$} & {} \\
\colhead{1} & \colhead{2} & \colhead{3} & \colhead{4} & \colhead{5} & \colhead{6} & \colhead{7} & \colhead{8} & \colhead{9} 
}

\startdata
\sidehead{2.311}
 2.310144 &     9.4 &   \nodata &    2.65 $\pm$    0.16 &    \nodata &    \nodata &    \nodata &    \nodata &  1  \\ 
 2.310436 &     8.4 &   \nodata &    3.10 $\pm$    0.16 &    \nodata &    \nodata &    \nodata &    \nodata &  2  \\ 
 2.310659 &     8.3 &   \nodata &    3.24 $\pm$    0.16 &    \nodata &    \nodata &    \nodata &    \nodata &  3  \\ 
 2.311034 &    16.4 &   \nodata &   20.1 $\pm$    0.3\phm{0} &    \nodata &    \nodata &    \nodata &    \nodata &  4  \\ 
 2.311266 &    12.1 &   \nodata &   15.3 $\pm$    0.2\phm{0} &    \nodata &    \nodata &    \nodata &    \nodata &  5  \\ 
 2.311701 &    22.4 &   \nodata &    9.06 $\pm$    0.27 &    \nodata &    \nodata &    \nodata &    \nodata &  6  \\ 
 2.312124 &     9.6 &   \nodata &    0.83 $\pm$    0.16 &    \nodata &    \nodata &    \nodata &    \nodata &  7  \\ 
 2.312480 &    25.7 &   \nodata &    3.94 $\pm$    0.29 &    \nodata &    \nodata &    \nodata &    \nodata &  8  \\ 
\sidehead{2.325}
 2.324375 &     4.4 &   \nodata &    0.25 $\pm$    0.08 &    \nodata &    \nodata &    \nodata &    \nodata &  1  \\ 
 2.324733 &     9.3 &   \nodata &    2.13 $\pm$    0.12 &    \nodata &    \nodata &    \nodata &    \nodata &  2  \\ 
 2.324870 &     7.1 &   \nodata &    4.49 $\pm$    0.12 &    \nodata &    \nodata &    \nodata &    \nodata &  3  \\ 
 2.325083 &    11.0 &   \nodata &    3.37 $\pm$    0.13 &    \nodata &    \nodata &    \nodata &    \nodata &  4  \\ 
 2.325250 &     6.8 &   \nodata &    0.26 $\pm$    0.10 &    \nodata &    \nodata &    \nodata &    \nodata &  5  \\ 
\sidehead{2.436}
 2.435488 &     3.2 &   \nodata &    0.69 $\pm$    0.09 &    \nodata &    \nodata &    \nodata &    \nodata &  1  \\ 
\sidehead{2.474\tablenotemark{a}}
 2.472728 &     4.8 &   \nodata &   11.1 $\pm$    0.3\phm{0} &    \nodata &    \nodata &    \nodata &    \nodata &  1  \\ 
 2.472757 &    13.4 &   \nodata &    5.75 $\pm$    0.38 &    \nodata &    \nodata &    \nodata &    \nodata &  2  \\ 
 2.473157 &     8.0 &   \nodata &    5.84 $\pm$    0.17 &    \nodata &    \nodata &    \nodata &    \nodata &  3  \\ 
 2.473406 &     5.8 &   \nodata &    3.64 $\pm$    0.15 &    \nodata &    \nodata &    \nodata &    \nodata &  4  \\ 
 2.473570 &     9.8 &   \nodata &    5.79 $\pm$    0.18 &    \nodata &    \nodata &    \nodata &    \nodata &  5  \\ 
 2.473849 &    19.4 &   \nodata &    5.74 $\pm$    0.25 &    \nodata &    \nodata &    \nodata &    \nodata &  6  \\ 
 2.474483 &     7.0 &   \nodata &   13.0 $\pm$    0.2\phm{0} &    \nodata &    \nodata &    \nodata &    \nodata &  7  \\ 
 2.474607 &    10.7 &   \nodata &    8.13 $\pm$    0.22 &    \nodata &    \nodata &    \nodata &    \nodata &  8  \\ 
 2.474945 &    15.2 &   \nodata &   11.0 $\pm$    0.2\phm{0} &    \nodata &    \nodata &    \nodata &    \nodata &  9  \\ 
 2.475229 &     6.8 &     (5.7) &   53.4 $\pm$    0.7\phm{0} &   77.6 $\pm$    4.4\phm{0} &    2.69 $\pm$    0.47 &    3.07 $\pm$    0.78 &    \nodata & 10  \\ 
 2.475353 &     5.6 &     3.8\phm{)} &   56.1 $\pm$    1.1\phm{0} &  113 $\pm$    7\phm{00} &   11.0 $\pm$    0.6\phm{0} &   10.8 $\pm$    0.9\phm{0} &    \nodata & 11  \\ 
 2.475462 &     9.1 &   \nodata &   17.2 $\pm$    0.3\phm{0} &    \nodata &    \nodata &    \nodata &    \nodata & 12  \\ 
 2.475853 &    28.9 &   \nodata &    5.96 $\pm$    0.40 &    \nodata &    \nodata &    \nodata &    \nodata & 13  \\ 
 2.475952 &     7.0 &   \nodata &    3.29 $\pm$    0.18 &    \nodata &    \nodata &    \nodata &    \nodata & 14  \\ 
\sidehead{2.595}
 2.594488 &     8.2 &   \nodata &    1.86 $\pm$    0.12 &    \nodata &    \nodata &    \nodata &    \nodata &  1  \\ 
 2.594723 &     4.6 &   \nodata &    0.79 $\pm$    0.11 &    \nodata &    \nodata &    \nodata &    \nodata &  2  \\ 
 2.594813 &     5.8 &   \nodata &    1.46 $\pm$    0.12 &    \nodata &    \nodata &    \nodata &    \nodata &  3  \\ 
\sidehead{2.621}
 2.620991 &    12.6 &   \nodata &    1.69 $\pm$    0.23 &    \nodata &    \nodata &    \nodata &    \nodata &  1  \\ 
 2.621408 &    17.2 &   \nodata &    4.90 $\pm$    0.27 &    \nodata &    \nodata &    \nodata &    \nodata &  2  \\ 
\sidehead{2.681\tablenotemark{b}}
 2.680206 &     8.7 &     (6.9) &    1.49 $\pm$    0.19 &  $<$    0.94 &    \nodata &    \nodata &    \nodata &  1  \\ 
 2.681560 &    16.8 &    (15.9) &    0.94 $\pm$    0.31 &  $<$    1.62 &    \nodata &    \nodata &    \nodata &  2  \\ 
 2.681707 &     8.0 &     (6.3) &    0.49 $\pm$    0.22 &    1.43 $\pm$    1.11 &    \nodata &    \nodata &    \nodata &  3  \\ 
 2.682112 &    16.9 &    (15.9) &    6.59 $\pm$    0.26 &  $<$    1.33 &    \nodata &    \nodata &    \nodata &  4  \\ 
\sidehead{2.690}
 2.690462 &     7.6 &     (6.3) &    1.67 $\pm$    0.21 &    1.38 $\pm$    0.44 &    \nodata &    \nodata &    \nodata &  1  \\ 
\sidehead{2.868\tablenotemark{c}}
 2.867163 &     9.3 &     (7.5) &    0.35 $\pm$    0.20 &  $<$    1.14 &  $<$    0.32 &    \nodata &    \nodata &  1  \\ 
 2.867425 &     6.5 &     (5.3) &    0.97 $\pm$    0.18 &  $<$    1.02 &  $<$    0.19 &    \nodata &    \nodata &  2  \\ 
 2.867898 &     9.4 &     (7.6) &    3.48 $\pm$    0.22 &  $<$    0.86 &  $<$    0.26 &    \nodata &    \nodata &  3  \\ 
 2.868403 &     8.6 &     (6.8) &    0.83 $\pm$    0.18 &  $<$    0.57 &  $<$    0.21 &    \nodata &    \nodata &  4  \\ 
 2.868776 &     9.1 &     (7.3) &    5.30 $\pm$    0.21 &    3.08 $\pm$    0.55 &  $<$    0.21 &    \nodata &    \nodata &  5  \\ 
\sidehead{2.892}
 2.890399 &     6.4 &     (5.2) &    0.81 $\pm$    0.12 &    1.05 $\pm$    0.46 &  $<$    0.24 &    \nodata &    \nodata &  1  \\ 
 2.890728 &    14.6 &    (13.5) &    1.18 $\pm$    0.19 &  $<$    1.00 &  $<$    0.35 &    \nodata &    \nodata &  2  \\ 
 2.890956 &     3.7 &     (2.4) &    0.43 $\pm$    0.11 &  $<$    0.49 &  $<$    0.33 &    \nodata &    \nodata &  3  \\ 
 2.891170 &     3.7 &     (2.4) &    0.79 $\pm$    0.11 &  $<$    0.38 &  $<$    0.20 &    \nodata &    \nodata &  4  \\ *
 2.891325 &     7.3 &     (6.0) &    3.44 $\pm$    0.14 &  $<$    0.52 &  $<$    0.26 &    \nodata &    \nodata &  5  \\ *
 2.891551 &    11.1 &     (9.6) &    1.18 $\pm$    0.16 &    2.05 $\pm$    0.61 &    0.71 $\pm$    0.30 &    \nodata &    \nodata &  6  \\ *
 2.891850 &     6.5 &     (5.7) &    7.40 $\pm$    0.21 &    1.29 $\pm$    0.61 &  $<$    0.50 &    \nodata &    \nodata &  7  \\ *
 2.891949 &    22.7 &    (22.0) &   11.0 $\pm$    0.4\phm{0} &  $<$    1.22 &  $<$    0.62 &    \nodata &    \nodata &  8  \\ 
 2.893129 &    11.2 &     (9.8) &    2.38 $\pm$    0.16 &  $<$    0.67 &  $<$    0.40 &    \nodata &    \nodata &  9  \\ 
 2.893908 &     3.7 &     (2.4) &    0.44 $\pm$    0.10 &  $<$    0.43 &  $<$    0.32 &    \nodata &    \nodata & 10  \\ 
\sidehead{2.904\tablenotemark{d}}
 2.902425 &    20.3 &   \nodata &    7.52 $\pm$    0.31 &    \nodata &    \nodata &    \nodata &    \nodata &  1  \\ 
 2.902592 &     8.2 &     (6.8) &    4.42 $\pm$    0.28 &   16.7 $\pm$    1.0\phm{0} &  $<$    0.58 &    \nodata &    \nodata &  2  \\ 
 2.902788 &    16.0 &    (15.1) &   21.8 $\pm$    0.7\phm{0} &   61.2 $\pm$    2.3\phm{0} &  $<$    0.90 &    \nodata &    \nodata &  3  \\ 
 2.902947 &    25.5 &    (24.9) &   28.5 $\pm$    0.9\phm{0} &   20.2 $\pm$    2.9\phm{0} &    4.14 $\pm$    1.21 &    \nodata &    \nodata &  4  \\ 
 2.903233 &     4.3 &     (3.3) &    1.46 $\pm$    0.24 &    7.27 $\pm$    0.84 &  $<$    0.43 &    \nodata &    \nodata &  5  \\ 
 2.903325 &    20.2 &    (19.5) &   28.7 $\pm$    0.5\phm{0} &   52.0 $\pm$    1.8\phm{0} &    2.20 $\pm$    0.77 &    \nodata &    \nodata &  6  \\ 
 2.903622 &     7.1 &     (5.8) &    1.91 $\pm$    0.18 &  $<$    0.67 &  $<$    0.56 &    \nodata &    \nodata &  7  \\ 
 2.903881 &     7.3 &     (6.3) &    2.05 $\pm$    0.41 &   28.1 $\pm$    1.4\phm{0} &    1.81 $\pm$    1.20 &    \nodata &    \nodata &  8  \\ 
 2.903969 &    15.8 &    (14.8) &   18.8 $\pm$    0.7\phm{0} &   12.3 $\pm$    2.1\phm{0} &  $<$    2.37 &    \nodata &    \nodata &  9  \\ 
 2.904118 &     3.8 &     (2.9) &    5.56 $\pm$    0.32 &   34.9 $\pm$    1.5\phm{0} &  [486 $\pm$   57]\phm{0} &  524 $\pm$   12\phm{0} &    \nodata & 10  \\ 
 2.904254 &     7.0 &     (5.8) &   31.5 $\pm$    0.6\phm{0} &  330 $\pm$    8\phm{00} &   79.1 $\pm$   13.5 &  167 $\pm$    5\phm{00} &    \nodata & 11  \\ 
 2.904380 &     7.7 &     (6.3) &    4.38 $\pm$    1.30 &  160 $\pm$    8\phm{00} &  213 $\pm$   31\phm{0} &  492 $\pm$    7\phm{00} &    \nodata & 12  \\ 
 2.904546 &    18.4 &    (17.5) &   94.6 $\pm$    3.5\phm{0} &  432 $\pm$   14\phm{0} &   35.2 $\pm$   17.6 &   17.4 $\pm$    7.4\phm{0} &    \nodata & 13  \\ 
 2.904656 &     9.2 &     (7.4) &   24.4 $\pm$    3.5\phm{0} &  182 $\pm$   17\phm{0} &  $<$   16.10 &   37.3 $\pm$    7.2\phm{0} &    \nodata & 14  \\ 
 2.904704 &     5.6 &     (4.6) &  $<$    1.79 &   40.5 $\pm$    8.9\phm{0} &  117 $\pm$    8\phm{00} &  182 $\pm$    6\phm{00} &    \nodata & 15  \\ 
 2.904924 &     5.7 &     (4.7) &    3.31 $\pm$    0.55 &    7.83 $\pm$    1.90 &    2.35 $\pm$    1.13 &    5.14 $\pm$    1.26 &    \nodata & 16  \\ 
 2.904928 &    25.6 &    (25.0) &   16.3 $\pm$    1.6\phm{0} &   65.2 $\pm$    5.8\phm{0} &  $<$    3.54 &    \nodata &    \nodata & 17  \\ 
 2.905130 &     6.4 &     4.6\phm{)} &    9.26 $\pm$    0.44 &   75.9 $\pm$    2.0\phm{0} &   12.8 $\pm$    1.0\phm{0} &   15.7 $\pm$    1.1\phm{0} &    \nodata & 18  \\ 
 2.905290 &     9.3 &     (7.6) &    3.27 $\pm$    0.25 &   20.8 $\pm$    1.1\phm{0} &    4.14 $\pm$    0.55 &    8.57 $\pm$    0.85 &    \nodata & 19  \\ 
 2.905463 &     4.7 &     2.7\phm{)} &  $<$    0.26 &    1.77 $\pm$    0.77 &    5.71 $\pm$    0.47 &   13.8 $\pm$    1.0\phm{0} &    \nodata & 20  \\ 
 2.905486 &    19.3 &    (18.5) &    3.27 $\pm$    0.40 &   16.1 $\pm$    1.8\phm{0} &    2.03 $\pm$    0.92 &    \nodata &    \nodata & 21  \\ 
\sidehead{3.009}
 3.007389 &     8.7 &     (6.8) &    1.02 $\pm$    0.25 &  $<$    0.57 &  $<$    0.39 &    \nodata &    \nodata &  1  \\ 
 3.010072 &    13.3 &    (12.1) &    4.65 $\pm$    0.31 &  $<$    0.67 &  $<$    0.49 &    \nodata &    \nodata &  2  \\ 
\sidehead{3.013}
 3.012866 &    17.5 &    (16.6) &    0.83 $\pm$    0.34 &  $<$    0.77 &  $<$    0.49 &    \nodata &    \nodata &  1  \\ 
 3.013227 &    18.5 &    (17.6) &    2.33 $\pm$    0.43 &  $<$    0.91 &  $<$    0.61 &    \nodata &    \nodata &  2  \\ 
 3.013444 &    12.7 &    (11.5) &    9.83 $\pm$    0.38 &    5.91 $\pm$    0.73 &  $<$    0.53 &    \nodata &    \nodata &  3  \\ 
 3.013706 &     3.2 &     (2.5) &    0.82 $\pm$    0.17 &  $<$    0.36 &  $<$    0.26 &    \nodata &    \nodata &  4  \\ 
 3.013909 &     9.3 &     (7.5) &    1.95 $\pm$    0.24 &  $<$    0.49 &  $<$    0.34 &    \nodata &    \nodata &  5  \\ 
\sidehead{3.018}
 3.017206 &     6.2 &     (5.4) &   22.1 $\pm$    0.4\phm{0} &    5.71 $\pm$    0.58 &    0.40 $\pm$    0.31 &    \nodata &    \nodata &  1  \\ 
 3.017367 &     6.4 &     (5.2) &   18.0 $\pm$    0.4\phm{0} &    2.96 $\pm$    0.57 &  $<$    0.37 &    \nodata &    \nodata &  2  \\ 
 3.017513 &     7.5 &     (5.1) &    3.51 $\pm$    0.26 &  $<$    0.62 &  $<$    0.34 &    \nodata &    \nodata &  3  \\ 
 3.017851 &     2.6 &     (1.7) &    0.79 $\pm$    0.23 &  $<$    0.44 &  $<$    0.24 &    \nodata &    \nodata &  4  \\ 
\sidehead{3.111\tablenotemark{e}}
 3.111291 &    10.3 &     (8.7) &    2.34 $\pm$    0.15 &  $<$    0.52 &  $<$    0.76 &  $<$    0.22 &  $<$    0.54 &  1  \\ 
\enddata

\tablenotetext{*}{See Figure 1.}

\tablenotetext{a}{In the strong, narrow components 10 and 11 \SiIV~$\lambda$1394, \CII\ and \SiII~$\lambda$1260, all in the forest, are 
relatively free of contamination and well-measured; additionally \SiII~$\lambda$1527, outside the forest, is used to supplement 
$\lambda$1260 in these components but elsewhere is too weak to give useful values.}

\tablenotetext{b}{\SiIV~$\lambda$1394 is blended with \CIV~$\lambda$1548 at $z = 2.311$ and overall is rather noisy, but $\lambda$1403 
is clear.}

\tablenotetext{c}{\SiIV~$\lambda$1394 in components 1 and 2 is blended with stronger \CIV~$\lambda$1551 at $z = 2.474$ but $\lambda$1403 
is clear. \CII\ is outside the forest.}

\tablenotetext{d}{This is a highly complex system with strong \CII\ and \SiII\ and relatively weak \CIV\ in components 10, 12 and 15: 
From the extent of the \HI\ Lyman $\alpha$ wings associated with these the limit on the combined column densities is 
log~$N$(\HI) $\lesssim 18.7$. In the broad component 1 \SiIV\ and \CII\ are weak and too uncertain to include. \CII\ in component 10 is 
close to saturation and the value (in square brackets) is approximate only. \SiII~$\lambda$1260 is in the forest but onward from component 
10 is relatively clear of contamination and is used together with $\lambda$1304 which is outside and relatively strong except in components 
1--9 (upper limits from the latter are too high to be useful). For component 20 $b$(C) comes from \CII.}

\tablenotetext{e}{\SiII~$\lambda$1260 is outside the forest. \NV\ is outside the forest; $\lambda$1239 is blended with strong 
\SiII~$\lambda$1304 at $z = 2.904$ and the upper limit is from $\lambda$1243.}

\end{deluxetable}

\clearpage
\begin{deluxetable}{crrcccccc}
\tabletypesize{\scriptsize}
\tablecaption{Heavy Element Absorption Lines: Q1425+6039 $z_{em}=3.18^*$}
\tablewidth{0pt}
\tablecolumns{9}
\tablehead{
\colhead{$z$} & \colhead{$b$(C)}  & \colhead{$b$(Si)} &
\colhead{$N$(\CIV)} & \colhead{$N$(\SiIV)} &
\colhead{$N$(\CII)} & \colhead{$N$(\SiII)} &
\colhead{$N$(\NV)} &  \colhead{n} \\
{} & \colhead{km~s$^{-1}$} & \colhead{km~s$^{-1}$} & \colhead{10$^{12}$cm$^{-2}$} &
\colhead{10$^{11}$cm$^{-2}$} & \colhead{10$^{12}$cm$^{-2}$} & \colhead{10$^{11}$cm$^{-2}$} & \colhead{10$^{12}$cm$^{-2}$} & {} \\
\colhead{1} & \colhead{2} & \colhead{3} & \colhead{4} & \colhead{5} & \colhead{6} & \colhead{7} & \colhead{8} & \colhead{9} 
}

\startdata
\sidehead{2.343}
 2.342638 &     9.9 &   \nodata &    4.83 $\pm$    0.07 &    \nodata &    \nodata &    \nodata &    \nodata &  1  \\ 
 2.342805 &    10.1 &   \nodata &    5.57 $\pm$    0.07 &    \nodata &    \nodata &    \nodata &    \nodata &  2  \\ 
 2.343092 &     8.3 &   \nodata &    2.50 $\pm$    0.06 &    \nodata &    \nodata &    \nodata &    \nodata &  3  \\ 
\sidehead{2.476}
 2.475425 &    13.2 &   \nodata &    2.16 $\pm$    0.12 &    \nodata &    \nodata &    \nodata &    \nodata &  1  \\ 
 2.475714 &    12.9 &   \nodata &    4.87 $\pm$    0.13 &    \nodata &    \nodata &    \nodata &    \nodata &  2  \\ 
 2.475989 &     7.4 &   \nodata &    8.01 $\pm$    0.12 &    \nodata &    \nodata &    \nodata &    \nodata &  3  \\ 
 2.476169 &     3.9 &   \nodata &    0.38 $\pm$    0.08 &    \nodata &    \nodata &    \nodata &    \nodata &  4  \\ 
 2.476537 &     2.1 &   \nodata &    0.24 $\pm$    0.07 &    \nodata &    \nodata &    \nodata &    \nodata &  5  \\ 
\sidehead{2.486}
 2.485329 &    11.7 &   \nodata &    2.04 $\pm$    0.10 &    \nodata &    \nodata &    \nodata &    \nodata &  1  \\ 
 2.485620 &    10.4 &   \nodata &    4.17 $\pm$    0.10 &    \nodata &    \nodata &    \nodata &    \nodata &  2  \\ 
 2.486001 &     8.1 &   \nodata &    2.27 $\pm$    0.09 &    \nodata &    \nodata &    \nodata &    \nodata &  3  \\ 
 2.486212 &     9.9 &   \nodata &    7.76 $\pm$    0.11 &    \nodata &    \nodata &    \nodata &    \nodata &  4  \\ 
 2.486445 &    10.6 &   \nodata &   10.5 $\pm$    0.1\phm{0} &    \nodata &    \nodata &    \nodata &    \nodata &  5  \\ 
 2.486614 &     7.8 &   \nodata &    9.60 $\pm$    0.12 &    \nodata &    \nodata &    \nodata &    \nodata &  6  \\ 
 2.486826 &    10.4 &   \nodata &   18.8 $\pm$    0.2\phm{0} &    \nodata &    \nodata &    \nodata &    \nodata &  7  \\ 
 2.487025 &    12.9 &   \nodata &   12.6 $\pm$    0.2\phm{0} &    \nodata &    \nodata &    \nodata &    \nodata &  8  \\ 
 2.487356 &    29.2 &   \nodata &    0.99 $\pm$    0.17 &    \nodata &    \nodata &    \nodata &    \nodata &  9  \\ 
\sidehead{2.513}
 2.513429 &     8.3 &   \nodata &    1.29 $\pm$    0.11 &    \nodata &    \nodata &    \nodata &    \nodata &  1  \\ 
\sidehead{2.606}
 2.605723 &     7.4 &   \nodata &    1.13 $\pm$    0.08 &    \nodata &    \nodata &    \nodata &    \nodata &  1  \\ 
\sidehead{2.624}
 2.624426 &    16.3 &   \nodata &    1.77 $\pm$    0.16 &    \nodata &    \nodata &    \nodata &    \nodata &  1  \\ 
\sidehead{2.646\tablenotemark{a}}
 2.645649 &    12.5 &   (11.2) &    1.13 $\pm$    0.11 &  $<$    0.58 &    \nodata &    \nodata &    \nodata &  1  \\ 
 2.646243 &    23.5 &   (22.8) &    1.03 $\pm$    0.16 &  $<$    1.52 &    \nodata &    \nodata &    \nodata &  2  \\ 
 2.646535 &    11.8 &   (10.4) &    4.08 $\pm$    0.12 &  $<$    0.61 &    \nodata &    \nodata &    \nodata &  3  \\ 
\sidehead{2.702\tablenotemark{b}}
 2.702156 &    20.9 &   (20.1) &    0.99 $\pm$    0.14 &  $<$    0.51 &    \nodata &    \nodata &    \nodata &  1  \\ 
 2.702441 &     8.0 &    (6.6) &    0.69 $\pm$    0.09 &    0.75 $\pm$    0.24 &    \nodata &    \nodata &    \nodata &  2  \\ 
\sidehead{2.726\tablenotemark{c}}
 2.726061 &    24.1 &   (23.5) &    2.98 $\pm$    0.17 &  $<$    0.75 &    \nodata &    \nodata &    \nodata &  1  \\ 
 2.726257 &    11.1 &    (9.7) &    3.07 $\pm$    0.12 &    3.71 $\pm$    0.37 &    \nodata &    \nodata &    \nodata &  2  \\ 
\sidehead{2.770\tablenotemark{d}}
 2.768764 &    14.6 &   (13.6) &    0.70 $\pm$    0.12 &    1.82 $\pm$    0.40 &    \nodata &  $<$    1.45 &    \nodata &  1  \\ 
 2.769092 &     8.4 &    (6.9) &    2.44 $\pm$    0.10 &    0.89 $\pm$    0.29 &    \nodata &  $<$    1.18 &    \nodata &  2  \\ 
 2.769310 &     7.3 &    (6.1) &    0.66 $\pm$    0.09 &    5.28 $\pm$    0.30 &    \nodata &    2.76 $\pm$    1.21 &    \nodata &  3  \\ 
 2.769581 &    16.9 &   (15.9) &    4.45 $\pm$    0.16 &    2.24 $\pm$    0.50 &    \nodata &   10.1 $\pm$    2.2\phm{0} &    \nodata &  4  \\ 
 2.769745 &     5.6 &    (4.7) &    2.13 $\pm$    0.11 &   42.4 $\pm$    0.5\phm{0} &    \nodata &  608 $\pm$    9\phm{00} &    \nodata &  5  \\ 
 2.769848 &     3.3 &    (2.2) &    0.57 $\pm$    0.13 &    8.31 $\pm$    0.43 &    \nodata &  548 $\pm$   54\phm{0} &    \nodata &  6  \\ 
 2.769925 &     7.2 &    (6.0) &    1.67 $\pm$    0.19 &   16.8 $\pm$    0.6\phm{0} &    \nodata &  459 $\pm$    10\phm{0} &    \nodata &  7  \\ 
 2.770007 &     3.7 &    (2.7) &    0.39 $\pm$    0.13 &    8.87 $\pm$    0.43 &    \nodata &  275 $\pm$   12\phm{0} &    \nodata &  8  \\ 
 2.770100 &     6.0 &    (4.9) &    0.94 $\pm$    0.12 &    8.21 $\pm$    0.38 &    \nodata &  192 $\pm$    3\phm{00} &    \nodata &  9  \\ 
 2.770230 &    30.1 &   (29.6) &    1.62 $\pm$    0.22 &  $<$    0.74 &    \nodata &  $<$    1.74 &    \nodata & 10  \\ 
 2.771032 &     6.9 &    (5.6) &    0.91 $\pm$    0.08 &    7.66 $\pm$    0.30 &    \nodata &    3.21 $\pm$    1.13 &    \nodata & 11  \\ 
\sidehead{2.773\tablenotemark{e}}
 2.772514 &    11.5 &   (10.1) &    3.31 $\pm$    0.25 &  $<$    0.31 &    \nodata &    \nodata &    \nodata &  1  \\ 
 2.772704 &     6.8 &    (3.9) &  [494 $\pm$    9]\phm{00} &    5.72 $\pm$    0.27 &    0.88 $\pm$    0.14 &    \nodata &  121 $\pm$    2\phm{00} &  2  \\ 
 2.772841 &     4.0 &    (2.7) &   56.8 $\pm$    1.6\phm{0} &    3.53 $\pm$    0.47 &    0.57 $\pm$    0.15 &     \nodata &   11.1 $\pm$    0.8\phm{0} &  3  \\ 
 2.772871 &     9.6 &    (7.9) &    2.72 $\pm$    0.47 &  $<$    0.68 &    \nodata &    \nodata &    \nodata &  4  \\ 
 2.773405 &    24.0 &   (23.3) &    2.86 $\pm$    0.24 &  $<$    0.45 &    \nodata &    \nodata &    \nodata &  5  \\ 
 2.774060 &    15.6 &   (14.6) &    1.14 $\pm$    0.19 &  $<$    0.36 &    \nodata &    \nodata &    \nodata &  6  \\ 
\sidehead{2.796}
 2.795950 &     5.6 &    (4.7) &    2.22 $\pm$    0.12 &  $<$    0.39 &    \nodata &    \nodata &    \nodata &  1  \\ *
 2.796160 &    10.2 &    (8.6) &    1.72 $\pm$    0.14 &  $<$    0.44 &    \nodata &    \nodata &    \nodata &  2  \\ 
\sidehead{2.827\tablenotemark{f}}
 2.825061 &     9.0 &    (7.1) &    1.01 $\pm$    0.12 &    1.29 $\pm$    0.45 &    2.15 $\pm$    0.14 &    2.04 $\pm$    1.19 &    \nodata &  1  \\ 
 2.825309 &     7.6 &    (6.1) &    2.12 $\pm$    0.14 &  $<$    0.45 &    9.82 $\pm$    0.20 &   12.6 $\pm$    0.4\phm{0} &    \nodata &  2  \\ 
 2.825426 &     4.2 &    (3.2) &   25.2 $\pm$    0.4\phm{0} &    4.32 $\pm$    0.59 &   20.1 $\pm$    0.4\phm{0} &   33.1 $\pm$    1.0\phm{0} &    \nodata &  3  \\ 
 2.825511 &     3.6 &    (2.5) &   27.7 $\pm$    0.7\phm{0} &   16.5 $\pm$    0.9\phm{0} &   27.8 $\pm$    0.8\phm{0} &    8.53 $\pm$    2.36 &    \nodata &  4  \\ 
 2.825598 &     5.5 &    (4.4) &   18.5 $\pm$    0.3\phm{0} &   15.0 $\pm$    0.9\phm{0} &   82.5 $\pm$    2.1\phm{0} &  539 $\pm$   67\phm{0} &    \nodata &  5  \\ 
 2.825740 &     7.0 &    (5.7) &    6.88 $\pm$    0.72 &    7.23 $\pm$    1.11 &  104 $\pm$    5\phm{00} &  166 $\pm$   14\phm{0} &    \nodata &  6  \\ 
 2.825808 &     (11) &    9.5\phm{)} &  $<$    1.94 &   25.0 $\pm$    1.6\phm{0} &  $<$    7.48 &  $<$   13.9 &    \nodata &  7  \\ 
 2.825894 &     5.8 &    (4.9) &    2.44 $\pm$    0.61 &  $<$    1.59 &   51.4 $\pm$    3.5\phm{0} &   97.7 $\pm$    7.6\phm{0} &    \nodata &  8  \\ 
 2.826001 &    13.8 &   (12.6) &   26.4 $\pm$    0.6\phm{0} &   78.5 $\pm$    1.6\phm{0} &   46.5 $\pm$    3.2\phm{0} &   22.7 $\pm$    2.4\phm{0} &    \nodata &  9  \\ 
 2.826111 &     5.7 &    (5.7) &  $<$    0.37 &  $<$    1.98 &   18.0 $\pm$    7.1\phm{0} &   52.1 $\pm$    3.1\phm{0} &    \nodata & 10  \\ 
 2.826217 &     9.6 &    (7.8) &    5.74 $\pm$    0.55 &    9.33 $\pm$    1.63 &   99.6 $\pm$   29.4 &   41.3 $\pm$    6.2\phm{0} &    \nodata & 11  \\ 
 2.826301 &     9.7 &    (8.0) &    9.87 $\pm$    0.49 &   26.8 $\pm$    1.4\phm{0} &   83.1 $\pm$   68.3 &  238 $\pm$    9\phm{00} &    \nodata & 12  \\ 
 2.826444 &     5.2 &    (3.8) &    3.16 $\pm$    0.29 &    5.30 $\pm$    0.77 &    \nodata &  348 $\pm$   30\phm{0} &    \nodata & 13  \\ 
 2.826544 &     7.7 &    (6.1) &    7.23 $\pm$    0.62 &   19.1 $\pm$    1.9\phm{0} &    \nodata &    \nodata &    \nodata & 14  \\ 
 2.826621 &    (11) &    9.5\phm{)} &  $<$    1.90 &    9.40 $\pm$    5.82 &    \nodata &    \nodata &    \nodata & 15  \\ 
 2.826708 &    15.5 &   (14.5) &    8.70 $\pm$    1.98 &   32.7 $\pm$    8.4\phm{0} &    \nodata &    \nodata &    \nodata & 16  \\ 
 2.826794 &    22.1 &   (21.4) &   40.0 $\pm$    3.0\phm{0} &   67.3 $\pm$    9.5\phm{0} &    \nodata &    \nodata &    \nodata & 17  \\ 
 2.826844 &    (13) &   12.0\phm{)} &  $<$    2.12 &   41.4 $\pm$    6.2\phm{0} &    \nodata &    \nodata &    \nodata & 18  \\ 
 2.827186 &     4.6 &    (3.7) &    0.65 $\pm$    0.35 &  $<$    1.03 &  183 $\pm$   40\phm{0} &  298 $\pm$   12\phm{0} &    \nodata & 19  \\ 
 2.827251 &    16.9 &   (16.0) &   23.6 $\pm$    1.1\phm{0} &   41.3 $\pm$    3.3\phm{0} &  $<$   26.8 &   37.5 $\pm$   14.7 &    \nodata & 20  \\ 
 2.827372 &     7.3 &    (6.0) &    3.49 $\pm$    0.48 &   12.8 $\pm$    1.4\phm{0} &  125 $\pm$    8\phm{00} &  201 $\pm$    6\phm{00} &    \nodata & 21  \\ 
 2.827548 &    14.2 &   (13.1) &    3.26 $\pm$    0.46 &  $<$    1.83 &    4.31 $\pm$    2.94 &   23.9 $\pm$    5.9\phm{0} &    \nodata & 22  \\ 
 2.827623 &     8.1 &    (6.5) &  $<$    0.38 &    3.94 $\pm$    1.02 &   99.5 $\pm$    1.8\phm{0} &  178 $\pm$    4\phm{00} &    \nodata & 23  \\ 
 2.827797 &     9.3 &    (7.6) &    1.71 $\pm$    0.19 &  $<$    0.64 &   31.6 $\pm$    0.5\phm{0} &   54.5 $\pm$    2.9\phm{0} &    \nodata & 24  \\ 
 2.827953 &    12.9 &   (11.7) &    5.12 $\pm$    0.36 &   11.8 $\pm$    1.3\phm{0} &    0.78 $\pm$    0.48 &  $<$    4.31 &    \nodata & 25  \\ 
 2.828063 &    16.4 &   (15.5) &    2.67 $\pm$    0.34 &  $<$    1.32 &    1.17 $\pm$    0.40 &    \nodata &    \nodata & 26  \\ 
\sidehead{2.830\tablenotemark{g}}
 2.828528 &    13.7 &   (12.6) &    2.44 $\pm$    0.14 &    2.93 $\pm$    0.49 &    0.29 $\pm$    0.14 &    \nodata &    \nodata &  1  \\*
 2.828808 &    11.0 &    (9.5) &    2.45 $\pm$    0.16 &    3.02 $\pm$    0.55 &    1.99 $\pm$    0.14 &    \nodata &    \nodata &  2  \\
 2.829030 &     3.3 &    (2.2) &  $<$    0.13 &    0.57 $\pm$    0.40 &    0.92 $\pm$    0.11 &    \nodata &    \nodata &  3  \\ 
 2.829348 &    15.2 &   (14.2) &    4.94 $\pm$    1.14 &   32.8 $\pm$    2.8\phm{0} &   18.3 $\pm$    0.8\phm{0} &   27.0 $\pm$    4.9\phm{0} &    \nodata &  4  \\ 
 2.829359 &     3.3 &    (2.2) &    1.42 $\pm$    0.17 &    7.41 $\pm$    0.64 &    5.28 $\pm$    0.19 &    7.70 $\pm$    1.56 &    \nodata &  5  \\ 
 2.829504 &    37.2 &   (36.8) &   35.8 $\pm$    2.4\phm{0} &  $<$    5.73 &  $<$    1.62 &    \nodata &    \nodata &  6  \\ 
 2.829548 &     5.9 &    (4.8) &   25.9 $\pm$    0.6\phm{0} &   51.8 $\pm$    1.6\phm{0} &    0.39 $\pm$    0.20 &    \nodata &    \nodata &  7  \\ 
 2.829648 &     7.2 &    (5.8) &   75.9 $\pm$    3.8\phm{0} &  168 $\pm$    3\phm{00} &   20.4 $\pm$    0.4\phm{0} &   22.7 $\pm$    2.0\phm{0} &    \nodata &  8  \\ 
 2.829749 &     5.2 &    (4.2) &   12.7 $\pm$    1.4\phm{0} &   93.7 $\pm$    3.6\phm{0} &   18.0 $\pm$    0.4\phm{0} &   27.9 $\pm$    2.1\phm{0} &    \nodata &  9  \\ 
 2.829807 &     8.1 &    (6.7) &   44.7 $\pm$    1.1\phm{0} &   66.9 $\pm$    2.3\phm{0} &    3.24 $\pm$    0.47 &  $<$    2.97 &    \nodata & 10  \\ 
 2.829898 &    14.3 &   (13.2) &    1.34 $\pm$    0.61 &   11.9 $\pm$    1.5\phm{0} &    4.76 $\pm$    0.45 &   12.6 $\pm$    3.3\phm{0} &    \nodata & 11  \\ 
 2.830133 &    10.2 &    (8.7) &    5.02 $\pm$    0.44 &   14.3 $\pm$    1.0\phm{0} &    6.05 $\pm$    0.37 &    7.85 $\pm$    2.40 &    \nodata & 12  \\ 
 2.830312 &    31.3 &   (30.8) &   55.6 $\pm$    1.1\phm{0} &   27.6 $\pm$    2.6\phm{0} &    4.61 $\pm$    0.98 &  $<$    5.32 &    \nodata & 13  \\ 
 2.830398 &     4.1 &    (3.3) &   21.5 $\pm$    0.8\phm{0} &   60.8 $\pm$    1.9\phm{0} &   44.8 $\pm$    1.0\phm{0} &   74.0 $\pm$    1.9\phm{0} &    \nodata & 14  \\ 
 2.830560 &     4.7 &    (3.5) &   19.0 $\pm$    1.7\phm{0} &   82.1 $\pm$    6.1\phm{0} &   43.4 $\pm$    2.5\phm{0} &  280 $\pm$    7\phm{00} &    \nodata & 15  \\ 
 2.830587 &    12.6 &   (11.3) &   41.8 $\pm$    2.6\phm{0} &  185 $\pm$    7\phm{00} &   20.6 $\pm$    6.0\phm{0} &  $<$    7.62 &    \nodata & 16  \\ 
 2.830655 &     4.5 &    (3.5) &    1.60 $\pm$    0.49 &   11.1 $\pm$    1.6\phm{0} &   26.3 $\pm$    1.1\phm{0} &   36.7 $\pm$    3.0\phm{0} &    \nodata & 17  \\ 
 2.830710 &     5.9 &    (5.0) &    3.29 $\pm$    0.62 &    3.95 $\pm$    1.46 &    1.64 $\pm$    1.12 &   \nodata &    \nodata & 18  \\ 
 2.830834 &    14.9 &   (13.8) &    2.10 $\pm$    0.23 &   11.0 $\pm$    0.8\phm{0} &    4.61 $\pm$    0.30 &   16.6 $\pm$    2.8\phm{0} &    \nodata & 19  \\ 
\sidehead{2.950}
 2.947815 &    11.9 &   (10.6) &    4.35 $\pm$    0.16 &  $<$    0.43 &  $<$    0.20 &    \nodata &    \nodata &  1  \\ 
 2.948133 &    13.6 &   (12.4) &    4.56 $\pm$    0.17 &    0.78 $\pm$    0.46 &  $<$    0.21 &    \nodata &    \nodata &  2  \\ 
 2.949936 &    10.8 &    (9.3) &    2.18 $\pm$    0.12 &  $<$    0.58 &  $<$    0.23 &    \nodata &    \nodata &  3  \\ 
 2.950319 &    14.5 &   (13.4) &    7.64 $\pm$    0.21 &    1.92 $\pm$    0.70 &  $<$    0.48 &    \nodata &    \nodata &  4  \\ 
 2.950432 &     6.8 &     5.0\phm{)} &    4.94 $\pm$    0.17 &    5.36 $\pm$    0.61 &  $<$    0.27 &    \nodata &    \nodata &  5  \\ 
 2.950655 &    11.1 &    (9.6) &   16.2 $\pm$    0.2\phm{0} &    5.53 $\pm$    0.42 &  $<$    0.24 &    \nodata &    \nodata &  6  \\ 
 2.951177 &    19.1 &   (18.3) &    0.74 $\pm$    0.14 &  $<$    0.59 &  $<$    0.28 &    \nodata &    \nodata &  7  \\ 
\sidehead{2.964\tablenotemark{h}}
 2.963274 &    14.4 &   (13.3) &    2.66 $\pm$    0.15 &  $<$    0.75 &  $<$    0.25 &    \nodata &    \nodata &  1  \\ *
 2.963415 &     8.9 &    (7.1) &    1.98 $\pm$    0.12 &    0.88 $\pm$    0.53 &  $<$    0.21 &    \nodata &    \nodata &  2  \\ *
 2.963608 &     7.6 &    (6.1) &    1.14 $\pm$    0.09 &  $<$    0.39 &  $<$    0.15 &    \nodata &    \nodata &  3  \\ 
 2.963946 &    13.0 &   (11.8) &    3.89 $\pm$    0.15 &    1.68 $\pm$    0.65 &  $<$    0.27 &    \nodata &    \nodata &  4  \\ 
 2.964076 &     8.8 &    (7.0) &    7.26 $\pm$    0.14 &    7.89 $\pm$    0.54 &    0.36 $\pm$    0.21 &    \nodata &    \nodata &  5  \\ 
\sidehead{3.025\tablenotemark{i}}
 3.024600 &     7.2 &    (6.0) &    0.66 $\pm$    0.13 &  $<$    0.54 &  $<$    0.36 &    \nodata &    \nodata &  1  \\ 
 3.024766 &     4.1 &    (2.7) &    0.62 $\pm$    0.11 &  $<$    0.35 &  $<$    0.36 &    \nodata &    \nodata &  2  \\ 
\sidehead{3.056\tablenotemark{j}}
 3.055455 &    23.9 &   (23.2) &    1.87 $\pm$    0.23 &  $<$    0.56 &  $<$    0.31 &    \nodata &    \nodata &  1  \\ 
 3.055859 &    16.4 &   (15.4) &    1.79 $\pm$    0.20 &  $<$    0.51 &    \nodata &    \nodata &    \nodata &  2  \\ 
 3.056111 &     6.0 &    (4.9) &    0.74 $\pm$    0.12 &  $<$    0.56 &  $<$    0.20 &    \nodata &    \nodata &  3  \\ 
\sidehead{3.067\tablenotemark{k}}
 3.066603 &    10.5 &    (8.9) &    0.97 $\pm$    0.20 &  $<$    0.34 &  $<$    0.19 &  $<$    0.15 &    \nodata &  1  \\ 
 3.066946 &    10.4 &    (8.9) &    6.69 $\pm$    0.26 &    8.61 $\pm$    0.39 &  $<$    0.23 &  $<$    0.21 &  $<$    0.43 &  2  \\ 
 3.067098 &     6.0 &     4.9\phm{)} &    3.33 $\pm$    0.22 &   15.1 $\pm$    0.4\phm{0} &  $<$    0.31 &  $<$    0.20 &  $<$    0.22 &  3  \\ 
\sidehead{3.074\tablenotemark{l}}
 3.073964 &     7.1 &    (5.8) &    2.60 $\pm$    0.17 &  $<$    0.55 &  $<$    0.24 &  $<$    0.23 &    \nodata &  1  \\ 
\sidehead{3.096\tablenotemark{m}}
 3.095166 &    10.9 &    (9.5) &    2.33 $\pm$    0.14 &  $<$    0.39 &  $<$    0.17 &  $<$    0.11 &    0.57 $\pm$    0.17 &  1  \\ 
 3.095394 &     6.3 &    (3.7) &    7.53 $\pm$    0.13 &   10.3 $\pm$    0.4\phm{0} &  $<$    0.15 &  $<$    0.09 &    0.43 $\pm$    0.14 &  2  \\ 
 3.095729 &    10.6 &    (9.1) &    1.31 $\pm$    0.13 &    0.60 $\pm$    0.42 &  $<$    0.35 &  $<$    0.12 &  $<$    0.32 &  3  \\ 
 3.095914 &     5.3 &    (4.4) &    1.46 $\pm$    0.11 &    0.87 $\pm$    0.32 &  $<$    0.14 &  $<$    0.09 &  $<$    0.22 &  4  \\ 
\sidehead{3.135\tablenotemark{n}}
 3.133446 &     4.8 &    (3.8) &    3.18 $\pm$    0.34 &    4.58 $\pm$    0.26 &  $<$    0.23 &  $<$    0.23 &    0.26 $\pm$    0.09 &  1  \\ 
 3.133621 &     9.9 &    (8.2) &    9.22 $\pm$    0.49 &   11.7 $\pm$    0.4\phm{0} &  $<$    0.48 &    0.47 $\pm$    0.18 &  $<$    0.13 &  2  \\ 
 3.133905 &    19.4 &   (18.6) &    6.50 $\pm$    0.62 &    4.93 $\pm$    0.50 &  $<$    0.42 &  $<$    0.26 &  $<$    0.36 &  3  \\ 
 3.134173 &     4.1 &    (3.0) &    1.08 $\pm$    0.31 &  $<$    0.25 &  $<$    0.27 &  $<$    0.12 &  $<$    0.09 &  4  \\ 
 3.134434 &    20.1 &   (19.4) &    3.35 $\pm$    0.29 &  $<$    0.48 &  $<$    0.38 &  $<$    0.24 &  $<$    0.18 &  5  \\ 
 3.135223 &     8.2 &    (6.7) &    0.89 $\pm$    0.16 &    0.75 $\pm$    0.29 &  $<$    0.25 &  $<$    0.26 &  $<$    0.09 &  6  \\ 
 3.135488 &     7.8 &    (7.1) &    2.06 $\pm$    0.18 &    7.12 $\pm$    0.33 &  $<$    0.37 &  $<$    0.18 &  $<$    0.10 &  7  \\ 
 3.135658 &     8.4 &    (6.9) &    3.96 $\pm$    0.22 &    3.26 $\pm$    0.36 &  $<$    0.29 &  $<$    0.22 &  $<$    0.22 &  8  \\ 
 3.135817 &     5.7 &    (4.5) &    0.85 $\pm$    0.26 &    5.88 $\pm$    0.44 &  $<$    0.35 &  $<$    0.26 &  $<$    0.15 &  9  \\ 
 3.135860 &    16.2 &   (15.3) &    7.18 $\pm$    0.41 &  $<$    0.74 &  $<$    0.57 &  $<$    0.46 &  $<$    0.23 & 10  \\ 
\sidehead{3.155\tablenotemark{o}}
 3.152956 &    14.6 &   (13.5) &    3.37 $\pm$    0.22 &  $<$    0.86 &  $<$    0.29 &  $<$    0.19 &  $<$    0.15 &  1  \\ 
 3.153323 &    11.7 &   (10.3) &    2.45 $\pm$    0.19 &  $<$    0.47 &  $<$    0.25 &  $<$    0.17 &    0.25 $\pm$    0.09 &  2  \\ 
 3.153658 &     4.1 &    (3.1) &    0.58 $\pm$    0.13 &  $<$    0.29 &  $<$    0.15 &  $<$    0.12 &  $<$    0.06 &  3  \\ 
 3.154100 &    14.9 &   (13.8) &    4.14 $\pm$    0.21 &    0.85 $\pm$    0.49 &  $<$    0.27 &  $<$    0.19 &  $<$    0.10 &  4  \\ 
 3.154541 &    12.2 &   (10.9) &    1.11 $\pm$    0.19 &    0.74 $\pm$    0.44 &  $<$    0.23 &  $<$    0.18 &  $<$    0.09 &  5  \\ 
 3.154991 &    13.6 &   (12.4) &    0.63 $\pm$    0.19 &  $<$    0.46 &  $<$    0.25 &  $<$    0.42 &  $<$    0.09 &  6  \\ 
 3.156022 &    27.4 &   (26.8) &    2.63 $\pm$    0.31 &  $<$    0.97 &  $<$    0.32 &  $<$    0.27 &    0.39 $\pm$    0.15 &  7  \\ 
\sidehead{3.174\tablenotemark{p}}
 3.172765 &     7.0 &    (5.6) &    0.34 $\pm$    0.09 &  $<$    0.40 &  $<$    0.13 &  $<$    0.12 &  $<$    0.07 &  1  \\ 
 3.173304 &     7.0 &    (5.7) &    0.59 $\pm$    0.09 &  $<$    0.43 &  $<$    0.18 &  $<$    0.12 &  $<$    0.07 &  2  \\ 
 3.174358 &     7.5 &    (6.1) &    0.40 $\pm$    0.10 &  $<$    0.28 &  $<$    0.13 &  $<$    0.13 &  $<$    0.07 &  3  \\ 
 3.174725 &     7.9 &    (6.3) &    1.18 $\pm$    0.10 &    0.78 $\pm$    0.28 &  $<$    0.14 &  $<$    0.13 &  $<$    0.12 &  4  \\ 
 3.175005 &     8.8 &     6.1\phm{)} &    3.17 $\pm$    0.11 &    3.05 $\pm$    0.49 &    0.49 $\pm$    0.15 &  $<$    0.13 &  $<$    0.09 &  5  \\ 
 3.175337 &    14.1 &   (13.0) &   14.3 $\pm$    0.3\phm{0} &   10.7 $\pm$    0.7\phm{0} &  $<$    0.33 &  $<$    0.29 &    0.57 $\pm$    0.17 &  6  \\ 
 3.175407 &     6.5 &     5.3\phm{)} &   11.9 $\pm$    0.2\phm{0} &   29.1 $\pm$    0.5\phm{0} &    1.08 $\pm$    0.23 &    0.77 $\pm$    0.21 &  $<$    0.12 &  7  \\ 
 3.175578 &     7.6 &     6.6\phm{)} &    6.02 $\pm$    0.13 &   14.9 $\pm$    0.3\phm{0} &    0.62 $\pm$    0.15 &    0.42 $\pm$    0.14 &  $<$    0.08 &  8  \\ 
 3.175781 &     3.7 &    (2.4) &    0.49 $\pm$    0.08 &    0.68 $\pm$    0.23 &  $<$    0.12 &  $<$    0.10 &  $<$    0.06 &  9  \\ 
\enddata

\tablenotetext{*}{See Figure 1.}

\tablenotetext{a}{\CIV~$\lambda$1548 is blended with very weak \SiIV~$\lambda$1403 at $z = 3.025$ and $\lambda$1551 with very weak 
\SiIV~$\lambda$1394 at $z = 3.056$, and are not significantly affected. \SiIV\ $\lambda$1394 is in the forest and the values are from 
$\lambda$1403 alone.}

\tablenotetext{b}{\SiIV~$\lambda$1403 is blended with \SiIV~$\lambda$1394 at $z = 2.726$ but $\lambda$1394 is clear.}

\tablenotetext{c}{\SiIV~$\lambda$1394 is blended with \SiIV~$\lambda$1403 at $z = 2.702$ but $\lambda$1403 is clear.}

\tablenotetext{d}{Spanning components 5--9 there is very strong \SiII\ and weak related \CIV: for these collectively a good fit to the 
\HI\ Lyman $\alpha$ damping wings is given by log~$N$(\HI) $\sim 19.4$. \CIV~$\lambda$1551 is blended with strong \SiII~$\lambda$1527 at 
$z = 2.830$ and is not used but $\lambda$1548 is clear. \SiII~$\lambda$1260 is in the forest and the values are from $\lambda$1527.}

\tablenotetext{e}{This system contains a region of unusually high ionization. \CIV~$\lambda$1548 is blended with strong \SiII~$\lambda$1527 
at $z = 2.827$ but $\lambda$1551 is clear; in component 2 both members of the doublet are saturated and the value (in square brackets) is 
highly approximate. \CII\ is in the forest: the values for components 2 and 3 are extracted from a complex blend with \SiIII~$\lambda$1207 
at $z = 3.174$ and \HI\ Lyman $\alpha$ but are useful. \NV\ is in the forest but gives good values for the two strong components; these 
appear slightly broader than is consistent with \CIV.}

\tablenotetext{f}{This is a highly complex system with very strong \CII\ and \SiII\ and relatively weak related \CIV\ centred on components 
13--19, as well as having several other distributed low ionization components: for these collectively a reasonable fit to the \HI\ 
Lyman $\alpha$ damping wings is given by log~$N$(\HI) $\sim 20.3$, reaching the ``damped Lyman $\alpha$'' classification. \CII\ is outside 
the forest. \SiII\ is obtained from appropriate sections of $\lambda$1527, which is partially blended with strong \CIV~$\lambda$1548 
at $z = 2.773$, and of $\lambda$1260 and $\lambda$1190, which are in the forest but useful in significant parts. \CII\ is heavily saturated 
in components 13--18, and \SiII~$\lambda$1527 in 14--18, and values for these are not included.}

\tablenotetext{g}{This is a complex system. \CII\ apparently requires an additional strong and narrow component 
(log~$N$(\CII) $=$ 13.9 $\pm$ 0.9 and $b$(C) = 3.3 km~s$^{-1}$) at $z = 2.830490$, not detected in the other available species and not listed, 
which probably is an interloper. \SiII\ is from $\lambda$1527, blended with \CIV~$\lambda$1551 at $z = 2.770$ which masks the first few 
components.}

\tablenotetext{h}{\CII\ in component 4 is blended with the outlying narrow component 11 of \SiIV~$\lambda$1403 in the $z = 2.770$ system.} 

\tablenotetext{i}{\SiIV~$\lambda$1403 is masked by \CIV~$\lambda$1548 at $z = 2.646$ but $\lambda$1394 is clear.}

\tablenotetext{j}{In component 3 \CIV~$\lambda$1551 at $z = 2.646$ encroaches on \SiIV~$\lambda$1394 and the value is from $\lambda$1403 
alone. \CII\ in component 2 seems contaminated and a value is not included.}

\tablenotetext{k}{\CIV~$\lambda$1551 is blended with \CIV~$\lambda$1548 at $z = 3.074$ but $\lambda$1548 is clear. \SiII~$\lambda$1260 is 
outside the forest The \NV\ values are from $\lambda$1243 which is in a clear region of the forest.}

\tablenotetext{l}{\CIV~$\lambda$1548 is blended with \CIV~$\lambda$1551 at $z = 3.067$ but $\lambda$1551 is clear.}

\tablenotetext{m}{\NV\ values are from $\lambda$1243 which is outside the forest.}

\tablenotetext{n}{\CIV~$\lambda$1548 in components 1--5 is blended with strong \AlII~$\lambda$1671 at $z = 2.830$ but $\lambda$1548 is clear. 
\NV\ is outside the forest.} 

\tablenotetext{o}{\CIV~$\lambda$1551 in component 7 is blended with \NiII~$\lambda$1710 at $z = 2.770$ but $\lambda$1548 is clear. 
\NV~$\lambda$1243 is blended with \NiII~$\lambda$1370 at $z = 2.770$ but $\lambda$1239 is clear.} 

\tablenotetext{p}{This is close to $z_{em}$. \SiIV~$\lambda$1394 in component 5 contains an unknown interloper and is not used but 
$\lambda$1403 is clear.}

\end{deluxetable}

\clearpage
\begin{deluxetable}{crrcccccc}
\tabletypesize{\scriptsize}
\tablecaption{Heavy Element Absorption Lines: Q1422+2309C $z_{em}=3.63^*$}
\tablewidth{0pt}
\tablecolumns{9}
\tablehead{
\colhead{$z$} & \colhead{$b$(C)}  & \colhead{$b$(Si)} &
\colhead{$N$(\CIV)} & \colhead{$N$(\SiIV)} &
\colhead{$N$(\CII)} & \colhead{$N$(\SiII)} &
\colhead{$N$(\NV)} &  \colhead{n} \\
{} & \colhead{km~s$^{-1}$} & \colhead{km~s$^{-1}$} & \colhead{10$^{12}$cm$^{-2}$} &
\colhead{10$^{11}$cm$^{-2}$} & \colhead{10$^{12}$cm$^{-2}$} & \colhead{10$^{11}$cm$^{-2}$} & \colhead{10$^{12}$cm$^{-2}$} & {} \\
\colhead{1} & \colhead{2} & \colhead{3} & \colhead{4} & \colhead{5} & \colhead{6} & \colhead{7} & \colhead{8} & \colhead{9} 
}

\startdata
\sidehead{2.666}
 2.664759 &    12.2 &   \nodata &    0.51 $\pm$    0.08 &    \nodata &    \nodata &    \nodata &    \nodata &  1  \\ 
 2.665200 &    11.0 &   \nodata &    2.32 $\pm$    0.08 &    \nodata &    \nodata &    \nodata &    \nodata &  2  \\ 
 2.665450 &    11.0 &   \nodata &    1.95 $\pm$    0.10 &    \nodata &    \nodata &    \nodata &    \nodata &  3  \\ 
 2.665655 &    19.0 &   \nodata &    0.99 $\pm$    0.12 &    \nodata &    \nodata &    \nodata &    \nodata &  4  \\ 
 2.666912 &    23.8 &   \nodata &    1.75 $\pm$    0.11 &    \nodata &    \nodata &    \nodata &    \nodata &  5  \\ 
\sidehead{2.683\tablenotemark{a}}
 2.682752 &    12.3 &   \nodata &    5.55 $\pm$    0.45 &    \nodata &    \nodata &    \nodata &    \nodata &  1  \\ 
 2.682904 &     5.9 &   \nodata &    0.98 $\pm$    0.20 &    \nodata &    \nodata &    \nodata &    \nodata &  2  \\ 
 2.683031 &     5.5 &   \nodata &    1.18 $\pm$    0.17 &    \nodata &    \nodata &    \nodata &    \nodata &  3  \\ 
\sidehead{2.698}
 2.696855 &     4.4 &   \nodata &    0.39 $\pm$    0.09 &    \nodata &    \nodata &    \nodata &    \nodata &  1  \\ 
 2.697200 &    16.3 &   \nodata &    3.73 $\pm$    0.17 &    \nodata &    \nodata &    \nodata &    \nodata &  2  \\ 
 2.697636 &    11.6 &   \nodata &    9.58 $\pm$    0.16 &    \nodata &    \nodata &    \nodata &    \nodata &  3  \\ 
 2.698077 &     9.1 &   \nodata &    1.11 $\pm$    0.12 &    \nodata &    \nodata &    \nodata &    \nodata &  4  \\ 
 2.698378 &     9.6 &   \nodata &    9.62 $\pm$    0.17 &    \nodata &    \nodata &    \nodata &    \nodata &  5  \\ 
 2.698564 &    15.3 &   \nodata &    7.62 $\pm$    0.19 &    \nodata &    \nodata &    \nodata &    \nodata &  6  \\ 
 2.698860 &    13.9 &   \nodata &    5.46 $\pm$    0.15 &    \nodata &    \nodata &    \nodata &    \nodata &  7  \\ 
\sidehead{2.720}
 2.719798 &    12.1 &   \nodata &    1.26 $\pm$    0.13 &    \nodata &    \nodata &    \nodata &    \nodata &  1  \\ 
 2.720051 &     7.9 &   \nodata &    1.74 $\pm$    0.12 &    \nodata &    \nodata &    \nodata &    \nodata &  2  \\ 
 2.720205 &     6.9 &   \nodata &    4.84 $\pm$    0.14 &    \nodata &    \nodata &    \nodata &    \nodata &  3  \\ 
 2.720681 &    27.5 &   \nodata &    0.98 $\pm$    0.16 &    \nodata &    \nodata &    \nodata &    \nodata &  4  \\ 
\sidehead{2.749\tablenotemark{b}}
 2.748373 &     7.4 &   \nodata &    0.78 $\pm$    0.20 &    \nodata &    \nodata &    \nodata &    \nodata &  1  \\ 
 2.748587 &     6.0 &   \nodata &    0.92 $\pm$    0.18 &    \nodata &    \nodata &    \nodata &    \nodata &  2  \\ 
 2.748750 &     6.6 &   \nodata &    1.64 $\pm$    0.20 &    \nodata &    \nodata &    \nodata &    \nodata &  3  \\ 
 2.748876 &     5.1 &     (3.6) &   11.8 $\pm$    0.3\phm{0} &   76.8 $\pm$    3.3\phm{0} &    \nodata &    \nodata &    \nodata &  4  \\ 
 2.748966 &     7.9 &     (6.5) &    9.43 $\pm$    0.32 &   27.5 $\pm$    1.7\phm{0} &    \nodata &    \nodata &    \nodata &  5  \\ 
 2.749170 &     6.5 &     (5.3) &    8.29 $\pm$    0.22 &    3.30 $\pm$    0.68 &    \nodata &    \nodata &    \nodata &  6  \\ 
 2.749402 &     8.4 &   \nodata &    1.65 $\pm$    0.26 &    \nodata &    \nodata &    \nodata &    \nodata &  7  \\ 
 2.749660 &    29.1 &   \nodata &    5.23 $\pm$    0.89 &    \nodata &    \nodata &    \nodata &    \nodata &  8  \\ 
 2.749847 &     6.6 &   \nodata &    3.36 $\pm$    0.27 &    \nodata &    \nodata &    \nodata &    \nodata &  9  \\ 
\sidehead{2.772}
 2.771655 &     8.2 &   \nodata &    1.36 $\pm$    0.10 &    \nodata &    \nodata &    \nodata &    \nodata &  1  \\ 
\sidehead{2.797}
 2.796525 &     3.9 &   \nodata &    0.61 $\pm$    0.16 &    \nodata &    \nodata &    \nodata &    \nodata &  1  \\ 
 2.796748 &    15.6 &   \nodata &    2.29 $\pm$    0.33 &    \nodata &    \nodata &    \nodata &    \nodata &  2  \\ 
\sidehead{2.810}
 2.809759 &     8.4 &   \nodata &    1.82 $\pm$    0.19 &    \nodata &    \nodata &    \nodata &    \nodata &  1  \\ 
\sidehead{2.895}
 2.895077 &    16.6 &   \nodata &    2.04 $\pm$    0.19 &    \nodata &    \nodata &    \nodata &    \nodata &  1  \\* 
 2.895430 &    13.4 &   \nodata &    7.51 $\pm$    0.27 &    \nodata &    \nodata &    \nodata &    \nodata &  2  \\* 
 2.895596 &    24.7 &   \nodata &    5.03 $\pm$    0.34 &    \nodata &    \nodata &    \nodata &    \nodata &  3  \\ 
\sidehead{2.908}
 2.907666 &     7.1 &   \nodata &    0.35 $\pm$    0.10 &    \nodata &    \nodata &    \nodata &    \nodata &  1  \\ 
 2.908035 &    15.6 &   \nodata &    0.62 $\pm$    0.15 &    \nodata &    \nodata &    \nodata &    \nodata &  2  \\ 
\sidehead{2.910}
 2.909673 &    11.1 &   \nodata &    2.54 $\pm$    0.23 &    \nodata &    \nodata &    \nodata &    \nodata &  1  \\ 
\sidehead{2.946\tablenotemark{c}}
 2.945131 &     4.1 &   \nodata &    0.67 $\pm$    0.21 &    \nodata &    \nodata &    \nodata &    \nodata &  1  \\ 
 2.945459 &    11.7 &   \nodata &    1.41 $\pm$    0.37 &    \nodata &    \nodata &    \nodata &    \nodata &  2  \\ 
 2.947497 &     8.3 &   \nodata &    3.11 $\pm$    0.26 &    \nodata &    \nodata &    \nodata &    \nodata &  3  \\ 
\sidehead{2.962\tablenotemark{d}}
 2.960717 &    21.5 &   \nodata &    2.15 $\pm$    0.26 &    \nodata &    \nodata &    \nodata &    \nodata &  1  \\ 
 2.961013 &    11.0 &   \nodata &    2.22 $\pm$    0.19 &    \nodata &    \nodata &    \nodata &    \nodata &  2  \\ 
 2.961504 &     8.3 &   \nodata &    3.03 $\pm$    0.16 &    \nodata &    \nodata &    \nodata &    \nodata &  3  \\ 
 2.961843 &    10.2 &   \nodata &   14.7 $\pm$    0.2\phm{0} &    \nodata &    \nodata &    \nodata &  $<$    0.99 &  4  \\ 
 2.962061 &     7.4 &   \nodata &    6.12 $\pm$    0.18 &    \nodata &    \nodata &    \nodata &    \nodata &  5  \\ 
 2.962344 &    10.1 &   \nodata &    6.82 $\pm$    0.19 &    \nodata &    \nodata &    \nodata &    \nodata &  6  \\ 
\sidehead{2.971\tablenotemark{e}}
 2.970982 &    17.0 &   \nodata &    0.58 $\pm$    0.50 &    \nodata &    \nodata &    \nodata &    \nodata &  1  \\ 
 2.971423 &    14.3 &   \nodata &    2.05 $\pm$    0.46 &    \nodata &    \nodata &    \nodata &    \nodata &  2  \\ 
 2.971613 &     3.5 &   \nodata &    0.43 $\pm$    0.12 &    \nodata &    \nodata &    \nodata &    \nodata &  3  \\ 
\sidehead{2.976}
 2.975799 &    23.4 &   \nodata &    2.55 $\pm$    0.24 &    \nodata &    \nodata &    \nodata &    \nodata &  1  \\ 
 2.976200 &    10.6 &   \nodata &    6.54 $\pm$    0.18 &    \nodata &    \nodata &    \nodata &    \nodata &  2  \\ 
\sidehead{2.999}
 2.999190 &     9.4 &   \nodata &    1.19 $\pm$    0.40 &    \nodata &    \nodata &    \nodata &    \nodata &  1  \\ 
 2.999204 &    25.1 &   \nodata &    5.03 $\pm$    0.63 &    \nodata &    \nodata &    \nodata &    \nodata &  2  \\ 
\sidehead{3.036}
 3.035037 &     4.7 &   \nodata &    0.95 $\pm$    0.13 &    \nodata &    \nodata &    \nodata &    \nodata &  1  \\ 
 3.036486 &    12.7 &   \nodata &    1.40 $\pm$    0.20 &    \nodata &    \nodata &    \nodata &    \nodata &  2  \\ 
\sidehead{3.064\tablenotemark{f}}
 3.063205 &     6.6 &     (5.4) &    1.52 $\pm$    0.26 &  $<$    0.20 &    \nodata &    \nodata &    \nodata &  1  \\ 
 3.063418 &     5.8 &     (4.8) &    2.57 $\pm$    0.38 &  $<$    0.27 &    \nodata &    \nodata &    \nodata &  2  \\ 
 3.063467 &    17.5 &    (16.6) &    3.98 $\pm$    0.63 &  $<$    0.48 &    \nodata &    \nodata &    \nodata &  3  \\ 
 3.063973 &    18.8 &    (18.0) &    2.49 $\pm$    0.43 &  $<$    0.32 &    \nodata &    \nodata &    \nodata &  4  \\ 
 3.064434 &     9.2 &     (7.4) &    1.71 $\pm$    0.49 &  $<$    0.22 &    \nodata &    \nodata &    \nodata &  5  \\ 
\sidehead{3.071}
 3.071076 &     7.2 &     (5.7) &    4.40 $\pm$    0.25 &  $<$    0.36 &    \nodata &    \nodata &    \nodata &  1  \\ 
\sidehead{3.086\tablenotemark{g}}
 3.086040 &     5.6 &     (4.4) &    1.06 $\pm$    0.28 &  $<$    0.39 &    \nodata &    \nodata &    \nodata &  1  \\ 
 3.086314 &     2.7 &     (2.2) &    0.52 $\pm$    0.15 &  $<$    0.28 &    \nodata &    \nodata &    \nodata &  2  \\ 
 3.086651 &     7.1 &     (5.9) &    5.97 $\pm$    0.31 &  $<$    0.39 &    \nodata &    \nodata &    \nodata &  3  \\ 
 3.086815 &     8.6 &     (6.8) &    2.10 $\pm$    0.31 &  $<$    0.45 &    \nodata &    \nodata &    \nodata &  4  \\ 
\sidehead{3.090\tablenotemark{h}}
 3.089876 &    13.0 &    (11.8) &   10.0 $\pm$    0.4\phm{0} &    2.27 $\pm$    0.99 &    \nodata &    \nodata &    \nodata &  1  \\* 
 3.090256 &    14.1 &    (13.0) &   14.1 $\pm$    0.3\phm{0} &   12.3 $\pm$    0.7\phm{0} &    \nodata &    \nodata &    \nodata &  2  \\* 
 3.090509 &     6.9 &     (5.5) &    4.22 $\pm$    0.23 &    4.00 $\pm$    0.45 &    \nodata &    \nodata &    \nodata &  3  \\ 
 3.090769 &     7.0 &     (5.6) &    2.94 $\pm$    0.23 &    3.12 $\pm$    0.80 &    \nodata &    \nodata &    \nodata &  4  \\ 
 3.091044 &     8.7 &     (6.9) &   10.6 $\pm$    0.2\phm{0} &    9.16 $\pm$    0.84 &    \nodata &    \nodata &    \nodata &  5  \\   
\sidehead{3.095}
 3.094865 &    34.0 &    (33.6) &    2.07 $\pm$    0.79 &  $<$    4.90 &    \nodata &    \nodata &    \nodata &  1  \\ 
\sidehead{3.135\tablenotemark{i}}
 3.132487 &    25.2 &    (24.6) &    2.76 $\pm$    0.46 &  $<$    2.92 &    \nodata &    \nodata &    \nodata &  1  \\ 
 3.133928 &    29.1 &    (28.6) &   10.4 $\pm$    0.8\phm{0} &   10.3 $\pm$    1.1\phm{0} &    \nodata &    \nodata &    \nodata &  2  \\ 
 3.134476 &    12.3 &    (11.0) &    7.34 $\pm$    0.66 &    5.86 $\pm$    0.78 &    \nodata &    \nodata &    \nodata &  3  \\ 
 3.136880 &    11.0 &     (9.6) &    1.47 $\pm$    0.98 &    5.73 $\pm$    2.40 &    \nodata &    \nodata &    \nodata &  4  \\ 
 3.137120 &    13.5 &    (12.3) &    5.37 $\pm$    0.97 &   10.8 $\pm$    2.5\phm{0} &    \nodata &    \nodata &    \nodata &  5  \\ 
\sidehead{3.191}
 3.191350 &     3.2 &     (2.2) &    0.44 $\pm$    0.12 &    0.91 $\pm$    0.47 &    \nodata &    \nodata &    \nodata &  1  \\ 
 3.191440 &     8.9 &     (7.1) &    1.09 $\pm$    0.20 &  $<$    0.74 &    \nodata &    \nodata &    \nodata &  2  \\ 
\sidehead{3.240\tablenotemark{j}}
 3.240169 &    19.6 &    (18.9) &    2.02 $\pm$    0.31 &  $<$    1.12 &  $<$    0.17 &    \nodata &    \nodata &  1  \\ 
 3.240658 &     6.0 &     (3.9) &    0.80 $\pm$    0.17 &  $<$    0.57 &  $<$    0.21 &    \nodata &    \nodata &  2  \\ 
\sidehead{3.257}
 3.256875 &     3.7 &     (2.4) &    0.33 $\pm$    0.13 &  $<$    0.41 &  $<$    0.15 &    \nodata &    \nodata &  1  \\ 
 3.257276 &    11.1 &     (9.6) &    0.64 $\pm$    0.21 &  $<$    1.79 &  $<$    1.06 &    \nodata &    \nodata &  2  \\ 
\sidehead{3.265\tablenotemark{k}}
 3.264775 &    24.1 &    (23.5) &    0.74 $\pm$    0.34 &    \nodata &  $<$    0.70 &    \nodata &    \nodata &  1  \\ 
 3.265398 &     3.7 &     (2.4) &    0.69 $\pm$    0.14 &    \nodata &  $<$    0.18 &    \nodata &    \nodata &  2  \\ 
 3.265741 &    14.8 &    (13.7) &    3.75 $\pm$    0.26 &  $<$    0.91 &  $<$    0.30 &    \nodata &    \nodata &  3  \\ 
\sidehead{\phm{0}}
\sidehead{3.381\tablenotemark{l}}
 3.380044 &    22.5 &    (21.9) &    6.52 $\pm$    0.61 &  $<$    1.64 &  $<$    0.54 &    \nodata &    \nodata &  1  \\ 
 3.380465 &     7.0 &     (5.7) &    0.30 $\pm$    0.28 &    \nodata &  $<$    0.30 &    \nodata &    \nodata &  2  \\ 
 3.380676 &    13.1 &    (11.9) &    2.53 $\pm$    0.42 &  $<$    1.01 &  $<$    0.31 &    \nodata &    \nodata &  3  \\ 
 3.381392 &    10.9 &     (9.4) &    4.10 $\pm$    0.38 &  $<$    0.87 &  $<$    0.28 &    \nodata &    \nodata &  4  \\ 
 3.381652 &     9.4 &     (7.6) &   14.0 $\pm$    0.4\phm{0} &  $<$    0.76 &  $<$    0.27 &    \nodata &    \nodata &  5  \\ 
 3.382017 &    11.5 &    (10.1) &    6.79 $\pm$    0.40 &  $<$    0.89 &  $<$    0.29 &    \nodata &    \nodata &  6  \\ 
 3.382285 &    10.0 &     (8.4) &    8.68 $\pm$    0.37 &    2.81 $\pm$    1.06 &  $<$    0.28 &    \nodata &    \nodata &  7  \\ 
 3.382676 &     8.2 &     (6.3) &    3.27 $\pm$    0.38 &  $<$    1.19 &  $<$    0.31 &    \nodata &    \nodata &  8  \\ 
 3.382871 &    34.1 &    (33.7) &    4.73 $\pm$    0.91 &  $<$    2.29 &  $<$    0.65 &    \nodata &    \nodata &  9  \\ 
\sidehead{3.411\tablenotemark{m}}
 3.410989 &    23.7 &    (23.0) &    1.85 $\pm$    0.35 &  $<$    2.55 &  $<$    0.52 &    \nodata &    \nodata &  1  \\ 
 3.411523 &    18.1 &    (17.2) &    7.21 $\pm$    0.33 &    5.11 $\pm$    2.17 &    \nodata &    \nodata &    \nodata &  2  \\ 
\sidehead{3.449\tablenotemark{n}}
 3.446200 &    14.8 &    (13.7) &    2.95 $\pm$    0.43 &  $<$    1.13 &  $<$    0.48 &    \nodata &    \nodata &  1  \\ 
 3.446574 &    12.6 &    (11.4) &    1.51 $\pm$    0.39 &    4.72 $\pm$    0.89 &  $<$    0.35 &    \nodata &    \nodata &  2  \\ 
 3.446894 &     6.4 &     5.6\phm{)} &    8.18 $\pm$    0.34 &   28.2 $\pm$    0.8\phm{0} &  $<$    0.32 &    \nodata &    \nodata &  3  \\ 
 3.447156 &     9.4 &     (7.8) &    7.56 $\pm$    0.41 &   16.8 $\pm$    0.8\phm{0} &  $<$    0.32 &    \nodata &    \nodata &  4  \\ 
 3.447410 &    10.9 &     (9.5) &   19.1 $\pm$    0.6\phm{0} &   36.9 $\pm$    1.0\phm{0} &  $<$    0.44 &    \nodata &    \nodata &  5  \\ 
 3.448211 &    12.4 &    (11.1) &    1.15 $\pm$    0.36 &    1.94 $\pm$    0.85 &  $<$    0.84 &    \nodata &    \nodata &  6  \\ 
 3.448625 &    13.9 &    (12.7) &    3.44 $\pm$    0.46 &   17.7 $\pm$    1.0\phm{0} &    2.38 $\pm$    0.36 &    \nodata &    \nodata &  7  \\ 
 3.449267 &    15.6 &    (14.6) &    2.48 $\pm$    0.59 &    6.39 $\pm$    1.00 &  $<$    0.37 &    \nodata &    \nodata &  8  \\ 
 3.450546 &     7.4 &     (6.0) &    0.89 $\pm$    0.31 &  $<$    0.96 &  $<$    0.59 &    \nodata &    \nodata &  9  \\ 
 3.451340 &    16.0 &    (15.0) &   10.5 $\pm$    0.7\phm{0} &    5.83 $\pm$    1.02 &  $<$    0.36 &    \nodata &    \nodata & 10  \\ 
 3.452078 &     7.9 &     (6.3) &    0.82 $\pm$    0.29 &    3.51 $\pm$    0.70 &  $<$    0.33 &    \nodata &    \nodata & 11  \\ 
\sidehead{3.480\tablenotemark{o}}
 3.479479 &    15.3 &    (14.2) &    1.75 $\pm$    0.45 &  $<$    1.10 &  $<$    0.70 &  $<$    0.46 &    \nodata &  1  \\ 
 3.479781 &     3.7 &     (2.4) &    0.95 $\pm$    0.25 &  $<$    0.48 &  $<$    0.32 &  $<$    0.09 &    \nodata &  2  \\ 
 3.480366 &     8.4 &     (6.5) &    0.88 $\pm$    0.33 &  $<$    0.68 &  $<$    0.44 &  $<$    0.22 &    \nodata &  3  \\ 
\sidehead{3.494\tablenotemark{p}}
 3.493290 &    14.6 &    (13.5) &    2.02 $\pm$    0.37 &  $<$    1.00 &  $<$    0.78 &  $<$    0.23 &    \nodata &  1  \\ 
 3.494847 &     6.6 &     (5.8) &    2.08 $\pm$    0.25 &  $<$    0.90 &  $<$    0.38 &  $<$    0.14 &    \nodata &  2  \\ 
\sidehead{3.515\tablenotemark{q}}
 3.514508 &    11.3 &     (9.9) &    0.71 $\pm$    0.27 &    2.73 $\pm$    1.60 &  $<$    0.30 &  $<$    0.38 &  $<$    0.65 &  1  \\ 
 3.514639 &     6.2 &     (5.0) &    5.79 $\pm$    0.28 &    5.69 $\pm$    1.28 &  $<$    0.52 &  $<$    0.33 &  $<$    0.52 &  2  \\ 
 3.514781 &    11.8 &    (10.5) &    0.85 $\pm$    0.33 &  $<$    3.74 &  $<$    0.52 &  $<$    0.42 &  $<$    0.67 &  3  \\ 
 3.514988 &    10.4 &     (8.9) &    1.65 $\pm$    0.25 &  $<$    1.21 &  $<$    0.31 &  $<$    0.33 &  $<$    0.56 &  4  \\ 
\sidehead{3.538\tablenotemark{r}}
 3.534573 &    22.7 &    (22.0) &    5.12 $\pm$    0.41 &  $<$    1.26 &  $<$    0.30 &  $<$    0.60 &  $<$    0.39 &  1  \\ 
 3.534873 &     3.3 &     (2.2) &    1.55 $\pm$    0.20 &  $<$    0.51 &  $<$    0.21 &  $<$    0.25 &  $<$    0.28 &  2  \\ 
 3.535056 &     7.4 &     (5.1) &    4.60 $\pm$    0.26 &  $<$    0.63 &    \nodata &  $<$    0.30 &  $<$    0.19 &  3  \\ 
 3.535562 &    19.5 &    (18.7) &    9.82 $\pm$    0.44 &  $<$    1.82 &    \nodata &  $<$    0.57 &  $<$    0.31 &  4  \\ 
 3.535939 &    12.8 &    (11.6) &   27.2 $\pm$    0.6\phm{0} &   38.2 $\pm$    1.3\phm{0} &  $<$    0.41 &  $<$    0.52 &  $<$    0.33 &  5  \\ 
 3.536132 &     5.5 &     (3.7) &    4.67 $\pm$    0.42 &   21.7 $\pm$    0.9\phm{0} &    1.27 $\pm$    0.20 &  $<$    0.31 &  $<$    0.28 &  6  \\ 
 3.536367 &    35.7 &    (35.3) &   32.1 $\pm$    1.0\phm{0} &   37.0 $\pm$    2.7\phm{0} &    1.27 $\pm$    0.63 &    \nodata &    \nodata &  7  \\ 
 3.536634 &     9.5 &     (8.9) &  $<$    0.45 &    6.48 $\pm$    1.18 &    2.75 $\pm$    0.27 &    2.12 $\pm$    0.51 &  $<$    0.40 &  8  \\ 
 3.537240 &     6.6 &     (5.5) &    5.93 $\pm$    0.61 &   56.4 $\pm$    1.9\phm{0} &    2.60 $\pm$    0.41 &    3.03 $\pm$    0.64 &  $<$    0.30 &  9  \\ 
 3.537303 &    19.1 &    (18.3) &    7.76 $\pm$    1.21 &   32.5 $\pm$    3.0\phm{0} &    5.87 $\pm$    0.82 &    2.26 $\pm$    1.25 &  $<$    0.72 & 10  \\ 
 3.537475 &     7.1 &     4.8\phm{)} &    4.70 $\pm$    0.49 &   12.0 $\pm$    1.1\phm{0} &  $<$    0.37 &    0.73 $\pm$    0.46 &  $<$    0.29 & 11  \\ 
 3.538010 &    20.6 &    (19.8) &    4.50 $\pm$    0.41 &    7.58 $\pm$    1.23 &  $<$    0.28 &  $<$    0.56 &  $<$    0.34 & 12  \\ 
 3.538598 &    22.2 &    (21.5) &   38.6 $\pm$    0.8\phm{0} &   69.3 $\pm$    2.0\phm{0} &    1.16 $\pm$    0.42 &    \nodata &  $<$    0.46 & 13  \\ 
 3.538702 &     6.1 &     4.5\phm{)} &   24.8 $\pm$    1.0\phm{0} &   74.3 $\pm$    2.0\phm{0} &    1.20 $\pm$    0.26 &    0.65 $\pm$    0.40 &  $<$    0.28 & 14  \\ 
 3.538794 &     5.2 &     (3.8) &    2.64 $\pm$    0.53 &   13.4 $\pm$    1.5\phm{0} &    0.98 $\pm$    0.21 &    1.16 $\pm$    0.33 &  $<$    0.31 & 15  \\ 
 3.539293 &     9.1 &     (7.5) &   22.2 $\pm$    0.6\phm{0} &   21.6 $\pm$    1.3\phm{0} &  $<$    0.32 &  $<$    0.53 &  $<$    0.43 & 16  \\ 
 3.539503 &     6.8 &     (5.2) &    8.83 $\pm$    0.50 &    8.23 $\pm$    1.31 &  $<$    0.32 &  $<$    0.43 &  $<$    0.32 & 17  \\ 
 3.539512 &    29.2 &    (28.7) &    7.72 $\pm$    1.19 &  $<$    3.71 &  $<$    0.84 &  $<$    1.28 &  $<$    0.66 & 18  \\ 
 3.539894 &     8.7 &     (6.8) &  $<$    0.52 &  $<$    1.14 &    1.21 $\pm$    0.32 &    1.18 $\pm$    0.42 &  $<$    0.39 & 19  \\ 
 3.540026 &     5.7 &     (4.6) &    0.54 $\pm$    0.27 &    5.44 $\pm$    0.73 &    1.08 $\pm$    0.22 &    0.96 $\pm$    0.29 &    \nodata & 20  \\ 
 3.540205 &     6.8 &     (5.9) &    0.97 $\pm$    0.32 &    2.27 $\pm$    0.88 &    2.55 $\pm$    0.27 &    3.11 $\pm$    0.38 &  $<$    0.27 & 21  \\ 
 3.540280 &    32.9 &    (32.4) &    5.22 $\pm$    0.74 &  $<$    7.86 &  $<$    1.67 &    \nodata &    \nodata & 22  \\ 
 3.541446 &     8.5 &     (6.9) &    2.60 $\pm$    0.26 &  $<$    0.84 &  $<$    0.22 &  $<$    4.15 &  $<$    0.51 & 23  \\ 
\sidehead{3.588\tablenotemark{s}}
 3.586827 &    15.6 &    (14.6) &    9.40 $\pm$    0.23 &   16.1 $\pm$    1.1\phm{0} &  $<$    0.48 &    \nodata &  $<$    0.41 &  1  \\ 
 3.587146 &    16.0 &    (15.0) &    7.07 $\pm$    0.21 &    4.72 $\pm$    1.09 &  $<$    0.58 &    \nodata &  $<$    0.48 &  2  \\ 
 3.588866 &    18.7 &    (17.9) &    2.53 $\pm$    0.19 &  $<$    1.09 &  $<$    0.49 &  $<$    0.55 &  $<$    0.86 &  3  \\ 
 3.589211 &     9.8 &     (8.1) &    2.97 $\pm$    0.15 &  $<$    1.19 &  $<$    0.36 &  $<$    0.40 &  $<$    0.48 &  4  \\ 
 3.589570 &    15.0 &    (14.0) &    0.48 $\pm$    0.17 &  $<$    2.13 &  $<$    0.49 &  $<$    0.80 &  $<$    0.44 &  5  \\ 
\sidehead{3.624\tablenotemark{t}}
 3.623361 &     7.2 &     (4.7) &    3.02 $\pm$    0.18 &  $<$    0.95 &  $<$    0.32 &  $<$    0.23 &    0.85 $\pm$    0.20 &  1  \\ 
 3.623519 &     5.7 &     (4.5) &   12.9 $\pm$    0.3\phm{0} &  $<$    0.58 &  $<$    0.36 &  $<$    0.24 &    4.84 $\pm$    0.35 &  2  \\ 
 3.623641 &     5.9 &     (4.7) &    9.83 $\pm$    0.24 &  $<$    0.57 &  $<$    0.34 &  $<$    0.25 &    8.21 $\pm$    0.28 &  3  \\ 
 3.623830 &     7.8 &     (6.6) &    7.45 $\pm$    0.19 &  $<$    0.62 &  $<$    0.48 &  $<$    0.26 &    2.11 $\pm$    0.23 &  4  \\ 
 3.624179 &    10.6 &     (9.1) &   13.1 $\pm$    0.3\phm{0} &  $<$    0.76 &  $<$    0.58 &  $<$    0.31 &    5.41 $\pm$    0.27 &  5  \\ 
 3.624333 &     4.4 &     (3.0) &    5.21 $\pm$    0.33 &    2.45 $\pm$    0.62 &  $<$    0.49 &  $<$    0.43 &    2.09 $\pm$    0.23 &  6  \\ 
 3.624406 &     5.7 &     (4.7) &   18.3 $\pm$    0.4\phm{0} &    2.72 $\pm$    0.69 &  $<$    0.42 &  $<$    0.29 &    1.76 $\pm$    0.30 &  7  \\ 
 3.624632 &     9.8 &     (8.1) &    3.99 $\pm$    0.20 &  $<$    0.67 &  $<$    0.65 &  $<$    0.29 &    0.71 $\pm$    0.22 &  8  \\ 
 3.624886 &     8.6 &     (6.7) &    1.95 $\pm$    0.17 &  $<$    0.61 &  $<$    0.40 &  $<$    0.26 &    0.52 $\pm$    0.19 &  9  \\ 
 3.625242 &    12.9 &    (11.6) &    3.48 $\pm$    0.21 &  $<$    0.84 &  $<$    0.41 &  $<$    0.33 &    0.67 $\pm$    0.27 & 10  \\ 
\enddata

\tablenotetext{*}{See Figure 1. For this object, in addition to the sky background including atmospheric line emission whose mean level 
is removed by the normal sky subtraction procedure (with related residual noise enhancement), both weak and strong atmospheric H$_{2}$O and 
O$_{2}$ absorption bands here are signifantly present in the observed wavelength range and have to be taken intricately into account while 
assessing the structures of the QSO absorber systems within the fitting process (see text). The lines in these bands are intrinsically very 
narrow compared with the QSO absorber components except in the most heavily absorbed regions where information is unavoidably obscured.
However, the QSO data have been reduced to vacuum, heliocentric redshifts so the apparent positions and widths of the atmospheric lines 
seen in this frame of reference reflect the integrated collective epochs of observation: here for Q1422+2309C the specific lengths and 
separation of observing runs become manifest as a mild broadening and pronounced spectral doubling (and therefore a related apparent 
weakening) of the atmospheric lines.}

\tablenotetext{a}{\CIV~$\lambda$1548 is blended with \SiIV~$\lambda$1394 at $z = 3.090$ but $\lambda$1551 is clear.}

\tablenotetext{b}{\CIV~$\lambda$1548 in components 1 and 2 is blended with \SiIV~$\lambda$1403 at $z = 3.135$ but $\lambda$1551 is clear. 
Although \SiIV\ is in the forest, for $\lambda$1394 components 4--6 are in a relatively clear region.}

\tablenotetext{c}{\CIV~$\lambda$1548 in components 1 and 2 is blended with \SiIV~$\lambda$1394 at $z = 3.381$ but $\lambda$1551 is clear.}

\tablenotetext{d}{The \NV\ upper limit is from $\lambda$1243 in a clear region of the forest.}

\tablenotetext{e}{\CIV~$\lambda$1548 is blended with \SiIV~$\lambda$1394 at $z = 3.411$ and \SiIV~$\lambda$1403 at $z = 3.381$ but 
$\lambda$1551 is clear.}

\tablenotetext{f}{\CIV~$\lambda$1548 is mildly contaminated by atmospheric line absorption and in component 5 is blended with 
\SiIV~$\lambda$1394 at $z = 3.515$ while $\lambda$1551 contains both atmospheric line absorption and emission and, also in component 5, is 
blended with \CIV~$\lambda$1548 at $z = 3.071$: acceptable values are obtained largely through $\lambda$1548. \SiIV~$\lambda$1403 is masked by 
\SiIV~$\lambda$1394 at $z = 3.090$ and \CIV~$\lambda$1548 at $z = 2.683$ and the upper limits are from $\lambda$1394.}

\tablenotetext{g}{\CIV~$\lambda$1548 is blended with \SiIV~$\lambda$1394 at $z = 3.538$ but $\lambda$1551 is clear. \SiIV~$\lambda$1403 in
component 4 is blended with component 1 of \CIV~$\lambda$1551 at $z = 2.698$ and here the upper limit is from $\lambda$1394.}

\tablenotetext{h}{\CIV~$\lambda$1548 contains \SiIV~$\lambda$1403 at $z = 3.515$ but $\lambda$1551 is clear. \SiIV~$\lambda$1394 is in a blend 
with \CIV~$\lambda$1548 at $z = 2.683$, very weak \SiIV~$\lambda$1403 at $z = 3.064$ and an unidentified interloper at observed wavelength 
$\lambda$5700.35 near component 1, while $\lambda$1403 also has some unknown contamination covering component 1, but from the available 
complementary regions acceptable values are obtained for all the other compoments.}

\tablenotetext{i}{\SiIV~$\lambda$1394 in component 1 is masked by \CIV~$\lambda$1548 at $z = 2.720$ while components 4 and 5 of $\lambda$1403 
are blended with \CIV~$\lambda$1548 at $z = 2.749$, but together acceptable values are obtained.}

\tablenotetext{j}{\CII\ is outside the forest.}

\tablenotetext{k}{\SiIV~$\lambda$1394 in the weak components 1 and 2 is contaminated by atmospheric line absorption and $\lambda$1403 is in a 
noisy region, making values here too uncertain to include.}

\tablenotetext{l}{This is nominally a partial Lyman limit system with total log~$N$(\HI) $\sim17.2$ but \CII\ is very weak. \SiIV~$\lambda$1394 
in component 2 is masked by an unidentified feature and in components 7 and 8 is blended with \CIV~$\lambda$1548 at $z = 2.946$ while 
$\lambda$1403 in components 1 and 2 is blended with \CIV~$\lambda$1551 at $z = 2.962$, in components 8 and 9 with \CIV~$\lambda$1548 at 
$z = 2.971$ and in component 8 also with very weak \SiIV~$\lambda$1394 at $z = 3.411$: nevertheless, acceptable values are obtained for all but 
component 2. \CII\ in components 8 and 9 is blended with weak \CIV~$\lambda$1551 at $z = 2.772$ but together with $\lambda$1548 of the latter 
the upper limits obtained are reliable.}

\tablenotetext{m}{\SiIV~$\lambda$1394 is blended with \CIV~$\lambda$1548 at $z = 2.971$ and borders on \SiIV~$\lambda$1403 at $z = 3.381$ but 
$\lambda$1403 is clear. \CII\ in component 2 is masked by \CIV~$\lambda$1551 at $z = 2.797$.} 

\tablenotetext{n}{\CIV\ is partially contaminated by strong atmospheric line absorption features but by including these in the fitting process 
for both lines of the doublet acceptable values are obtained in all components.} 

\tablenotetext{o}{\SiIV~$\lambda$1403 is contaminated by atmospheric line absorption and the upper limits are from $\lambda$1394. 
\SiII~$\lambda$1260 is outside the forest.}

\tablenotetext{p}{\SiIV~$\lambda$1403 is blended with \CIV~$\lambda$1548 at $z = 3.071$ and \CIV~$\lambda$1551 at $z = 3.064$ and there is also 
contamination by atmospheric line absorption: the upper limits are from $\lambda$1394.}

\tablenotetext{q}{\CIV\ is contaminated by atmospheric line absorption but good values are obtained in all components from complementary regions  
of the doublet. \SiIV~$\lambda$1394 is mildly affected by atmospheric line absorption and is blended with \CIV~$\lambda$1548 at $z = 3.064$ while 
$\lambda$1403 is blended with strong \CIV~$\lambda$1548 at $z = 3.090$ but taken together useful values are obtained. \NV\ upper limits are from 
$\lambda$1243 in a clear region of the forest.}

\tablenotetext{r}{This is a highly complex system. \CIV\ is mildly contaminated by atmospheric line absorption but good values are obtained 
overall. \SiIV~$\lambda$1394 in components 14--17 is blended with \CIV~$\lambda$1548 at $z = 3.086$ and $\lambda$1403 in component 12 has some 
interference from atmospheric line emission, but together good values are obtained. \CII\ values in components 3 and 4 are too uncertain to use, 
due to blending with \CIV~$\lambda$1548 at $z = 2.910$. \NV~$\lambda$1243 is outside the forest and yields good upper limits.}

\tablenotetext{s}{\CII\ in component 2 is blended with \CIV~$\lambda$1551 at $z = 2.946$ but a useful upper limit is obtained from close analysis
of the \CIV\ doublet. \SiII\ in components 1 and 2 seems too strong relative to \CII\ and probably contains an unidentified interloper, and is 
not used. \NV\ is outside the forest; $\lambda$1239 in components 3--5 is blended with \CIV~$\lambda$1551 at $z = 2.666$ and $\lambda$1243 in 
components 1 and 2  is masked by \SiIV~$\lambda$1394 at $z = 3.090$ and \CIV~$\lambda$1548 at $z = 2.683$ but taken together useful upper limits 
are obtained.}
 
\tablenotetext{t}{This is close to $z_{em}$. \CIV~$\lambda$1551 has some contamination by atmospheric line absorption but $\lambda$1548 is not 
significantly affected. \SiIV~$\lambda$1403 is mildly contaminated by atmospheric line absorption but $\lambda$1394 is clear.}

\end{deluxetable}

\clearpage
\begin{deluxetable}{crrcccccc}
\tabletypesize{\scriptsize}
\tablecaption{Heavy Element Absorption Lines: Q1645+5520  $z_{em} = 4.10^*$}
\tablewidth{0pt}
\tablecolumns{9}
\tablehead{
\colhead{$z$} & \colhead{$b$(C)}  & \colhead{$b$(Si)} &
\colhead{$N$(\CIV)} & \colhead{$N$(\SiIV)} &
\colhead{$N$(\CII)} & \colhead{$N$(\SiII)} &
\colhead{$N$(\NV)} &  \colhead{n} \\
{} & \colhead{km~s$^{-1}$} & \colhead{km~s$^{-1}$} & \colhead{10$^{12}$cm$^{-2}$} &
\colhead{10$^{11}$cm$^{-2}$} & \colhead{10$^{12}$cm$^{-2}$} & \colhead{10$^{11}$cm$^{-2}$} & \colhead{10$^{12}$cm$^{-2}$} & {} \\
\colhead{1} & \colhead{2} & \colhead{3} & \colhead{4} & \colhead{5} & \colhead{6} & \colhead{7} & \colhead{8} & \colhead{9} 
}

\startdata
\sidehead{3.015\tablenotemark{a}}
 3.015026 &     7.6 &   \nodata &    0.33 $\pm$    0.14 &    \nodata &    \nodata &    \nodata &    \nodata &  1  \\ 
\sidehead{3.077\tablenotemark{b}}
 3.076957 &     9.4 &   \nodata &    1.87 $\pm$    0.17 &    \nodata &    \nodata &    \nodata &    \nodata &  1  \\ 
 3.077197 &     7.5 &   \nodata &    6.25 $\pm$    0.20 &    \nodata &    \nodata &    \nodata &    \nodata &  2  \\ 
\sidehead{3.129}
 3.128686 &     9.6 &   \nodata &    5.63 $\pm$    0.18 &    \nodata &    \nodata &    \nodata &    \nodata &  1  \\ 
 3.128941 &     8.8 &   \nodata &    3.32 $\pm$    0.17 &    \nodata &    \nodata &    \nodata &    \nodata &  2  \\ 
 3.129349 &    12.3 &   \nodata &    0.58 $\pm$    0.17 &    \nodata &    \nodata &    \nodata &    \nodata &  3  \\ 
 3.129694 &     3.9 &   \nodata &    0.39 $\pm$    0.12 &    \nodata &    \nodata &    \nodata &    \nodata &  4  \\ 
\sidehead{3.162}
 3.161200 &    10.7 &   \nodata &    1.53 $\pm$    0.18 &    \nodata &    \nodata &    \nodata &    \nodata &  1  \\ 
 3.161467 &    11.1 &   \nodata &    2.19 $\pm$    0.19 &    \nodata &    \nodata &    \nodata &    \nodata &  2  \\ 
 3.161877 &     6.9 &   \nodata &    0.86 $\pm$    0.15 &    \nodata &    \nodata &    \nodata &    \nodata &  3  \\ 
 3.162148 &     8.2 &   \nodata &    3.78 $\pm$    0.18 &    \nodata &    \nodata &    \nodata &    \nodata &  4  \\ 
 3.162442 &    17.0 &   \nodata &    4.51 $\pm$    0.23 &    \nodata &    \nodata &    \nodata &    \nodata &  5  \\ 
 3.162982 &    20.4 &   \nodata &    2.00 $\pm$    0.23 &    \nodata &    \nodata &    \nodata &    \nodata &  6  \\ 
 3.163669 &    14.8 &   \nodata &    1.42 $\pm$    0.20 &    \nodata &    \nodata &    \nodata &    \nodata &  7  \\ 
\sidehead{3.193\tablenotemark{c}}
 3.191938 &    12.3 &   \nodata &    1.57 $\pm$    0.16 &    \nodata &    \nodata &    \nodata &    \nodata &  1  \\ 
 3.192421 &     6.7 &   \nodata &    2.94 $\pm$    0.15 &    \nodata &    \nodata &    \nodata &    \nodata &  2  \\ 
 3.192914 &    54.9 &   \nodata &   11.8 $\pm$    0.6\phm{0} &    \nodata &    \nodata &    \nodata &    \nodata &  3  \\ 
 3.193257 &    15.8 &   \nodata &   22.1 $\pm$    0.6\phm{0} &    \nodata &    \nodata &    \nodata &    \nodata &  4  \\ 
 3.193301 &    26.4 &   \nodata &   10.5 $\pm$    0.8\phm{0} &    \nodata &    \nodata &    \nodata &    \nodata &  5  \\ 
\sidehead{3.197}
 3.197438 &    21.8 &   \nodata &    2.38 $\pm$    0.17 &    \nodata &    \nodata &    \nodata &    \nodata &  1  \\ 
\sidehead{3.217}
 3.216331 &     9.5 &   \nodata &    2.34 $\pm$    0.17 &    \nodata &    \nodata &    \nodata &    \nodata &  1  \\ 
 3.216788 &    24.6 &   \nodata &    3.71 $\pm$    0.26 &    \nodata &    \nodata &    \nodata &    \nodata &  2  \\ 
\sidehead{3.262}
 3.261969 &     7.3 &   \nodata &    1.58 $\pm$    0.25 &    \nodata &    \nodata &    \nodata &    \nodata &  1  \\ 
 3.262136 &    12.3 &   \nodata &    0.69 $\pm$    0.19 &    \nodata &    \nodata &    \nodata &    \nodata &  2  \\ 
\sidehead{3.289}
 3.287686 &    27.0 &   \nodata &    3.72 $\pm$    0.28 &    \nodata &    \nodata &    \nodata &    \nodata &  1  \\ 
 3.287971 &     4.8 &   \nodata &    0.96 $\pm$    0.15 &    \nodata &    \nodata &    \nodata &    \nodata &  2  \\ 
 3.288192 &    13.6 &   \nodata &   10.5 $\pm$    0.2\phm{0} &    \nodata &    \nodata &    \nodata &    \nodata &  3  \\ 
 3.288662 &    16.6 &   \nodata &   14.1 $\pm$    0.3\phm{0} &    \nodata &    \nodata &    \nodata &    \nodata &  4  \\ 
 3.288957 &     6.2 &   \nodata &    3.93 $\pm$    0.29 &    \nodata &    \nodata &    \nodata &    \nodata &  5  \\ 
 3.289017 &    18.0 &   \nodata &   10.3 $\pm$    0.4\phm{0} &    \nodata &    \nodata &    \nodata &    \nodata &  6  \\ 
 3.290072 &    21.9 &   \nodata &    4.97 $\pm$    0.28 &    \nodata &    \nodata &    \nodata &    \nodata &  7  \\ 
\sidehead{3.322\tablenotemark{d}}
 3.321727 &    21.4 &   \nodata &    1.74 $\pm$    0.22 &    \nodata &    \nodata &    \nodata &    \nodata &  1  \\ 
 3.322327 &    10.0 &   \nodata &    2.34 $\pm$    0.18 &    \nodata &    \nodata &    \nodata &    \nodata &  2  \\ 
 3.322627 &    11.5 &   \nodata &    3.72 $\pm$    0.19 &    \nodata &    \nodata &    \nodata &    \nodata &  3  \\* 
 3.323030 &    19.4 &   \nodata &    2.38 $\pm$    0.25 &    \nodata &    \nodata &    \nodata &    \nodata &  4  \\ 
\sidehead{3.341\tablenotemark{e}}
 3.340588 &    10.5 &   \nodata &    1.06 $\pm$    0.18 &    \nodata &    \nodata &    \nodata &    \nodata &  1  \\ 
 3.340847 &     7.7 &   \nodata &    1.08 $\pm$    0.19 &    \nodata &    \nodata &    \nodata &    \nodata &  2  \\ 
 3.341160 &     9.0 &   \nodata &    1.35 $\pm$    0.23 &    \nodata &    \nodata &    \nodata &    \nodata &  3  \\ 
 3.341462 &    19.6 &   \nodata &    1.43 $\pm$    0.35 &    \nodata &    \nodata &    \nodata &    \nodata &  4  \\ 
 3.341963 &     6.0 &   \nodata &    0.60 $\pm$    0.26 &    \nodata &    \nodata &    \nodata &    \nodata &  5  \\ 
\sidehead{3.348}
 3.347684 &    11.7 &   \nodata &    0.29 $\pm$    0.16 &    \nodata &    \nodata &    \nodata &    \nodata &  1  \\ 
 3.348334 &     3.0 &   \nodata &    0.20 $\pm$    0.11 &    \nodata &    \nodata &    \nodata &    \nodata &  2  \\ 
 3.348930 &     4.2 &   \nodata &    0.15 $\pm$    0.10 &    \nodata &    \nodata &    \nodata &    \nodata &  3  \\ 
 3.349228 &     3.3 &   \nodata &    0.17 $\pm$    0.13 &    \nodata &    \nodata &    \nodata &    \nodata &  4  \\ 
\sidehead{\phm{0}}
\sidehead{3.355}
 3.354884 &    38.2 &   \nodata &    2.04 $\pm$    0.30 &    \nodata &    \nodata &    \nodata &    \nodata &  1  \\ 
 3.355420 &     4.2 &   \nodata &    0.57 $\pm$    0.13 &    \nodata &    \nodata &    \nodata &    \nodata &  2  \\ 
 3.355849 &    13.3 &   \nodata &    1.75 $\pm$    0.19 &    \nodata &    \nodata &    \nodata &    \nodata &  3  \\ 
\sidehead{3.421\tablenotemark{f}}
 3.420509 &     6.9 &     (5.7) &    4.12 $\pm$    0.24 &    7.72 $\pm$    0.37 &    \nodata &    \nodata &    \nodata &  1  \\ 
 3.420657 &     5.5 &     (4.4) &    2.20 $\pm$    0.13 &   15.8 $\pm$    0.4\phm{0} &    \nodata &    \nodata &    \nodata &  2  \\ 
 3.420838 &     7.8 &     (6.3) &    2.90 $\pm$    0.12 &    6.15 $\pm$    0.38 &    \nodata &    \nodata &    \nodata &  3  \\ 
 3.421212 &    15.4 &    (14.4) &    1.09 $\pm$    0.14 &  $<$    3.24 &    \nodata &    \nodata &    \nodata &  4  \\ 
\sidehead{3.468\tablenotemark{g}}
 3.467141 &    20.2 &    (19.5) &    7.25 $\pm$    0.44 &    4.44 $\pm$    0.74 &    \nodata &    \nodata &    \nodata &  1  \\ 
 3.467446 &     9.1 &     (7.3) &    0.77 $\pm$    0.26 &    3.40 $\pm$    0.52 &    \nodata &    \nodata &    \nodata &  2  \\ 
 3.467751 &    12.4 &    (11.1) &    7.70 $\pm$    0.32 &    9.72 $\pm$    0.60 &    \nodata &    \nodata &    \nodata &  3  \\ 
 3.467968 &     9.1 &     (7.3) &    4.75 $\pm$    0.27 &    6.65 $\pm$    0.49 &    \nodata &    \nodata &    \nodata &  4  \\ 
\sidehead{3.514\tablenotemark{h}}
 3.513426 &     3.8 &   \nodata &    0.18 $\pm$    0.11 &    \nodata &    \nodata &    \nodata &    \nodata &  1  \\ 
 3.513691 &     7.9 &   \nodata &    0.57 $\pm$    0.15 &    \nodata &    \nodata &    \nodata &    \nodata &  2  \\ 
 3.514141 &    12.0 &   \nodata &    2.10 $\pm$    0.21 &    \nodata &    \nodata &    \nodata &    \nodata &  3  \\ 
 3.514405 &     8.0 &   \nodata &    6.32 $\pm$    0.20 &    \nodata &    \nodata &    \nodata &    \nodata &  4  \\ 
\sidehead{3.544\tablenotemark{i}}
 3.541717 &    23.7 &    (23.1) &    4.70 $\pm$    0.41 &    7.15 $\pm$    1.42 &    \nodata &    \nodata &    \nodata &  1  \\ 
 3.542056 &    10.5 &     (8.9) &    5.18 $\pm$    0.31 &   12.8 $\pm$    1.1\phm{0} &    \nodata &    \nodata &    \nodata &  2  \\ 
 3.542427 &    13.6 &    (12.5) &   26.7 $\pm$    0.5\phm{0} &   79.7 $\pm$    1.6\phm{0} &    \nodata &    \nodata &    \nodata &  3  \\ 
 3.542665 &    11.9 &    (10.6) &   27.2 $\pm$    0.7\phm{0} &   43.5 $\pm$    1.8\phm{0} &    \nodata &    \nodata &    \nodata &  4  \\ 
 3.542967 &    17.2 &    (16.3) &   29.7 $\pm$    0.7\phm{0} &   27.6 $\pm$    1.9\phm{0} &    \nodata &    \nodata &    \nodata &  5  \\ 
 3.543164 &     5.7 &     (4.8) &    6.26 $\pm$    0.40 &  $<$    1.26 &    \nodata &    \nodata &    \nodata &  6  \\ 
 3.543341 &     8.5 &     (7.0) &   13.8 $\pm$    0.4\phm{0} &   14.4 $\pm$    1.2\phm{0} &    \nodata &    \nodata &    \nodata &  7  \\ 
 3.543622 &     9.6 &     (7.3) &   10.2 $\pm$    0.3\phm{0} &   21.1 $\pm$    1.0\phm{0} &    \nodata &    \nodata &    \nodata &  8  \\ 
 3.544162 &    37.0 &    (36.6) &   22.5 $\pm$    0.6\phm{0} &   10.7 $\pm$    2.1\phm{0} &    \nodata &    \nodata &    \nodata &  9  \\ 
 3.544469 &     5.9 &      4.0\phm{)} &   13.2 $\pm$    0.4\phm{0} &   37.3 $\pm$    0.9\phm{0} &    \nodata &    \nodata &    \nodata & 10  \\ 
 3.544663 &     3.7 &     (2.4) &    2.49 $\pm$    0.22 &    3.88 $\pm$    0.53 &    \nodata &    \nodata &    \nodata & 11  \\ 
 3.544837 &     6.5 &      4.3\phm{)} &   17.1 $\pm$    0.4\phm{0} &   18.3 $\pm$    0.7\phm{0} &    \nodata &    \nodata &    \nodata & 12  \\ 
 3.545112 &    12.2 &    (10.9) &    2.20 $\pm$    0.30 &    3.21 $\pm$    0.85 &    \nodata &    \nodata &    \nodata & 13  \\ 
 3.545386 &    13.5 &    (12.4) &    0.98 $\pm$    0.38 &    6.37 $\pm$    1.06 &    \nodata &    \nodata &    \nodata & 14  \\ 
 3.545645 &    20.6 &    (19.9) &    1.22 $\pm$    0.44 &    6.95 $\pm$    1.34 &    \nodata &    \nodata &    \nodata & 15  \\ 
 3.546023 &     9.6 &     (8.1) &    0.84 $\pm$    0.24 &    2.80 $\pm$    0.71 &    \nodata &    \nodata &    \nodata & 16  \\ 
\sidehead{3.567}
 3.566854 &    18.6 &    (17.8) &    4.43 $\pm$    0.30 &  $<$    0.76 &    \nodata &    \nodata &    \nodata &  1  \\ 
\sidehead{3.690\tablenotemark{j}}
 3.689699 &     3.7 &     (2.5) &    0.71 $\pm$    0.13 &  $<$    0.80 &  $<$    0.17 &    \nodata &    \nodata &  1  \\ 
 3.689963 &     9.9 &     (8.2) &    1.89 $\pm$    0.19 &  $<$    0.98 &  $<$    0.23 &    \nodata &    \nodata &  2  \\ 
 3.690267 &    10.2 &     (8.6) &    4.35 $\pm$    0.20 &    1.41 $\pm$    0.98 &  $<$    0.24 &    \nodata &    \nodata &  3  \\ 
 3.690600 &     9.7 &     (8.0) &    1.57 $\pm$    0.18 &  $<$    0.97 &    \nodata &    \nodata &    \nodata &  4  \\ 
\sidehead{3.752\tablenotemark{k}}  
 3.751237 &     4.8 &     (3.8) &    2.13 $\pm$    0.27 &  $<$    0.78 &    \nodata &    \nodata &    \nodata &  1  \\ 
 3.751557 &    30.1 &    (29.6) &    6.29 $\pm$    0.60 &  $<$    1.40 &    \nodata &    \nodata &    \nodata &  2  \\ 
 3.751980 &     6.0 &      4.4\phm{)} &   18.1 $\pm$    0.5\phm{0} &    4.25 $\pm$    0.61 &  $<$    0.32 &    \nodata &    \nodata &  3  \\ 
 3.752309 &    11.2 &     (9.8) &    7.66 $\pm$    0.43 &    2.12 $\pm$    0.82 &  $<$    0.49 &    \nodata &    \nodata &  4  \\ 
 3.752474 &     3.3 &     (2.2) &    3.17 $\pm$    1.04 &   10.7 $\pm$    1.2\phm{0} &  $<$    0.43 &    \nodata &    \nodata &  5  \\ 
 3.752535 &     6.6 &     (5.1) &   28.2 $\pm$    1.3\phm{0} &   23.0 $\pm$    1.3\phm{0} &  $<$    0.53 &    \nodata &    \nodata &  6  \\ 
 3.752742 &     5.4 &     (4.4) &   10.6 $\pm$    0.6\phm{0} &   18.5 $\pm$    1.0\phm{0} &  $<$    0.43 &    \nodata &    \nodata &  7  \\ 
 3.752798 &    17.8 &    (17.0) &    7.86 $\pm$    0.76 &    4.31 $\pm$    1.64 &    \nodata &    \nodata &    \nodata &  8  \\ 
 3.753312 &     5.9 &     (4.9) &    2.74 $\pm$    0.34 &    2.13 $\pm$    0.59 &  $<$    0.27 &    \nodata &    \nodata &  9  \\ 
 3.753501 &     4.9 &     3.6\phm{)} &    5.19 $\pm$    0.64 &   12.3 $\pm$    1.0\phm{0} &  $<$    0.45 &    \nodata &    \nodata & 10  \\ 
 3.753573 &    11.1 &     (9.6) &   13.6 $\pm$    0.7\phm{0} &    2.35 $\pm$    1.21 &  $<$    0.86 &    \nodata &    \nodata & 11  \\ 
\sidehead{3.759\tablenotemark{l}}
 3.758220 &    13.2 &    (12.0) &   10.2 $\pm$    0.5\phm{0} &   14.5 $\pm$    0.9\phm{0} &    \nodata &    \nodata &    \nodata &  1  \\ *
 3.758825 &     6.9 &     (5.8) &    1.07 $\pm$    0.28 &  $<$    0.67 &  $<$    0.35 &    \nodata &    \nodata &  2  \\ 
\sidehead{3.779\tablenotemark{m}}
 3.779232 &     9.3 &     (7.5) &    2.45 $\pm$    0.27 &  $<$    0.99 &  $<$    0.29 &    \nodata &    \nodata &  1  \\ 
\sidehead{3.811}
 3.810675 &    15.3 &    (14.3) &    1.34 $\pm$    0.28 &  $<$    1.40 &  $<$    0.76 &    \nodata &    \nodata &  1  \\ 
 3.811041 &     7.2 &     (6.2) &    3.78 $\pm$    0.22 &  $<$    1.39 &  $<$    0.43 &    \nodata &    \nodata &  2  \\ 
\sidehead{3.863\tablenotemark{n}}
 3.863018 &     8.6 &     (6.8) &    0.80 $\pm$    0.25 &  $<$    1.52 &    \nodata &    \nodata &    \nodata &  1  \\ 
 3.863389 &    12.8 &    (11.6) &    2.35 $\pm$    0.31 &    2.21 $\pm$    0.84 &    \nodata &    \nodata &    \nodata &  2  \\ 
\sidehead{3.881\tablenotemark{o}}
 3.880046 &    49.7 &    (49.4) &   12.1 $\pm$    0.6\phm{0} &    9.92 $\pm$    1.49 &  $<$    0.72 &    \nodata &    \nodata &  1  \\ 
 3.880840 &     5.3 &     (4.3) &    1.24 $\pm$    0.22 &  $<$    0.50 &    \nodata &    \nodata &    \nodata &  2  \\ 
\sidehead{3.921\tablenotemark{p}}
 3.920918 &    10.5 &     (9.7) &    2.93 $\pm$    0.40 &   19.0\phm{0} $\pm$    0.60 &    0.59 $\pm$    0.34 &  $<$    0.27 &    \nodata &  1  \\ 
 3.921157 &    16.2 &    (15.7) &   14.9\phm{0} $\pm$    0.69 &    3.92 $\pm$    0.92 &  $<$    0.51 &  $<$    0.41 &    \nodata &  2  \\ 
 3.921199 &     7.9 &     (6.8) &    5.66 $\pm$    0.42 &   21.3\phm{0} $\pm$    0.54 &    0.91 $\pm$    0.30 &  $<$    0.23 &    \nodata &  3  \\ 
 3.921509 &     7.7 &     (6.6) &    3.70 $\pm$    0.35 &   14.3\phm{0} $\pm$    0.54 &  $<$    0.32 &  $<$    0.24 &    \nodata &  4  \\ 
 3.921682 &     6.4 &     (5.1) &    3.47 $\pm$    0.32 &    2.77 $\pm$    0.45 &  $<$    0.29 &  $<$    0.22 &    \nodata &  5  \\ 
\sidehead{3.976\tablenotemark{q}}
 3.974237 &    34.1 &    (33.6) &    2.60 $\pm$    0.53 &  $<$    2.37 &    \nodata &    \nodata &    \nodata &  1  \\* 
 3.975244 &    12.9 &    (11.6) &    1.49 $\pm$    0.30 &  $<$    1.24 &    \nodata &    \nodata &    \nodata &  2  \\* 
 3.976158 &    32.9 &   \nodata &    4.62 $\pm$    0.51 &    \nodata &    \nodata &    \nodata &    \nodata &  3  \\ 
\sidehead{3.994}
 3.993795 &    11.6 &    (10.3) &    3.40 $\pm$    0.25 &  $<$    0.91 &  $<$    0.44 &    \nodata &    \nodata &  1  \\ 
 3.994076 &     8.8 &     (7.1) &    2.00 $\pm$    0.22 &  $<$    0.74 &  $<$    0.43 &    \nodata &    \nodata &  2  \\ 
\sidehead{4.024\tablenotemark{r}}
 4.022713 &    18.2 &   \nodata &    1.78 $\pm$    0.41 &    \nodata &  $<$    0.41 &    \nodata &  $<$    0.28 &  1  \\ 
 4.023240 &     9.4 &     (7.7) &    3.28 $\pm$    0.36 &    5.39 $\pm$    0.80 &  $<$    0.33 &    \nodata &  $<$    0.23 &  2  \\ 
 4.023394 &     4.1 &     (2.9) &    1.57 $\pm$    0.32 &    6.25 $\pm$    0.69 &  $<$    0.25 &    \nodata &  $<$    0.19 &  3  \\ 
 4.023609 &    11.7 &    (10.4) &    8.46 $\pm$    0.42 &   25.7 $\pm$    0.9\phm{0} &    0.96 $\pm$    0.34 &    \nodata &  $<$    0.28 &  4  \\ 
 4.023930 &     7.5 &     (6.3) &    3.70 $\pm$    0.31 &    1.74 $\pm$    0.66 &  $<$    0.53 &    \nodata &  $<$    0.20 &  5  \\ 
 4.024154 &     4.3 &     (3.2) &    1.20 $\pm$    0.23 &  $<$    0.76 &  $<$    0.50 &    \nodata &  $<$    0.16 &  6  \\ 
 4.024557 &    10.4 &     (8.9) &    1.62 $\pm$    0.30 &  $<$    0.71 &  $<$    0.55 &    \nodata &  $<$    0.21 &  7  \\ 
\sidehead{4.031\tablenotemark{s}}
 4.029082 &     6.1 &     (5.1) &    1.29 $\pm$    0.35 &  $<$    0.85 &  $<$    0.30 &  $<$    0.40 &  $<$    0.19 &  1  \\ 
 4.029235 &     8.3 &     (7.1) &    1.78 $\pm$    0.42 &    2.29 $\pm$    0.99 &  $<$    0.38 &  $<$    0.48 &  $<$    0.27 &  2  \\ 
 4.029449 &     8.2 &     (6.8) &    7.84 $\pm$    0.48 &    2.31 $\pm$    0.95 &  $<$    0.36 &    \nodata &  $<$    0.26 &  3  \\ 
 4.029637 &     6.7 &     (5.5) &   17.3 $\pm$    0.8\phm{0} &   10.2 $\pm$    1.5\phm{0} &  $<$    0.66 &    \nodata &  $<$    0.24 &  4  \\ 
 4.029792 &     7.2 &     (6.0) &    9.66 $\pm$    1.85 &   84.4 $\pm$    4.9\phm{0} &  $<$    2.77 &  $<$   12.30 &  $<$    0.60 &  5  \\ 
 4.029875 &     8.0 &     (6.6) &    2.66 $\pm$    2.52 &    5.68 $\pm$    5.29 &   67.3 $\pm$    4.1\phm{0} &  102 $\pm$   13\phm{0} &  $<$    0.80 &  6  \\ 
 4.029986 &     8.4 &     (6.7) &   19.4 $\pm$    1.5\phm{0} &   17.9 $\pm$    2.1\phm{0} &    8.40 $\pm$    1.98 &  $<$   15.2\phm{0} &  $<$    0.43 &  7  \\ 
 4.030165 &     5.1 &     (4.1) &    2.74 $\pm$    0.49 &    4.03 $\pm$    0.96 &   38.8 $\pm$    1.1\phm{0} &  119 $\pm$    7\phm{00} &  $<$    0.24 &  8  \\ 
 4.030371 &     9.3 &     (7.6) &   13.3 $\pm$    0.6\phm{0} &   62.1 $\pm$    1.8\phm{0} &   33.3 $\pm$    0.8\phm{0} &   72.5 $\pm$    4.7\phm{0} &  $<$    0.25 &  9  \\ 
 4.030507 &     4.3 &     (3.5) &    0.19 $\pm$    0.53 &    2.64 $\pm$    1.54 &   42.8 $\pm$    1.6\phm{0} &   68.4 $\pm$    3.7\phm{0} &  $<$    0.23 & 10  \\ 
 4.030608 &     7.4 &     (6.1) &    5.24 $\pm$    0.64 &   47.4 $\pm$    2.0\phm{0} &    4.28 $\pm$    0.82 &    7.58 $\pm$    1.49 &  $<$    0.29 & 11  \\ 
 4.030774 &    11.7 &    (10.3) &    7.53 $\pm$    0.53 &   25.0 $\pm$    1.8\phm{0} &    0.80 $\pm$    0.44 &  $<$    0.96 &  $<$    0.33 & 12  \\ 
 4.031528 &    26.0 &    (25.4) &    5.34 $\pm$    1.05 &    6.69 $\pm$    2.96 &  $<$    0.89 &    \nodata &  $<$    0.67 & 13  \\ 
 4.031661 &    15.1 &    (14.1) &    3.60 $\pm$    0.85 &  $<$    2.90 &  $<$    0.69 &  $<$    1.00 &  $<$    0.43 & 14  \\ 
 4.031999 &    16.2 &    (15.3) &    1.05 $\pm$    0.47 &  $<$    1.08 &  $<$    0.76 &    \nodata &  $<$    0.26 & 15  \\ 
 4.032692 &     6.4 &     (5.5) &    0.46 $\pm$    0.26 &  $<$    0.78 &  $<$    0.22 &  $<$    0.35 &  $<$    0.16 & 16  \\ 
 4.033157 &    18.1 &    (17.3) &    3.04 $\pm$    0.42 &  $<$    2.47 &  $<$    0.38 &    \nodata &  $<$    0.36 & 17  \\ 
\sidehead{4.037\tablenotemark{t}}
 4.035726 &    11.3 &     (9.9) &    4.98 $\pm$    0.48 &    3.77 $\pm$    1.14 &  $<$    0.50 &  $<$    0.34 &  $<$    0.20 &  1  \\ 
 4.036301 &    11.8 &    (10.5) &   14.1 $\pm$    0.4\phm{0} &   28.6 $\pm$    1.6\phm{0} &    1.30 $\pm$    0.56 &  $<$    0.46 &  $<$    0.20 &  2  \\ 
 4.036650 &     6.2 &     (5.0) &    6.61 $\pm$    0.34 &    6.57 $\pm$    1.23 &  $<$    0.46 &  $<$    0.27 &  $<$    0.25 &  3  \\ 
 4.037141 &    13.9 &    (12.8) &    6.62 $\pm$    0.41 &  $<$    0.52 &  $<$    0.78 &    $<$    0.85 &    $<$    0.54 &  4  \\ 
 4.037506 &    11.5 &    (10.1) &    2.00 $\pm$    0.40 &  $<$    2.31 &  $<$    1.91 &  $<$    0.35 &  $<$    0.34 &  5  \\ 
\enddata
 
\tablenotetext{*}{See Figure 1. For this object refer to the corresponding footnote to Table 7: here for Q1645+5520 the specific 
lengths and separation of observing runs become manifest simply as a mild broadening of the atmospheric emission and absorption lines.}

\tablenotetext{a}{\CIV~$\lambda$1551 is blended with \SiIV~$\lambda$1394 at $z = 3.468$ but $\lambda$1548 is clear.}

\tablenotetext{b}{\CIV~$\lambda$1548 has mild contamination by atmospheric line absorption but $\lambda$1551 is clear.}

\tablenotetext{c}{\CIV~$\lambda$1551 in component 1 has weak atmospheric line absorption but $\lambda$1548 is clear.}

\tablenotetext{d}{\CIV~$\lambda$1551 is blended with weak \CII\ at $z = 4.024$ but $\lambda$1548 is clear.}

\tablenotetext{e}{\CIV~$\lambda$1548 is blended with weak \CII\ at $z = 4.037$ but $\lambda$1551 is clear. This system marks the start 
of a sequence of three weak \CIV\ systems: $\lambda$1551 blended with $\lambda$1548 at $z = 3.348$ and $\lambda$1551 of the latter 
system blended with $\lambda$1548 at $z = 3.355$.}

\tablenotetext{f}{\CIV~$\lambda$1548 is closely adjacent to weak \SiIV~$\lambda$1403 at $z = 3.881$ and in components 1 and 2 contains 
weak atmospheric line emission but $\lambda$1551 is clear. The values for \SiIV\ are from $\lambda$1403 which is outside the forest.}

\tablenotetext{g}{\CIV~$\lambda$1548 in component 1 has strong contamination by atmospheric line absorption but $\lambda$1551 is clear. 
\SiIV~$\lambda$1394 in component 2  contains weak \CIV~$\lambda$1551 at $z = 3.015$ but $\lambda$1403 is clear.}

\tablenotetext{h}{\CIV~$\lambda$1551 is contaminated by atmospheric line absorption and blended with \SiIV~$\lambda$1394 at $z = 4.024$ 
but $\lambda$1548 is sufficiently clear of interference for definition of the recorded components. \SiIV~$\lambda$1394 is too 
contaminated by atmospheric line absorption to be useful and $\lambda$1403 is masked by \SiIV~$\lambda$1394 at $z = 3.544$.}

\tablenotetext{i}{\CIV~$\lambda$1551 is blended with \SiIV~$\lambda$1403 at $z = 4.024$ but $\lambda$1548 is clear. \SiIV~$\lambda$1394 is 
partially blended with weak \SiIV~$\lambda$1403 at $z = 3.514$ and components 3 and 4 include atmospheric line emission but $\lambda$1403 
is clear.}

\tablenotetext{j}{CIV~$\lambda$1548 has mild contamination by atmospheric line absorption but $\lambda$1551 is clear. \SiIV~$\lambda$1403 
has strong interference by atmospheric line emission but $\lambda$1394 is clear. \CII\ is outside the forest; in component 4 it there is 
mild interference by atmospheric line emission.}

\tablenotetext{k}{\CIV~$\lambda$1551 in components 9--11 is excluded by strong atmospheric line emission but here 
$\lambda$1548 is clear, and additional mild atmospheric line absorption is present in both, but taken together good values are obtained 
for all components. \CII\ in components 1 and 2 is masked by strong \SiII~$\lambda$1260 at $z = 4.031$ and in further components overlaps 
with extended weak \SiII\ in the same system but useful upper limits are obtained as listed.}

\tablenotetext{l}{\CII\ in component 1 is too contaminated by atmospheric line emission to be useful.}

\tablenotetext{m}{\SiIV~$\lambda$1403 is close in wavelength to \CIV~$\lambda$1551 at $z = 3.322$ and the upper limit is from 
$\lambda$1394.}

\tablenotetext{n}{\CII\ is obscured by \CIV~$\lambda$1548 at $z = 3.193$.}

\tablenotetext{o}{There is weak atmospheric line emission in both \CIV\ doublet members but taken together the interference is not 
significant. \SiIV~$\lambda$1403 is close in wavelength to \CIV~$\lambda$1548 at $z = 3.421$ but $\lambda$1394 is clear. \CII\ in 
component 2 is excluded by weak atmospheric line absorption.} 

\tablenotetext{p}{\CIV~$\lambda$1551 is largely immersed in strong atmospheric line absorption and not used but $\lambda$1548 is clear 
except for component 1 which is only mildly affected. \SiIV~$\lambda$1403 has similar interference in component 1 and in 4 and 5 is 
rather noisy but $\lambda$1394 is clear. \SiII~$\lambda$1260 is outside the forest.}

\tablenotetext{q}{\CIV~$\lambda$1551 has interference by strong line emission but $\lambda$1548 is relatively clear and taken together 
the listed values are acceptable. Weak atmospheric line absorption occurs in component 1 of \SiIV~$\lambda$1394, component 2 of 
$\lambda$1403 and component 3 in both and the listed values are from the available complementary regions. \CII\ is obscured by 
\CIV~$\lambda$1548 at $z = 3.289$. \SiII~$\lambda$1260 is too "noisy" to include.}

\tablenotetext{r}{\CIV~$\lambda$1551 is blended with \CIV~$\lambda$1548 at $z = 4.031$ but $\lambda$1548 is clear of significant 
atmospheric interference and good values are obtained. \SiIV~$\lambda$1403 is immersed in strong \CIV~$\lambda$1551 at $z =3.544$ while 
$\lambda$1394 in components 1 and 2 is blended with \CIV~$\lambda$1551 at $z = 3.514$ and also in component 1 is contaminated by 
atmospheric line absorption, but together good values are obtained for components 2--7. \CII\ in components 1--4 is blended with 
\CIV~$\lambda$1551 at $z = 3.322$ and in 7 with weak \SiIV~$\lambda$1394 at $z = 3.811$ but acceptable values are obtained for all 
components. \SiII~$\lambda$1260 is masked by \SiIV~$\lambda$1394 at $z = 3.544$. \NV\ is outside the forest.}

\tablenotetext{s}{This complex system has strong \CII\ and \SiII\ and relatively weak related \CIV\ in components 6--10: for these 
collectively a reasonable fit to the \HI\ Lyman $\alpha$ damping wings is given by log~$N$(\HI) $\sim19.7$. Onward from component 12 
\CIV~$\lambda$1548 is blended with \CIV~$\lambda$1551 at $z = 4.024$ and in components 1 and 2 $\lambda$1551 is blended with 
\CIV~$\lambda$1548 at $z = 4.037$ but taken together good values are obtained. Although \SiIV\ has some contamination 
by atmospheric line absorption, good values are obtained for all components. \CII\ is in a fully clear region. In the strong 
components \SiII~$\lambda$1260 is mildly saturated but together with $\lambda$1304 good values are obtained ($\lambda$1527 is heavily 
contaminated by atmospheric line absorption); in components 12--17 weak $\lambda$1260 is blended with weak \CII\ at $z = 3.752$ but 
some useful upper limits are listed. All of \NV\ is clear.}  

\tablenotetext{t}{\CIV~$\lambda$1548 in component 1 coincides with strong atmospheric line emission and in 5 encroaches on 
\CIV~$\lambda$1551 at $z = 4.031$, while $\lambda$1551 in components 2 and 3 also has interference from strong atmospheric line emission, 
but together good values are obtained. \SiIV~$\lambda$1394 is partly contaminated by atmospheric line absorption but the values extracted 
are acceptable, while \SiIV~$\lambda$1403 is too "noisy" to be useful. \CII\ is blended with weak \CIV~$\lambda$1548 at $z = 3.341$ but 
by including \CIV~$\lambda$1551 useful values are obtained. All of \NV\ is clear.}

\end{deluxetable}

\clearpage
\begin{deluxetable}{crrcccccc}
\tabletypesize{\scriptsize}
\tablecaption{Heavy Element Absorption Lines: Q1055+4611  $z_{em} = 4.15^*$}
\tablewidth{0pt}
\tablecolumns{9}
\tablehead{
\colhead{$z$} & \colhead{$b$(C)}  & \colhead{$b$(Si)} &
\colhead{$N$(\CIV)} & \colhead{$N$(\SiIV)} &
\colhead{$N$(\CII)} & \colhead{$N$(\SiII)} &
\colhead{$N$(\NV)} &  \colhead{n} \\
{} & \colhead{km~s$^{-1}$} & \colhead{km~s$^{-1}$} & \colhead{10$^{12}$cm$^{-2}$} &
\colhead{10$^{11}$cm$^{-2}$} & \colhead{10$^{12}$cm$^{-2}$} & \colhead{10$^{11}$cm$^{-2}$} & \colhead{10$^{12}$cm$^{-2}$} & {} \\
\colhead{1} & \colhead{2} & \colhead{3} & \colhead{4} & \colhead{5} & \colhead{6} & \colhead{7} & \colhead{8} & \colhead{9} 
}

\startdata
\sidehead{3.063\tablenotemark{a}}
 3.061530 &    11.8 &   \nodata &   11.7 $\pm$    0.5\phm{0} &    \nodata &    \nodata &    \nodata &    \nodata &  1  \\ 
 3.062074 &     5.1 &   \nodata &    3.92 $\pm$    0.33 &    \nodata &    \nodata &    \nodata &    \nodata &  2  \\ 
 3.062485 &     8.5 &   \nodata &   85.8 $\pm$    1.5\phm{0} &    \nodata &    \nodata &    \nodata &    \nodata &  3  \\ 
 3.062607 &     5.1 &   \nodata &   17.4 $\pm$    0.8\phm{0} &    \nodata &    \nodata &    \nodata &    \nodata &  4  \\ 
 3.063773 &    11.3 &   \nodata &    7.72 $\pm$    0.49 &    \nodata &    \nodata &    \nodata &    \nodata &  5  \\ 
\sidehead{3.090\tablenotemark{b}}
 3.090376 &    21.3 &   \nodata &    1.89 $\pm$    0.28 &    \nodata &    \nodata &    \nodata &    \nodata &  1  \\ 
\sidehead{3.122}
 3.121821 &     9.7 &   \nodata &    3.06 $\pm$    0.17 &    \nodata &    \nodata &    \nodata &    \nodata &  1  \\ 
 3.122077 &    19.5 &   \nodata &    1.99 $\pm$    0.30 &    \nodata &    \nodata &    \nodata &    \nodata &  2  \\ 
 3.122197 &     4.2 &   \nodata &   12.9 $\pm$    0.2\phm{0} &    \nodata &    \nodata &    \nodata &    \nodata &  3  \\ 
 3.122825 &    17.8 &   \nodata &    0.80 $\pm$    0.18 &    \nodata &    \nodata &    \nodata &    \nodata &  4  \\ 
\sidehead{3.154\tablenotemark{c}}
 3.153131 &    26.9 &   \nodata &    1.11 $\pm$    0.50 &    \nodata &    \nodata &    \nodata &    \nodata &  1  \\ 
 3.153408 &    13.2 &   \nodata &    2.44 $\pm$    0.36 &    \nodata &    \nodata &    \nodata &    \nodata &  2  \\ 
 3.154130 &    28.7 &   \nodata &    1.44 $\pm$    0.48 &    \nodata &    \nodata &    \nodata &    \nodata &  3  \\ 
 3.154489 &    16.0 &   \nodata &    2.83 $\pm$    0.40 &    \nodata &    \nodata &    \nodata &    \nodata &  4  \\ 
 3.155057 &    16.5 &   \nodata &    2.28 $\pm$    0.35 &    \nodata &    \nodata &    \nodata &    \nodata &  5  \\ 
\sidehead{3.167}
 3.163730 &     7.1 &   \nodata &    3.68 $\pm$    0.22 &    \nodata &    \nodata &    \nodata &    \nodata &  1  \\ 
 3.163925 &     7.6 &   \nodata &    3.92 $\pm$    0.23 &    \nodata &    \nodata &    \nodata &    \nodata &  2  \\ 
 3.165066 &    12.4 &   \nodata &    2.68 $\pm$    0.25 &    \nodata &    \nodata &    \nodata &    \nodata &  3  \\ 
 3.165533 &     4.6 &   \nodata &    1.17 $\pm$    0.17 &    \nodata &    \nodata &    \nodata &    \nodata &  4  \\ 
 3.166536 &    29.1 &   \nodata &    6.33 $\pm$    0.41 &    \nodata &    \nodata &    \nodata &    \nodata &  5  \\ 
 3.166717 &     7.1 &   \nodata &    6.39 $\pm$    0.24 &    \nodata &    \nodata &    \nodata &    \nodata &  6  \\ 
 3.167199 &     8.4 &   \nodata &    9.48 $\pm$    0.22 &    \nodata &    \nodata &    \nodata &    \nodata &  7  \\ 
 3.167448 &     7.3 &   \nodata &    1.98 $\pm$    0.18 &    \nodata &    \nodata &    \nodata &    \nodata &  8  \\ 
 3.168484 &     8.1 &   \nodata &    3.73 $\pm$    0.21 &    \nodata &    \nodata &    \nodata &    \nodata &  9  \\ 
 3.168668 &     7.2 &   \nodata &    5.53 $\pm$    0.23 &    \nodata &    \nodata &    \nodata &    \nodata & 10  \\ 
 3.168806 &     5.0 &   \nodata &    8.57 $\pm$    0.24 &    \nodata &    \nodata &    \nodata &    \nodata & 11  \\ 
 3.169067 &    11.5 &   \nodata &    4.87 $\pm$    0.22 &    \nodata &    \nodata &    \nodata &    \nodata & 12  \\ 
 3.169356 &    12.7 &   \nodata &    2.72 $\pm$    0.22 &    \nodata &    \nodata &    \nodata &    \nodata & 13  \\ 
\sidehead{3.228\tablenotemark{d}}
 3.227567 &     9.8 &   \nodata &   11.3 $\pm$    0.4\phm{0} &    \nodata &    \nodata &    \nodata &    \nodata &  1  \\ 
 3.227794 &    10.2 &   \nodata &    5.61 $\pm$    0.38 &    \nodata &    \nodata &    \nodata &    \nodata &  2  \\ 
 3.228220 &    21.5 &   \nodata &    2.87 $\pm$    0.51 &    \nodata &    \nodata &    \nodata &    \nodata &  3  \\ 
\sidehead{3.234\tablenotemark{e}}
 3.233204 &     6.1 &   \nodata &    3.34 $\pm$    0.42 &    \nodata &    \nodata &    \nodata &    \nodata &  1  \\ 
 3.233413 &    37.7 &   \nodata &    5.88 $\pm$    1.07 &    \nodata &    \nodata &    \nodata &    \nodata &  2  \\ 
 3.233769 &    11.5 &   \nodata &   22.8 $\pm$    0.7\phm{0} &    \nodata &    \nodata &    \nodata &    \nodata &  3  \\ 
 3.235253 &    16.1 &   \nodata &    3.79 $\pm$    0.38 &    \nodata &    \nodata &    \nodata &    \nodata &  4  \\ 
\sidehead{3.238}
 3.236618 &    13.2 &   \nodata &   16.7 $\pm$    0.6\phm{0} &    \nodata &    \nodata &    \nodata &    \nodata &  1  \\* 
 3.236759 &     4.1 &   \nodata &   17.1 $\pm$    1.1\phm{0} &    \nodata &    \nodata &    \nodata &    \nodata &  2  \\
 3.236866 &     6.6 &   \nodata &   24.0 $\pm$    1.0\phm{0} &    \nodata &    \nodata &    \nodata &    \nodata &  3  \\ 
 3.237038 &    11.9 &   \nodata &   16.8 $\pm$    0.6\phm{0} &    \nodata &    \nodata &    \nodata &    \nodata &  4  \\ 
 3.237794 &     9.1 &   \nodata &    1.34 $\pm$    0.37 &    \nodata &    \nodata &    \nodata &    \nodata &  5  \\ 
 3.238040 &     9.2 &   \nodata &   10.1 $\pm$    0.4\phm{0} &    \nodata &    \nodata &    \nodata &    \nodata &  6  \\ 
 3.238369 &     9.9 &   \nodata &   39.8 $\pm$    0.8\phm{0} &    \nodata &    \nodata &    \nodata &    \nodata &  7  \\ 
 3.238679 &     9.9 &   \nodata &   43.5 $\pm$    0.9\phm{0} &    \nodata &    \nodata &    \nodata &    \nodata &  8  \\ 
 3.238910 &     8.1 &   \nodata &   17.7 $\pm$    0.6\phm{0} &    \nodata &    \nodata &    \nodata &    \nodata &  9  \\ 
 3.239162 &     9.9 &   \nodata &   10.4 $\pm$    0.4\phm{0} &    \nodata &    \nodata &    \nodata &    \nodata & 10  \\ 
 3.239429 &     8.6 &   \nodata &    9.49 $\pm$    0.44 &    \nodata &    \nodata &    \nodata &    \nodata & 11  \\ 
 3.239661 &    13.1 &   \nodata &    5.14 $\pm$    0.48 &    \nodata &    \nodata &    \nodata &    \nodata & 12  \\ 
\sidehead{3.269}
 3.268610 &    11.6 &   \nodata &    4.45 $\pm$    0.19 &    \nodata &    \nodata &    \nodata &    \nodata &  1  \\ *
 3.269033 &    27.5 &   \nodata &    2.47 $\pm$    0.27 &    \nodata &    \nodata &    \nodata &    \nodata &  2  \\ 
\sidehead{3.317\tablenotemark{f}}
 3.314915 &     2.9 &     (1.9) &    0.83 $\pm$    0.18 &    \nodata &    \nodata &  $<$    2.37 &    \nodata &  1  \\ 
 3.315923 &     6.4 &     (5.2) &    6.89 $\pm$    0.27 &    \nodata &    \nodata &  $<$    2.84 &    \nodata &  2  \\ 
 3.316131 &     5.2 &     (4.3) &   25.6 $\pm$    0.6\phm{0} &    \nodata &    \nodata &  $<$    3.47 &    \nodata &  3  \\ 
 3.316262 &     4.6 &     (3.7) &   13.8 $\pm$    0.4\phm{0} &    \nodata &    \nodata &  $<$    2.95 &    \nodata &  4  \\ 
 3.316486 &     2.9 &     (1.9) &    1.66 $\pm$    0.20 &    \nodata &    \nodata &    3.66 $\pm$    2.43 &    \nodata &  5  \\ 
 3.316717 &     3.3 &     (2.2) &    1.20 $\pm$    0.20 &    \nodata &    \nodata &  $<$    4.07 &    \nodata &  6  \\ 
 3.316851 &     7.1 &     (5.7) &    1.28 $\pm$    0.42 &    \nodata &    \nodata &  344 $\pm$   13\phm{0} &    \nodata &  7  \\ 
 3.317070 &    (17) &  16.4\phm{)} &  $<$    1.38 &    \nodata &    \nodata &   70.1 $\pm$   26.0 &    \nodata &  8  \\ 
 3.317183 &    10.7 &     (9.1) &    3.92 $\pm$    0.76 &    \nodata &    \nodata &  986 $\pm$   46\phm{0} &    \nodata &  9  \\ 
 3.317346 &     5.8 &     (4.4) &    5.43 $\pm$    0.32 &    \nodata &    \nodata &  262 $\pm$   16\phm{0} &    \nodata & 10  \\ 
 3.317476 &     6.0 &     (5.5) &    4.96 $\pm$    0.29 &    \nodata &    \nodata &   44.1 $\pm$    4.4\phm{0} &    \nodata & 11  \\ 
 3.317710 &    11.8 &    (10.4) &   10.5 $\pm$    0.4\phm{0} &    \nodata &    \nodata &   19.8 $\pm$    4.6\phm{0} &    \nodata & 12  \\ 
 3.317878 &     (7) &   5.6\phm{)} &  $<$    0.34 &    \nodata &    \nodata &   21.0 $\pm$    4.2\phm{0} &    \nodata & 13  \\ 
 3.318006 &    11.1 &     (9.7) &    8.33 $\pm$    0.37 &    \nodata &    \nodata &  $<$    6.78 &    \nodata & 14  \\ 
 3.318554 &    13.4 &    (12.3) &    3.38 $\pm$    0.28 &    \nodata &    \nodata &  $<$    4.47 &    \nodata & 15  \\ 
 3.318961 &     9.4 &     (7.7) &    1.90 $\pm$    0.23 &    \nodata &    \nodata &  $<$    3.28 &    \nodata & 16  \\ 
 3.319446 &     5.7 &     (4.7) &    1.68 $\pm$    0.21 &    \nodata &    \nodata &  $<$    2.69 &    \nodata & 17  \\ 
\sidehead{3.332\tablenotemark{g}}
 3.331308 &     6.6 &   \nodata &    0.69 $\pm$    0.17 &    \nodata &    \nodata &    \nodata &    \nodata &  1  \\ 
 3.331784 &     7.1 &   \nodata &    0.99 $\pm$    0.16 &    \nodata &    \nodata &    \nodata &    \nodata &  2  \\ 
 3.332094 &     7.6 &   \nodata &    1.53 $\pm$    0.18 &    \nodata &    \nodata &    \nodata &    \nodata &  3  \\ 
 3.332283 &     6.7 &   \nodata &    0.76 $\pm$    0.18 &    \nodata &    \nodata &    \nodata &    \nodata &  4  \\ 
 3.332733 &     7.4 &   \nodata &    0.43 $\pm$    0.18 &    \nodata &    \nodata &    \nodata &    \nodata &  5  \\ 
\sidehead{3.359\tablenotemark{h}}
 3.358324 &    10.3 &   \nodata &    1.25 $\pm$    0.24 &    \nodata &    \nodata &    \nodata &    \nodata &  1  \\ 
 3.358781 &    15.4 &   \nodata &    5.23 $\pm$    0.31 &    \nodata &    \nodata &    \nodata &    \nodata &  2  \\ 
 3.359286 &    18.1 &   \nodata &    7.89 $\pm$    0.32 &    \nodata &    \nodata &    \nodata &    \nodata &  3  \\ 
\sidehead{3.406\tablenotemark{i}}
 3.404510 &    15.1 &   \nodata &    1.93 $\pm$    0.39 &    \nodata &    \nodata &    \nodata &    \nodata &  1  \\ 
 3.404899 &     4.2 &   \nodata &    1.42 $\pm$    0.25 &    \nodata &    \nodata &    \nodata &    \nodata &  2  \\*
 3.405261 &    12.4 &   \nodata &    7.54 $\pm$    0.40 &    \nodata &    \nodata &    \nodata &    \nodata &  3  \\ 
 3.405677 &    16.4 &   \nodata &    5.83 $\pm$    0.42 &    \nodata &    \nodata &    \nodata &    \nodata &  4  \\ 
 3.406767 &    20.0 &   \nodata &    1.83 $\pm$    0.36 &    \nodata &    \nodata &    \nodata &    \nodata &  5  \\ 
\sidehead{3.477\tablenotemark{j}}
 3.477253 &     9.6 &     (7.9) &    8.39 $\pm$    0.46 &   16.5 $\pm$    0.6\phm{0} &    \nodata &    \nodata &    \nodata &  1  \\ 
\sidehead{3.521\tablenotemark{k}}
 3.520863 &    11.6 &    (10.2) &    5.34 $\pm$    0.23 &  $<$    1.56 &    \nodata &    \nodata &    \nodata &  1  \\ 
 3.521115 &     7.2 &     (6.1) &    2.28 $\pm$    0.19 &  $<$    1.44 &    \nodata &    \nodata &    \nodata &  2  \\ 
\sidehead{3.533\tablenotemark{l}}
 3.530522 &     7.3 &     (5.8) &    3.81 $\pm$    0.22 &    8.68 $\pm$    0.78 &    \nodata &  \nodata &    \nodata &  1  \\ 
 3.530699 &     5.2 &     (3.6) &    3.15 $\pm$    0.34 &   14.4 $\pm$    1.2\phm{0} &    \nodata &  \nodata &    \nodata &  2  \\ 
 3.530866 &    15.6 &    (14.6) &   20.8 $\pm$    0.6\phm{0} &   92.4 $\pm$    1.9\phm{0} &    \nodata &  \nodata &    \nodata &  3  \\ 
 3.531125 &     9.1 &     (6.4) &   38.7 $\pm$    1.1\phm{0} &   78.9 $\pm$    2.2\phm{0} &    \nodata &  \nodata &    \nodata &  4  \\ 
 3.531322 &     6.3 &     (5.7) &   19.6 $\pm$    2.0\phm{0} &  215 $\pm$   11\phm{0} &    \nodata &   43.0 $\pm$   15.4 &    \nodata &  5  \\ 
 3.531380 &     7.6 &     (5.8) &   17.4 $\pm$    1.5\phm{0} &  $<$    4.19 &    \nodata &  \nodata &   \nodata &  6  \\ 
 3.531594 &    27.8 &    (27.3) &   35.3 $\pm$    0.8\phm{0} &   55.1 $\pm$    2.3\phm{0} &    \nodata &  \nodata &    \nodata &  7  \\ 
 3.532027 &    13.8 &    (12.7) &   12.9 $\pm$    0.5\phm{0} &    4.31 $\pm$    1.38 &    \nodata &    \nodata &    \nodata &  8  \\ 
 3.531908 &     5.1 &     (4.6) &    2.18 $\pm$    0.38 &   13.0 $\pm$    1.1\phm{0} &    \nodata &    \nodata &    \nodata &  9  \\ 
 3.532459 &    15.8 &    (14.8) &    7.21 $\pm$    0.30 &   21.6 $\pm$    1.1\phm{0} &    \nodata &    \nodata &    \nodata & 10  \\ 
 3.532985 &     8.7 &     (5.3) &    2.80 $\pm$    0.37 &   18.8 $\pm$    1.2\phm{0} &    \nodata &    \nodata &    \nodata & 11  \\ 
 3.533056 &     4.1 &     (3.8) &    1.72 $\pm$    0.30 &    1.41 $\pm$    0.97 &    \nodata &    \nodata &    \nodata & 12  \\ 
 3.533513 &    20.7 &    (19.9) &    2.07 $\pm$    0.31 &    5.88 $\pm$    1.16 &    \nodata &    \nodata &    \nodata & 13  \\ 
 3.534188 &    15.6 &    (14.6) &    5.78 $\pm$    0.34 &    1.66 $\pm$    1.08 &    \nodata &    \nodata &    \nodata & 14  \\ 
 3.534422 &     8.5 &     (7.0) &   15.1 $\pm$    0.4\phm{0} &   24.4 $\pm$    1.0\phm{0} &    \nodata &  \nodata &    \nodata & 15  \\ 
 3.534598 &     4.3 &     (3.8) &    1.70 $\pm$    0.22 &    8.72 $\pm$    0.73 &    \nodata &  \nodata &    \nodata & 16  \\ 
 3.534792 &     6.0 &     (4.7) &    6.78 $\pm$    0.69 &    4.14 $\pm$    2.41 &    \nodata &  \nodata &    \nodata & 17  \\ 
 3.535123 &    11.3 &     (9.9) &    3.26 $\pm$    0.25 &  $<$    0.98 &    \nodata &  \nodata &   \nodata & 18  \\ 
 3.535494 &     7.4 &     (6.4) &   67.7 $\pm$    2.0\phm{0} &   56.5 $\pm$    1.4\phm{0} &    \nodata &    \nodata &    \nodata & 19  \\ 
 3.535615 &     3.1 &     (2.4) &   19.4 $\pm$    3.5\phm{0} &  $<$    0.95 &    \nodata &    \nodata &    \nodata & 20  \\ 
 3.535714 &    11.1 &     (9.6) &   26.9 $\pm$    0.6\phm{0} &   14.4 $\pm$    1.2\phm{0} &    \nodata &    \nodata &    \nodata & 21  \\ 
\sidehead{3.644\tablenotemark{m}}
 3.644101 &     3.7 &     (2.4) &    7.49 $\pm$    0.46 &    5.96 $\pm$    0.46 &    \nodata &    \nodata &    \nodata &  1  \\ 
\sidehead{3.646\tablenotemark{n}}
 3.645768 &     5.4 &     (4.7) &    1.98 $\pm$    0.30 &  $<$    0.53 &    \nodata &    \nodata &    \nodata &  1  \\ 
 3.645995 &     4.2 &     (3.2) &    2.16 $\pm$    0.34 &    2.17 $\pm$    0.47 &    \nodata &    \nodata &    \nodata &  2  \\ 
\sidehead{3.754\tablenotemark{o}}
 3.753044 &    13.3 &    (12.2) &    7.04 $\pm$    1.05 &  $<$    1.11 &  $<$    0.52 &    \nodata &    \nodata &  1  \\ 
 3.753395 &     3.1 &     (2.0) &    1.63 $\pm$    0.66 &  $<$    0.65 &  $<$    0.29 &    \nodata &    \nodata &  2  \\ 
 3.753693 &     3.4 &     (2.4) &   11.2 $\pm$    2.6\phm{0} &   10.4 $\pm$    0.9\phm{0} &  $<$    0.22 &    \nodata &    \nodata &  3  \\ 
 3.753847 &    14.1 &    (13.0) &   64.6 $\pm$    3.2\phm{0} &   19.3 $\pm$    1.3\phm{0} &  $<$    0.39 &    \nodata &    \nodata &  4  \\ 
 3.754563 &    12.5 &    (11.2) &   13.5 $\pm$    1.0\phm{0} &    8.41 $\pm$    1.23 &  $<$    0.28 &    \nodata &    \nodata &  5  \\ 
\sidehead{3.789\tablenotemark{p}}
 3.787638 &     9.6 &     (7.9) &    3.48 $\pm$    0.43 &  $<$    1.47 &  $<$    0.29 &    \nodata &    \nodata &  1  \\ 
 3.788301 &     5.0 &     (4.0) &    0.97 $\pm$    0.32 &  $<$    1.38 &  $<$    0.24 &    \nodata &    \nodata &  2  \\ 
 3.789095 &     4.1 &     (2.7) &    0.91 $\pm$    0.30 &  $<$    1.66 &  $<$    0.44 &    \nodata &    \nodata &  3  \\* 
 3.789357 &     5.6 &     (4.1) &    0.62 $\pm$    0.32 &  $<$    1.28 &  $<$    0.68 &    \nodata &    \nodata &  4  \\ 
\sidehead{3.822}  
 3.822015 &     9.9 &     (8.2) &    3.82 $\pm$    0.57 &  $<$    0.66 &  $<$    0.52 &    \nodata &    \nodata &  1  \\ 
\sidehead{3.826\tablenotemark{q}}
 3.824474 &    24.5 &    (23.9) &    9.64 $\pm$    1.05 &    8.51 $\pm$    1.15 &  $<$    0.71 &    \nodata &    \nodata &  1  \\ 
 3.824886 &     8.8 &     (7.4) &    6.96 $\pm$    0.91 &   21.8 $\pm$    0.9\phm{0} &  $<$    0.53 &    \nodata &    \nodata &  2  \\ 
 3.825085 &    12.7 &    (11.5) &   15.3 $\pm$    1.0\phm{0} &   31.1 $\pm$    1.1\phm{0} &    3.41 $\pm$    0.60 &    \nodata &    \nodata &  3  \\ 
 3.825562 &    10.6 &     (9.1) &   11.8 $\pm$    1.0\phm{0} &   37.3 $\pm$    1.1\phm{0} &  $<$    0.74 &    \nodata &    \nodata &  4  \\ 
 3.825680 &     4.0 &     (3.2) &    3.09 $\pm$    0.81 &   56.9 $\pm$    2.0\phm{0} &    5.07 $\pm$    0.51 &    7.20 $\pm$    0.72 &    \nodata &  5  \\ 
 3.825841 &     6.2 &     (5.3) &    6.72 $\pm$    0.93 &   64.7 $\pm$    1.5\phm{0} &    9.60 $\pm$    0.56 &   16.30 $\pm$    0.73 &    \nodata &  6  \\ 
 3.826099 &    15.7 &    (14.7) &   36.9 $\pm$    1.7\phm{0} &   97.5 $\pm$    1.6\phm{0} &    1.03 $\pm$    0.77 &    \nodata &    \nodata &  7  \\ 
 3.826380 &    10.6 &     (9.4) &    8.90 $\pm$    1.57 &    3.01 $\pm$    1.17 &  $<$    0.70 &    \nodata &    \nodata &  8  \\ 
 3.826596 &    11.3 &     (9.9) &   15.9 $\pm$    1.5\phm{0} &   52.2 $\pm$    1.3\phm{0} &    1.48 $\pm$    0.71 &    \nodata &    \nodata &  9  \\ 
 3.826796 &     6.1 &     (5.0) &    6.65 $\pm$    1.21 &   31.1 $\pm$    1.1\phm{0} &  $<$    0.60 &    \nodata &    \nodata & 10  \\ 
 3.826924 &     8.5 &     (7.0) &    9.42 $\pm$    0.99 &   11.6 $\pm$    0.9\phm{0} &  $<$    0.56 &    \nodata &    \nodata & 11  \\ 
\sidehead{3.850\tablenotemark{r}}
 3.848913 &     4.8 &     (3.7) &  $<$    0.81 &    8.09 $\pm$    0.83 &    0.74 $\pm$    0.25 &    \nodata &    \nodata &  1  \\ 
 3.849088 &     4.2 &     (2.8) &    1.80 $\pm$    0.61 &   13.4 $\pm$    0.9\phm{0} &    6.00 $\pm$    0.30 &    \nodata &    \nodata &  2  \\ 
 3.849375 &    11.8 &    (10.4) &    7.68 $\pm$    0.91 &   27.0 $\pm$    1.3\phm{0} &    3.03 $\pm$    0.35 &    \nodata &    \nodata &  3  \\ 
 3.849792 &     8.5 &     (7.1) &    3.06 $\pm$    0.89 &   23.5 $\pm$    1.4\phm{0} &    1.92 $\pm$    0.41 &    \nodata &    \nodata &  4  \\ 
 3.849910 &     3.7 &     (2.4) &  $<$    0.71 &    2.51 $\pm$    1.15 &    1.21 $\pm$    0.31 &    \nodata &    \nodata &  5  \\ 
 3.851317 &    15.9 &    (14.9) &    7.85 $\pm$    1.06 &    6.17 $\pm$    1.60 &  $<$    0.40 &    \nodata &    \nodata &  6  \\ 
\sidehead{3.884}
 3.884106 &     6.6 &     (5.3) &    4.29 $\pm$    0.59 &    4.20 $\pm$    1.15 &  $<$    1.18 &    \nodata &    \nodata &  1  \\ 
 3.884255 &     8.4 &     (7.1) &   12.5 $\pm$    0.7\phm{0} &   11.8 $\pm$    1.3\phm{0} &  $<$    1.08 &    \nodata &    \nodata &  2  \\ 
 3.884492 &     8.0 &     (7.1) &    1.39 $\pm$    0.46 &    5.73 $\pm$    1.20 &  $<$    0.82 &    \nodata &    \nodata &  3  \\ 
\sidehead{3.934\tablenotemark{s}}
 3.932549 &    10.8 &     (9.4) &    6.86 $\pm$    1.59 &   14.0 $\pm$    0.8\phm{0} &  $<$    0.60 &    \nodata &    \nodata &  1  \\ 
 3.932807 &     4.8 &     (4.0) &    7.31 $\pm$    2.06 &  $<$    1.17 &  $<$    0.43 &    \nodata &    \nodata &  2  \\ 
 3.932907 &    10.2 &     (8.6) &    6.02 $\pm$    1.60 &    9.92 $\pm$    1.14 &  $<$    0.58 &    \nodata &    \nodata &  3  \\ 
 3.933551 &    15.5 &    (14.5) &   14.6 $\pm$    1.9\phm{0} &    4.73 $\pm$    1.12 &  $<$    0.47 &    \nodata &    \nodata &  4  \\ 
 3.934134 &    12.8 &    (11.6) &    9.08 $\pm$    1.44 &   24.1 $\pm$    1.6\phm{0} &  $<$    0.52 &    \nodata &    \nodata &  5  \\ 
 3.934314 &     8.4 &     (6.9) &    8.06 $\pm$    1.20 &   23.0 $\pm$    1.1\phm{0} &  $<$    0.53 &    \nodata &    \nodata &  6  \\ 
 3.934635 &     8.2 &     (6.9) &    7.46 $\pm$    0.88 &   16.4 $\pm$    0.8\phm{0} &    0.44 $\pm$    0.34 &    \nodata &    \nodata &  7  \\ 
\enddata
 
\tablenotetext{*}{See Figure 1. For this object refer to the corresponding footnote to Table 8: here for Q1055+4611 the specific 
lengths and separation of observing runs again become manifest as a mild broadening of the atmospheric emission and absorption lines. 
Furthermore, in addition to the normal sky subtraction procedure which is common to all objects, the atmospheric line absorption is 
here "divided out" by use of a related spectrum of a calibration object nearby on the sky. This gives the advantage of restoring the 
QSO spectrum to a normalized continuum and directly makes available a fraction more spectral coverage without the complication of 
unraveling the atmospheric absorption. Whether so processed or not, information is of course unavoidably compromised through the 
intrinsically enhanced noise and ultimately obscured in the most heavily absorbed regions where these lines are saturated.}

\tablenotetext{a}{\CIV~$\lambda$1548 is blended with \FeII~$\lambda$1608 at $z = 2.912$ but $\lambda$1551 is relatively clear with 
effectively only minor interference by atmospheric line absorption and emission.} 

\tablenotetext{b}{\CIV~$\lambda$1551 contains very weak \CII\ at $z = 3.754$ but $\lambda$1548 is clear.}

\tablenotetext{c}{\CIV~$\lambda$1551 is blended with \CII\ at $z = 3.826$ but $\lambda$1548 is clear.} 

\tablenotetext{d}{\CIV~$\lambda$1551 is blended with \CIV~$\lambda$1548 at $z = 3.234$ but $\lambda$1548 is clear although with mild 
interference by atmospheric line absorption and emission.}

\tablenotetext{e}{\CIV~$\lambda$1548 is blended with \CIV~$\lambda$1551 at $z = 3.228$ but $\lambda$1551 is clear.}

\tablenotetext{f}{This is a complex system with very strong \SiII~$\lambda$1527 ($\lambda$1260 is in the forest) and relatively weak 
related \CIV\ in components 7--10: for these, log~$N$(\HI) totalling $\sim20.3$ gives a reasonable fit to the \HI\ Lyman $\alpha$ 
damping wings, verging on the ``damped Lyman $\alpha$'' classification. Both of \CIV~$\lambda\lambda$1548,1551 are partially blended 
with \MgII~$\lambda$2804 in a strong, extended system at $z = 1.386$ but taken together good values are obtained. Si $b$-values here 
relate to \SiII.}

\tablenotetext{g}{\CIV~$\lambda$1551 is blended with weak \SiIV~$\lambda$1403 at $z = 3.789$ but $\lambda$1548 is clear.}

\tablenotetext{h}{\CIV~$\lambda$1551 is blended with \SiIV~$\lambda$1394 at $z = 3.850$ but $\lambda$1548 is clear.}

\tablenotetext{i}{\CIV~$\lambda$1551 has partial interference by atmospheric line emission but $\lambda$1548 is clear.} 

\tablenotetext{j}{\CIV~$\lambda$1551 is blended with strong \FeII~$\lambda$1608 at $z = 3.317$ but $\lambda$1548 is relatively clear. 
\SiIV~$\lambda$1403 is outside the forest and not significantly affected by mild atmospheric line absorption which is present.}

\tablenotetext{k}{\SiIV~$\lambda$1394 is heavily contaminated by atmospheric line emission and also is blended with \CIV~$\lambda$1551 
at $z = 3.063$ but $\lambda$1403 is clear.}

\tablenotetext{l}{This is a highly complex system. In \SiIV~$\lambda$1403 there is an unidentified interloper at observed wavelength 
$\lambda$6359.9 near component 13, but in a relatively clear region within the complex, and \SiIV~$\lambda$1394 is clear. 
\SiII~$\lambda$1260 is in the forest; \SiII~$\lambda$1527 is generally relatively weak and partially blended with much stronger 
\SiIV~$\lambda$1403 at $z = 3.934$ and, with the exception of the strong component 5, yields only upper limits which are too high to 
be useful.}

\tablenotetext{m}{\CIV~$\lambda$1551 has strong atmospheric line absorption but $\lambda$1548 is clear. \SiIV~$\lambda$1394 resides in 
\CII\ at $z = 3.850$ but $\lambda$1403 is clear although in a rather noisy region.}

\tablenotetext{n}{\CIV~$\lambda$1548 has strong atmospheric line absorption but $\lambda$1551 is clear. \SiIV~$\lambda$1403 is in a 
noisy region and not used but $\lambda$1394 is clear.} 

\tablenotetext{o}{\CIV~$\lambda$1548 near component 4 and $\lambda$1551 in component 1 have strong atmospheric line emission and both 
are in a rather noisy region, but taken together acceptable values are obtained in all components. \SiIV~$\lambda$1403 in component 5  
is blended with \MgII~$\lambda$2796 at $z = 1.386$ but $\lambda$1394 is clear. \CII\ is outside the forest; components 1 and 2 are 
blended with weak \CIV~$\lambda$1551 at $z = 3.090$ but taken with $\lambda$1548 reasonable upper limits are obtained.}

\tablenotetext{p}{\SiIV~$\lambda$1394 is masked by very strong \MgII~$\lambda$2796 at $z = 1.386$ and $\lambda$1403 in components 2--4  
is blended with weak \CIV~$\lambda$1551 at $z = 3.332$, but together with $\lambda$1548 in that system reasonable upper limits are 
obtained. \CII\ in component 4 is close in wavelength to \CIV~$\lambda$1551 at $z = 3.122$ but the upper limit is useful.}

\tablenotetext{q}{\CIV~$\lambda$1551 has significant interference by atmospheric line emission but $\lambda$1548 is relatively clear
and together good values are obtained. \CII\ is partially blended with \CIV~$\lambda$1551 at $z = 3.154$ but taken with $\lambda$1548 
in that system good values are obtained. \SiII~$\lambda$1260 is in the forest but the two narrow and strong components 5 and 6 are in 
a clear region and give good values consistent with the weak $\lambda$1527.}

\tablenotetext{r}{\CIV\ is in a rather noisy region. \SiIV~$\lambda$1394 includes weaker \CIV~$\lambda$1551 at $z = 3.359$ but 
$\lambda$1403 is clear. \CII\ contains narrow \SiIV~$\lambda$1394 at $z = 3.644$ which however is in a clear part of the complex.}

\tablenotetext{s}{\CIV\ is in a region with very strong atmospheric line absorption but over the doublet the values obtained are 
acceptable. \SiIV~$\lambda$1403 contains very weak \SiII~$\lambda$1527 at $z = 3.533$ but $\lambda$1394 is relatively clear and good 
values are obtained.}

\end{deluxetable}

\clearpage
\begin{deluxetable}{crrcccccc}
\tabletypesize{\scriptsize}
\tablecaption{Heavy Element Absorption Lines: Q2237-0607  $z_{em} = 4.56^*$}
\tablewidth{0pt}
\tablecolumns{9}
\tablehead{
\colhead{$z$} & \colhead{$b$(C)}  & \colhead{$b$(Si)} &
\colhead{$N$(\CIV)} & \colhead{$N$(\SiIV)} &
\colhead{$N$(\CII)} & \colhead{$N$(\SiII)} &
\colhead{$N$(\NV)} &  \colhead{n} \\
{} & \colhead{km~s$^{-1}$} & \colhead{km~s$^{-1}$} & \colhead{10$^{12}$cm$^{-2}$} &
\colhead{10$^{11}$cm$^{-2}$} & \colhead{10$^{12}$cm$^{-2}$} & \colhead{10$^{11}$cm$^{-2}$} & \colhead{10$^{12}$cm$^{-2}$} & {} \\
\colhead{1} & \colhead{2} & \colhead{3} & \colhead{4} & \colhead{5} & \colhead{6} & \colhead{7} & \colhead{8} & \colhead{9} 
}

\startdata
\sidehead{3.433\tablenotemark{a}}
 3.432137 &    18.1 &   \nodata &    1.38 $\pm$    0.22 &    \nodata &    \nodata &    \nodata &    \nodata &  1  \\ 
 3.432510 &    12.5 &   \nodata &    5.06 $\pm$    0.19 &    \nodata &    \nodata &    \nodata &    \nodata &  2  \\ 
 3.432872 &    15.4 &   \nodata &    4.17 $\pm$    0.20 &    \nodata &    \nodata &    \nodata &    \nodata &  3  \\ 
 3.433346 &    14.4 &   \nodata &    8.11 $\pm$    0.20 &    \nodata &    \nodata &    \nodata &    \nodata &  4  \\ 
 3.433767 &    15.7 &   \nodata &    2.27 $\pm$    0.20 &    \nodata &    \nodata &    \nodata &    \nodata &  5  \\ 
 3.434190 &     5.9 &   \nodata &    0.38 $\pm$    0.12 &    \nodata &    \nodata &    \nodata &    \nodata &  6  \\ 
\sidehead{3.448}
 3.447979 &     6.1 &   \nodata &    1.78 $\pm$    0.14 &    \nodata &    \nodata &    \nodata &    \nodata &  1  \\ 
\sidehead{3.454}
 3.454071 &     6.6 &   \nodata &    1.88 $\pm$    0.15 &    \nodata &    \nodata &    \nodata &    \nodata &  1  \\ 
\sidehead{3.482\tablenotemark{b}}
 3.480721 &     6.1 &   \nodata &    5.83 $\pm$    0.33 &    \nodata &    \nodata &    \nodata &    \nodata &  1  \\ 
 3.480900 &     7.2 &   \nodata &    1.57 $\pm$    0.39 &    \nodata &    \nodata &    \nodata &    \nodata &  2  \\ 
 3.481150 &     3.7 &   \nodata &    0.73 $\pm$    0.21 &    \nodata &    \nodata &    \nodata &    \nodata &  3  \\ 
 3.481795 &     3.2 &   \nodata &    0.58 $\pm$    0.21 &    \nodata &    \nodata &    \nodata &    \nodata &  4  \\ 
 3.482812 &     8.4 &   \nodata &    2.35 $\pm$    0.51 &    \nodata &    \nodata &    \nodata &    \nodata &  5  \\ 
\sidehead{3.591}
 3.590142 &     9.0 &   \nodata &    0.98 $\pm$    0.19 &    \nodata &    \nodata &    \nodata &    \nodata &  1  \\ 
 3.591022 &    14.4 &   \nodata &    3.81 $\pm$    0.24 &    \nodata &    \nodata &    \nodata &    \nodata &  2  \\ 
\sidehead{3.604\tablenotemark{c}}
 3.603319 &     6.2 &   \nodata &    0.78 $\pm$    0.25 &    \nodata &    \nodata &    \nodata &    \nodata &  1  \\ 
 3.603784 &     8.5 &   \nodata &    2.20 $\pm$    0.21 &    \nodata &    \nodata &    \nodata &    \nodata &  2  \\ 
 3.604063 &     9.3 &   \nodata &    0.44 $\pm$    0.18 &    \nodata &    \nodata &    \nodata &    \nodata &  3  \\ 
 3.604742 &    21.6 &   \nodata &    4.87 $\pm$    0.25 &    \nodata &    \nodata &    \nodata &    \nodata &  4  \\ 
 3.605166 &     9.7 &   \nodata &    3.18 $\pm$    0.22 &    \nodata &    \nodata &    \nodata &    \nodata &  5  \\ 
 3.605526 &     5.0 &   \nodata &    0.93 $\pm$    0.16 &    \nodata &    \nodata &    \nodata &    \nodata &  6  \\ 
\sidehead{3.627\tablenotemark{d}}
 3.626887 &    17.7 &   \nodata &    7.42 $\pm$    0.86 &    \nodata &    \nodata &    \nodata &    \nodata &  1  \\ 
 3.626972 &     7.8 &   \nodata &    5.65 $\pm$    0.65 &    \nodata &    \nodata &    \nodata &    \nodata &  2  \\ 
\sidehead{3.681\tablenotemark{e}}
 3.681058 &    15.6 &   \nodata &    3.29 $\pm$    0.40 &    \nodata &    \nodata &    \nodata &    \nodata &  1  \\ 
 3.681358 &    10.4 &   \nodata &    1.56 $\pm$    0.24 &    \nodata &    \nodata &    \nodata &    \nodata &  2  \\ 
 3.681789 &     9.7 &   \nodata &    0.91 $\pm$    0.33 &    \nodata &    \nodata &    \nodata &    \nodata &  3  \\ 
\sidehead{3.748\tablenotemark{f}}
 3.746527 &     9.4 &   \nodata &    4.92 $\pm$    0.37 &    \nodata &    \nodata &    \nodata &    \nodata &  1  \\ 
 3.746813 &     8.8 &   \nodata &    1.41 $\pm$    0.34 &    \nodata &    \nodata &    \nodata &    \nodata &  2  \\ 
 3.748577 &    10.5 &   \nodata &    3.14 $\pm$    0.38 &    \nodata &    \nodata &    \nodata &    \nodata &  3  \\ 
\sidehead{3.756\tablenotemark{g}}
 3.756282 &     6.1 &   \nodata &    1.56 $\pm$    0.31 &    \nodata &    \nodata &    \nodata &    \nodata &  1  \\ 
\sidehead{3.794}
 3.793849 &    17.1 &   \nodata &    4.78 $\pm$    0.32 &    \nodata &    \nodata &    \nodata &    \nodata &  1  \\* 
 3.794448 &    26.0 &   \nodata &    1.45 $\pm$    0.40 &    \nodata &    \nodata &    \nodata &    \nodata &  2  \\ 
\sidehead{3.811\tablenotemark{h}}
 3.808354 &    28.0 &   \nodata &    4.17 $\pm$    0.63 &    \nodata &    \nodata &    \nodata &    \nodata &  1  \\ 
 3.809181 &     9.4 &   \nodata &    1.60 $\pm$    0.38 &    \nodata &    \nodata &    \nodata &    \nodata &  2  \\ 
 3.810379 &     6.7 &   \nodata &    4.47 $\pm$    0.38 &    \nodata &    \nodata &    \nodata &    \nodata &  3  \\ 
 3.810655 &     7.1 &   \nodata &    1.44 $\pm$    0.36 &    \nodata &    \nodata &    \nodata &    \nodata &  4  \\ 
 3.810928 &    13.1 &   \nodata &    2.62 $\pm$    0.48 &    \nodata &    \nodata &    \nodata &    \nodata &  5  \\ 
 3.811434 &     9.3 &   \nodata &    2.98 $\pm$    0.39 &    \nodata &    \nodata &    \nodata &    \nodata &  6  \\ 
 3.811845 &     7.6 &   \nodata &    0.52 $\pm$    0.34 &    \nodata &    \nodata &    \nodata &    \nodata &  7  \\ 
 3.812758 &    16.9 &   \nodata &    3.42 $\pm$    0.50 &    \nodata &    \nodata &    \nodata &    \nodata &  8  \\ 
 3.813257 &     5.9 &   \nodata &    1.02 $\pm$    0.32 &    \nodata &    \nodata &    \nodata &    \nodata &  9  \\ 
\sidehead{3.825\tablenotemark{i}}
 3.824580 &     8.6 &     (7.0) &    4.08 $\pm$    0.63 &   17.3 $\pm$    0.9\phm{0} &    \nodata &    \nodata &    \nodata &  1  \\ 
 3.824747 &     7.2 &     (4.7) &    2.80 $\pm$    0.98 &    1.64 $\pm$    0.72 &    \nodata &    \nodata &    \nodata &  2  \\ 
\sidehead{\phm{0}}
\sidehead{3.832\tablenotemark{j}}
 3.829879 &    28.5 &   \nodata &   17.4 $\pm$    2.1\phm{0} &    \nodata &    \nodata &    \nodata &    \nodata &  1  \\ 
 3.830918 &    13.8 &   \nodata &    8.88 $\pm$    1.31 &    \nodata &    \nodata &    \nodata &    \nodata &  2  \\ 
 3.831427 &    15.0 &    (13.9) &   17.4 $\pm$    1.5\phm{0} &   42.4 $\pm$   37.6 &    \nodata &    \nodata &    \nodata &  3  \\ 
 3.831933 &    12.8 &    (11.5) &   41.7 $\pm$    1.9\phm{0} &  108 $\pm$   33\phm{0} &    \nodata &    \nodata &    \nodata &  4  \\ 
 3.832314 &    12.4 &    (11.1) &    5.18 $\pm$    1.28 &   39.9 $\pm$   18.4 &    \nodata &    \nodata &    \nodata &  5  \\ 
 3.833277 &     8.3 &   \nodata &    4.51 $\pm$    1.27 &    \nodata &    \nodata &    \nodata &    \nodata &  6  \\ 
\sidehead{3.978\tablenotemark{k}}
 3.976775 &     9.1 &     (7.3) &    1.80 $\pm$    0.52 &  $<$    0.93 &    \nodata &    \nodata &    \nodata &  1  \\ 
 3.977090 &     5.2 &     (3.4) &    1.39 $\pm$    0.44 &  $<$    0.72 &    \nodata &    \nodata &    \nodata &  2  \\ 
 3.977349 &     8.1 &     (6.6) &    3.07 $\pm$    0.61 &    4.99 $\pm$    1.97 &    \nodata &    \nodata &    \nodata &  3  \\ 
 3.977591 &    15.9 &    (14.9) &    3.47 $\pm$    0.73 &  $<$    1.86 &    \nodata &    \nodata &    \nodata &  4  \\ 
 3.978222 &    18.4 &    (17.5) &   10.2 $\pm$    0.7\phm{0} &    6.03 $\pm$    1.13 &    \nodata &    \nodata &    \nodata &  5  \\ 
 3.978890 &     8.6 &     (6.8) &    1.85 $\pm$    0.64 &    6.27 $\pm$    1.11 &    \nodata &    \nodata &    \nodata &  6  \\ 
 3.979087 &    19.4 &    (18.6) &    5.10 $\pm$    0.89 &  $<$    1.53 &    \nodata &    \nodata &    \nodata &  7  \\ 
\sidehead{4.030}
 4.028668 &    10.5 &     (9.1) &    2.19 $\pm$    0.42 &    1.23 $\pm$    0.87 &    \nodata &    \nodata &    \nodata &  1  \\ 
 4.029160 &    25.1 &    (24.5) &    2.45 $\pm$    0.61 &  $<$    1.34 &    \nodata &    \nodata &    \nodata &  2  \\ 
 4.029586 &     9.4 &     (7.8) &    4.56 $\pm$    0.40 &    4.87 $\pm$    0.79 &    \nodata &    \nodata &    \nodata &  3  \\ 
 4.030407 &     7.4 &     (6.6) &    0.76 $\pm$    0.35 &  $<$    0.69 &    \nodata &    \nodata &    \nodata &  4  \\ 
 4.031057 &    17.0 &    (16.1) &    5.37 $\pm$    0.54 &  $<$    1.03 &    \nodata &    \nodata &    \nodata &  5  \\ 
\sidehead{4.080\tablenotemark{l}}
 4.078068 &     7.8 &     (6.3) &    1.66 $\pm$    0.57 &    1.49 $\pm$    1.01 &    3.10 $\pm$    0.33 &    \nodata &    \nodata &  1  \\ 
 4.078373 &     3.6 &     (2.5) &  $<$    1.59 &  $<$    2.01 &   31.2 $\pm$    3.7\phm{0} &   18.1 $\pm$   10.0 &    \nodata &  2  \\ 
 4.078401 &     9.1 &     (7.2) &    3.94 $\pm$    1.86 &    4.75 $\pm$    2.69 &   55.2 $\pm$    6.9\phm{0} &   78.5 $\pm$   12.5 &    \nodata &  3  \\ 
 4.078669 &     9.8 &     (8.2) &    1.03 $\pm$    0.84 &    5.36 $\pm$    1.40 &  109 $\pm$    4\phm{00} &  164 $\pm$    7\phm{00} &    \nodata &  4  \\ 
 4.078874 &     8.4 &     (6.6) &  $<$    0.81 &   12.5 $\pm$    1.3\phm{0} &   66.1 $\pm$    4.5\phm{0} &   92.5 $\pm$    6.2\phm{0} &    \nodata &  5  \\ 
 4.079041 &     3.3 &     (2.2) &    1.06 $\pm$    0.51 &    2.88 $\pm$    0.92 &   16.9 $\pm$    2.4\phm{0} &   23.8 $\pm$    4.2\phm{0} &    \nodata &  6  \\ 
 4.079178 &     4.3 &     (3.2) &    1.39 $\pm$    0.56 &    5.69 $\pm$    1.02 &   39.4 $\pm$    1.7\phm{0} &   74.1 $\pm$    5.0\phm{0} &    \nodata &  7  \\ 
 4.079364 &     8.8 &     (7.1) &    5.36 $\pm$    0.79 &   17.9 $\pm$    1.3\phm{0} &  $<$    3.46 &   10.3 $\pm$    4.2\phm{0} &    \nodata &  8  \\ 
 4.079609 &     7.5 &     (5.9) &   18.0 $\pm$    1.7\phm{0} &   16.0 $\pm$    2.6\phm{0} &    4.65 $\pm$    1.24 &   11.4 $\pm$    5.5\phm{0} &    \nodata &  9  \\ 
 4.079721 &     9.2 &     (7.4) &   13.4 $\pm$    1.8\phm{0} &   42.1 $\pm$    2.5\phm{0} &    2.76 $\pm$    0.96 &  \nodata &    \nodata & 10  \\ 
 4.079979 &    10.6 &     (9.1) &    9.12 $\pm$    1.03 &   45.8 $\pm$    1.8\phm{0} &   35.0 $\pm$    1.3\phm{0} &   64.7 $\pm$    6.1\phm{0} &    \nodata & 11  \\ 
 4.080205 &     8.1 &     (6.5) &    3.01 $\pm$    0.80 &   28.9 $\pm$    1.5\phm{0} &  277 $\pm$   15\phm{0} &  468 $\pm$   17\phm{0} &    \nodata & 12  \\ 
 4.080359 &     3.4 &     (2.3) &  $<$    0.93 &   24.0 $\pm$    2.8\phm{0} &   28.5 $\pm$    6.8\phm{0} &   87.2 $\pm$   10.7 &    \nodata & 13  \\ 
 4.080462 &     6.6 &     (5.4) &    3.04 $\pm$    0.87 &   31.7 $\pm$    2.6\phm{0} &   39.3 $\pm$    1.5\phm{0} &   61.1 $\pm$    6.0\phm{0} &    \nodata & 14  \\ 
 4.080655 &     7.4 &     (5.8) &    1.70 $\pm$    0.63 &   10.9 $\pm$    1.2\phm{0} &   69.0 $\pm$    1.7\phm{0} &  111 $\pm$    6\phm{00} &    \nodata & 15  \\ 
 4.080852 &     4.9 &     (3.9) &    1.35 $\pm$    0.55 &    7.28 $\pm$    1.03 &  281 $\pm$   18\phm{0} &  385 $\pm$   189 &    \nodata & 16  \\ 
 4.081019 &     7.4 &     (6.0) &  $<$    0.80 &    4.13 $\pm$    1.22 &    7.26 $\pm$    0.56 &   14.3 $\pm$    4.8\phm{0} &    \nodata & 17  \\ 
 4.081189 &     8.4 &     (6.9) &  $<$    0.92 &    3.85 $\pm$    1.22 &    2.05 $\pm$    0.29 &    \nodata &    \nodata & 18  \\ 
 4.081509 &     9.7 &     (8.1) &    3.26 $\pm$    0.65 &    4.17 $\pm$    1.12 &    0.34 $\pm$    0.20 &    \nodata &    \nodata & 19  \\ 
\sidehead{4.086\tablenotemark{m}}
 4.085645 &    12.1 &    (10.8) &    3.31 $\pm$    0.68 &    6.03 $\pm$    1.26 &  $<$    0.21 &    \nodata &    \nodata &  1  \\ 
\sidehead{4.089\tablenotemark{n}}
 4.088900 &     7.8 &     (6.9) &    6.27 $\pm$    0.78 &    3.54 $\pm$    1.01 &    0.43 $\pm$    0.19 &    \nodata &    \nodata &  1  \\ 
\sidehead{4.094\tablenotemark{o}}
 4.093972 &    16.0 &    (15.1) &    6.50 $\pm$    0.77 &    4.68 $\pm$    1.42 &  $<$    0.34 &    \nodata &    \nodata &  1  \\ 
\sidehead{4.200\tablenotemark{p}}
 4.199860 &     9.2 &     (7.4) &    1.38 $\pm$    0.68 &  $<$    1.16 &    \nodata &    \nodata &    \nodata &  1  \\ 
 4.200502 &    12.9 &    (11.7) &   11.6 $\pm$    0.9\phm{0} &    6.22 $\pm$    1.53 &    \nodata &    \nodata &    \nodata &  2  \\ 
 4.200809 &     8.4 &     (6.8) &    6.77 $\pm$    0.87 &  $<$    1.06 &    $<$    0.12 &    \nodata &    \nodata &  3  \\ 
\sidehead{4.231\tablenotemark{q}}
 4.229886 &     8.3 &     (6.7) &    0.79 $\pm$    0.41 &  $<$    2.30 &  $<$    0.38 &    \nodata &    \nodata &  1  \\ 
 4.230182 &     5.8 &     (5.1) &    3.16 $\pm$    0.41 &  $<$    1.65 &    \nodata &    \nodata &    \nodata &  2  \\ 
 4.230435 &     7.9 &     (6.2) &   15.0 $\pm$    0.7\phm{0} &    4.66 $\pm$    0.78 &    \nodata &    \nodata &    \nodata &  3  \\ 
 4.230771 &     7.8 &     (6.1) &    3.06 $\pm$    0.45 &  $<$    0.73 &    \nodata &    \nodata &    \nodata &  4  \\ 
 4.231091 &     5.8 &     (4.9) &    3.06 $\pm$    0.43 &  $<$    0.68 &  $<$    0.29 &    \nodata &    \nodata &  5  \\ 
 4.231385 &     8.9 &     (7.2) &    1.23 $\pm$    0.47 &  $<$    0.79 &  $<$    0.36 &    \nodata &    \nodata &  6  \\ 
 4.231826 &     6.5 &     (5.3) &    1.15 $\pm$    0.37 &  $<$    0.86 &  $<$    0.47 &    \nodata &    \nodata &  7  \\ 
\sidehead{4.239\tablenotemark{r}}
 4.237075 &     8.4 &     (6.8) &    3.19 $\pm$    0.44 &  $<$    1.31 &  $<$    0.38 &    \nodata &    \nodata &  1  \\ 
 4.238095 &     7.2 &     (4.7) &    1.10 $\pm$    0.40 &  $<$    0.73 &  $<$    0.48 &    \nodata &    \nodata &  2  \\ 
 4.239683 &    13.5 &    (12.4) &    2.98 $\pm$    0.64 &   29.9 $\pm$    1.4\phm{0} &    3.61 $\pm$    0.57 &    \nodata &    \nodata &  3  \\ 
 4.239891 &     4.1 &     (3.0) &    1.38 $\pm$    0.42 &    6.40 $\pm$    0.90 &  $<$    0.50 &    \nodata &    \nodata &  4  \\ 
 4.240122 &    10.5 &     (9.0) &    3.29 $\pm$    0.54 &   10.2 $\pm$    1.0\phm{0} &  $<$    0.45 &    \nodata &    \nodata &  5  \\ 
\sidehead{4.247\tablenotemark{s}}
 4.245677 &     7.0 &     (5.8) &    1.16 $\pm$    0.52 &    3.61 $\pm$    1.01 &  $<$    0.37 &    \nodata &    \nodata &  1  \\ 
 4.246104 &     3.1 &     (2.1) &    3.38 $\pm$    0.56 &  $<$    0.77 &  $<$    0.39 &    \nodata &    \nodata &  2  \\ 
 4.247022 &    29.2 &    (28.7) &   12.2 $\pm$    1.5\phm{0} &   11.6 $\pm$    3.0\phm{0} &  $<$    0.96 &    \nodata &    \nodata &  3  \\ 
 4.247807 &    21.2 &    (20.4) &    9.09 $\pm$    1.02 &   25.3 $\pm$    2.1\phm{0} &  $<$    0.72 &    \nodata &    \nodata &  4  \\ 
\sidehead{4.286\tablenotemark{t}}
 4.284727 &    10.1 &     (8.5) &    1.30 $\pm$    0.71 &    5.78 $\pm$    1.58 &  $<$    0.54 &    \nodata &    \nodata &  1  \\ 
 4.285258 &     8.5 &     (7.0) &    7.82 $\pm$    0.93 &   18.4 $\pm$    1.5\phm{0} &  $<$    0.49 &    \nodata &    \nodata &  2  \\ 
 4.285650 &    10.4 &     (8.8) &   10.3 $\pm$    1.3\phm{0} &   46.2 $\pm$    2.0\phm{0} &  $<$    0.81 &    \nodata &    \nodata &  3  \\ 
 4.286004 &    14.4 &    (13.3) &    6.00 $\pm$    4.07 &   13.6 $\pm$    2.1\phm{0} &  $<$    0.70 &    \nodata &    \nodata &  4  \\ 
 4.286205 &     3.9 &     (2.8) &    5.60 $\pm$    1.44 &  $<$    1.24 &  $<$    0.50 &    \nodata &    \nodata &  5  \\ 
 4.286412 &    13.4 &    (12.2) &    5.97 $\pm$    0.99 &  $<$    2.76 &  $<$    0.70 &    \nodata &    \nodata &  6  \\ 
\sidehead{4.361\tablenotemark{u}}
 4.359466 &    15.3 &    (14.3) &   11.5 $\pm$    0.9\phm{0} &   55.8 $\pm$    4.6\phm{0} &    1.14 $\pm$    0.58 &  $<$    0.96 &    \nodata &  1  \\ 
 4.359949 &    14.4 &    (13.3) &   17.6 $\pm$    1.6\phm{0} &   68.6 $\pm$    3.6\phm{0} &  $<$    0.69 &  $<$    3.00 &    \nodata &  2  \\ 
 4.360144 &     5.8 &     (4.6) &    8.50 $\pm$    1.15 &   50.9 $\pm$    3.8\phm{0} &    1.42 $\pm$    0.45 &    2.03 $\pm$    0.49 &    \nodata &  3  \\ 
 4.360303 &     4.4 &     (3.5) &    4.48 $\pm$    0.91 &   71.3 $\pm$    5.1\phm{0} &    3.68 $\pm$    0.46 &    4.96 $\pm$    0.54 &    \nodata &  4  \\ 
 4.360496 &     9.5 &     (7.8) &   29.6 $\pm$    1.2\phm{0} &   34.2 $\pm$    2.3\phm{0} &  $<$    0.76 &  $<$    0.66 &    \nodata &  5  \\ 
 4.360899 &     6.4 &     (5.2) &   14.7 $\pm$    0.7\phm{0} &   34.0 $\pm$    2.0\phm{0} &    1.16 $\pm$    0.41 &  $<$    0.44 &    \nodata &  6  \\ 
 4.361351 &    10.2 &     (8.7) &    5.57 $\pm$    1.37 &  $<$    7.27 &  $<$    0.90 &  $<$    0.41 &    \nodata &  7  \\ 
 4.361474 &     7.7 &     (5.3) &    3.56 $\pm$    1.80 &   65.9 $\pm$    6.8\phm{0} &    3.29 $\pm$    0.78 &    1.66 $\pm$    0.29 &    \nodata &  8  \\ 
 4.361758 &    13.7 &    (10.2) &   13.5 $\pm$    1.2\phm{0} &  161 $\pm$    9\phm{00} &   12.0 $\pm$    0.7\phm{0} &   10.2 $\pm$    0.4\phm{0} &    \nodata &  9  \\ 
 4.362331 &     9.9 &     (8.2) &    9.02 $\pm$    0.73 &   37.8 $\pm$    2.5\phm{0} &    3.05 $\pm$    0.59 &    0.74 $\pm$    0.40 &    \nodata & 10  \\ 
 4.362589 &    10.4 &     (8.9) &   13.0 $\pm$    0.9\phm{0} &   90.4 $\pm$    4.1\phm{0} &   19.2 $\pm$    0.8\phm{0} &   18.6 $\pm$    0.4\phm{0} &    \nodata & 11  \\ 
 4.362803 &     6.9 &     (5.7) &    2.40 $\pm$    0.75 &   65.9 $\pm$    4.8\phm{0} &    8.78 $\pm$    0.69 &    9.98 $\pm$    0.32 &    \nodata & 12  \\ 
 4.363034 &    22.4 &    (21.7) &   14.5 $\pm$    1.1\phm{0} &   15.9 $\pm$    4.5\phm{0} &  $<$    1.09 &    \nodata &    \nodata & 13  \\ 
\sidehead{4.389\tablenotemark{v}}
 4.388714 &     7.2 &     (5.7) &    3.59 $\pm$    0.82 &    \nodata &  $<$    0.56 &  $<$    0.33  &    \nodata &  1  \\ 
 4.389280 &     3.7 &     (2.4) &    0.90 $\pm$    0.48 &  $<$    3.91 &  $<$    0.45 &  $<$    0.17 &    \nodata &  2  \\ 
 4.390043 &     6.1 &     (5.2) &    2.78 $\pm$    1.11 &  $<$    3.01 &  $<$    0.56 &  $<$    0.21 &    \nodata &  3  \\ 
\enddata
 
\tablenotetext{*}{See Figure 1. For this object refer to the corresponding footnote to Table 9: here for Q2237-0607 the atmospheric line 
absorption is treated similarly; and the specific lengths and separation of observing runs again become manifest as a mild broadening of the 
atmospheric emission and absorption lines.}

\tablenotetext{a}{\CIV~$\lambda$1551 has mild interference by atmospheric line absorption but $\lambda$1548 is clear.}

\tablenotetext{b}{\CIV~$\lambda$1548 is blended with weak \SiIV~$\lambda$1394 at $z = 3.978$ and weak \CII\ at $z = 4.200$ and also has mild 
interference by atmospheric line absorption while $\lambda$1551 has both absorption and emission, but taken together with both the \SiIV\ 
doublet members at $z = 3.978$ good values are obtained.}

\tablenotetext{c}{\CIV~$\lambda$1548 is partially blended with \SiIV~$\lambda$1403 at $z = 4.080$ while $\lambda$1551 in component 1 is blended 
with \SiIV~$\lambda$1403 at $z = 4.089$ but together acceptable values are obtained.}

\tablenotetext{d}{\CIV~$\lambda$1551 has mild interference by atmospheric line absorption but $\lambda$1548 is clear.}

\tablenotetext{e}{\CIV~$\lambda$1548 has interference by both atmospheric line absorption and emission while $\lambda$1551 has absorption alone 
but taken together acceptable values are obtained.}

\tablenotetext{f}{\CIV~$\lambda$1548 is blended with \SiIV~$\lambda$1403 at $z = 4.239$ and $\lambda$1551 is blended with \SiIV~$\lambda$1403 
at $z = 4.247$ and \CIV~$\lambda$1548 at $z = 3.756$, and both have some interference by atmospheric line absorption and emission, but taking 
all together acceptable values are obtained.}

\tablenotetext{g}{\CIV~$\lambda$1548 is blended with $\lambda$1551 at $z = 3.748$ but $\lambda$1551 is clear.}

\tablenotetext{h}{\CIV~$\lambda$1551 in components 8 and 9 is mildly affected by atmospheric line emission but $\lambda$1548 is clear. This is 
a weak extended system and may represent possibly three separate close regions.} 

\tablenotetext{i}{\CIV~$\lambda$1551 is blended with \CIV~$\lambda$1548 at $z = 3.832$ and heavily contaminated by atmospheric line emission and 
not used; $\lambda$1548 is blended with component 1 in \SiIV~$\lambda$1394 at $z = 4.361$ but good values are obtained by use of both \SiIV\ 
doublet members of that system. \SiIV~$\lambda$1394 is in the forest but $\lambda$1403 is clear.}

\tablenotetext{j}{\CIV~$\lambda$1548 is partially blended with \AlIII~$\lambda$1855 at $z = 3.032$ and strongly contaminated by atmospheric line 
emission while $\lambda$1551 is clear of this except in component 6 but together with \AlIII~$\lambda$1863, itself blended with very weak 
\SiIV~$\lambda$1394 at $z = 4.389$, good values are obtained for all components. \SiIV~$\lambda$1403 is outside the forest and immersed in very 
strong \CII~$\lambda$1335 at $z = 4.080$ but with the help of corresponding \CII~$\lambda$1036, partially available in the forest, the values 
obtained are useful indicators despite the large relative uncertainties.}

\tablenotetext{k}{\CIV~$\lambda$1551 has strong contamination by atmospheric line emission but $\lambda$1548 is clear although in a rather 
noisy region. \SiIV~$\lambda$1394 is blended with \CIV~$\lambda$1548 at $z = 3.482$ and weak \CII\ at $z = 4.200$ while $\lambda$1403 is 
partially blended with very weak \CII\ at $z = 4.231$ but together acceptable values are obtained.}

\tablenotetext{l}{This is a complex system with \CII\ and \SiII\ very strong and \CIV\ relatively weak in many of the components: collectively 
log~$N$(\HI) $\sim20.3$ gives a good fit to the outer parts of the \HI\ Lyman~$\alpha$ damping wings, at the ``damped Lyman $\alpha$'' 
classification. \CIV~$\lambda$1548 in component 2 is contaminated by atmospheric line emission while $\lambda$1551 in component 14 is blended 
with \CIV~$\lambda$1548 at $z = 4.089$ but together good values are obtained. \SiIV~$\lambda$1403 is blended with \CIV~$\lambda$1548 at 
$z = 3.604$ but $\lambda$1394 is clear. \CII~$\lambda$1335, outside the forest, contains much weaker \SiIV~$\lambda$1403 at $z = 3.832$ which is 
approximately determined with use of \CII~$\lambda$1036 in the forest; in components 12 and 16 \CII\ is mildly saturated, making the values 
listed for these less certain than the rest; there is weak associated \CII{$^\star$}~$\lambda$1336: for components 4, 12 and 16 
\CII{$^\star$}$/$\CII\ averages (3.4~$\pm$~1.1)~$\times~10^{-3}$. The \SiII\ values are from $\lambda$1527: component 1 has strong 
contamination by atmospheric line emission and values in components 10, 18 and 19, which have only upper limits, are too high to be useful.}

\tablenotetext{m}{\CIV~$\lambda$1551 is blended with \CIV~$\lambda$1548 at $z = 4.094$ but $\lambda$1548 is clear.}

\tablenotetext{n}{\CIV~$\lambda$1548 is blended with \CIV~$\lambda$1551 at $z = 4.080$ but is strong and together with both \CIV\ doublet members 
at $z = 4.080$ is well determined while $\lambda$1551 is strongly contaminated by atmospheric line emission and not used. \SiIV~$\lambda$1403 is 
blended with \CIV~$\lambda$1551 at $z = 3.604$ but $\lambda$1394 is clear although in a noisy region.}

\tablenotetext{o}{\CIV~$\lambda$1548 is blended with \CIV~$\lambda$1551 at $z = 4.086$ but $\lambda$1551 is clear.}

\tablenotetext{p}{\CIV~$\lambda$1551 in components 1 and 3 has strong contamination by atmospheric line emission but $\lambda$1548 is clear. 
\SiIV~$\lambda$1394 is weak and dominated by \CIV~$\lambda$1548 at $z = 3.681$ but $\lambda$1403 is clear. \CII\ in components 1 and 2 is masked by 
\CIV~$\lambda$1548 at $z = 3.482$ and \SiIV~$\lambda$1394 at $z = 3.978$.}

\tablenotetext{q}{\CIV~$\lambda$1551 is blended with \CIV~$\lambda$1548 at $z = 4.239$ but $\lambda$1548 is clear. \SiIV~$\lambda$1394 in components 
1 and 2 contains some unidentified narrow features but $\lambda$1403 is clear. \CII\ in components 2--4 is in a very noisy region and in 5--7 is 
blended with \SiIV~$\lambda$1403 at $z = 3.978$ but some useful upper limits are obtained.}

\tablenotetext{r}{\CIV~$\lambda$1548 is blended with \CIV~$\lambda$1551 at $z = 4.231$ while $\lambda$1551 is blended with \CIV~$\lambda$1548 at 
$z = 4.247$, but taking all associated doublets together good values are obtained. \SiIV~$\lambda$1403 is blended with \CIV~$\lambda$1548 at 
$z = 3.748$ but $\lambda$1394 appears clear apart from a strong feature at observed wavelength $\lambda$5908.40, at apparent redshift 4.239181 if  
$\lambda$1394, where however no features are detected in \CIV\ and \CII, so this is probably an unidentified interloper and not entered in the 
table. \CII\ is blended with \SiII~$\lambda$1304 at $z = 4.361$ but this is very weak and only the value in component 3 is significantly affected.}  

\tablenotetext{s}{\CIV~$\lambda$1548 is blended with \CIV~$\lambda$1551 at $z = 4.239$ while $\lambda$1551 is strongly affected by atmospheric line 
absorption but together with the associated \CIV\ doublets nearby in redshift good values are obtained. \SiIV~$\lambda$1403 is partially blended 
with \CIV~$\lambda$1551 at $z = 3.748$, while component 2 in $\lambda$1394 has an interloper indicated by strong inconsistency with $\lambda$1403 
and is not used, but taken collectively the listed values are acceptable.}

\tablenotetext{t}{\CIV~$\lambda$1548 contains very weak \SiII~$\lambda$1527 at $z = 4.361$ and in component 3 includes strong atmospheric line 
absorption while $\lambda$1551 in components 2 and 4 includes line emission, and both are in a rather noisy region, but together acceptable 
values are obtained.} 

\tablenotetext{u}{\CIV~$\lambda$1548 in components 2, 7 and 8 and $\lambda$1551 in components 3 and 4 are contaminated by strong atmospheric line 
emission but together good values are obtained. \SiIV~$\lambda$1394 in component 1 is blended with \CIV~$\lambda$1548 at $z = 3.825$ and has mild 
interference by atmospheric line emission but $\lambda$1403 is clear although in a rather noisy region. \SiII\ is derived from a combination of 
complementary portions of $\lambda$1260 (in the forest but useful) and $\lambda$1304 (partially blended with \CII~$\lambda$1335 at $z = 4.239$), 
while $\lambda$1527 is partially blended with \CIV~$\lambda$1548 at $z = 4.286$ and rather noisy and is not used; errors and upper limits are from 
$\lambda\lambda$1260,1304.}

\tablenotetext{v}{\CIV\ values are dominantly from $\lambda$1551 as $\lambda$1548 is heavily contaminated by atmospheric line emission and has other 
apparent interference. \SiIV~$\lambda$1394 is masked by strong \AlIII~$\lambda$1863 at $z = 3.032$ but $\lambda$1403 is clear except for component 1 
which is obscured by atmospheric line emission. \SiII~$\lambda$1260 is outside the forest.}

\end{deluxetable}


\clearpage
\begin{figure*}
\figurenum{\scriptsize 1}
\epsscale{1.9}
\plotone{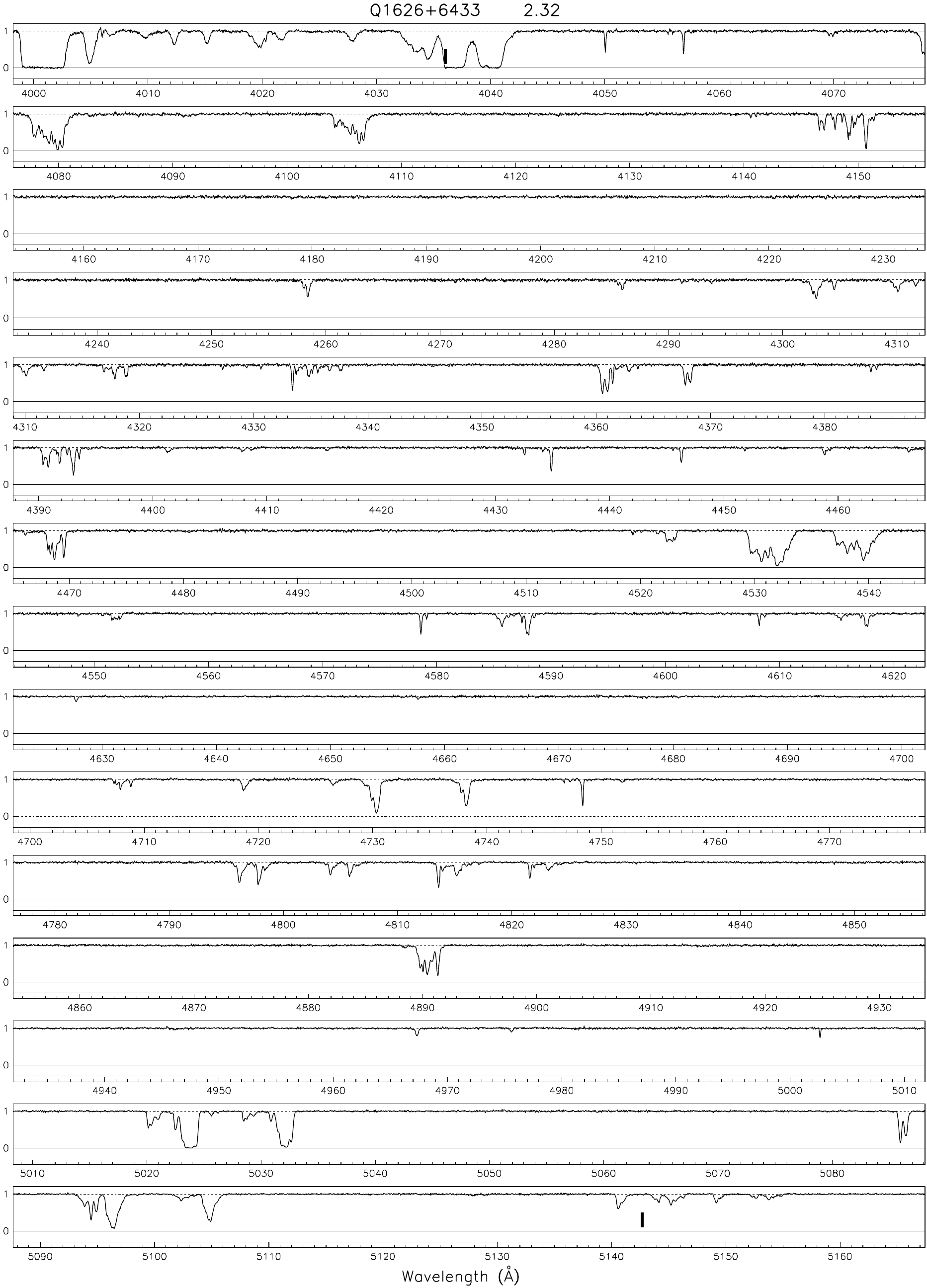}
\caption{\scriptsize Continuum-normalized spectra 
of the nine QSOs in observed wavelengths reduced to heliocentric vacuum 
values, scaled to cover the same rest wavelength range: \emph{thick vertical 
ticks} place the QSO broad Lyman~$\alpha$ and \CIV\ lines. Atmospheric line 
features are indicated by \emph{graded thin vertical ticks} (emission) and 
\emph{inset spectra below} (absorption); both reflect the heliocentric 
corrections over the runs (see text and Tables): the QSO spectra are 
``cleaned'' of resultant interference for display.} 
\end{figure*}

\clearpage
\begin{figure*}
\figurenum{\scriptsize 1}
\epsscale{1.9}
\plotone{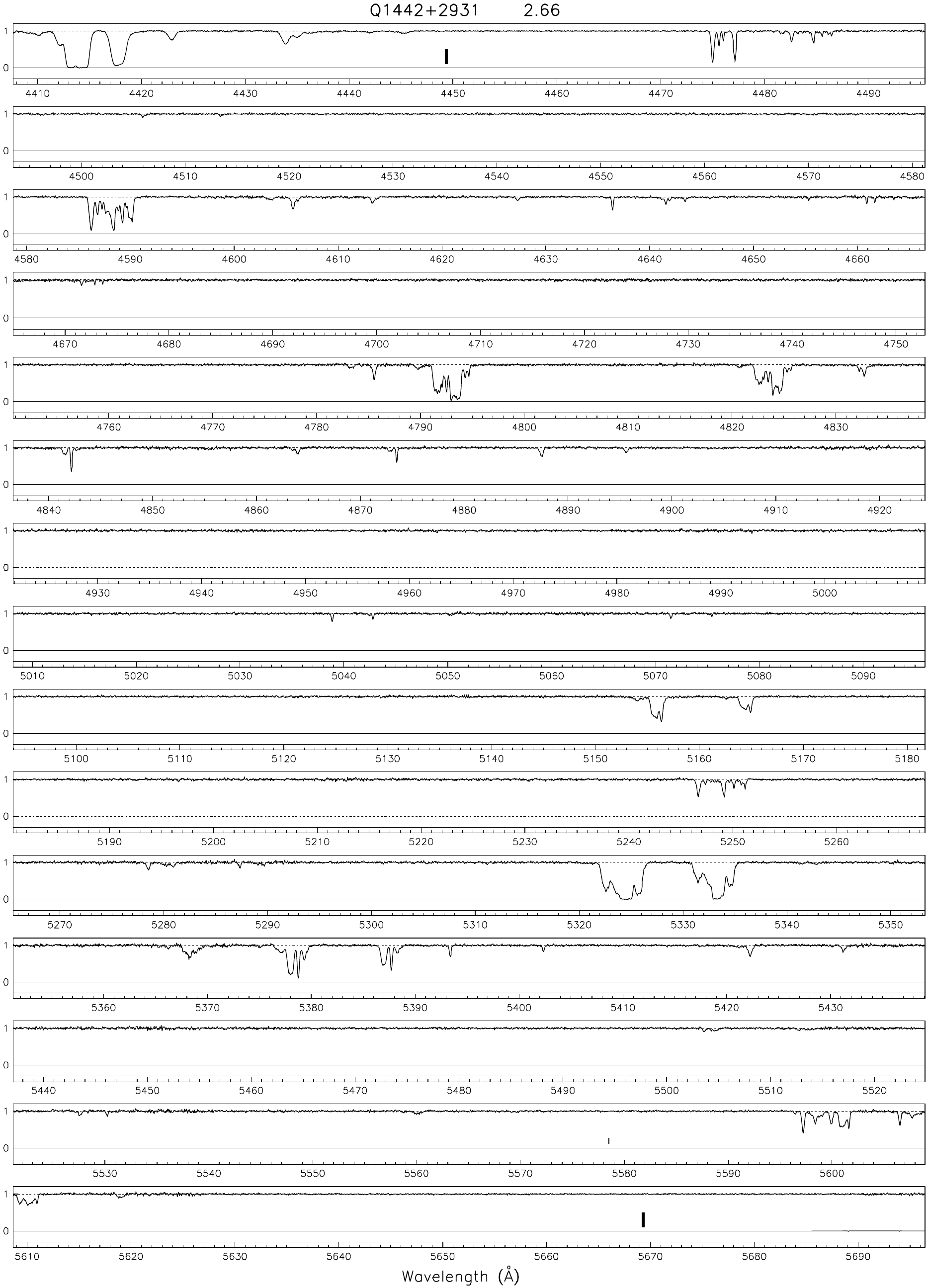}
\caption{\scriptsize Continued.}
\end{figure*}

\clearpage
\begin{figure*}
\figurenum{\scriptsize 1}
\epsscale{1.9}
\plotone{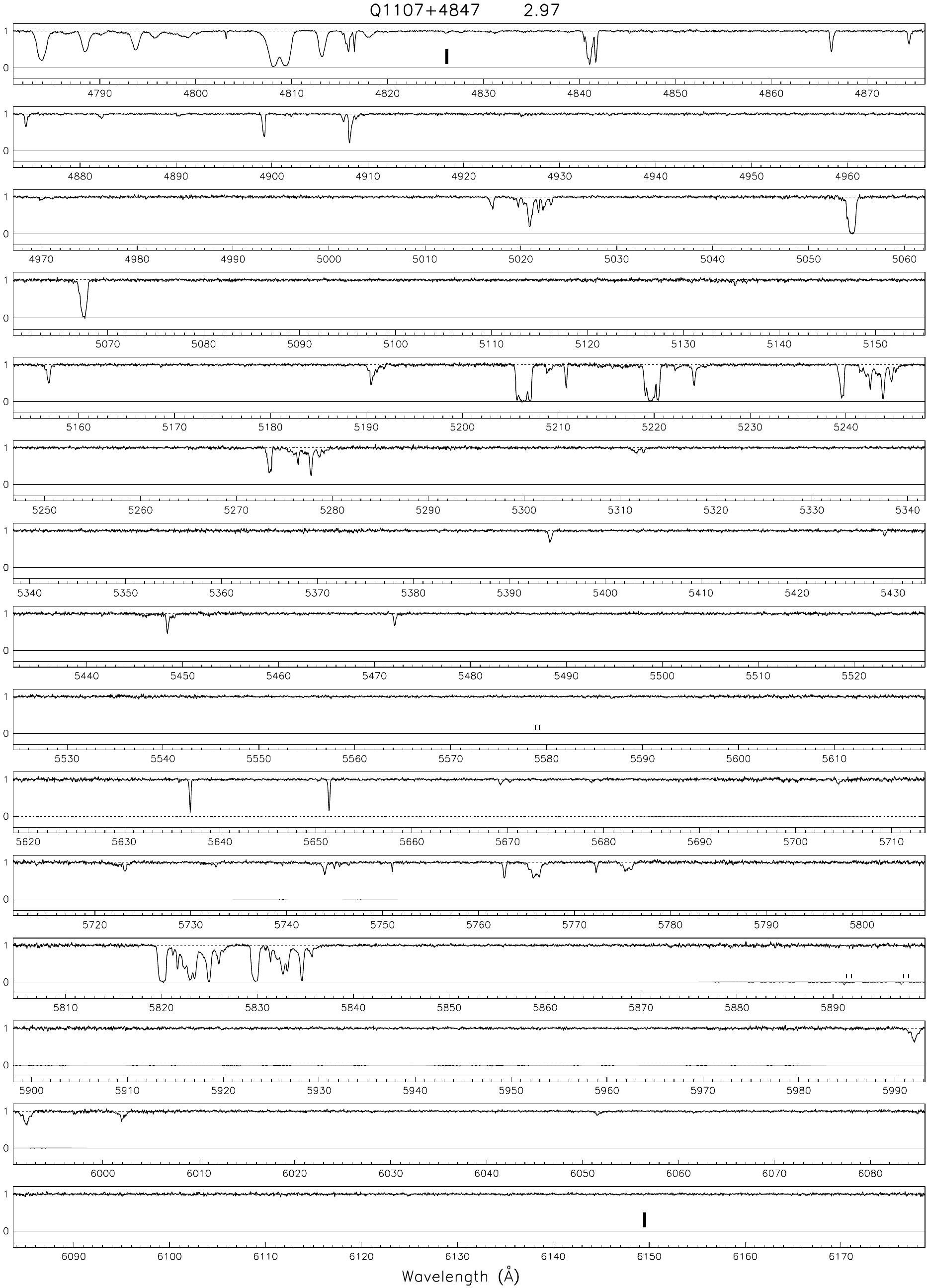}
\caption{\scriptsize Continued.}
\end{figure*}

\clearpage
\begin{figure*}
\figurenum{\scriptsize 1}
\epsscale{1.9}
\plotone{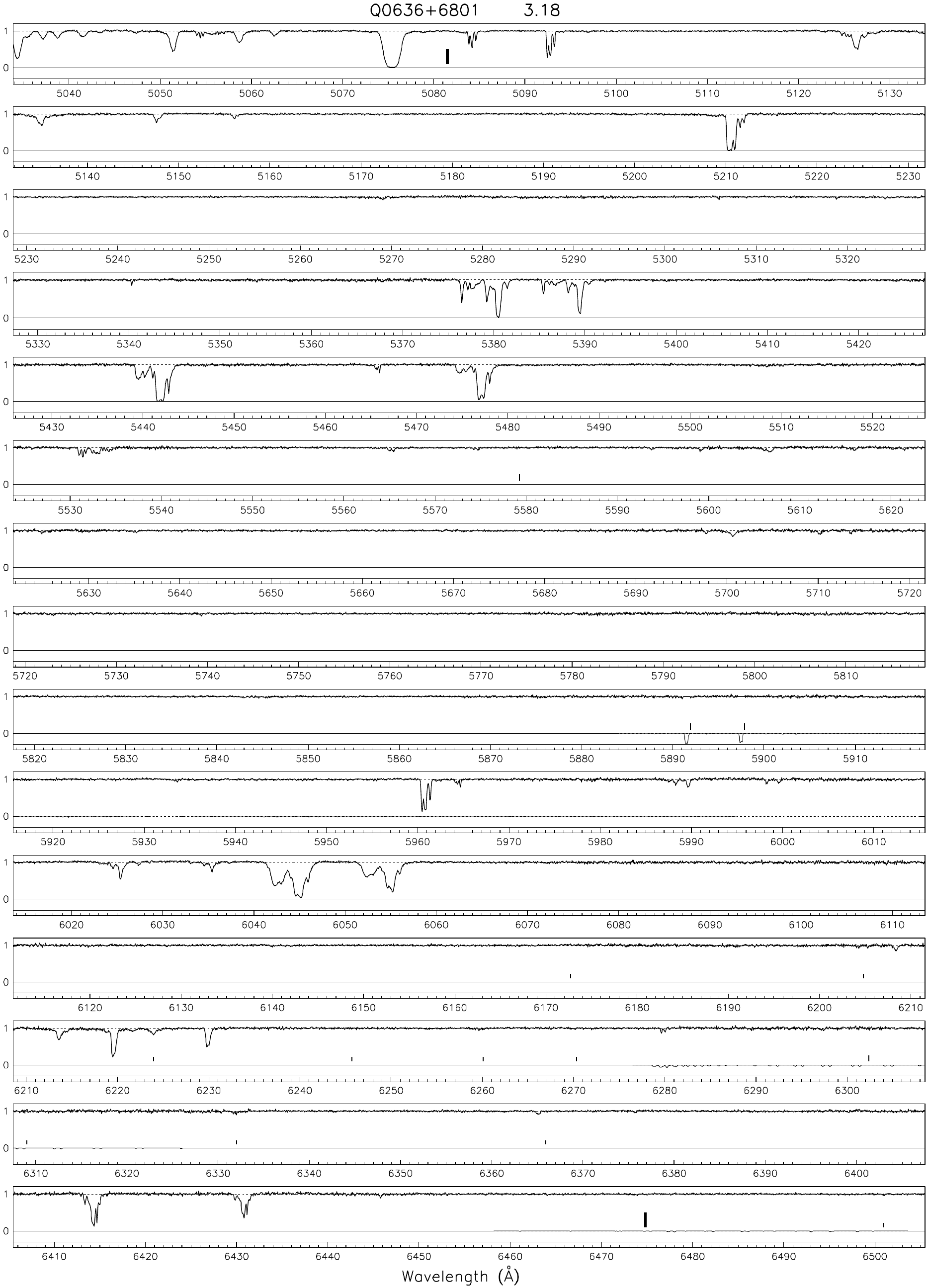}
\caption{\scriptsize Continued.}
\end{figure*}

\clearpage
\begin{figure*}
\figurenum{\scriptsize 1}
\epsscale{1.9}
\plotone{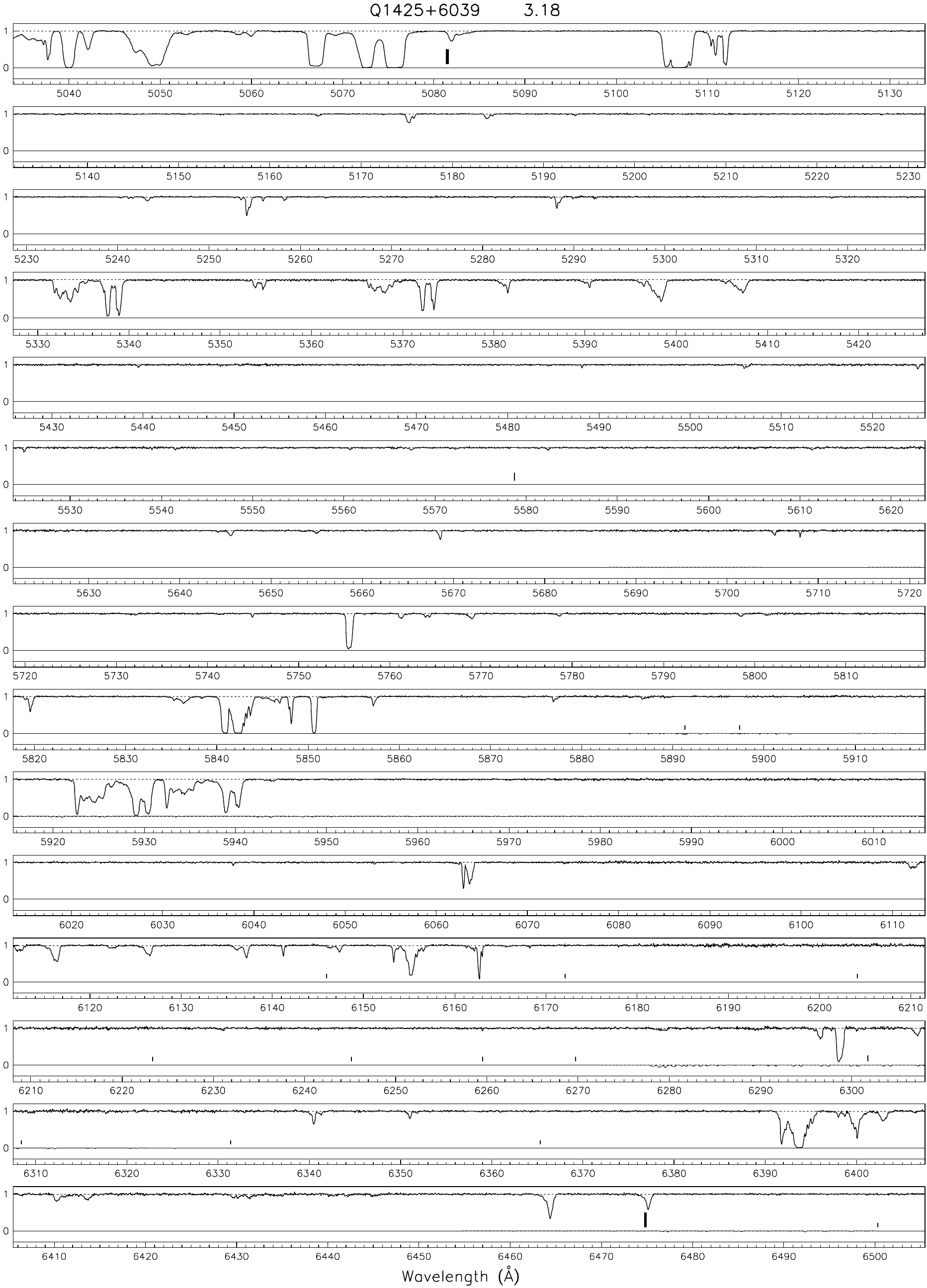}
\caption{\scriptsize Continued.}
\end{figure*}

\clearpage
\begin{figure*}
\figurenum{\scriptsize 1}
\epsscale{1.9}
\plotone{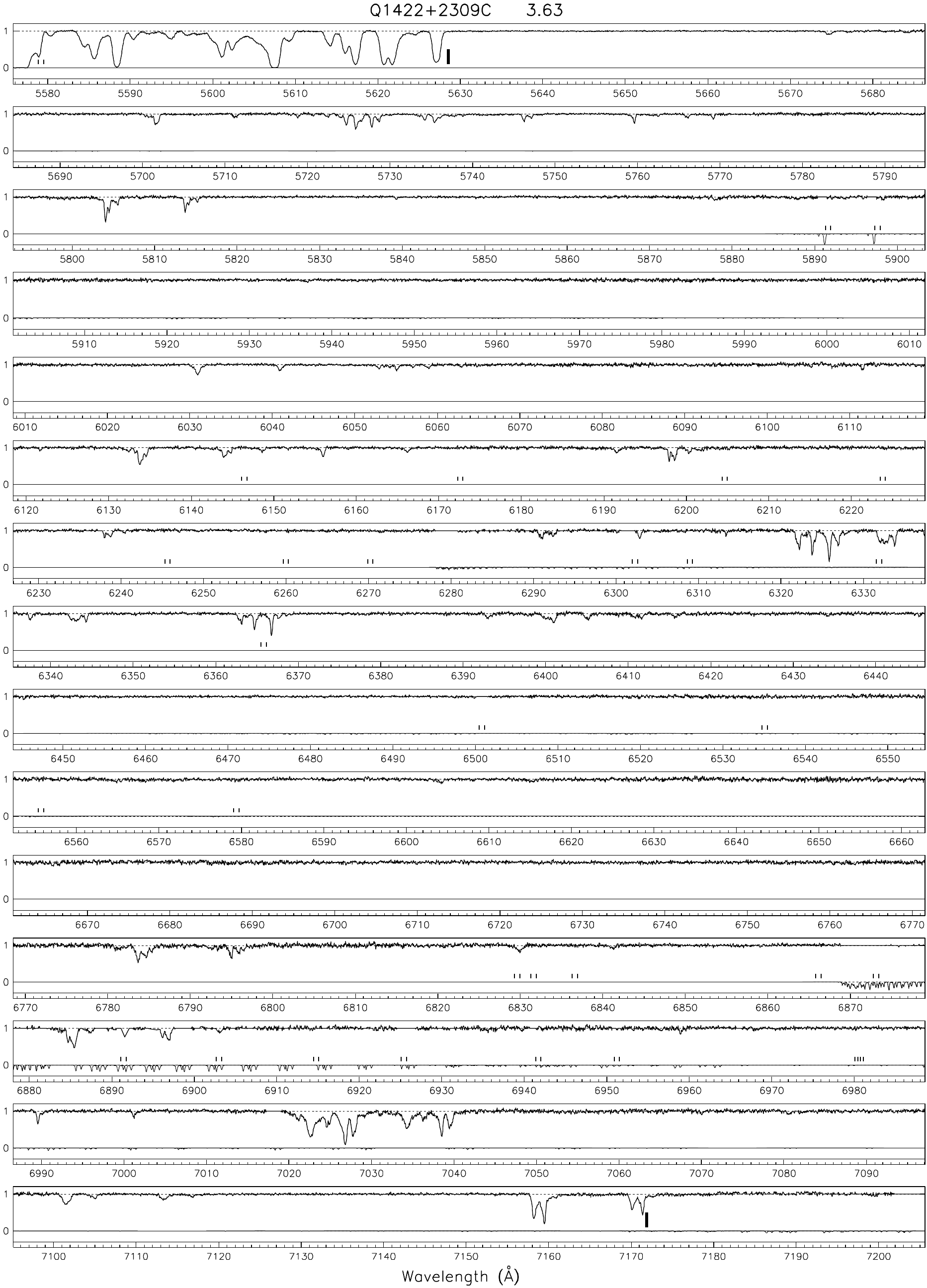}
\caption{\scriptsize Continued.}
\end{figure*}

\clearpage
\begin{figure*}
\figurenum{\scriptsize 1}
\epsscale{1.9}
\plotone{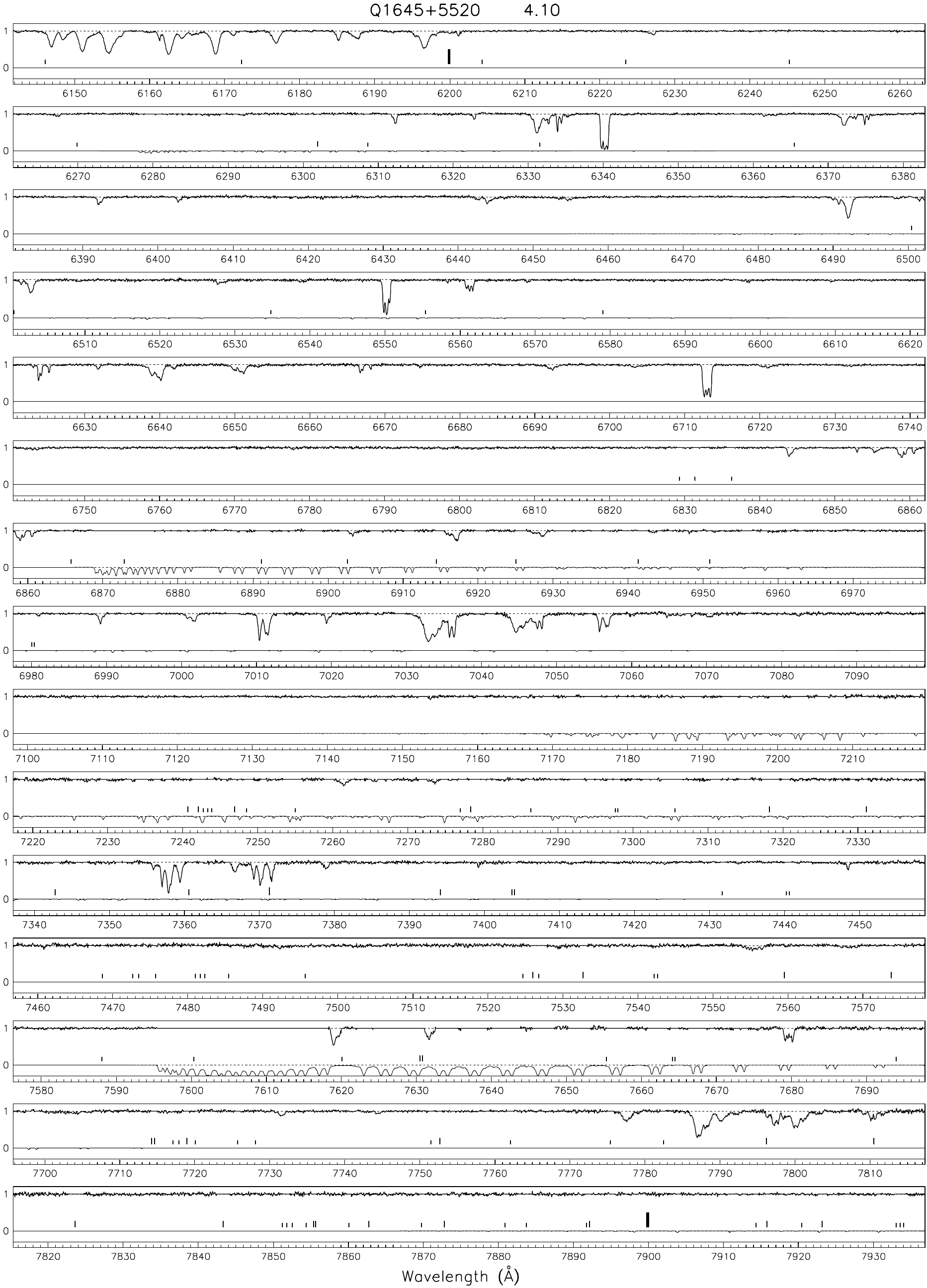}
\caption{\scriptsize Continued.}
\end{figure*}

\clearpage
\begin{figure*}
\figurenum{\scriptsize 1}
\epsscale{1.9}
\plotone{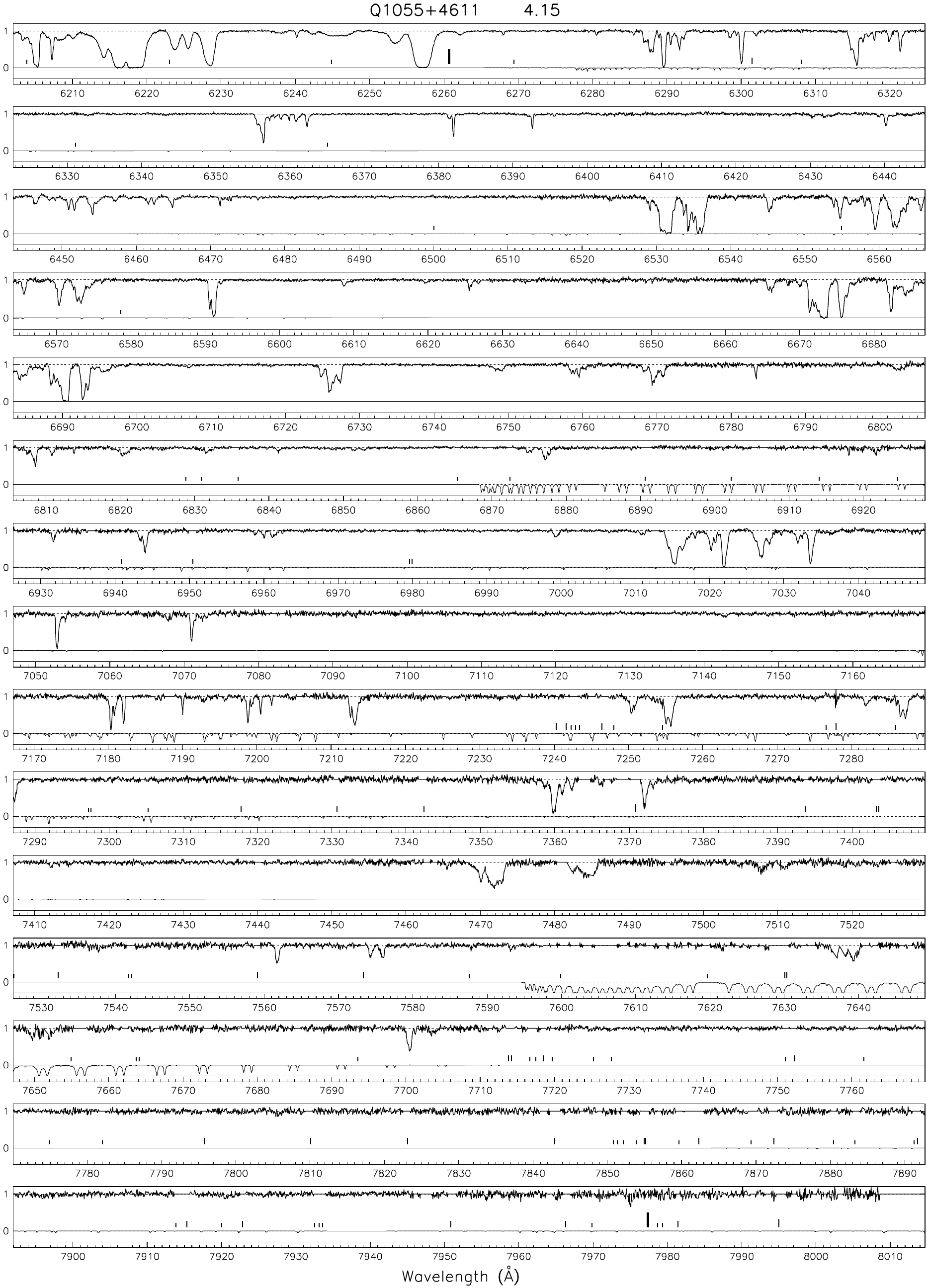}
\caption{\scriptsize Continued.}
\end{figure*}

\clearpage
\begin{figure*}
\figurenum{\scriptsize 1}
\epsscale{1.9}
\plotone{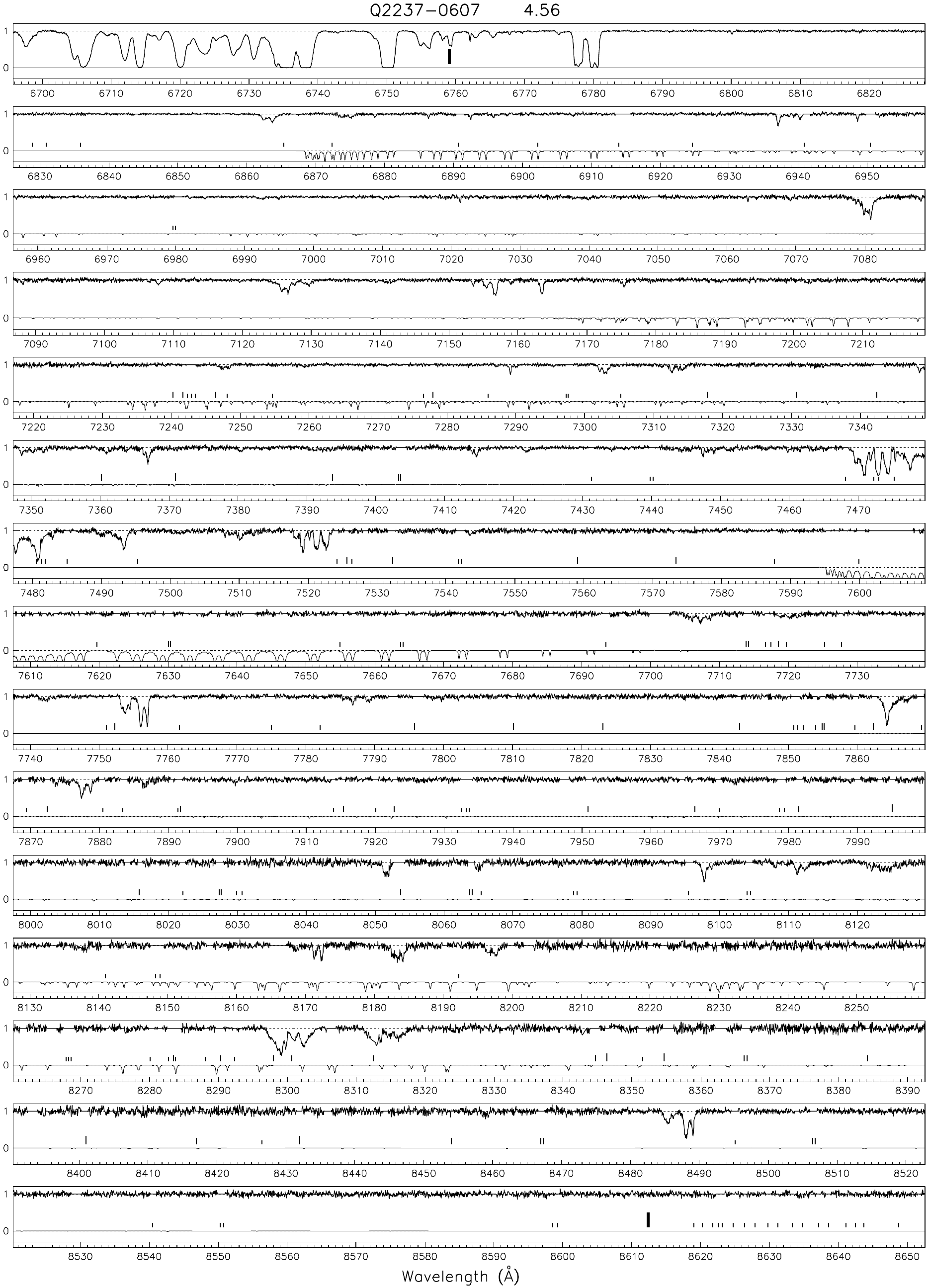}
\caption{\scriptsize Continued.}
\end{figure*}

\clearpage
\begin{figure*}
\figurenum{\scriptsize 2}
\epsscale{2.0}
\plotone{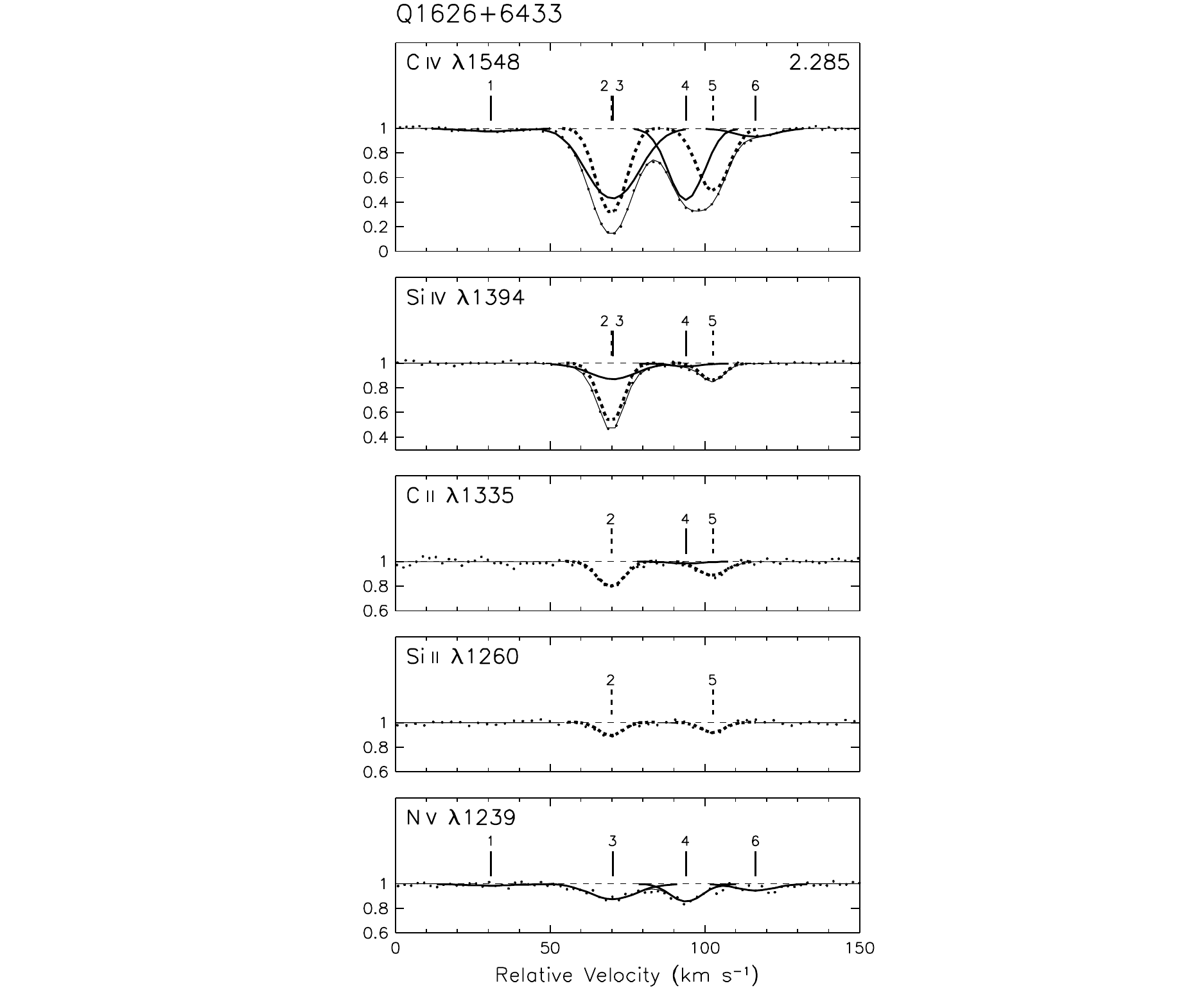}
\caption{\scriptsize \CIV~$\lambda$1548, 
\SiIV~$\lambda$1394, \CII~$\lambda$1335, \SiII~$\lambda$1260 and 
\NV~$\lambda$1239 continuum-normalized spectra for the $z = 2.285$ 
absorption system in Q1626$+$6433, demonstrating the separation into 
\emph{single-phase-ionization components} by means of the VPFIT analysis 
(see \S3.2 in the text for a full description). The complete data set 
for this system (including all observed transitions) is shown in Figure 
4. For each profile the composite fit through the system data points is 
shown by a \emph{thin continuous line} where not covered by individual 
component fits. Parameters of the constituent components indicated by 
the \emph{numbered vertical ticks} are given in Table 2. Ticks 
are shown only at the positions of \emph{detected} components, not upper 
limits. The six components shown separated with individual fits in the 
\CIV\ profile are reflected in partial fashion among the \SiIV, \CII, 
\SiII\ and \NV\ profiles, with the components that are dominant in \CII\ 
(low ionization) identified by \emph{dashed lines and ticks} and those 
dominant in \NV\ (high ionization) by \emph{thick continuous lines and 
ticks}. Note the two nearly coincident (in redshift) components 2 and 3 
having markedly different widths: both are strong in \CIV\ but differ 
greatly among the other species (see text). Component 4 which is strong 
in both \CIV\ and \NV\ still is weakly evident in the \SiIV\ and \CII\ 
profiles.} 
\end{figure*}

\clearpage
\begin{figure*}
\figurenum{\scriptsize 3}
\epsscale{2.0}
\plotone{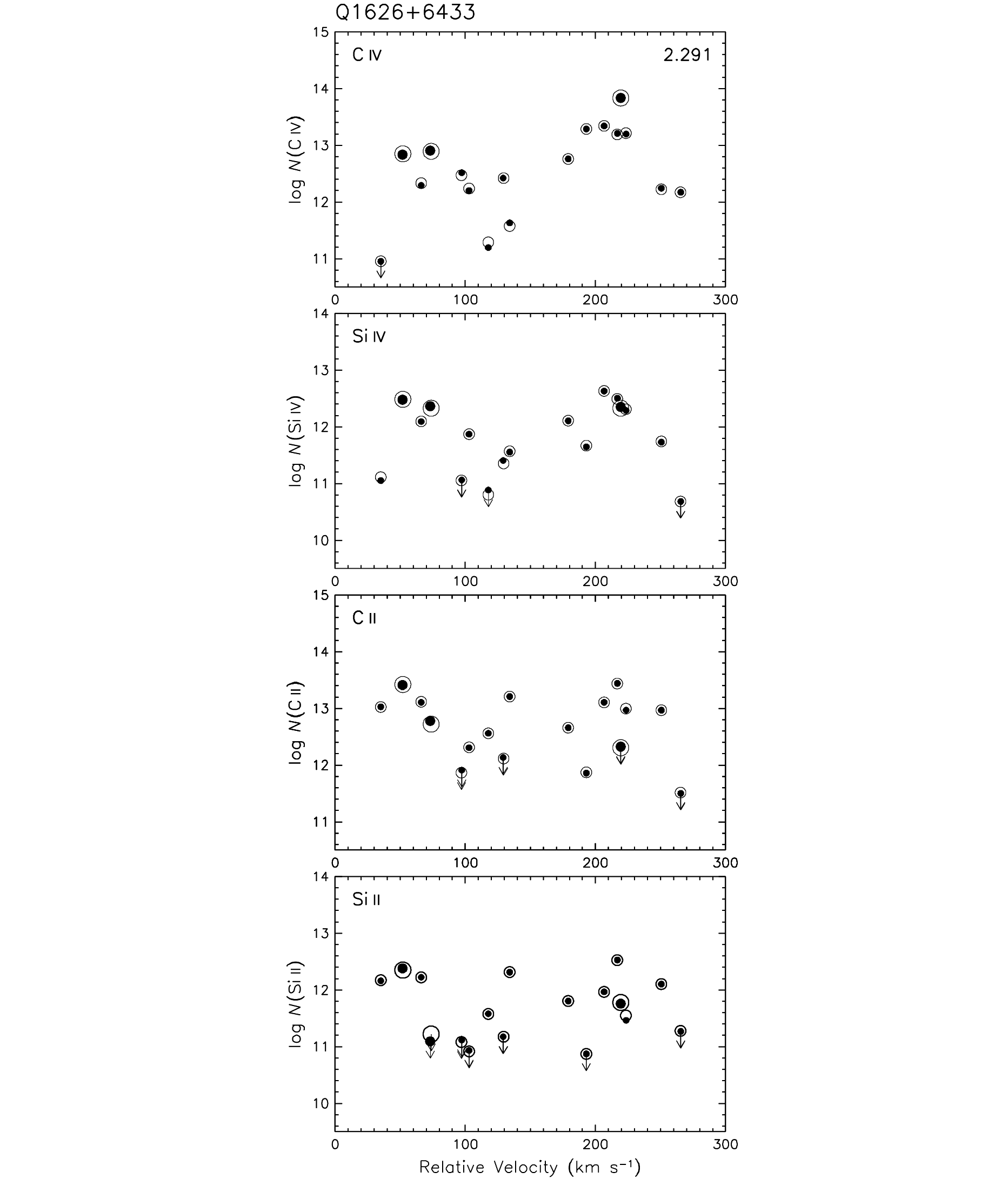}
\caption{\scriptsize \CIV, \SiIV, \CII\ and \SiII\ 
component column densities $N$ (cm$^{-2}$) for the complex absorption 
system at $z = 2.291$ in Q1626+6433 (see Figures 4 and 6 and Table 2 for 
identification). Two individual runs in the VPFIT analysis show the 
differential effect of fixing the set of three broadest components 
(indicated by \emph{enlarged symbols}) at the two widely separated 
temperatures $1 \times 10^{4}$ K and $1 \times 10^{5}$ K, but otherwise 
having the same mix of specific starting values for the remainder of the 
components (see text). All component values yielded in the first case are 
shown by \emph{filled circles} and in the second by \emph{open circles}.}
\end{figure*}

\clearpage
\begin{figure*}
\figurenum{\scriptsize 4}
\epsscale{2.0}
\plotone{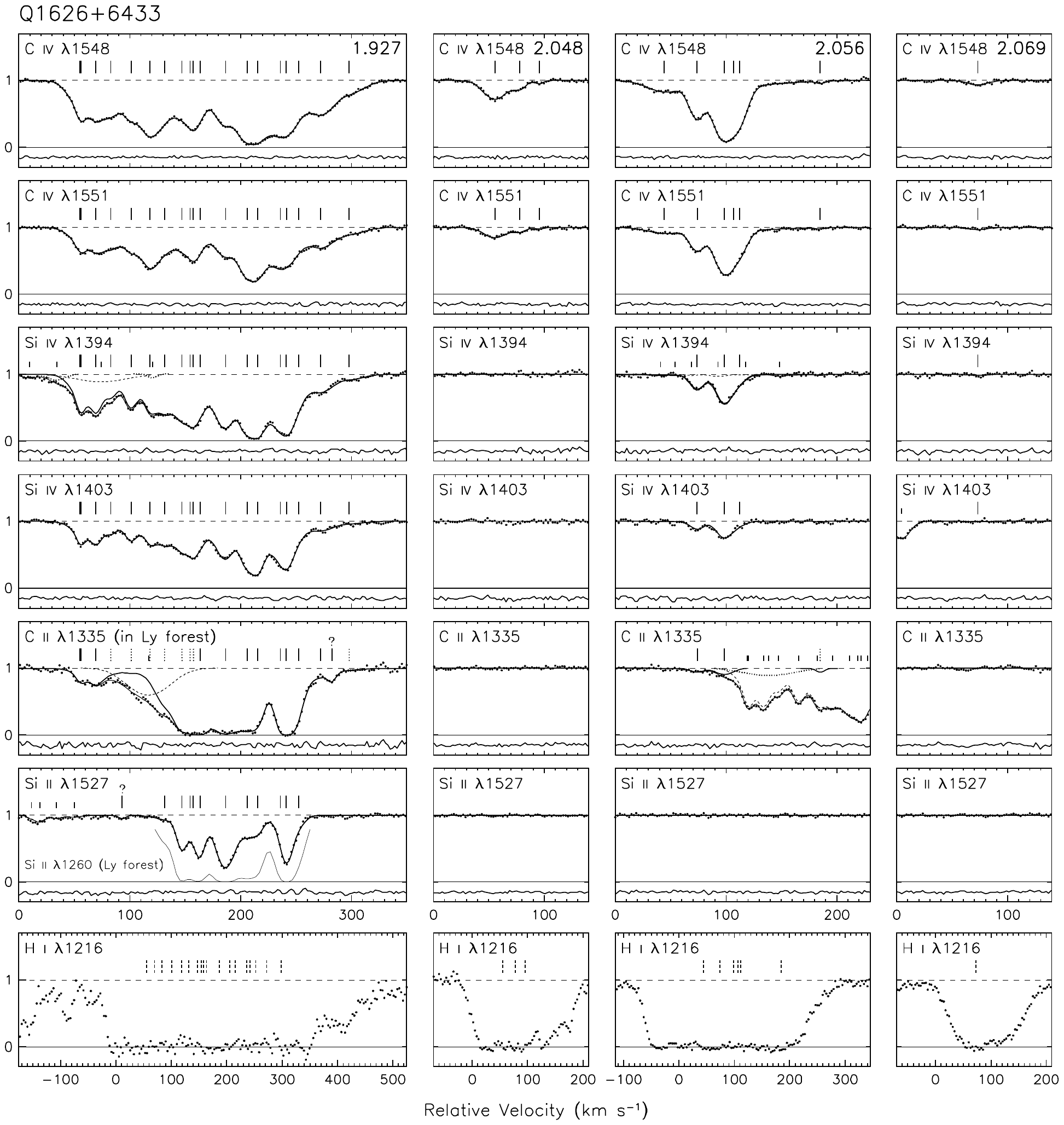}
\caption{\scriptsize Continuum-normalized spectra for 
all systems detected in Q1626+6433 ($z_{em} = 2.32$) having at least \CIV\ 
and \SiIV\ outside the Lyman forest, as an example of the outcome of the 
line-fitting procedure. For each system the available metal species of 
interest in this paper are compared on a common velocity scale. The system 
components identified using VPFIT are marked with \emph{long vertical ticks}; 
a few questionable features are so indicated. Components yielding only upper 
limits are unmarked. Component parameter values and upper limits are listed 
in Table 2; in the few cases where components have values which are too 
uncertain to be included in the table, as indicated in associated notes, they 
are marked with \emph{dotted ticks}. Blended or nearby interloper species from 
other systems, also indicated in the notes, are marked with \emph{short ticks}.
The observed data values are given as points and the overall fits obtained to 
these are shown as \emph{continuous lines}; overall residuals (i.e., 
[data] $-$ [fit]) are shown on the same scale beneath the profiles. When 
blending is present the deconvolved fits to the species under consideration 
again are shown as \emph{continuous lines} and fits to individual interloper 
species by \emph{short-dashed or dotted lines}. \SiII~$\lambda$1527 at 
$z = 1.927$ and $z = 2.110$ have superimposed fits to $\lambda1260$ (in these 
cases found in relatively clear regions of the forest and matching the data 
well) shown as \emph{continuous thin lines}. Lyman $\alpha$ observations, 
covering twice the velocity range, are shown unfitted, but with the positions 
of all components detected in \CIV\ indicated by \emph{broken ticks}.}
\end{figure*}

\clearpage
\begin{figure*}
\figurenum{\scriptsize 4}
\epsscale{2.0}
\plotone{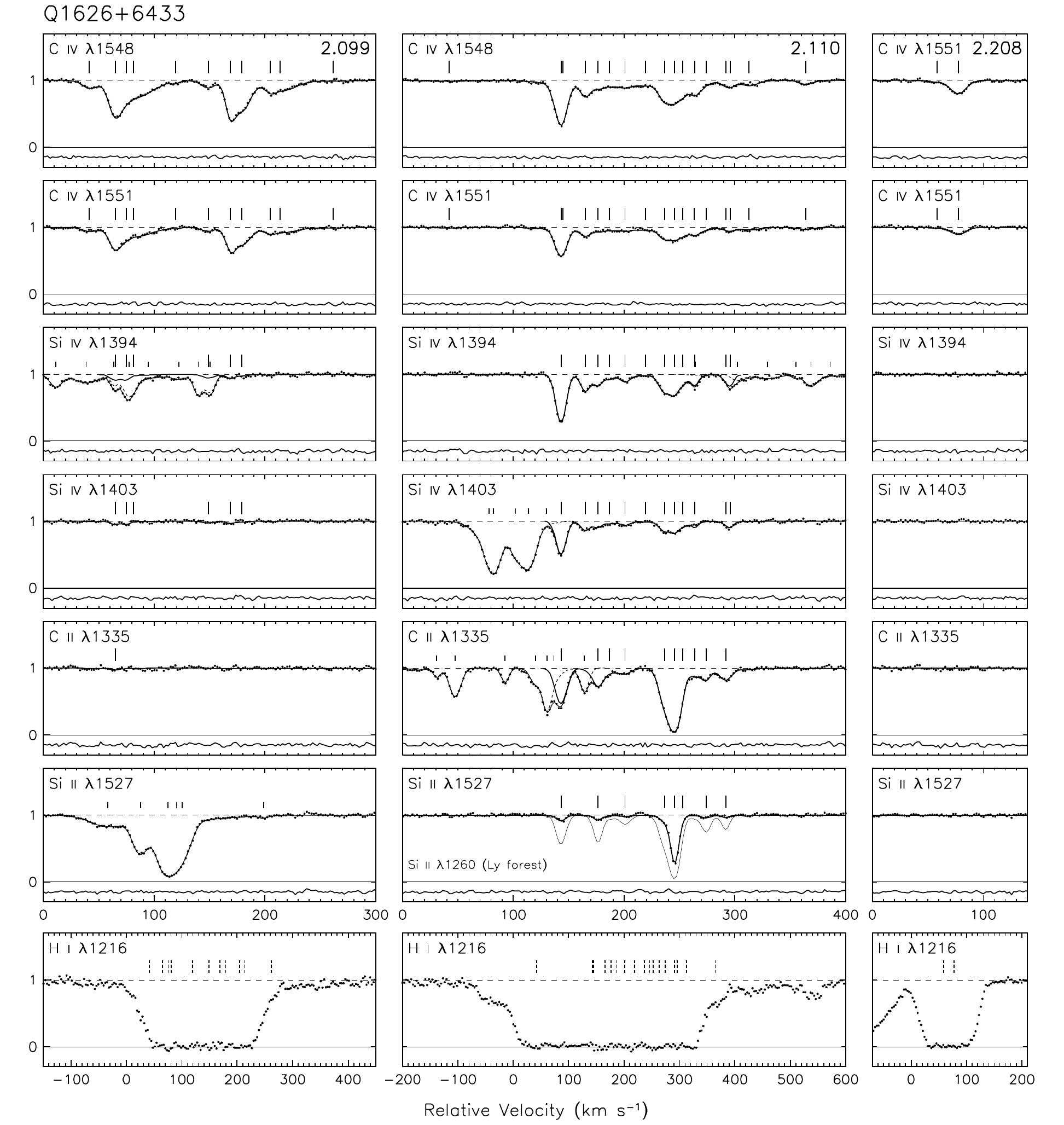}
\caption{\scriptsize Continued.}
\end{figure*}

\clearpage
\begin{figure*}
\figurenum{\scriptsize 4}
\epsscale{2.0}
\plotone{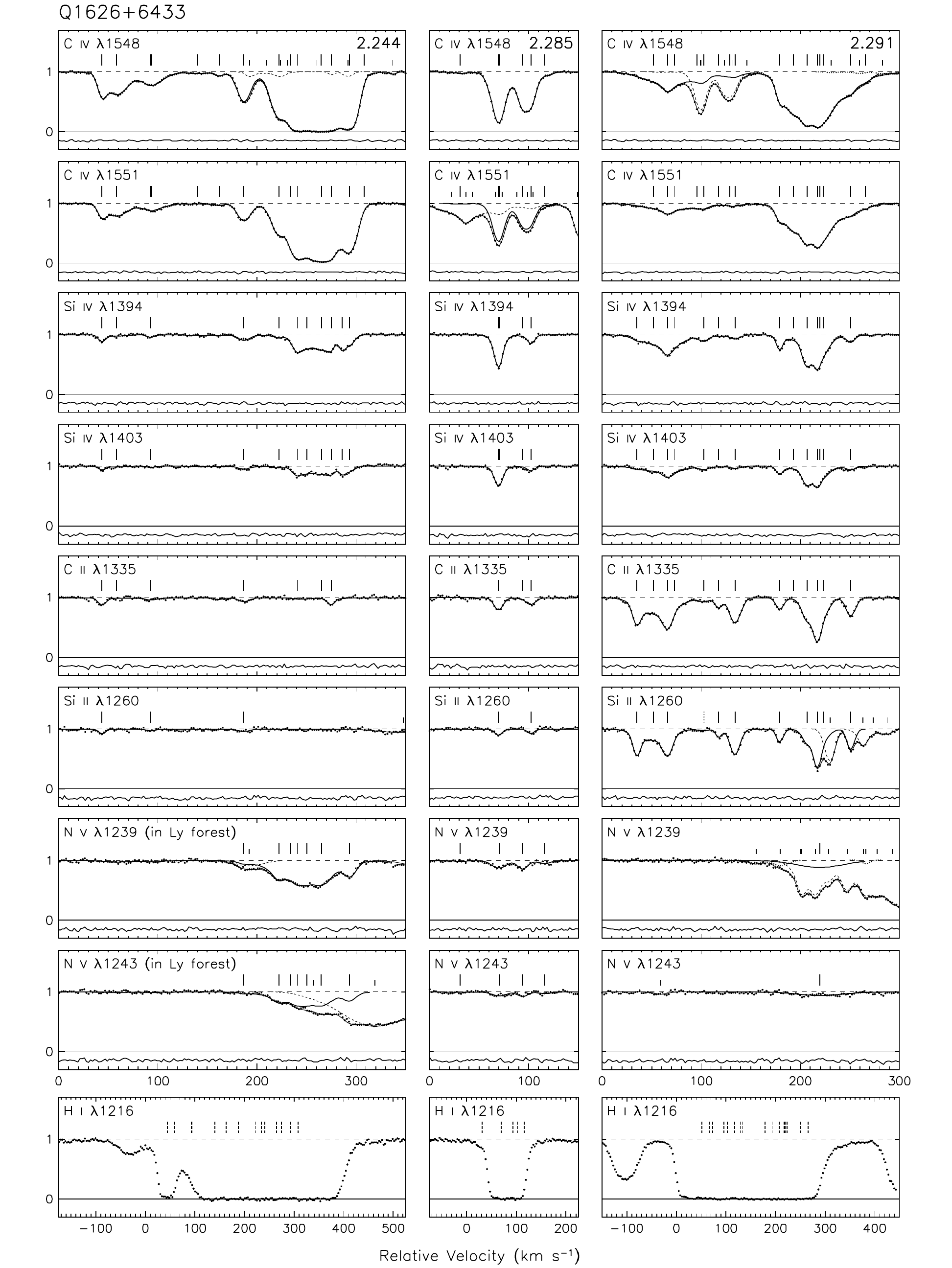}
\caption{\scriptsize Continued.}
\end{figure*}

\clearpage
\begin{figure*}
\figurenum{\scriptsize 4}
\epsscale{2.0}
\plotone{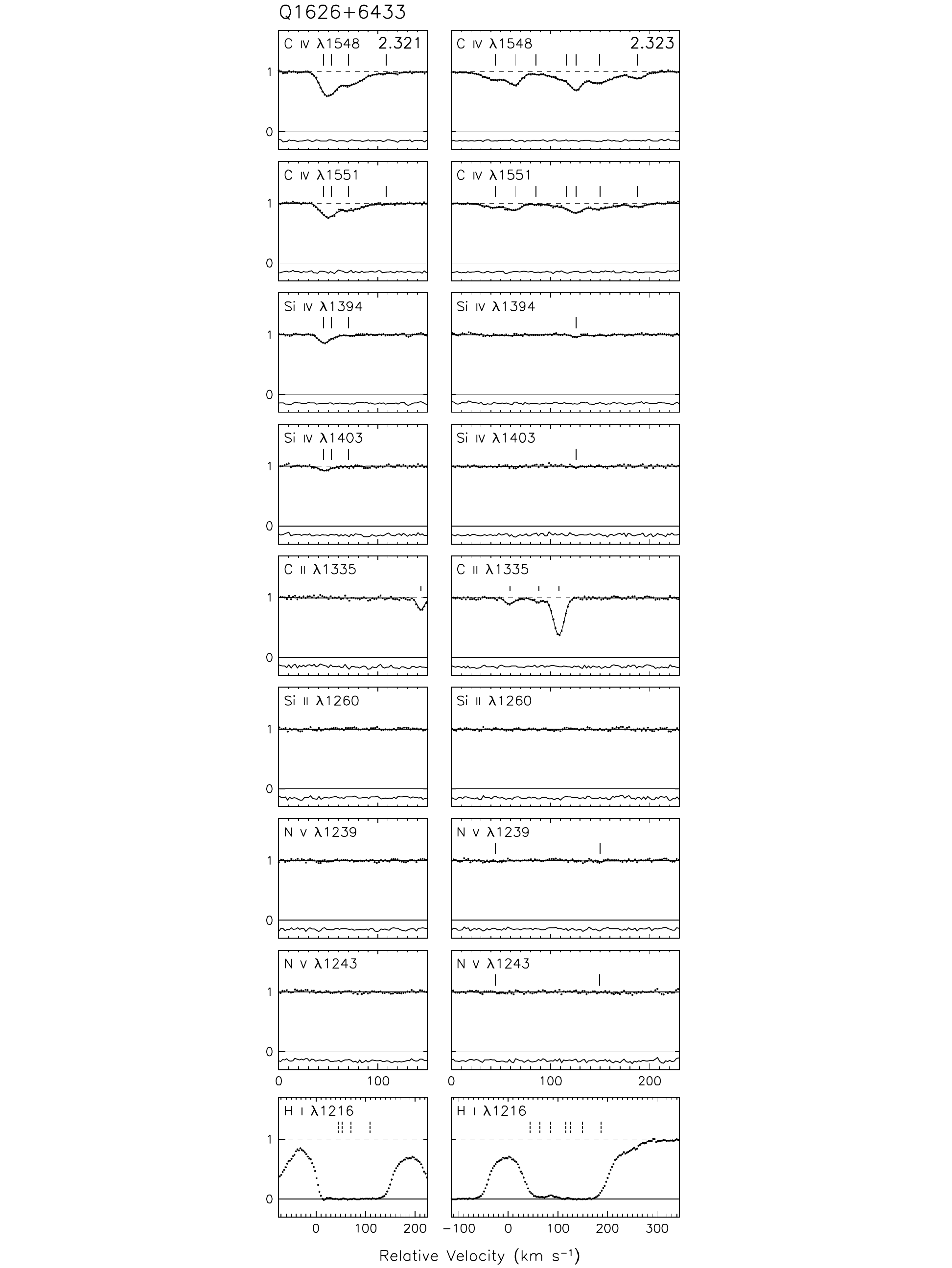}
\caption{\scriptsize Continued.}
\end{figure*}

\clearpage
\begin{figure*}
\figurenum{\scriptsize 5}
\epsscale{1.96}
\plotone{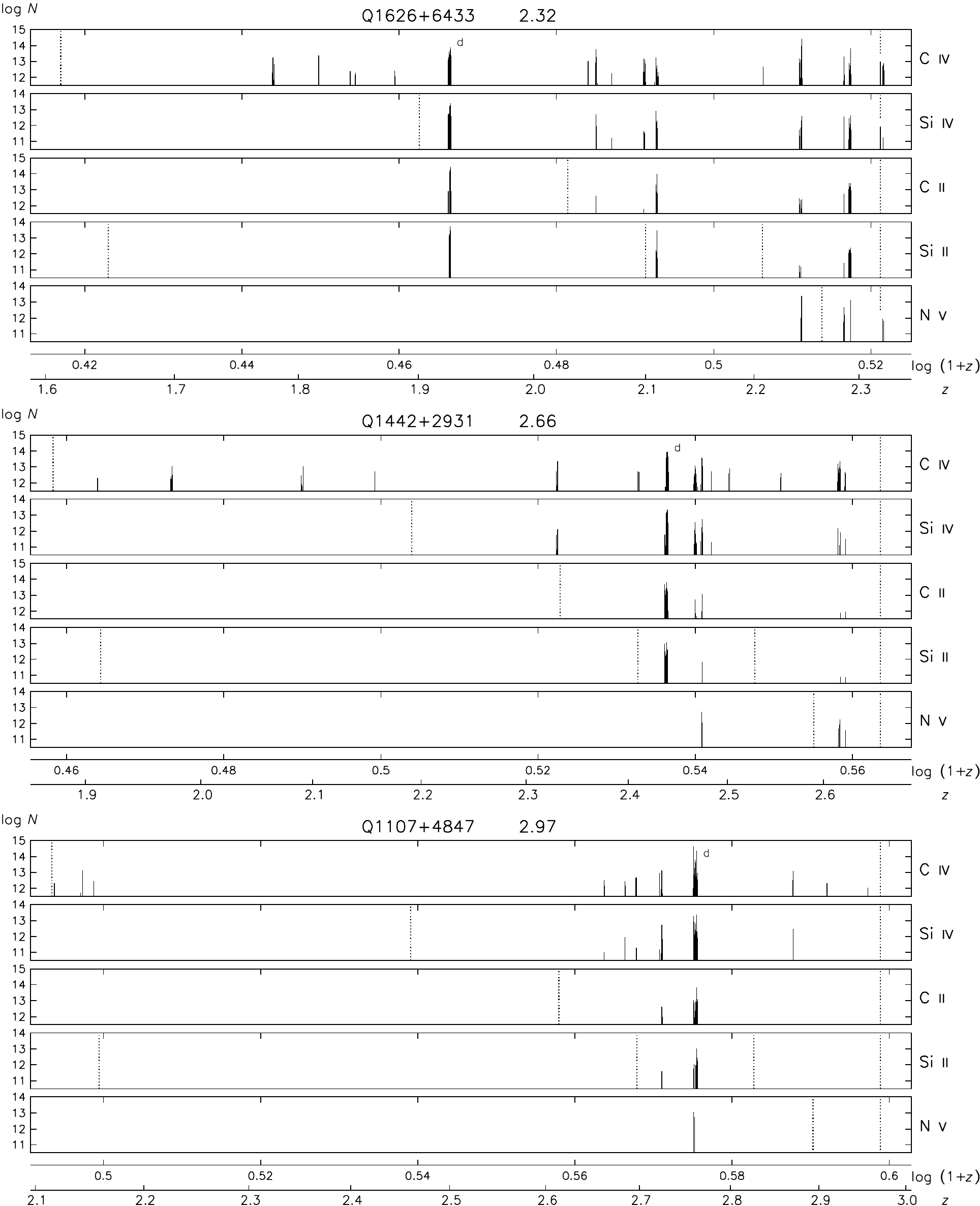}
\caption{\scriptsize \CIV, \SiIV, \CII, \SiII\ and \NV\ 
column densities $N$ (cm$^{-2}$) plotted against redshift $z$ and log ($1+z$) 
for the absorber components \emph{detected} in the nine QSOs, including the 
few bracketted values (see Tables 2--10). Errors and upper limits are not 
shown. Systems containing one or more components with significant 
Lyman~$\alpha$ damping wings are marked d. Note, the vertical scales for 
\SiIV, \SiII\ and \NV\ are shifted lower by 1 dex than the other two. All 
frames cover the same extent in log ($1+z$), which for \CIV\ encompasses the 
region between the QSO Lyman~$\alpha$ and \CIV\ emission lines with some 
margin. The \emph{dotted vertical line} at the right of each frame is at the 
emission redshift indicated in the heading; the similar line to the left of 
this marks the limit where a given ion falls in the Lyman forest 
(for \SiII\ the limits for $\lambda$1260, $\lambda$1304 and 
$\lambda$1527 are included; the \emph{strong} components appearing to the 
left of the $\lambda$1260 and $\lambda$1304 boundaries are invariably 
$\lambda$1527). Reliable values for \SiIV, \CII, \SiII\ and \NV\ appear at 
redshifts in the forest where clear (such cases are few), as clarified 
in the table footnotes.} 
\end{figure*}

\clearpage
\begin{figure*}
\figurenum{\scriptsize 5}
\epsscale{1.96}
\plotone{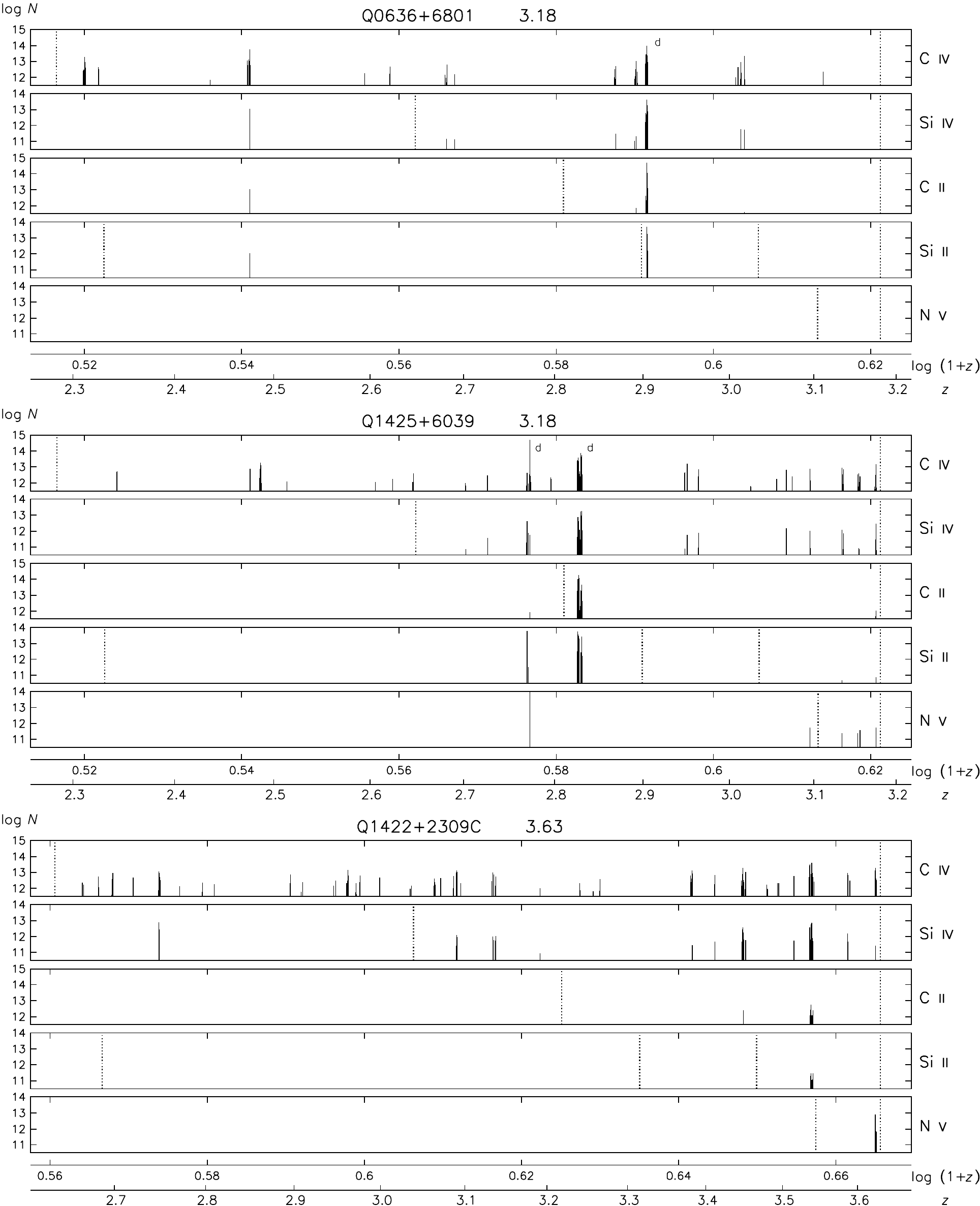}
\caption{\scriptsize Continued.}
\end{figure*}

\clearpage
\begin{figure*}
\figurenum{\scriptsize 5}
\epsscale{1.96}
\plotone{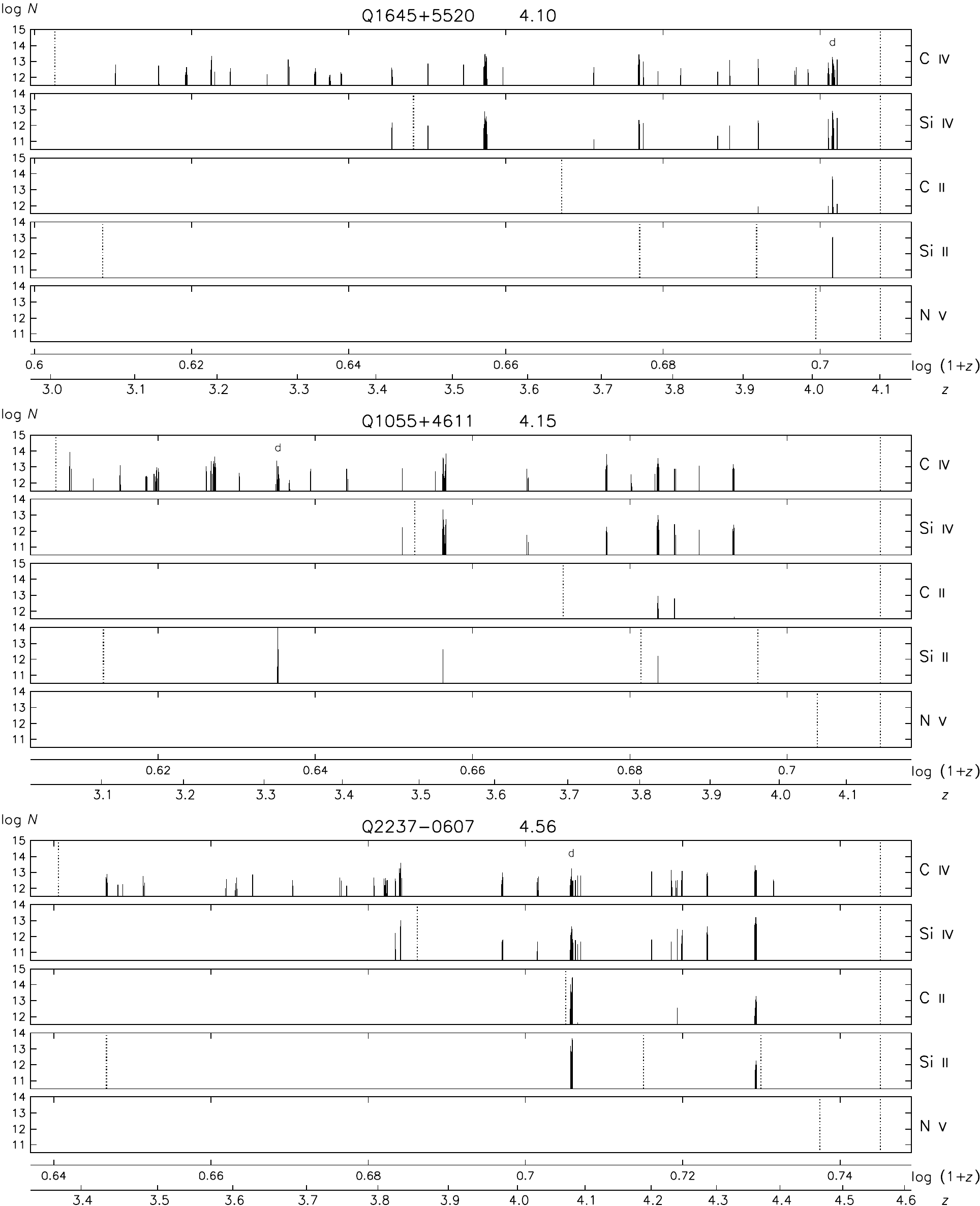}
\caption{\scriptsize Continued.}
\end{figure*}

\clearpage
\begin{figure*}
\figurenum{\scriptsize 6}
\epsscale{2.0}
\plotone{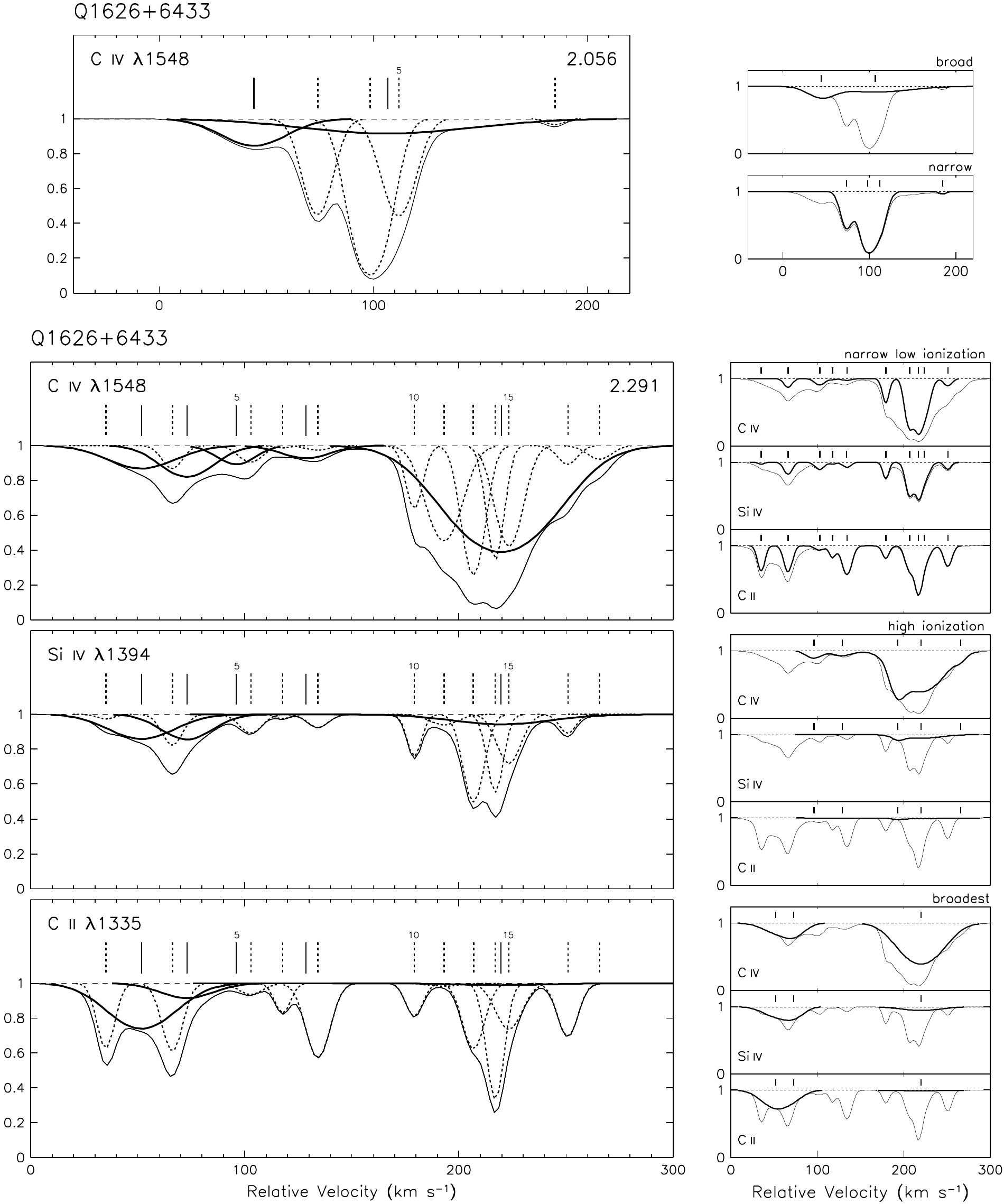}
\caption{\scriptsize Two examples from Q1626+6433 
displayed in Figure 4, showing details of the derived VPFIT components of 
systems containing \emph{broad} components (\emph{continuous thick lines 
and ticks}) among the more numerous \emph{narrow} components, 
{\it b}(\CIV) $\lesssim$ 10 km~s$^{-1}$ (\emph{dashed lines and ticks}), 
with numbering as in Table 2. Unlike in Figure 4 the ticks here 
\emph{include} positions of upper limits. The overall composite fits to the 
pure system profiles ({\it i.e.,} excluding any interloper species) are 
shown in \emph{continuous thin lines}. \emph{Upper panels:} 
\emph{left}, \CIV\ profile in a simple system with exposed broad components; 
\emph{right}, separately highlighting the composite profiles of the 
\emph{broad} and \emph{narrow} components. \emph{Lower panels:} 
\emph{left}, \CIV, \SiIV\ and \CII\ profiles in a complex system with 
immersed broad components; \emph{right}, separately highlighting the 
composite profiles of \emph{narrow components} which are strong in \CII, 
\emph{all high ionization components} and the \emph{broadest components}.}
\end{figure*}

\clearpage
\begin{figure*}
\figurenum{\scriptsize 7(a)}
\epsscale{2.0}
\plotone{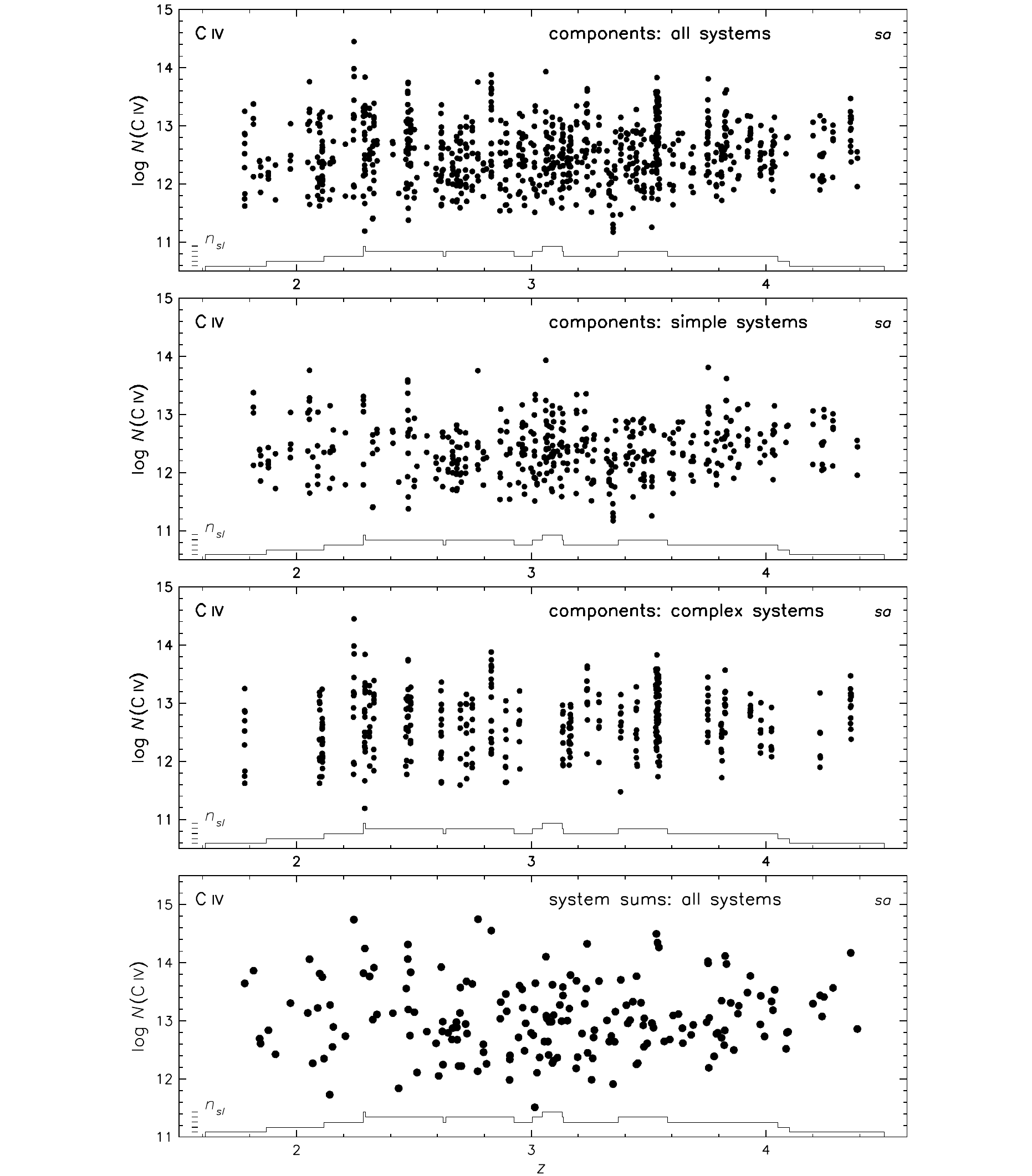}
\caption{\scriptsize Detected \CIV\ component and 
system column densities $N$ (cm$^{-2}$) in sample {\it sa} for systems with 
velocity $\gtrsim$ 3000 km s$^{-1}$ from the nominal redshift $z_{em}$ of 
their sightline QSO, plotted against redshift $z$. To avoid clutter, errors 
(see Tables 2--10) are not shown. The \emph{thin continuous histograms} 
display (in unit steps) the redshift distribution of the number of 
contributing sightlines ($n_{sl}$) from the nine QSOs of the sample within 
the applied redshift constraints. \emph{Upper three panels:} 
\emph{component} values in \emph{all} systems, \emph{simple} systems 
($\leqslant 6$ identified components) and \emph{complex} systems 
($\geqslant 7$ identified components). The vertical distributions manifest 
membership of systems. \emph{Bottom panel:} \emph{summed} column densities 
of the components within \emph{all} systems.} 
\end{figure*}

\clearpage
\begin{figure*}
\figurenum{\scriptsize 7(b)}
\epsscale{2.0}
\plotone{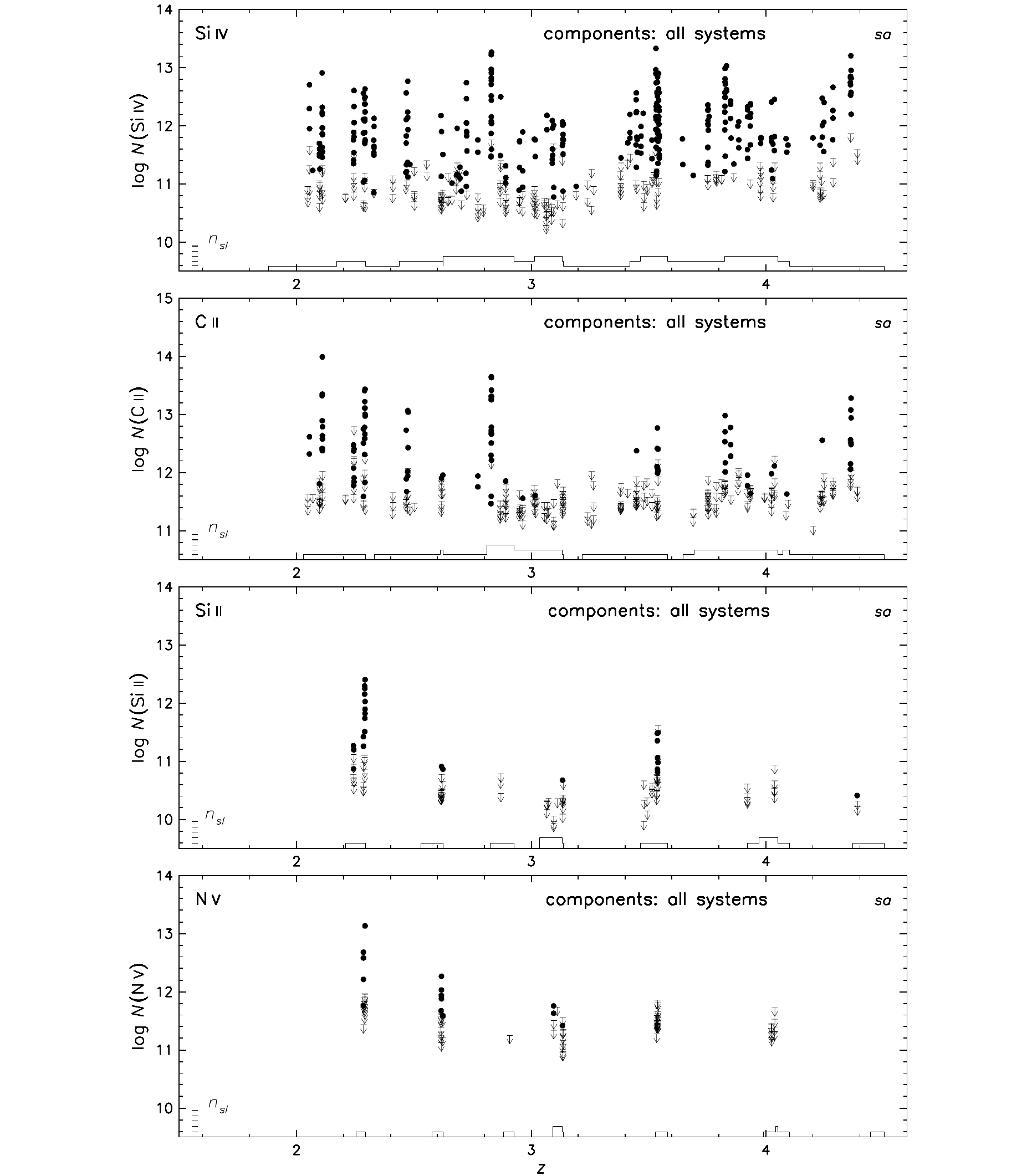}
\caption{\scriptsize \SiIV, \CII, \SiII\ and \NV\ 
\emph{component} column densities $N$ (cm$^{-2}$) for all \CIV\ detections 
in sample {\it sa}, following the style of Figure 7(a) but here for 
\emph{all} systems only; upper limits related to these are $1\sigma$ values.}
\end{figure*}

\clearpage
\begin{figure*}
\figurenum{\scriptsize 8}
\epsscale{2.0}
\plotone{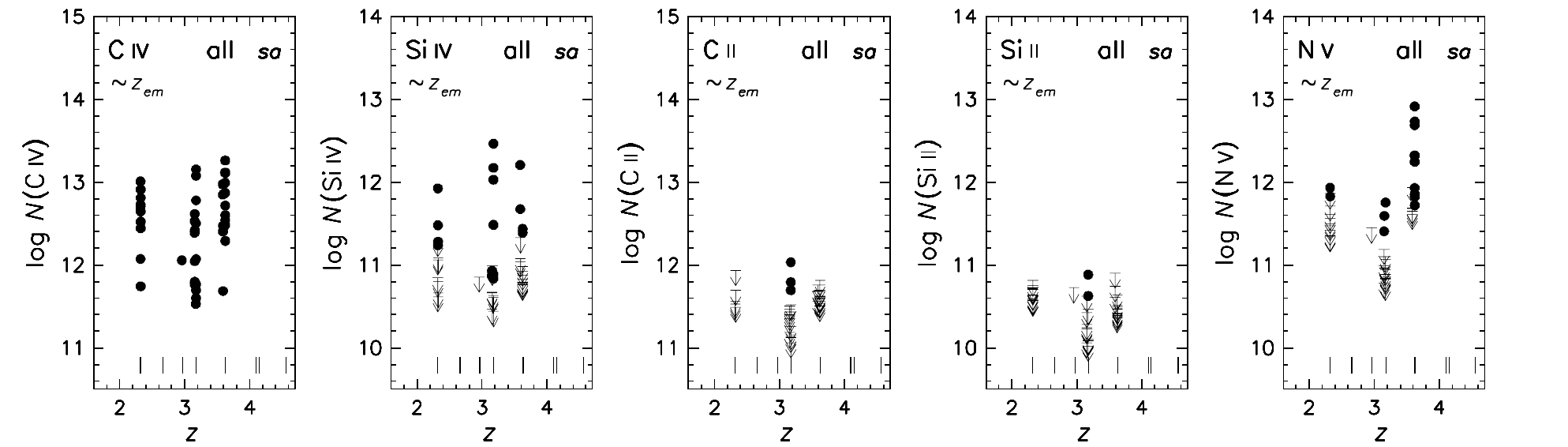}
\caption{\scriptsize \CIV, \SiIV, \CII, \SiII\ and 
\NV\ \emph{component} column densities $N$ (cm$^{-2}$) in sample {\it sa} 
for systems with velocity $\lesssim$ 3000 km s$^{-1}$ from the nominal 
redshift $z_{em}$ of their sightline QSO ($z_{em}$ for all nine QSOs are 
indicated by \emph{vertical ticks}).}
\end{figure*}

\clearpage
\begin{figure*}
\figurenum{\scriptsize 9(a)}
\epsscale{1.96}
\plotone{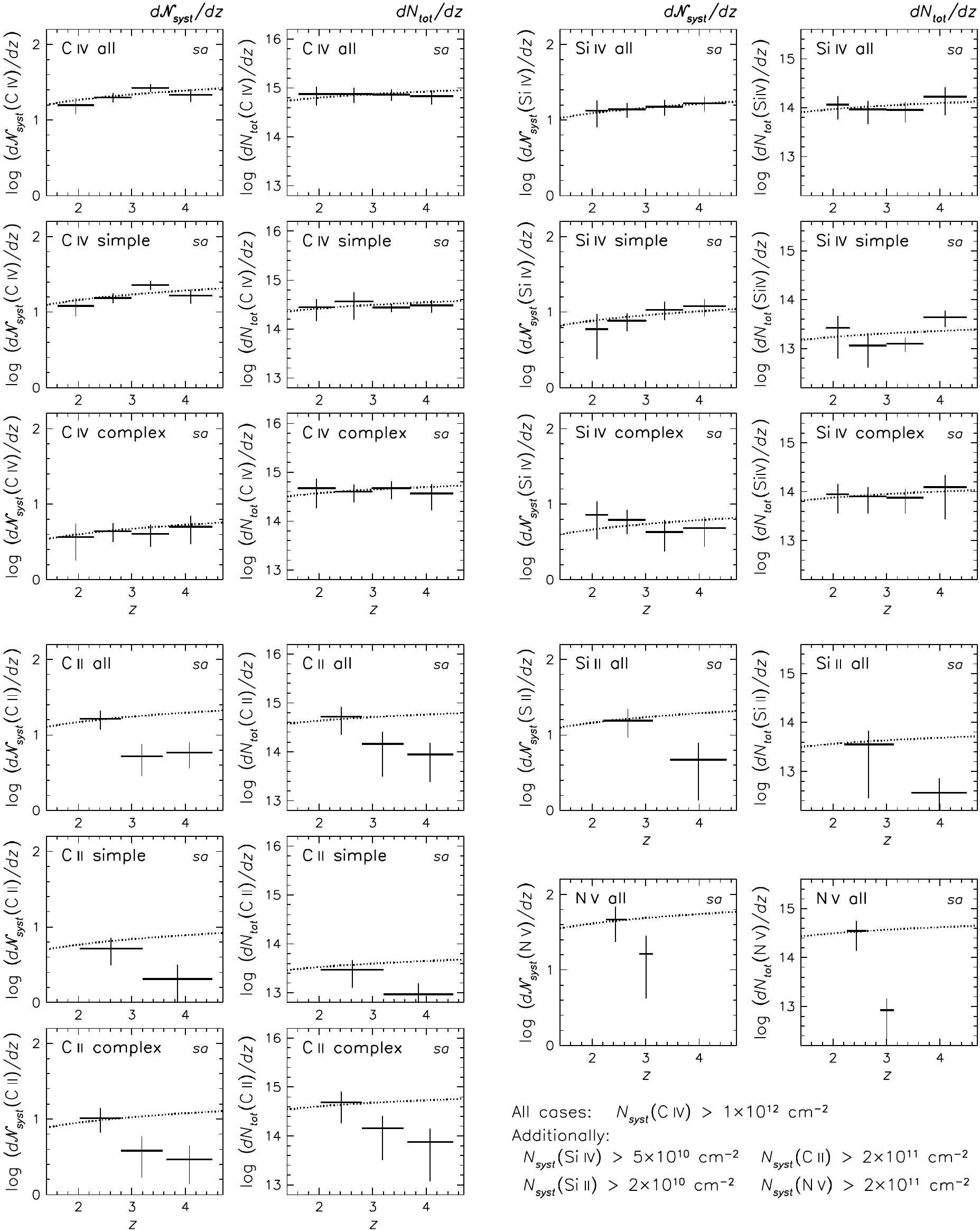}
\caption{\scriptsize \emph{Upper left panels:} Redshift 
evolution of \CIV\ total number of \emph{systems} per unit redshift interval, 
$d{\cal N}_{syst}/dz$, and total column density (cm$^{-2}$) per unit redshift 
interval, $dN_{tot}/dz$, accounted over the redshift range shown by each bar 
and including correction for multiple sightlines (indicated in Fig (a) and 
(b)), for systems in sample {\it sa} having 
$N_{syst}$(\CIV) $> 1 \times 10^{12}$ cm$^{-2}$, showing \emph{all} systems, 
\emph{simple} systems ($\leqslant 6$ \CIV\ components) and \emph{complex} 
systems ($\geqslant 7$ \CIV\ components); for errors see text.
\emph{Upper right and lower left panels:} Corresponding data for the same 
systems also having \emph{detected} components in \SiIV\ and \CII\ above the 
indicated thresholds. \emph{Lower right panels:} Similarly for \emph{all} 
systems having \emph{detected} components in \SiII\ and \NV. Note the various 
changes in vertical scale for $dN_{tot}/dz$. The \emph{dotted curves} 
indicate non-evolving quantities (see text) vertically scaled for ready 
comparison with the data.}
\end{figure*}

\clearpage
\begin{figure*}
\figurenum{\scriptsize 9(b)}
\epsscale{2.0}
\plotone{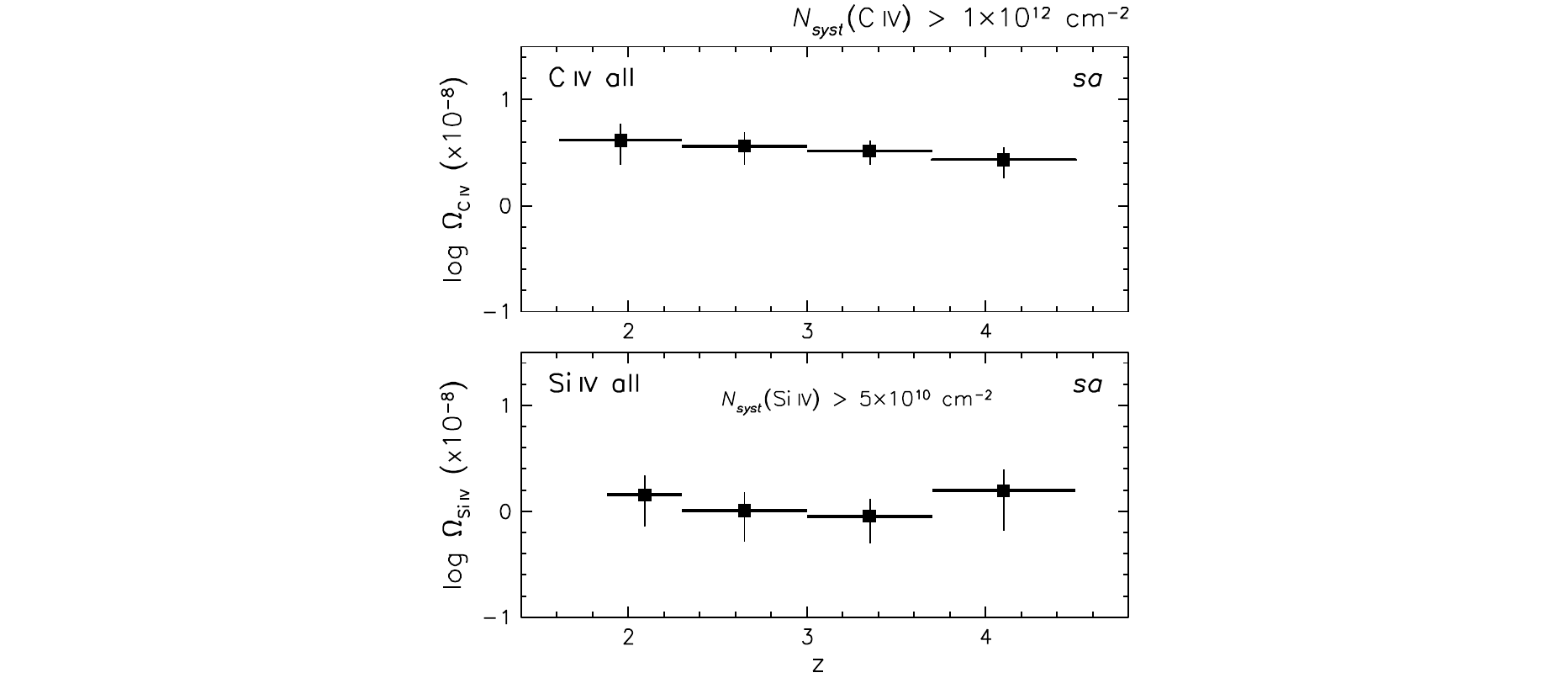}
\caption{\scriptsize Redshift evolution of \CIV\ and 
\SiIV\  cosmological mass densities for \emph{all} systems in sample {\it sa} 
having $N_{syst}$(\CIV) $> 1 \times 10^{12}$ cm$^{-2}$ and in the 
\emph{lower} panel \emph{additionally} 
$N_{syst}$(\SiIV) $> 5 \times 10^{10}$ cm$^{-2}$, from the 
corresponding data for $dN_{tot}/dz$ shown in the uppermost panels of 
Figure 9(a) and here similarly accounted over the redshift ranges indicated 
by the horizontal bars (see text).}
\end{figure*}

\clearpage
\begin{figure*}
\figurenum{\scriptsize 10}
\epsscale{2.0}
\plotone{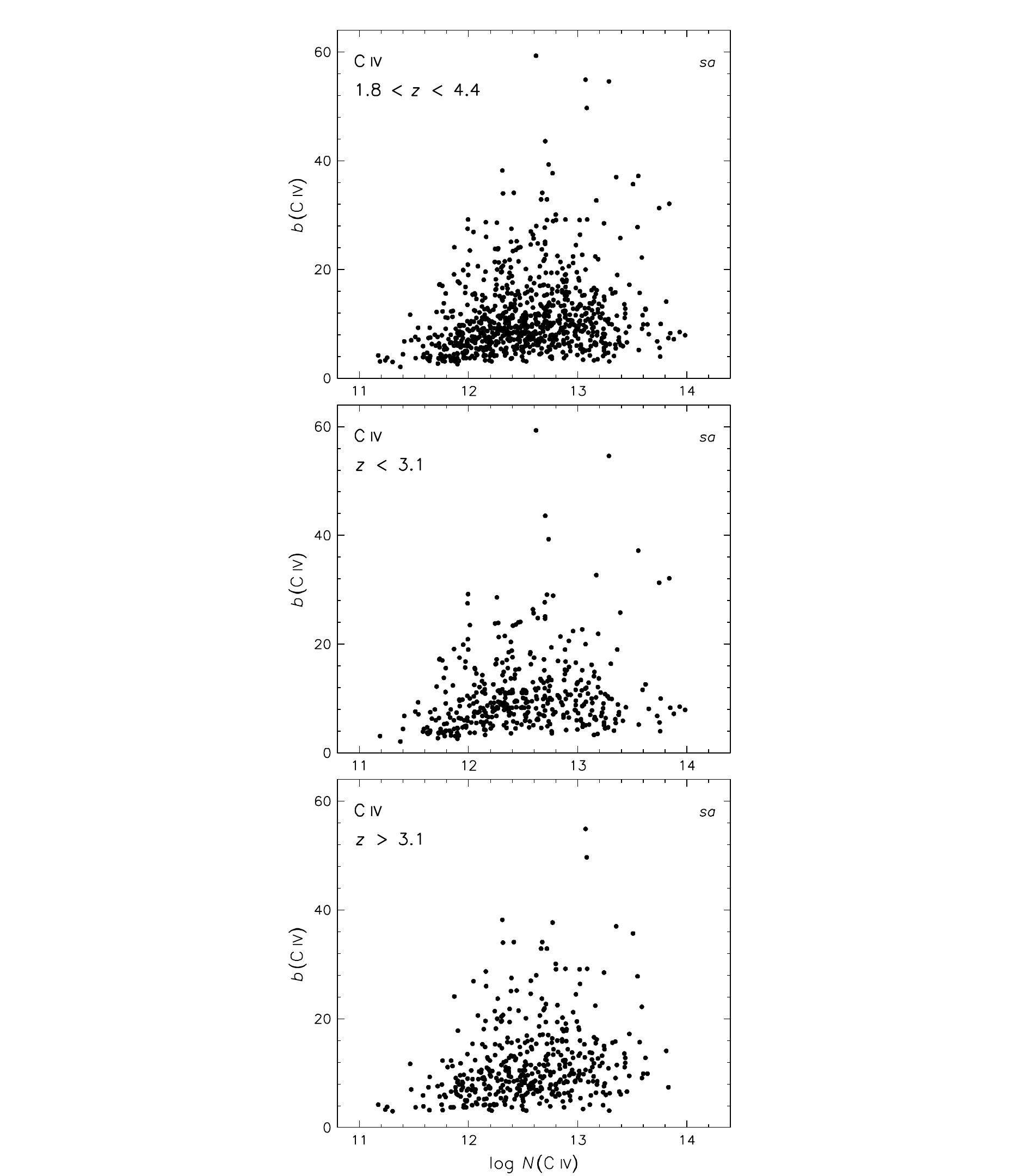}
\caption{\scriptsize \CIV\ Doppler parameter {\it b} 
(km s$^{-1}$) versus column density $N$ (cm$^{-2}$) for all detected  
\emph{components} of sample {\it sa}, also separately showing those with 
$z < 3.1$ and $z > 3.1$.}
\end{figure*}

\clearpage
\begin{figure*}
\figurenum{\scriptsize 11}
\epsscale{2.0}
\plotone{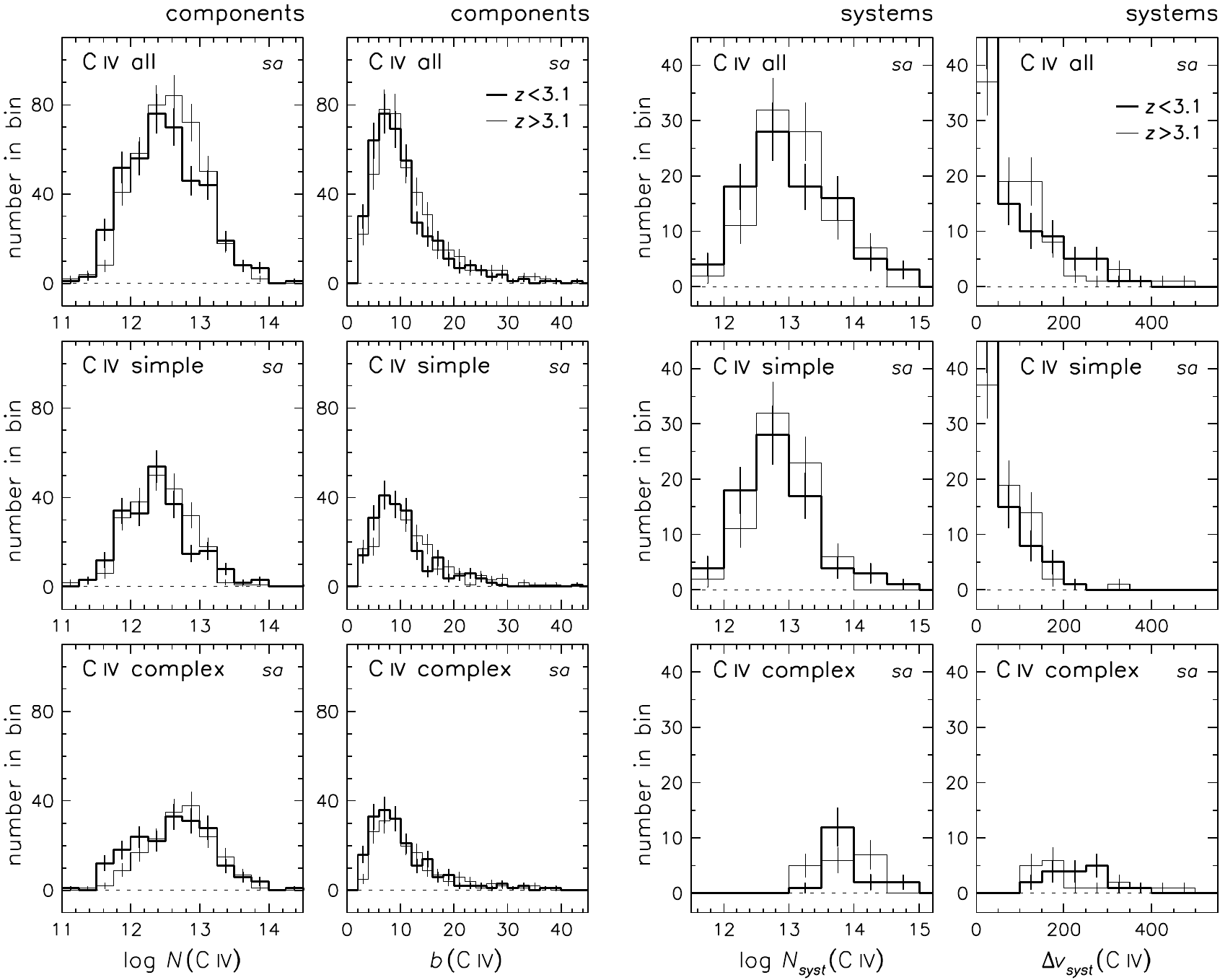}
\caption{\scriptsize \emph{Left panels:} Histograms of 
\CIV\ column density $N$ (cm$^{-2}$) and Doppler parameter 
{\it b} (km s$^{-1}$) for all detected \emph{components} of sample {\it sa}, 
comparing values for $z < 3.1$ (\emph{thick lines}) and $z > 3.1$ 
(\emph{thin lines}) and showing the data for \emph{all} systems, 
\emph{simple} systems ($\leqslant 6$ identified components) and 
\emph{complex} systems ($\geqslant 7$ identified components). \emph{Right 
panels:} Histograms, imitating \emph{left panels}, of \emph{system} summed 
\CIV\ column density $N_{syst}$ (cm$^{-2}$) and velocity spread of the 
components within a system $\Delta v_{syst}$ (km s$^{-1}$).}
\end{figure*}

\clearpage
\begin{figure*}
\figurenum{\scriptsize 12}
\epsscale{2.0}
\plotone{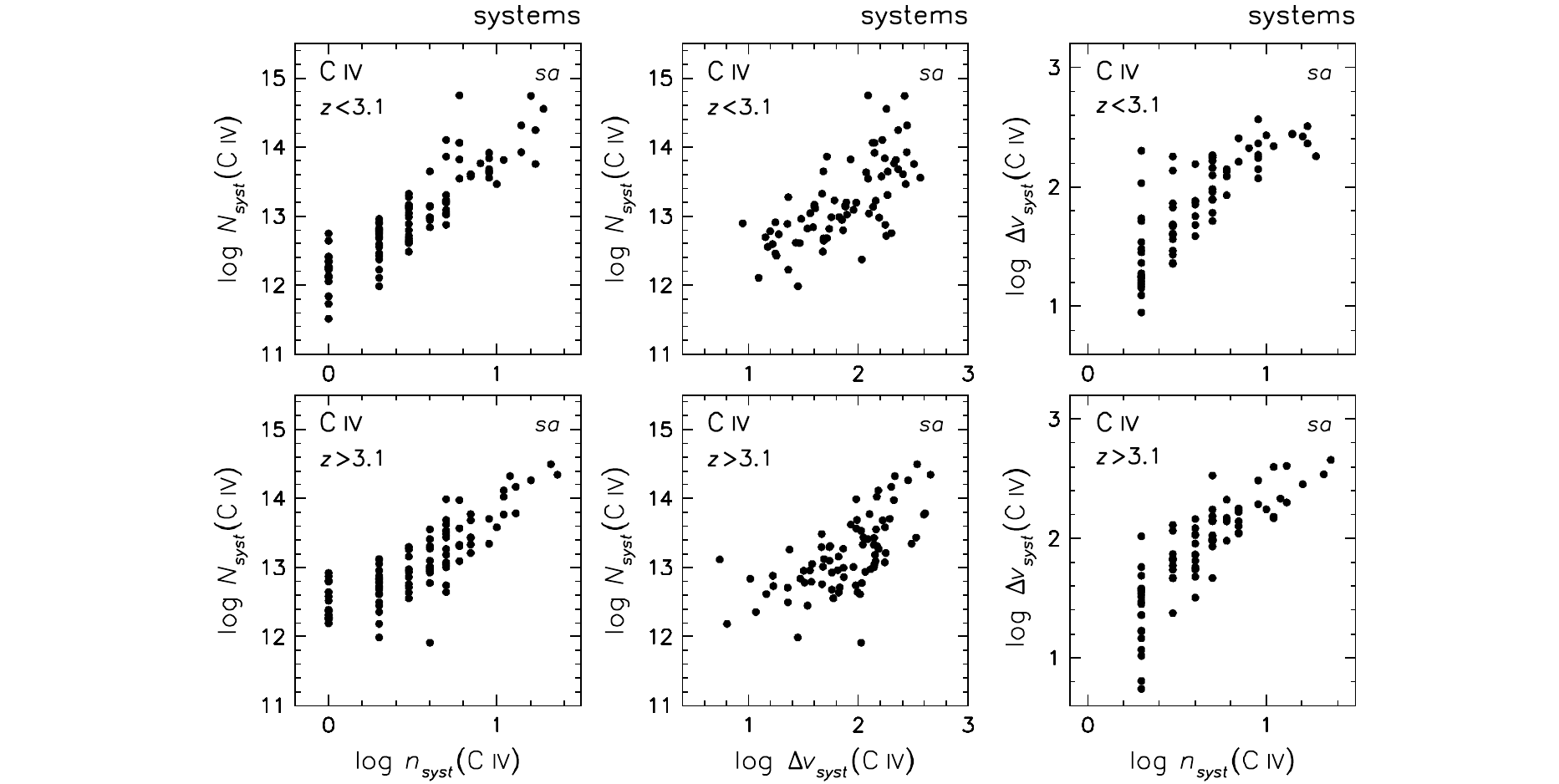}
\caption{\scriptsize Relationships for all \CIV\ 
\emph{systems} of sample {\it sa} at redshifts $z < 3.1$ and $z > 3.1$, 
showing: \emph{left panels}, system summed \CIV\ column density $N_{syst}$ 
(cm$^{-2}$) and number of components within a system $n_{syst}$; \emph{middle 
panels}, $N_{syst}$ and velocity spread of system components 
$\Delta v_{syst}$ (km s$^{-1}$); \emph{right panels}, $\Delta v_{syst}$ and 
$n_{\it syst}$.}
\end{figure*}

\clearpage
\begin{figure*}
\figurenum{\scriptsize 13}
\epsscale{2.0}
\plotone{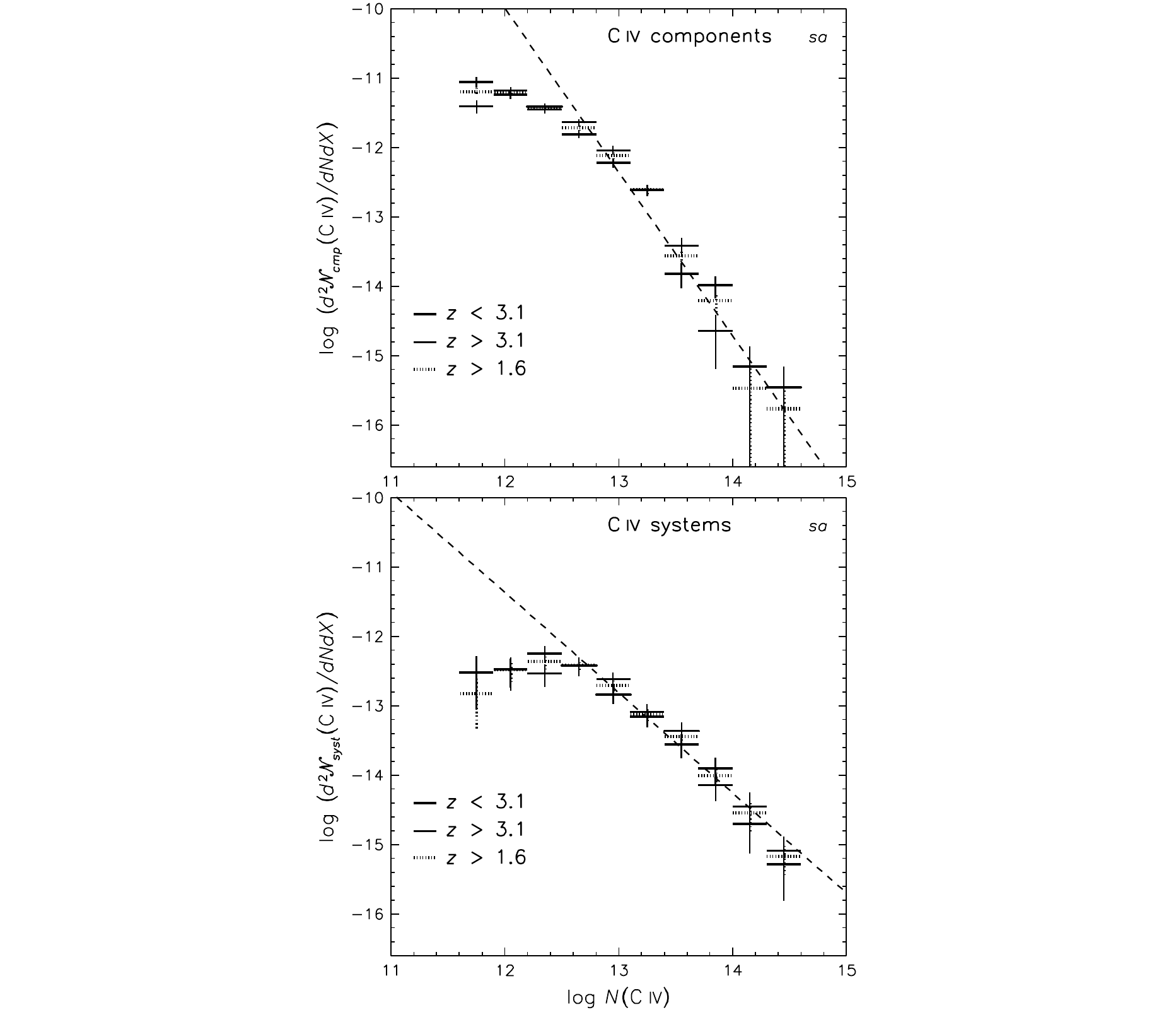}
\caption{\scriptsize Differential column density 
distribution of \CIV\ \emph{components} (\emph{upper}) and \emph{systems} 
(\emph{lower}) for sample {\it sa} covering redshifts $1.6 < z < 4.4$ and 
shown for ranges $z < 3.1$ (\emph{thick continuous lines}), $z > 3.1$ 
(\emph{thin continuous lines}) and the full extent $z > 1.6$ (\emph{dotted 
lines}). The bin size (shown by \emph{horizontal bars}) is 
$10^{0.3}N$ where $N$ is the column density (cm$^{-2}$); errors 
(\emph{vertical bars}) are $\pm1\sigma$ 
values based on the number of absorbers ${\cal N}$ in each bin. The few 
densely-solid looking horizontal bars are manifestations of closely similar 
results from the different redshift selections. The \emph{dashed lines} are 
power-law fits to the full data set as described in the text, with indices 
$\beta = 2.36\pm0.07$ (components) and $\beta = 1.68\pm0.09$ (systems).}
\end{figure*}

\clearpage
\begin{figure*}
\figurenum{\scriptsize 14}
\epsscale{2.0}
\plotone{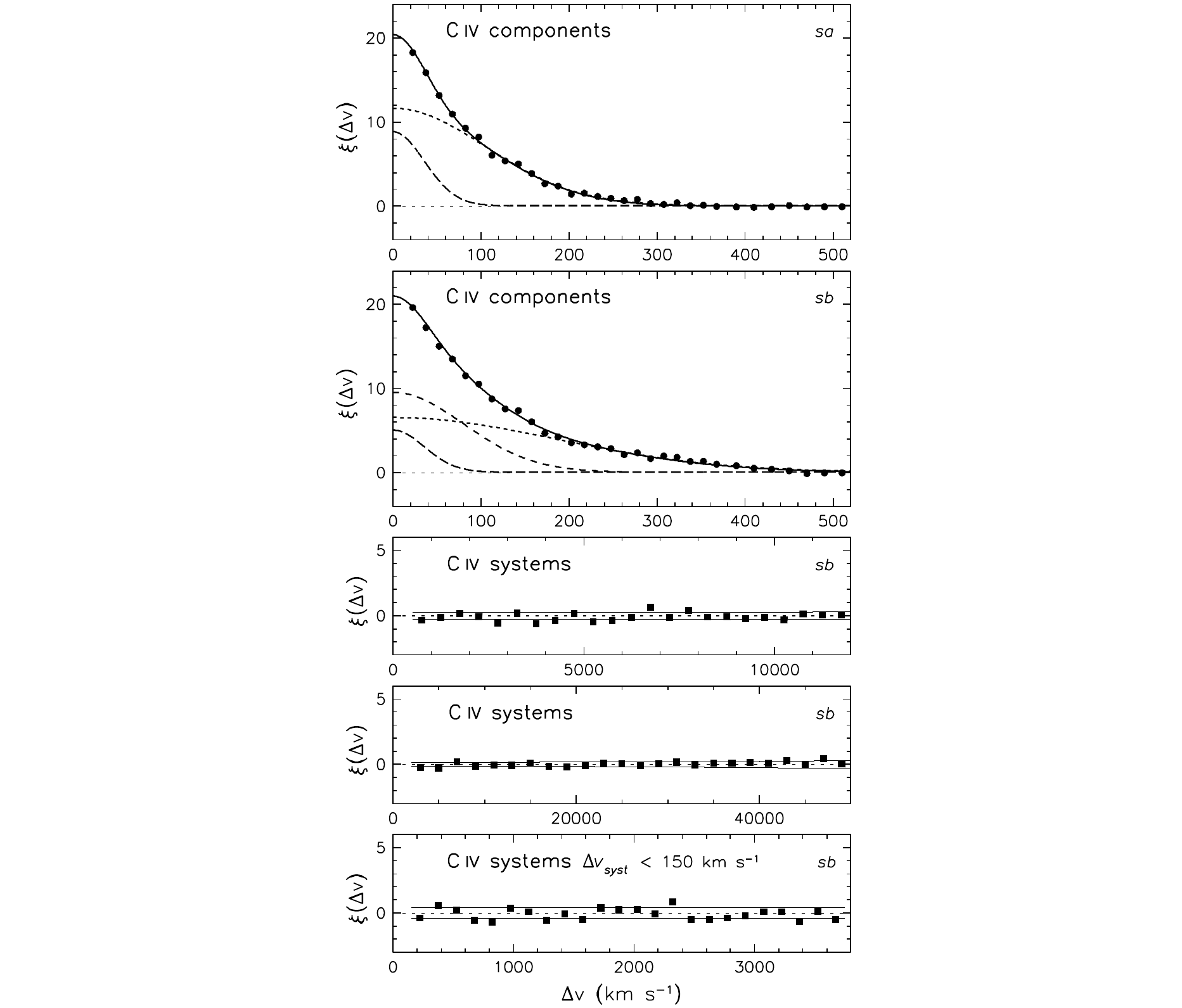}
\caption{\scriptsize Two-point correlation functions 
$\xi(\Delta v)$ versus velocity separations $\Delta v$ for \CIV\ absorber 
redshifts spanning $1.6 \lesssim z \lesssim 4.4$ and separated by 
$\gtrsim$ 3000 km s$^{-1}$ from the QSO nominal emission redshifts. 
\emph{First panel:} Points give $\xi(\Delta v)$ for the individual 
\emph{components} of sample {\it sa}, binned over 15 km s$^{-1}$ for 
$\Delta v \leqslant 370$ km s$^{-1}$ and 20 km s$^{-1}$ for 
$\Delta v \geqslant 370$; $\pm1\sigma$ errors in the random distribution are 
smaller than the symbol size. A two-component Gaussian fit and the separate 
components of the fit are shown with parameters as given in the text. 
\emph{Second panel:} Same as top panel, now including all components of the 
nine systems with significant Lyman~$\alpha$ damping wings, using sample 
{\it sb}; a three-component Gaussian fit is shown. 
\emph{Third panel:} Result for the \emph{system} redshifts of sample 
{\it sb}, binned over 500 km s$^{-1}$ and extending to 
$\Delta v = 12000$ km s$^{-1}$; $\pm1\sigma$ errors in the random 
distribution are shown by \emph{bounding thin lines}. 
\emph{Forth panel:} Same as third panel but binned over 2000 km s$^{-1}$ and 
extending to $\Delta v = 50000$ km s$^{-1}$. 
\emph{Fifth panel:} Same as third panel but using only systems having velocity 
spread $\Delta v_{syst} < 150$ km s$^{-1}$, binned over 150 km s$^{-1}$ and 
extending to $\Delta v = 3800$ km s$^{-1}$.}
\end{figure*}

\clearpage
\begin{figure*}
\figurenum{\scriptsize 15}
\epsscale{2.0}
\plotone{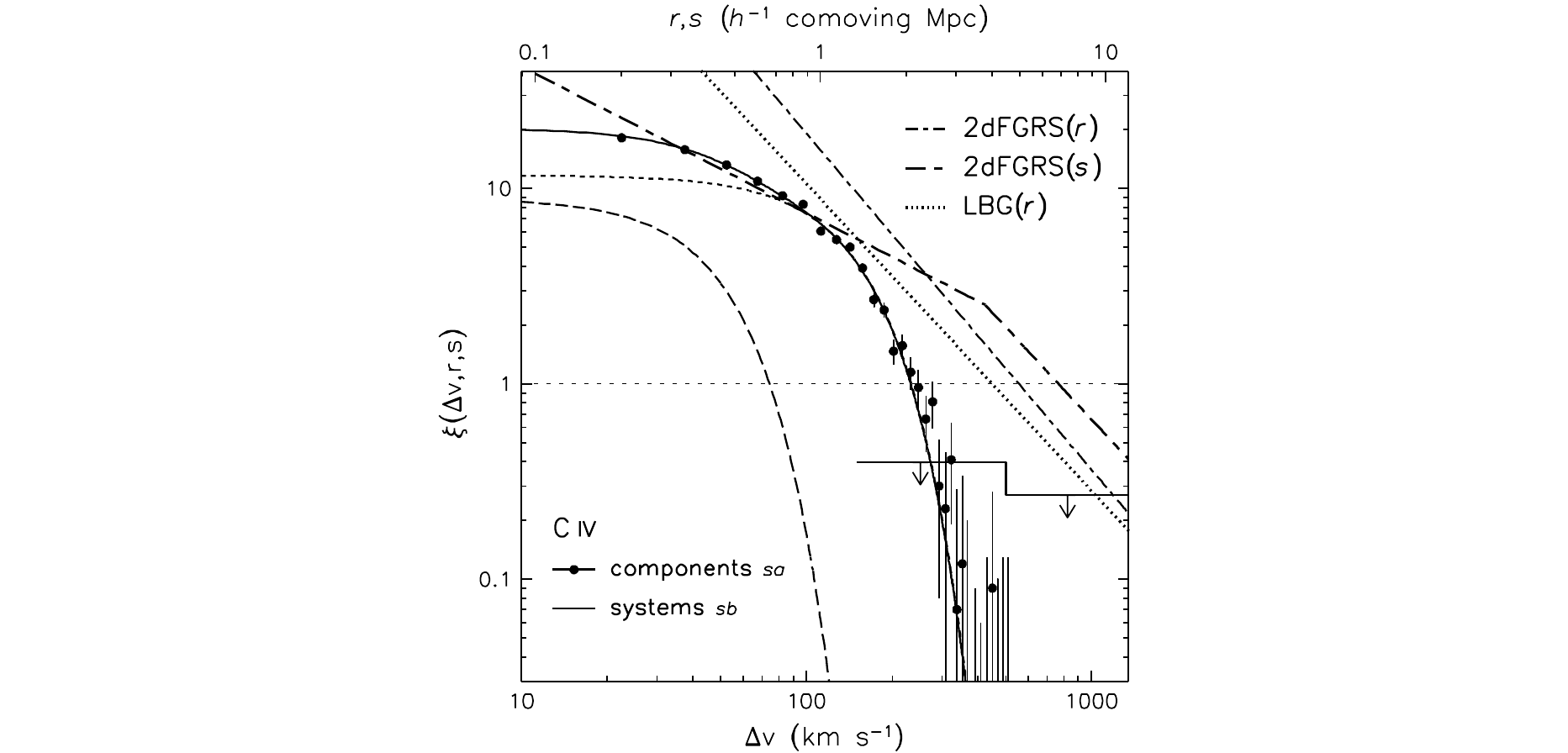}
\caption{\scriptsize Comparison of \CIV\ absorber and 
galaxy two-point correlation functions in logarithmic form. The \CIV\ sample 
{\it sa} data for individual \emph{components} are shown with the 
two-component Gaussian fit as in Figure 14; $\pm1\sigma$ errors in the random 
distributions are given with the data points. Sample {\it sb} data for 
\emph{systems} as described in the text are shown here as upper limits using 
$+1\sigma$ errors from Figure 14. Fits to results from the 2dF Galaxy 
Redshift Survey in real-space ($r$) and redshift-space ($s$) and in 
real-space for a sample of Lyman-break galaxies (LBG) use the upper axis, 
all as described in the text.}
\end{figure*}

\clearpage
\begin{figure*}
\figurenum{\scriptsize 16(a)}
\epsscale{2.0}
\plotone{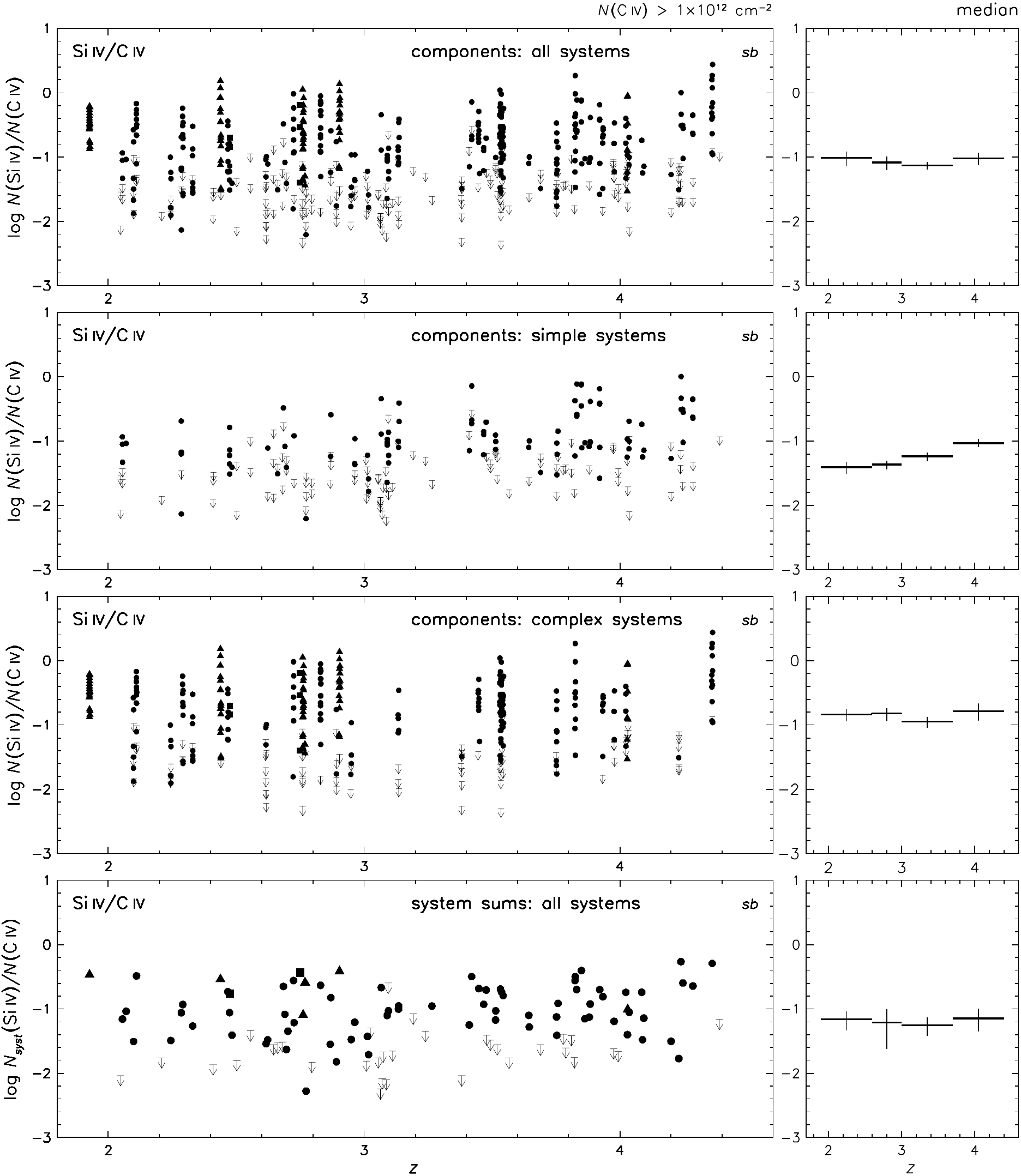}
\caption{\scriptsize \emph{Left panels}: Redshift 
evolution of \SiIV/\CIV\ column density ratios. Upper limits are $1\sigma$ 
values. The \emph{upper three panels} show individual \emph{component} 
values having $N$(\CIV) $> 1\times10^{12}$ cm$^{-2}$ from sample {\it sb} 
for \emph{all} systems, \emph{simple} systems ($\leqslant 6$ \CIV\ 
components) and \emph{complex} systems ($\geqslant 7$ \CIV\ components). 
\emph{Circles} show values obtained outside the Lyman forest; reliable 
values from lines in the forest are identified by \emph{squares}; selected 
components clear of regions of high $N$(\HI) in systems containing 
significant Lyman~$\alpha$ damping wings are shown by \emph{triangles}, and 
by \emph{diamonds} if also in the forest. The \emph{bottom panel} gives 
values obtained from \emph{summed} column densities for all available 
\emph{systems} in sample {\it sb} having 
$N_{syst}$(\CIV) $> 1\times10^{12}$ cm$^{-2}$. \emph{Right panels}: 
Redshift evolution of corresponding median values, obtained over the extent 
of each horizontal bar, indicated with $1\sigma$ error bars (see text).}
\end{figure*}

\clearpage
\begin{figure*}
\figurenum{\scriptsize 16(b)}
\epsscale{2.0}
\plotone{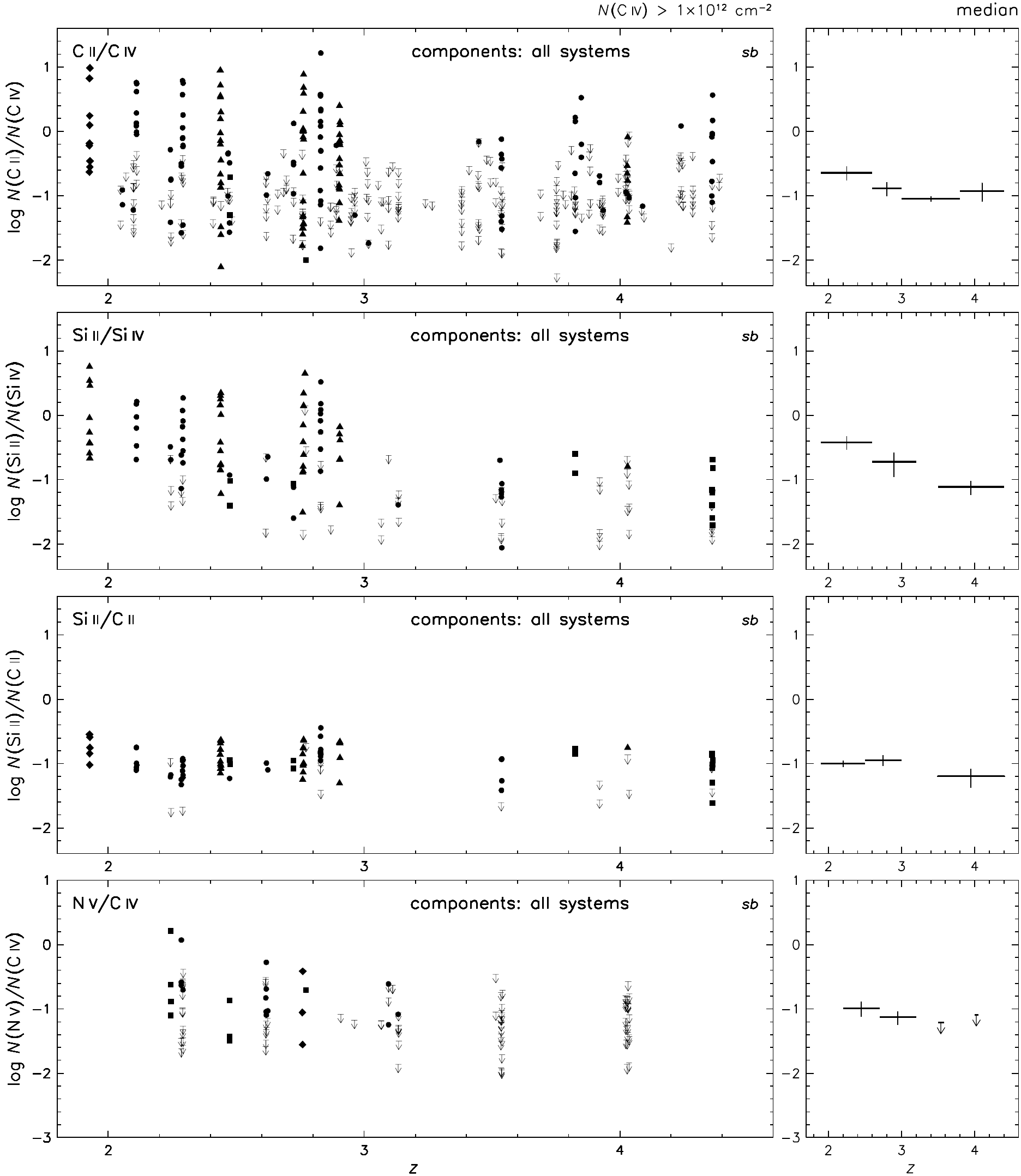}
\caption{\scriptsize Same as for the \emph{top} panels 
in Figure 16(a), here for the \emph{component} column density ratios: 
\CII/\CIV, \SiII/\SiIV\ (having \emph{detected} \SiIV\ with 
$N$(\SiIV) $> 5 \times 10^{10}$ cm$^{-2}$); 
\SiII/\CII\ (having \emph{detected} \CII\ with 
$N$(\CII) $> 5 \times 10^{11}$ cm$^{-2}$); \NV/\CIV. All cases arise from 
components having $N$(\CIV) $> 1\times10^{12}$ cm$^{-2}$.} 
\end{figure*}

\clearpage
\begin{figure*}
\figurenum{\scriptsize 17(a)}
\epsscale{1.92}
\plotone{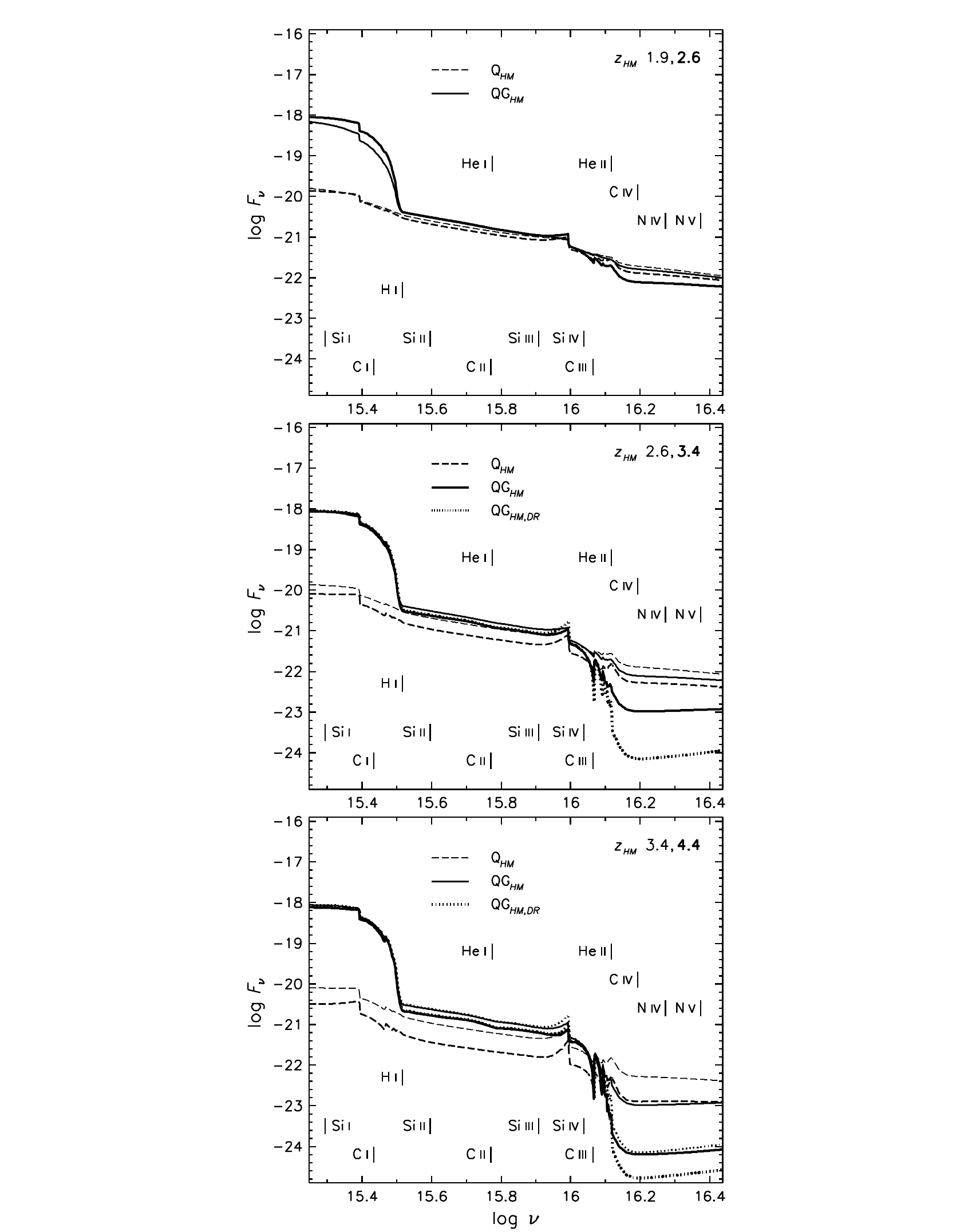}
\caption{\scriptsize Spectral energy distributions of 
the metagalactic ionizing radiation background \citep{ham12} plotted as flux 
$F_{\nu}$ erg cm$^{-2}$ s$^{-1}$ Hz$^{-1}$ (see text) against frequency 
$\nu$ Hz in the range effective for photoionization of the species studied 
in this paper (relevant ionization thresholds are shown), at bounding 
redshifts $z_{HM}$ of the inervals selected for the ionic displays in Figure 
17(b). Three spectral cases are included (see text): (i) partial model using 
QSO sources alone (Q$_{HM}$), in \emph{dashed lines}; (ii) full model with 
QSOs and star-forming galaxies (QG$_{HM}$), in \emph{continuous lines}; 
(iii) model simulating delayed reionization of \HeII\ (see text) at 
$z_{HM}$ = 3.4 and 4.4 (QG$_{HM,DR}$), in \emph{dotted lines}. Note the 
\emph{thin-thick} coded in correspondence with the indicated $z_{HM}$ 
boundaries.}
\end{figure*}

\clearpage
\begin{figure*}
\figurenum{\scriptsize 17(b)}
\epsscale{1.92}
\plotone{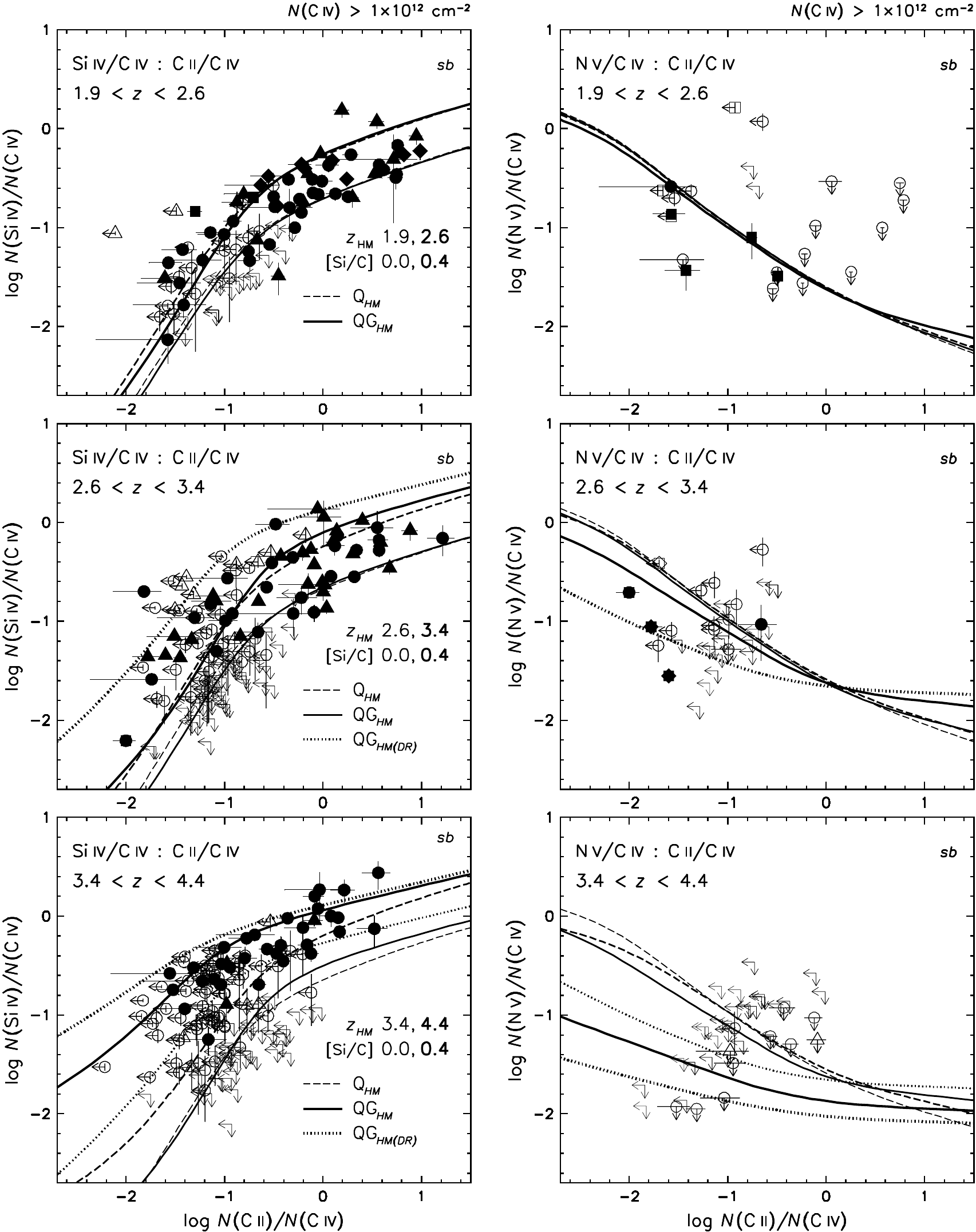}
\caption{\scriptsize Two-dimensional column density 
ratios \SiIV/\CIV~:~\CII/\CIV, \NV/\CIV~:~\CII/\CIV, 
\SiII/\SiIV~:~\CII/\CIV\ and \SiII/\CII~:~\CII/\CIV\ for all 
\emph{components} in sample {\it sb} having 
$N$(\CIV) $> 1\times10^{12}$ cm$^{-2}$. Data symbols are defined in Figure 
16(a). Error bars give $\pm1\sigma$ uncertainties. Upper limit arrows point 
from $+1\sigma$ values; cases where one of the two ionic ratios is an upper 
limit have an {\it open} symbol. The curves are model predictions of the 
\Cloudy\ code at the bounding redshifts $z_{HM}$ for the spectral models 
in Figure 17(a) (see text) with fiducial absorbers (see text) and 
[Si/C] = 0.0, 0.4 (see text) applied respectively at the lower and upper 
redshift for each pair containing \SiIV/\CIV\ or \SiII/\CII. Similarly, 
\NV/\CIV\ has [N/C] = 0.0, $-1.4$ (see text) but is shown only for the solar 
case; the progression down to [N/C] = $-1.4$ remains implied here and in all 
similar following figures. All curves are \emph{thin-thick} redshift coded
in correspondence with the $z_{HM}$ boundaries.} 
\end{figure*}

\clearpage
\begin{figure*}
\figurenum{\scriptsize 17(b)}
\epsscale{1.92}
\plotone{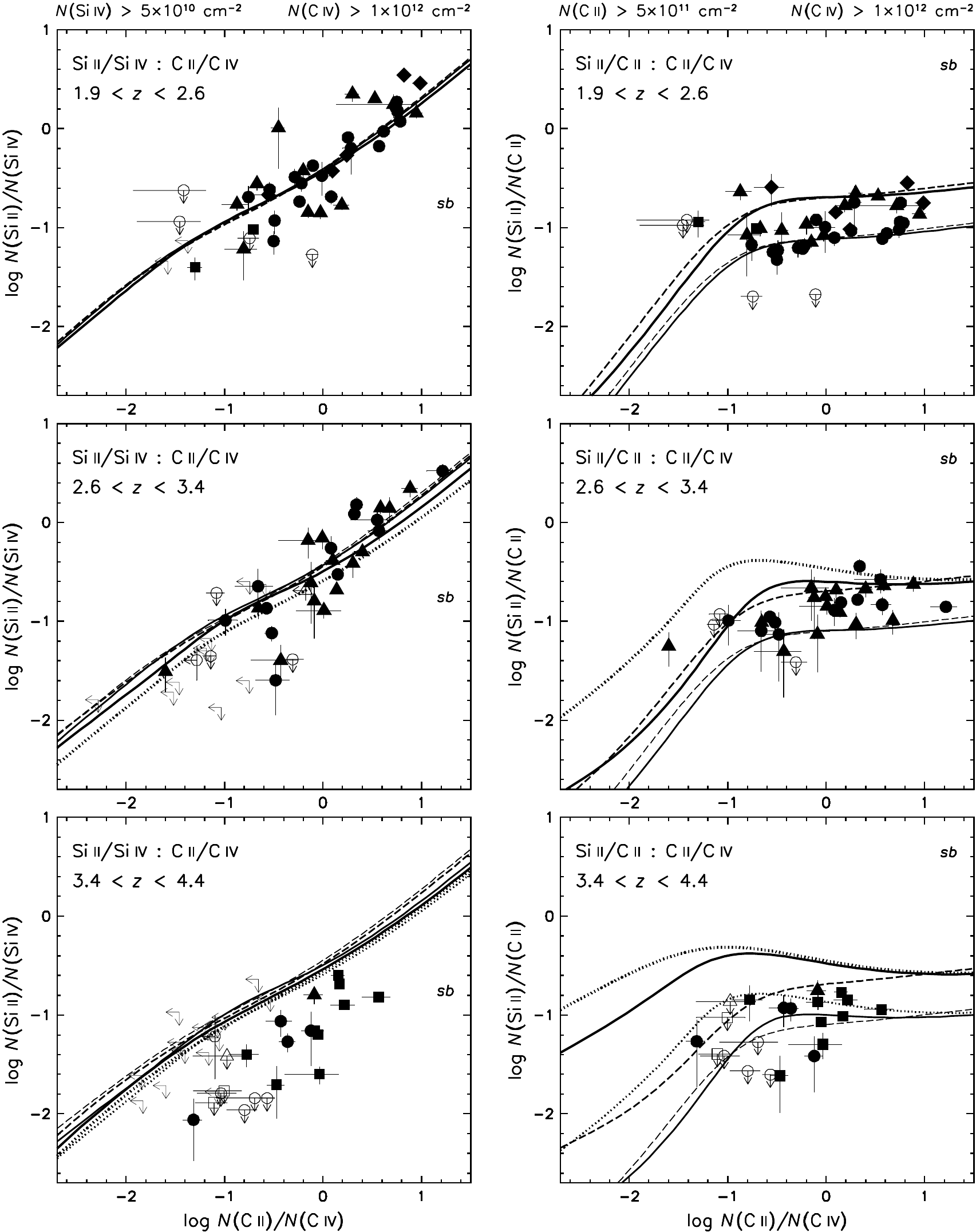}
\caption{\scriptsize Continued. Note the additional
imposed requirements of \emph{detected} 
$N$(\SiIV) $> 5\times10^{10}$ cm$^{-2}$ and 
$N$(\CII) $> 5\times10^{11}$ cm$^{-2}$ as respective bases for \SiII/\SiIV\ 
and \SiII/\CII.}
\end{figure*}

\clearpage
\begin{figure*}
\figurenum{\scriptsize 18}
\epsscale{2.0}
\plotone{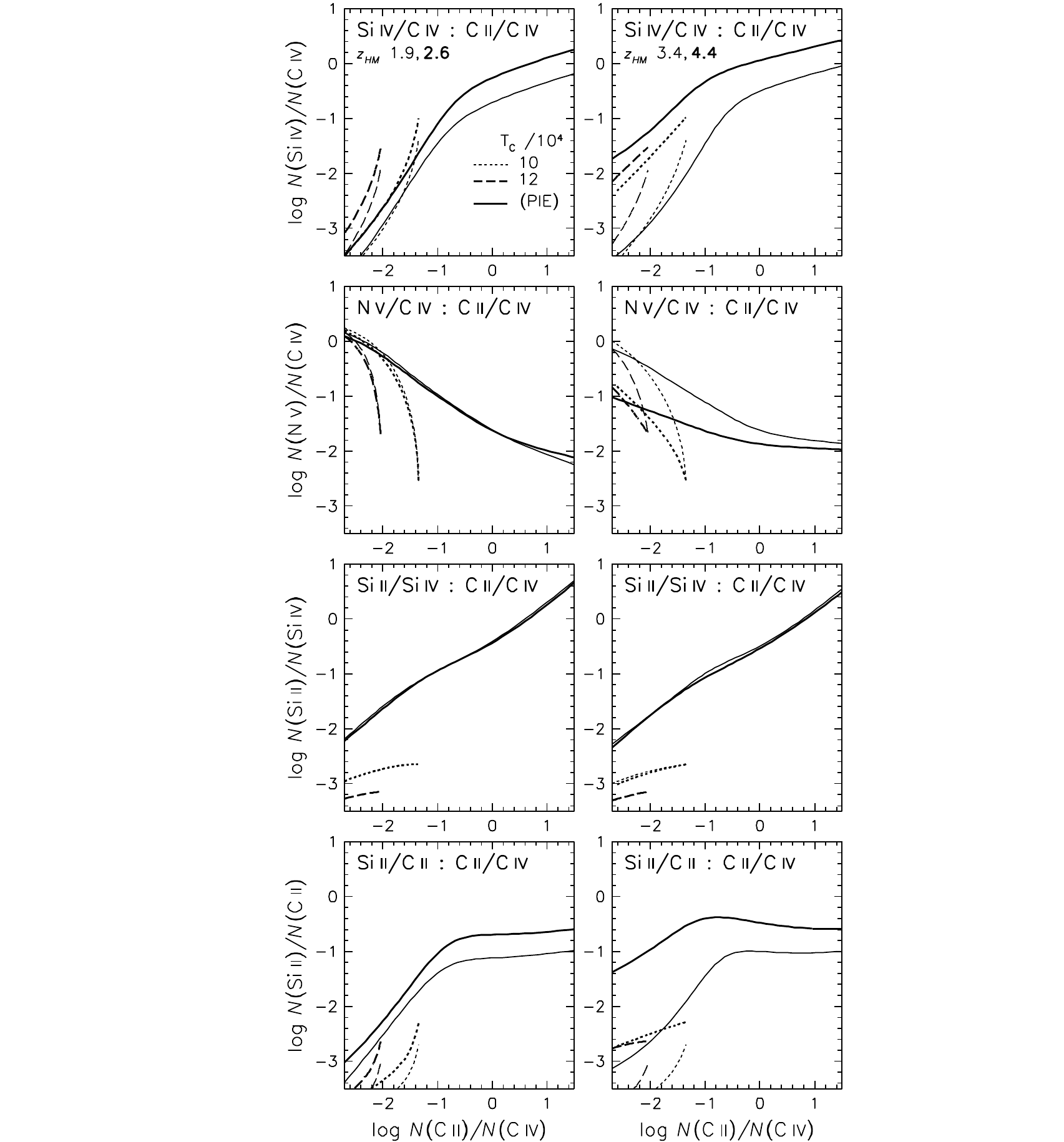}
\caption{\scriptsize \Cloudy\ models for 
\SiIV/\CIV~:~\CII/\CIV, \NV/\CIV~:~\CII/\CIV, \SiII/\SiIV~:~\CII/\CIV\ and 
\SiII/\CII~:~\CII/\CIV\ at the bounding redshifts $z_{HM} =$ 1.9, 2.6 and 
3.4, 4.4, following the style of the displays in the corresponding panels of
Figure 17(b) but with downward extended axes. The coded \emph{dashed lines} 
show results for collisional ionization equilibrium models at the two fixed 
temperatures near $10^5$ K as indicated, in the presence of the \citet{ham12} 
QG$_{HM}$ metagalactic ionizing radiation background and the cosmic microwave 
background (see text) and using the fiducial absorber properties including 
both values of [Si/C]) as in Figure 17(b); the curves terminate within each 
diagram where the process becomes independent of density. For comparison the 
corresponding models in pure photoionization equilibrium (PIE) from Figure 
17(b) are included separately in \emph{continuous lines}. For each case note 
the respective \emph{thin-thick} redshift coding of the model curves.}
\end{figure*}

\clearpage
\begin{figure*}
\figurenum{\scriptsize 19}
\epsscale{2.0}
\plotone{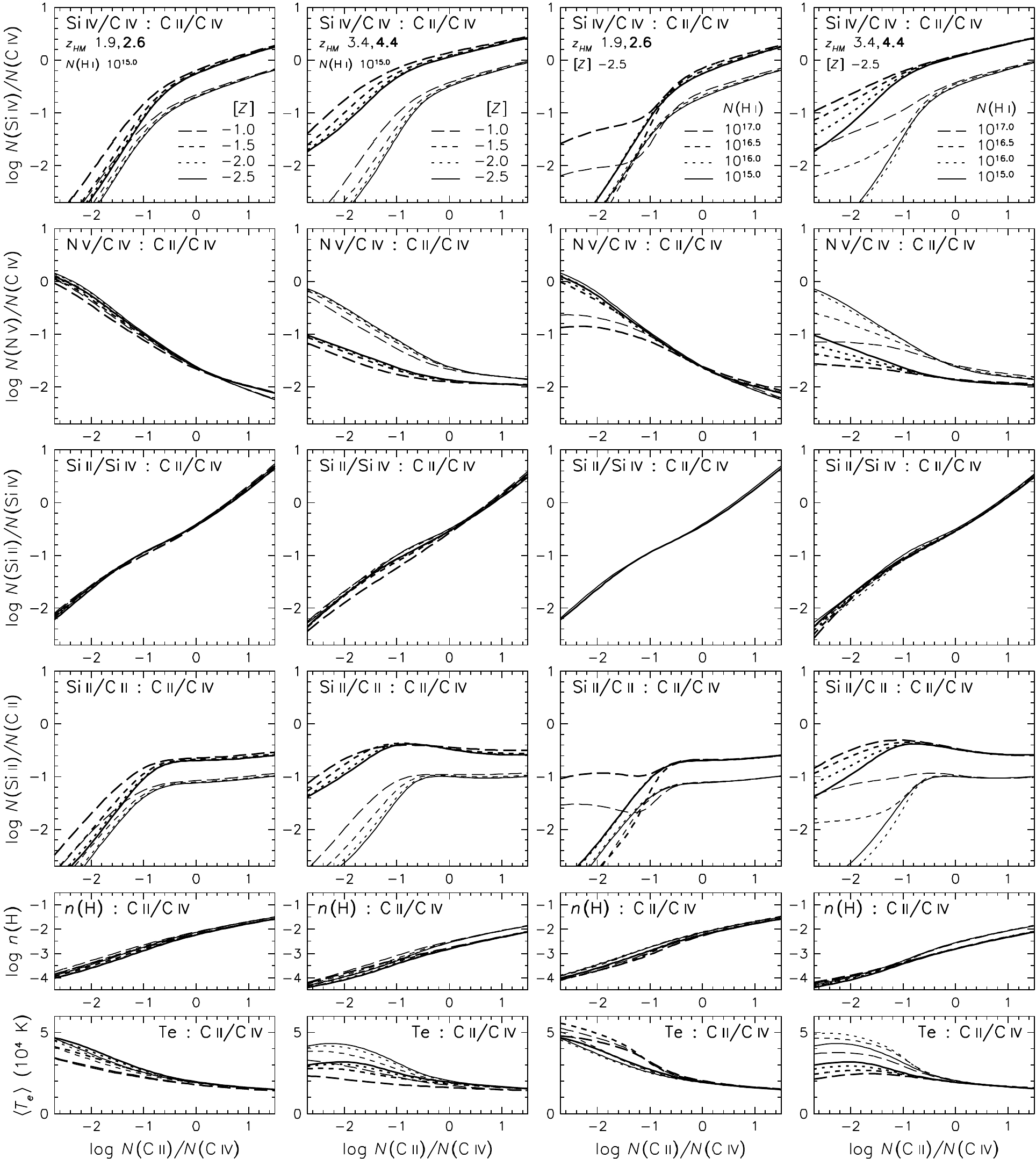}
\caption{\scriptsize \Cloudy\ models in photoionization 
equilibrium using the \citet{ham12} QG$_{HM}$ metagalactic ionizing radiation 
background and the cosmic microwave background (see text) for the bounding 
redshifts $z_{HM} =$ 1.9, 2.6 and 3.4, 4.4, following the corresponding 
displays in Figure 17(b) including both values of [Si/C], but here with 
absorber parameters arbitrarily differing in metallicity, $[Z]$, and neutral 
hydrogen column density, $N$(\HI) cm$^{-2}$, in coded \emph{dashed lines} as 
indicated; the fiducial case with $[Z] = -2.5$ and 
$N$(\HI) $=10^{15.0}$ cm$^{-2}$ used in Figure 17(b) is shown in all panels 
by \emph{continuous lines}. The apparent trends are discussed in the text. In 
the \emph{bottom two sets of  panels} are related displays for total hydrogen 
volume density, $n$(H), this indicating broad proportionality with \CII/\CIV, 
and mean temperature within the absorber gaseous column, 
$\langle T_e \rangle$. For each case note the respective \emph{thin-thick} 
redshift coding of the model curves.} 
\end{figure*}

\clearpage
\begin{figure*}
\figurenum{\scriptsize 20(a)}
\epsscale{1.92}
\plotone{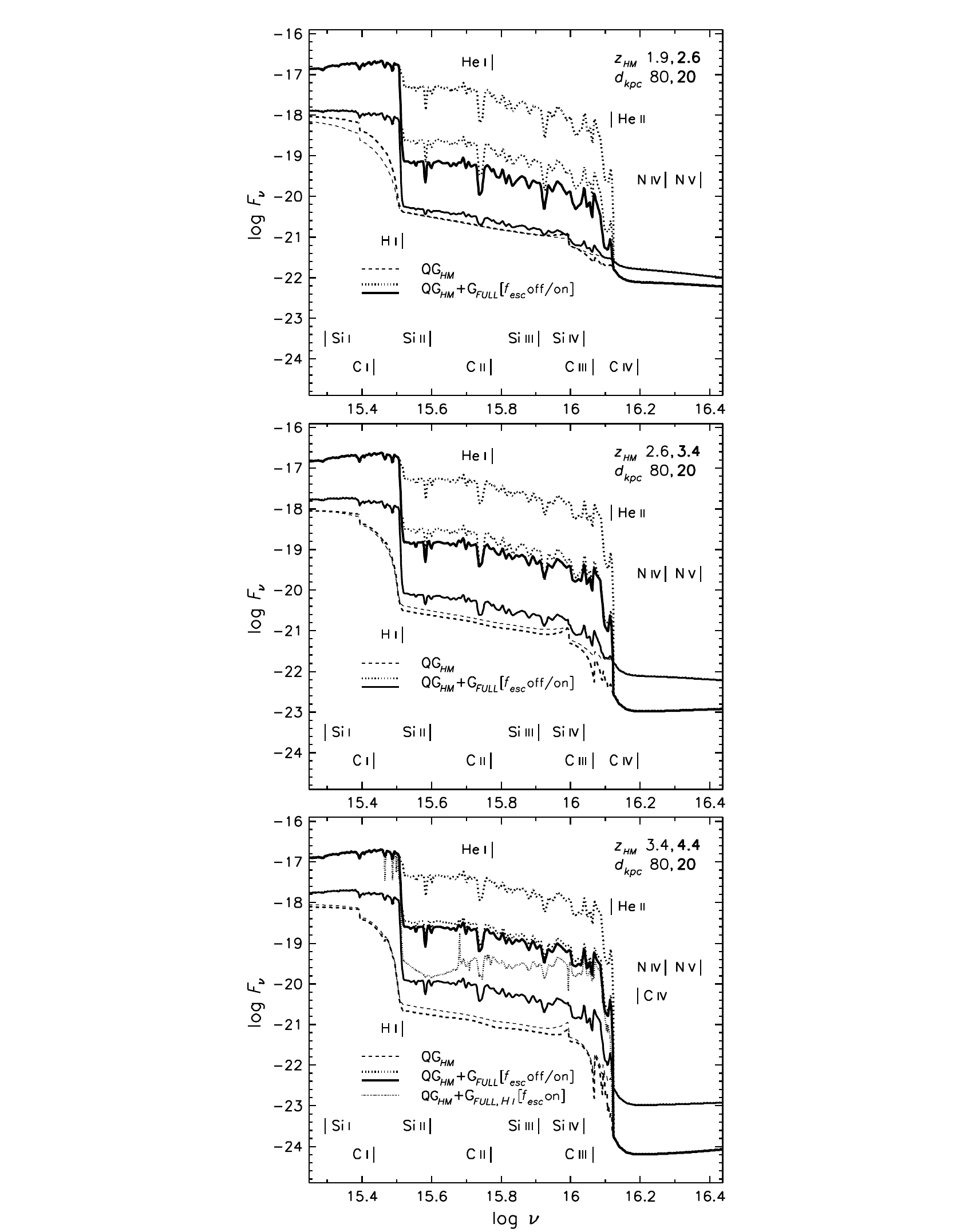}
\caption{\scriptsize Spectral energy distributions 
plotted as in Figure 17(a), here for the \citet{ham12} QG$_{HM}$ background 
augmented by radiative contributions from local template galaxies at 80 and 
20 physical kpc from the absorbers (QG$_{HM}$+G$_{FULL}$) (see text), in 
\emph{continuous lines}. For reference, corresponding cases having the 
galactic spectra without applied escape fraction are shown in \emph{thick 
dotted lines}, and the basic QG$_{HM}$ background from Figure 17(a), in 
\emph{dashed lines}. All are \emph{thin-thick} redshift-coded in 
correspondence with the $z_{HM}$ boundaries. Included in the \emph{bottom 
panel} at $z_{HM}$ = 4.4 is a spectrum with the local galaxy contribution 
G$_{FULL}$ attenuated by intrinsic \HI\ having 
$N$(\HI) $=10^{18}$ cm$^{-2}$ (QG$_{HM}$+G$_{FULL,HI}$), in \emph{fine 
dotted line}.}
\end{figure*}

\clearpage
\begin{figure*}
\figurenum{\scriptsize 20(b)}
\epsscale{1.92}
\plotone{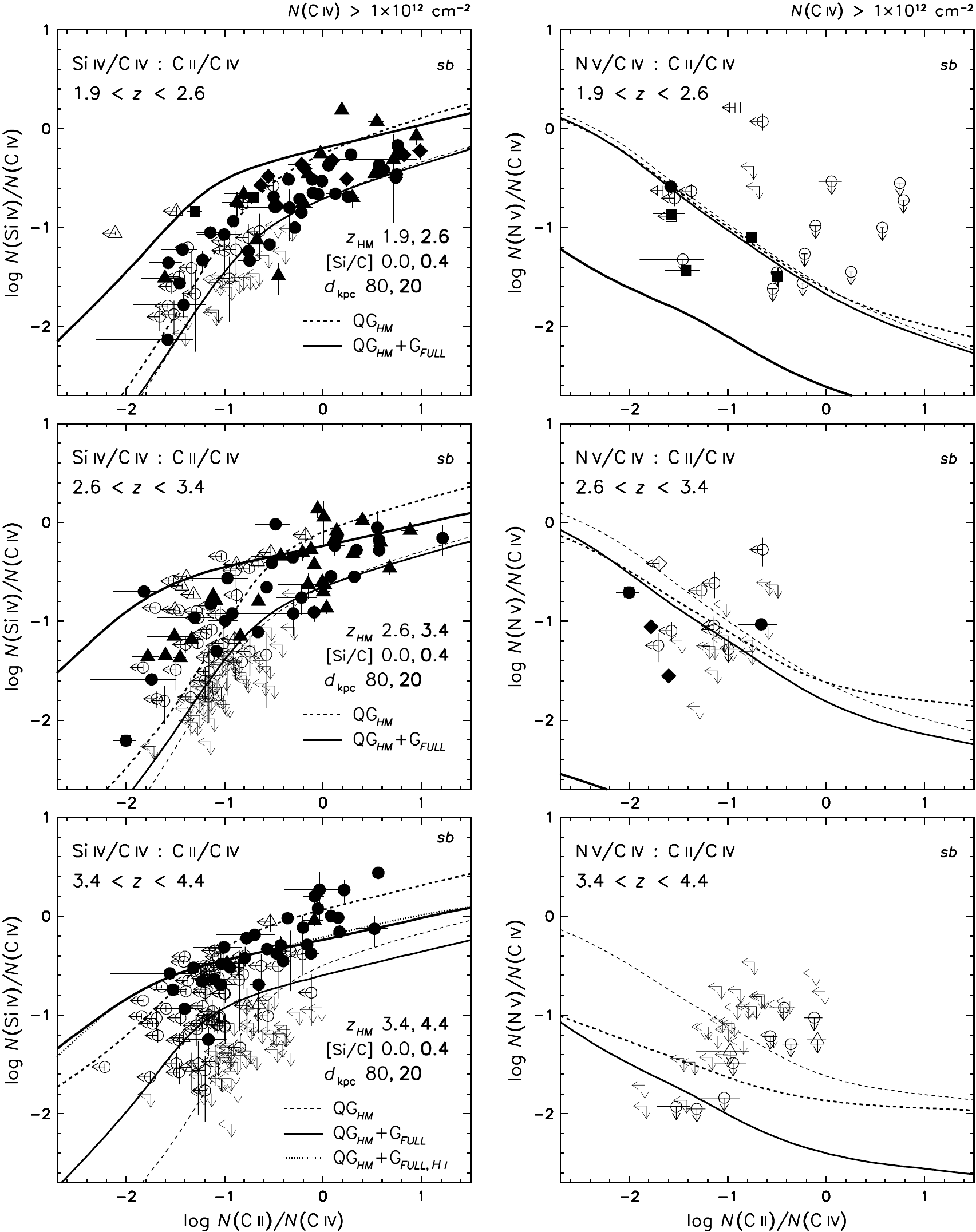}
\caption{\scriptsize \Cloudy\ models of 
two-dimensional column density ratios at the bounding redshifts $z_{HM}$ 
for the radiative cases in Figure 20(a) and similarly coded, with fiducial 
absorbers (see text), and associated data points, all plotted as in Figure 
17(b) and described in its caption. The basic \citet{ham12} QG$_{HM}$ 
background from Figure 17(b) is shown in \emph{thin dashed lines}. All in 
the figure except the single case QG$_{HM}$+G$_{FULL,HI}$ are 
\emph{thin-thick} redshift-coded in correspondence with the $z_{HM}$ 
boundaries.}
\end{figure*}

\clearpage
\begin{figure*}
\figurenum{\scriptsize 20(b)}
\epsscale{1.92}
\plotone{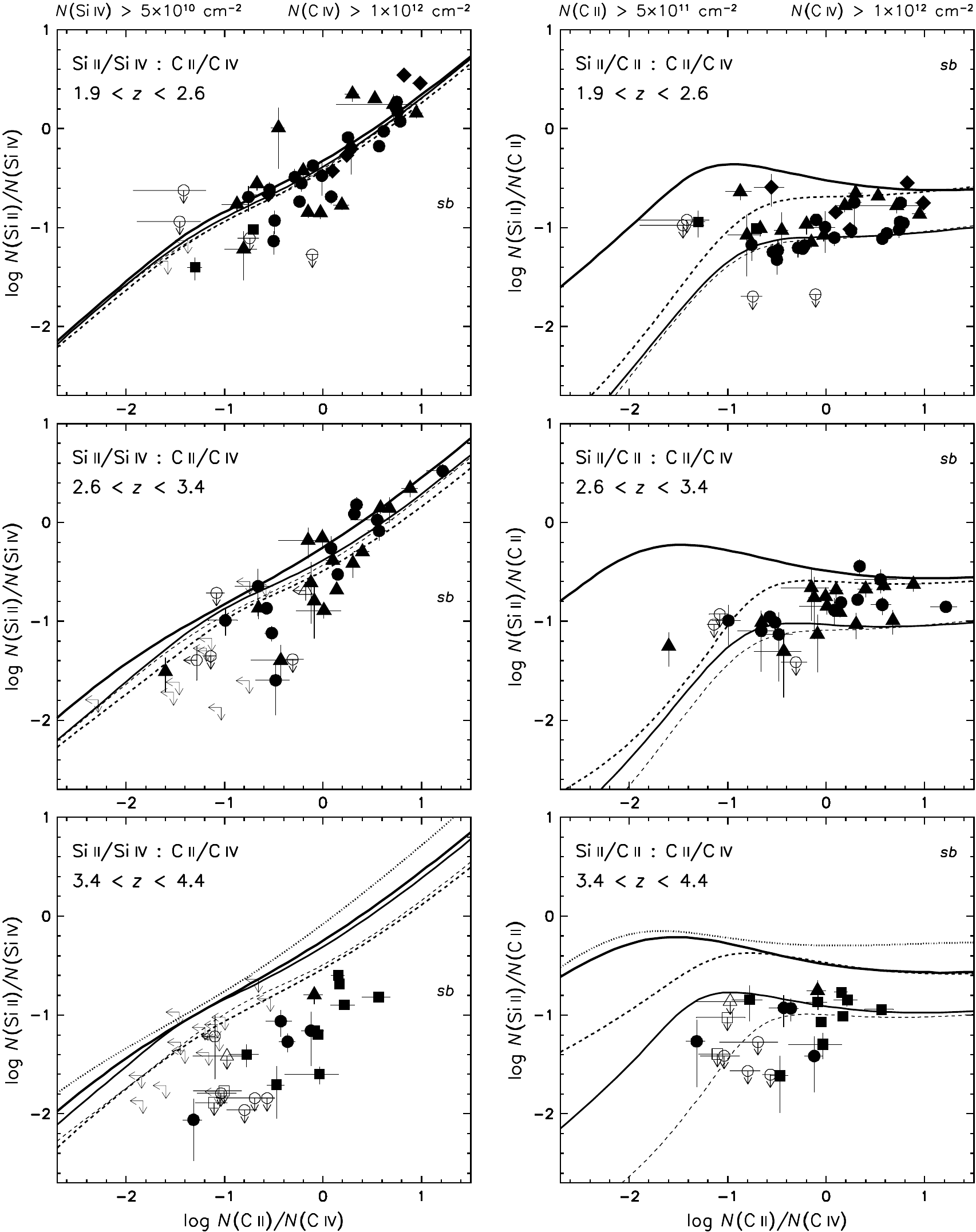}
\caption{\scriptsize Continued. Note the caption to
the equivalent part of Figure 17(b).} 
\end{figure*}

\clearpage
\begin{figure*}
\figurenum{\scriptsize 21}
\epsscale{2.0}
\plotone{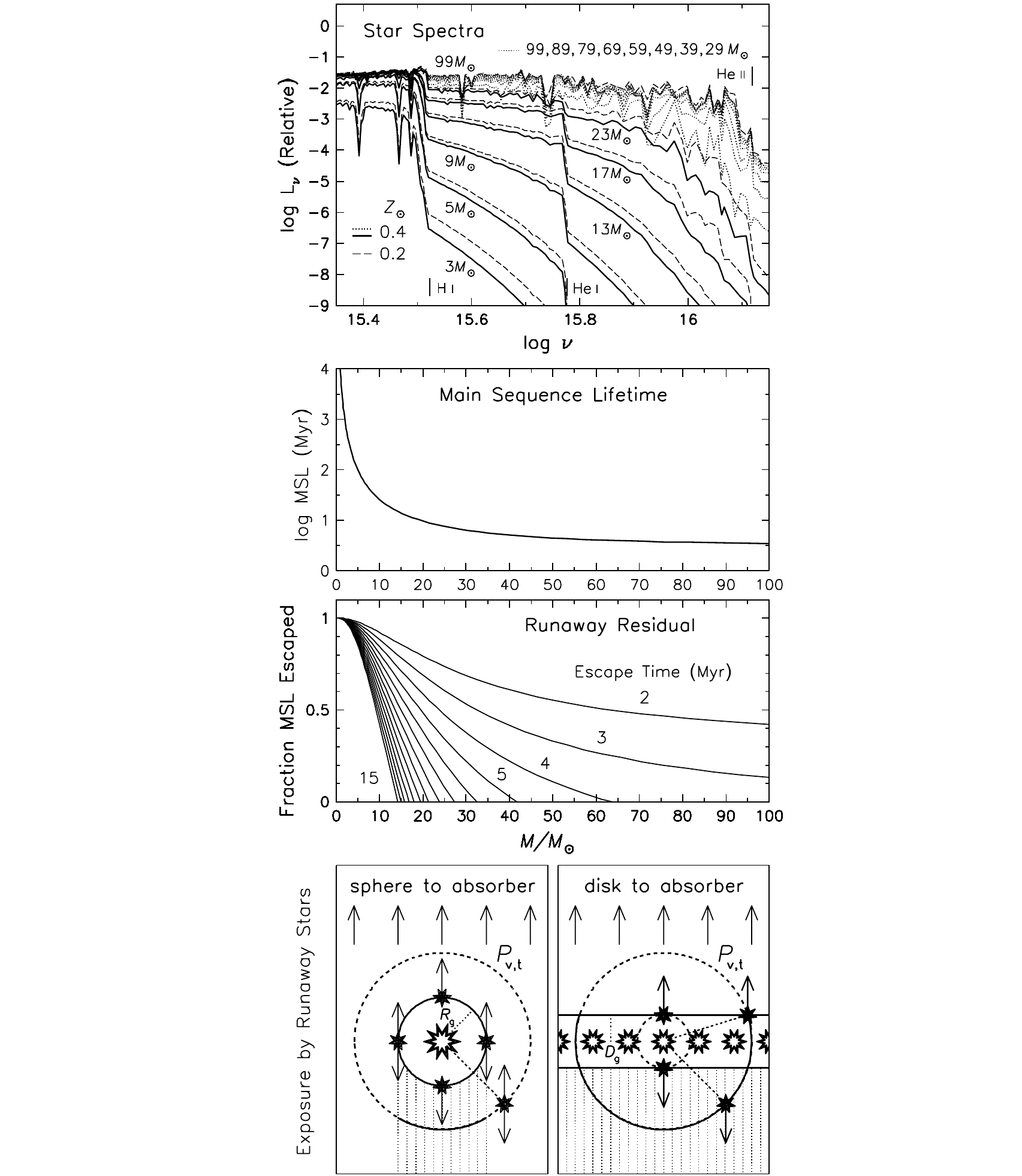}
\caption{\scriptsize \emph{First panel:} Intrinsic 
spectra of component stellar 2 $M_{\Sun}$ slices selected from the sets of 
metallicity 0.2 and 0.4 $Z_{\Sun}$ obtained using Starburst99 as described in 
the text, here highlighting the region containing the \HI, \HeI\ and \HeII\ 
ionization thresholds (marked). The pairs of components labeled 99 and over 
3--23 $M_{\Sun}$ indicate the relatively small spectral differences beween 
the two metallicities (the rest shown here are of the higher metallicity only). 
\emph{Second panel:} Stellar main sequence lifetime (MSL) shown against mass 
($M$/$M_{\Sun}$) for metallicity 0.3 $Z_{\Sun}$, from Hurley at al. (2000). 
\emph{Third panel:} For runaway stars over the mass range and a set of trial 
\emph{potential} intervals for escape from obscuration of ionizing radiation, 
showing the fraction of MSL remaining after reaching the effective boundary of 
the galaxy. \emph{Fourth panel:} Depictions of a central section through a 
spheroidal galaxy and a central vertical section through a disk-like galaxy, 
indicating compact star-forming regions. Absorptive boundaries are shown 
respectively by a \emph{continuous circle} and \emph{continuous horizontal 
lines}. Scenarios for illumination of a relatively distant absorber by runaway 
stars are described in the text. Circles of \emph{partially dashed and 
continuous lines} indicate positions of stars at a time after ejection at a 
given velocity (see text). Arrows indicate specific directions of radiation.}
\end{figure*}

\clearpage
\begin{figure*}
\figurenum{\scriptsize 22}
\epsscale{2.0}
\plotone{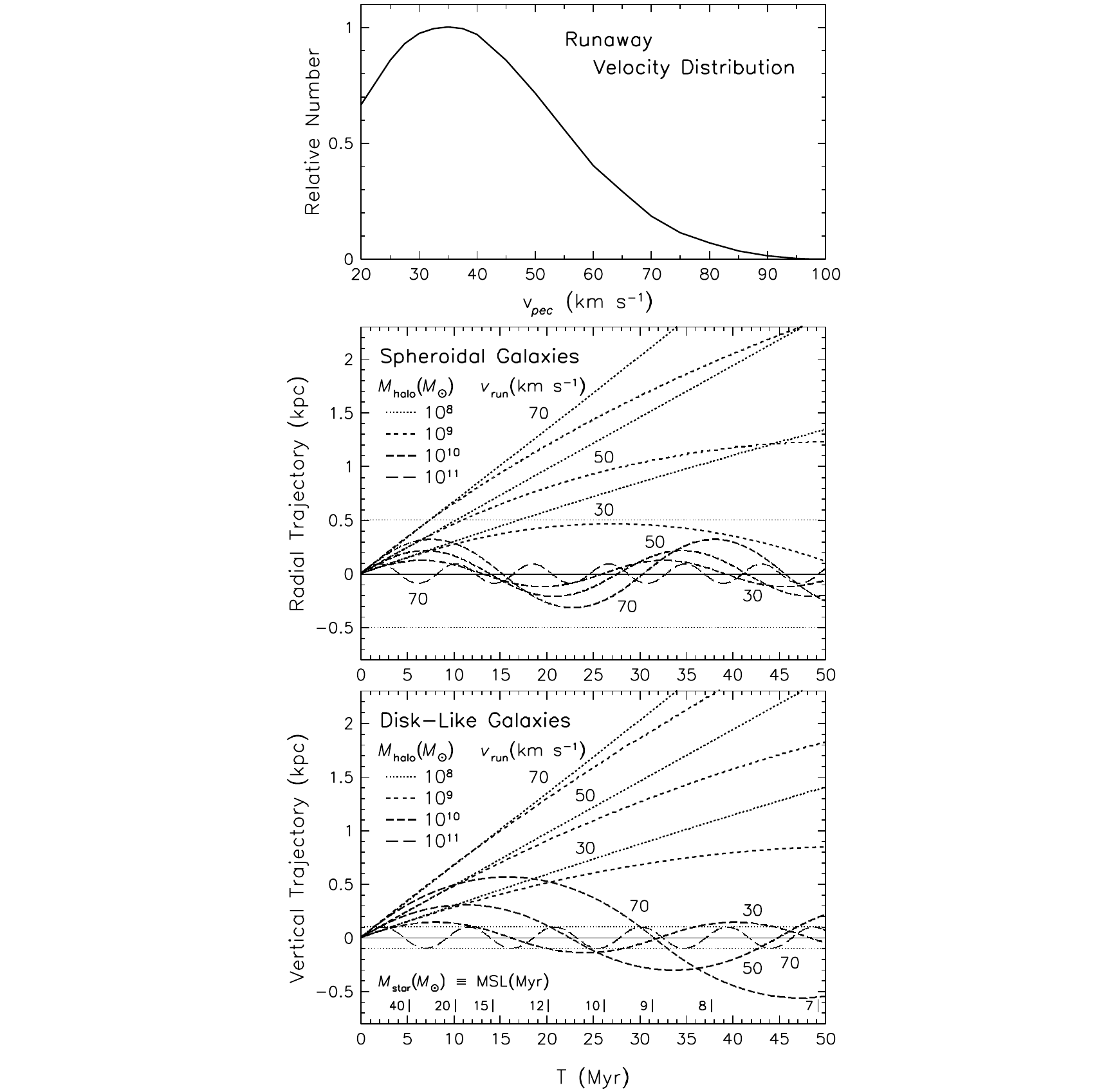}
\caption{\scriptsize \emph{Top panel:} Runaway star 
velocity distribution from \citet{tet11}, common to all stellar masses. 
\emph{Middle panel:} Temporal displays of a selection of typical 
\emph{radial} trajectories of runaway stars with indicated velocities 
$v_{\it{run}}$ initiated at the centre of model \emph{spheroidal} galaxies 
having spheroidal dark matter halos of indicated mass and baryonic content 
0.17 of halo mass (see text for details). The \emph{dotted horizontal lines} 
indicate the baryonic radial softening length applied in the example shown 
here. \emph{Bottom panel:} Similar presentation for a selection of 
\emph{vertical} trajectories initiated at the central plane of model 
\emph{disk-like} galaxies having spheroidal dark matter halos (see text for 
details). Notice in both that stars with trajectories that return to the 
galaxy pass through and out again on the other side. Marked along the 
bottom time axis but applicable to both displays are indicative positions 
of main sequence lifetimes of stars of shown mass in the active range of 
significance here (refer also to the \emph{top panel} of Figure 21).} 
\end{figure*}

\clearpage
\begin{figure*}
\figurenum{\scriptsize 23(a)}
\epsscale{1.92}
\plotone{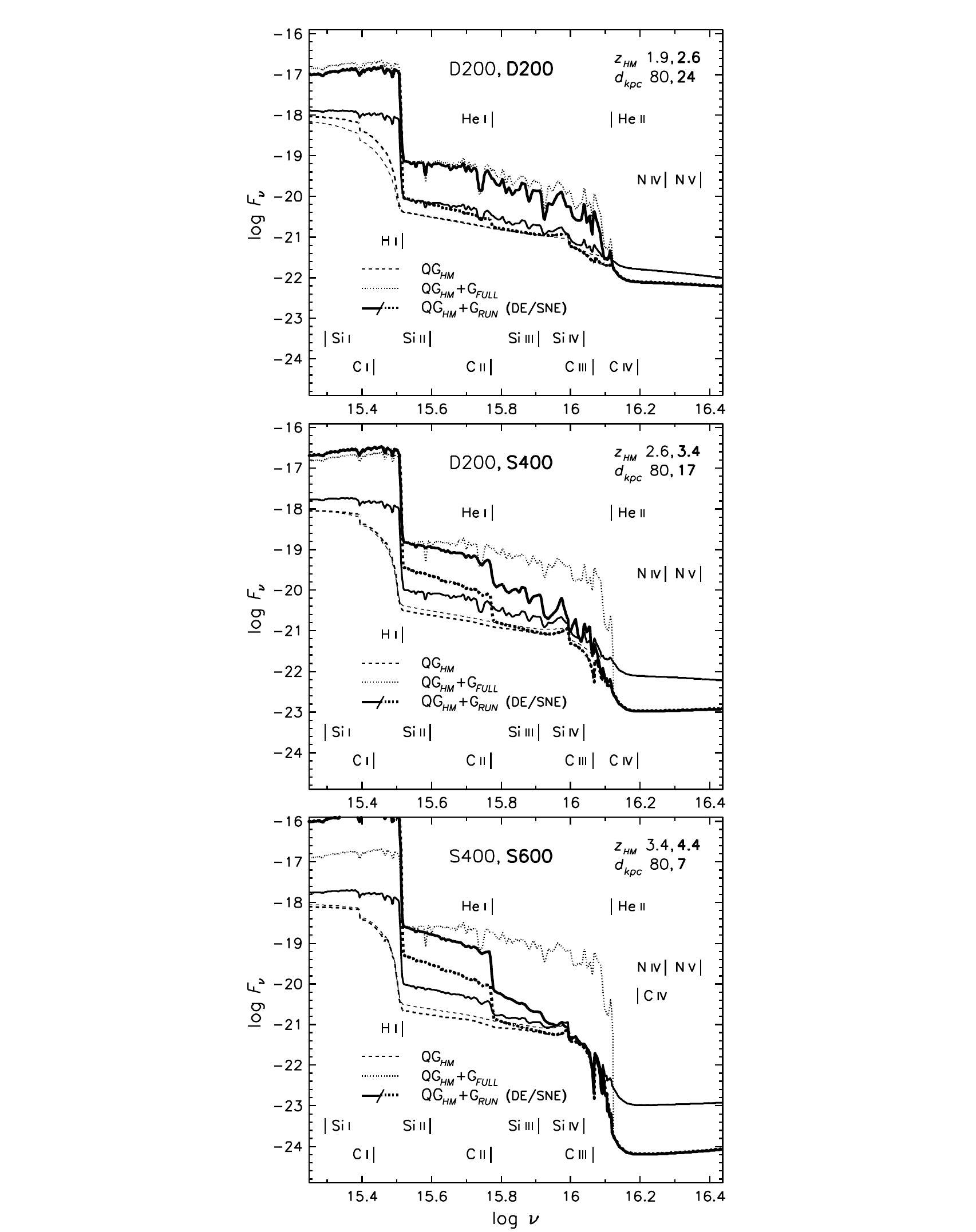}
\caption{\scriptsize Spectral energy distributions 
plotted as in Figure 17(a), here for the \citet{ham12} QG$_{HM}$ background  
augmented by radiative contributions from local template galaxies due to 
runaway stars (identified G$_{RUN}$) over ranges of internal escape 
intervals to the gaseous boundaries of the galaxies as labeled (e.g.: D200, 
indicating a disk-like galaxy with escape height 200 pc; S400, a spheroidal 
galaxy with escape radius 400 pc) (see text) and positioned at indicated 
physical kpc from the absorbers. Included are \emph{dynamically ejected} 
stars (DE), in \emph{continuous lines}, and \emph{supernova ejected} stars 
(SNE), in \emph{bold dotted lines} (shown only at the higher redshift in 
each panel). For comparison, QG$_{HM}$+G$_{FULL}$) from Figure 20 are 
included at the higher redshift in each panel, in \emph{fine dotted lines}. 
The basic QG$_{HM}$ background from Figure 17(a) is in \emph{dashed lines}. 
\emph{Thin-thick} redshift coding in correspondence with the $z_{HM}$ 
boundaries is applied where necessary.} 
\end{figure*}

\clearpage
\begin{figure*}
\figurenum{\scriptsize 23(b)}
\epsscale{1.92}
\plotone{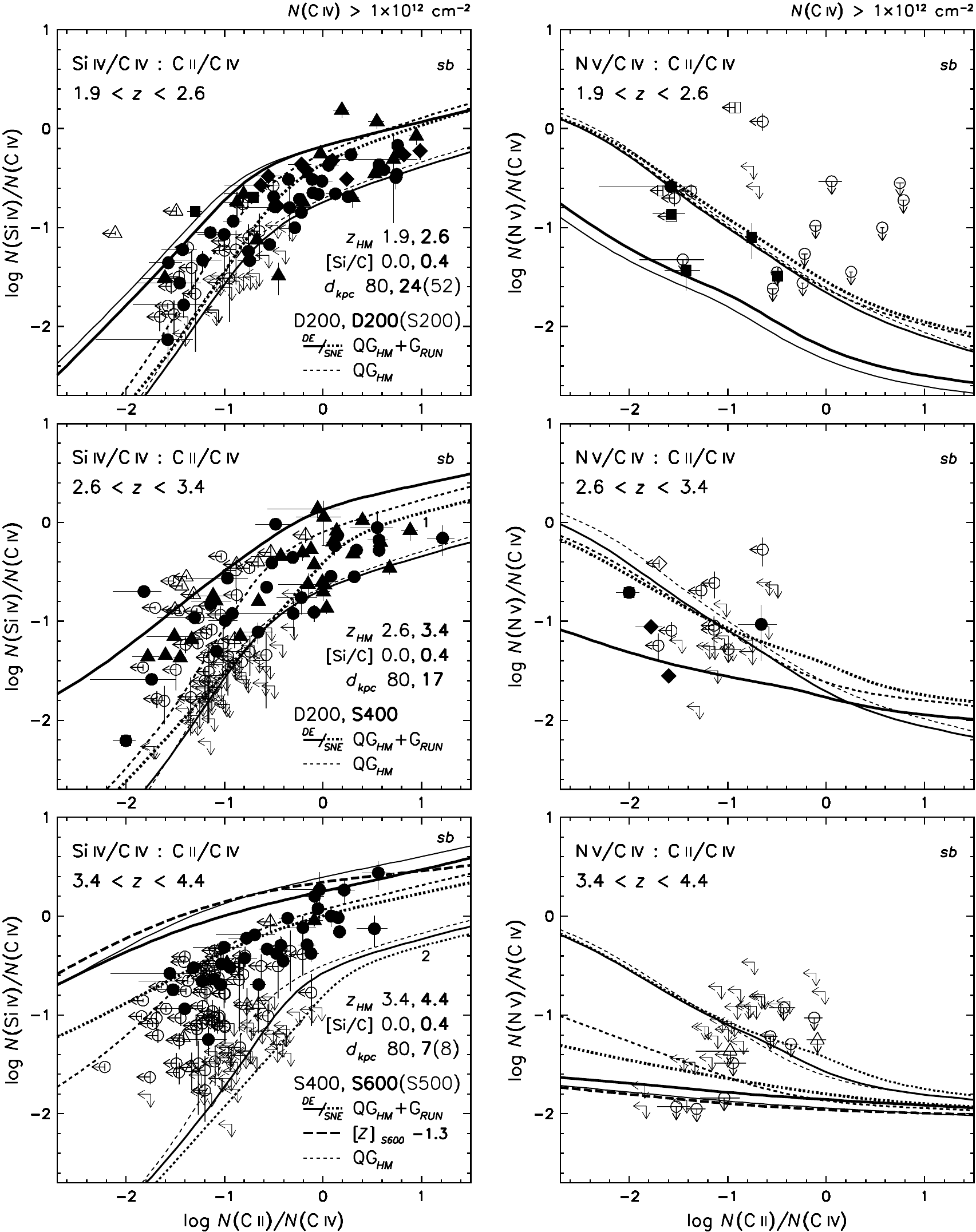}
\caption{\scriptsize \Cloudy\ models of 
two-dimensional column density ratios at the bounding redshifts $z_{HM}$ 
for the radiative cases in Figure 23(a) and similarly coded, with fiducial 
absorbers (see text), and associated data points, all plotted as in Figure 
17(b) and described in its caption. In addition, in the \emph{top panels} 
there is a case at $z_{HM}$ = 2.6 for a spheroidal model S200 and in the 
\emph{bottom panels} a case at $z_{HM}$ = 4.4 for S500, both in \emph{thin 
continuous lines}; also in the \emph{bottom panels} the SNE runaway case 
from the \emph{middle panels} at $z_{HM}$ = 3.4 (indicated 1) is now 
displayed for the lower boundary [Si/C] = 0.0 (indicated 2); and at 
$z_{HM}$ = 4.4 is a case for S600 with the metallicity $[Z]$ of the 
absorber raised from the fiducial value $(-2.5)$ to $-1.3$, in \emph{thick 
dashed lines}); see text for all these cases. The basic QG$_{HM}$ 
background from Figure 17(b) is in \emph{thin dashed lines}. 
\emph{Thin-thick} redshift coding in correspondence with the $z_{HM}$ 
boundaries is applied where necessary.}
\end{figure*}

\clearpage
\begin{figure*}
\figurenum{\scriptsize 23(b)}
\epsscale{1.92}  
\plotone{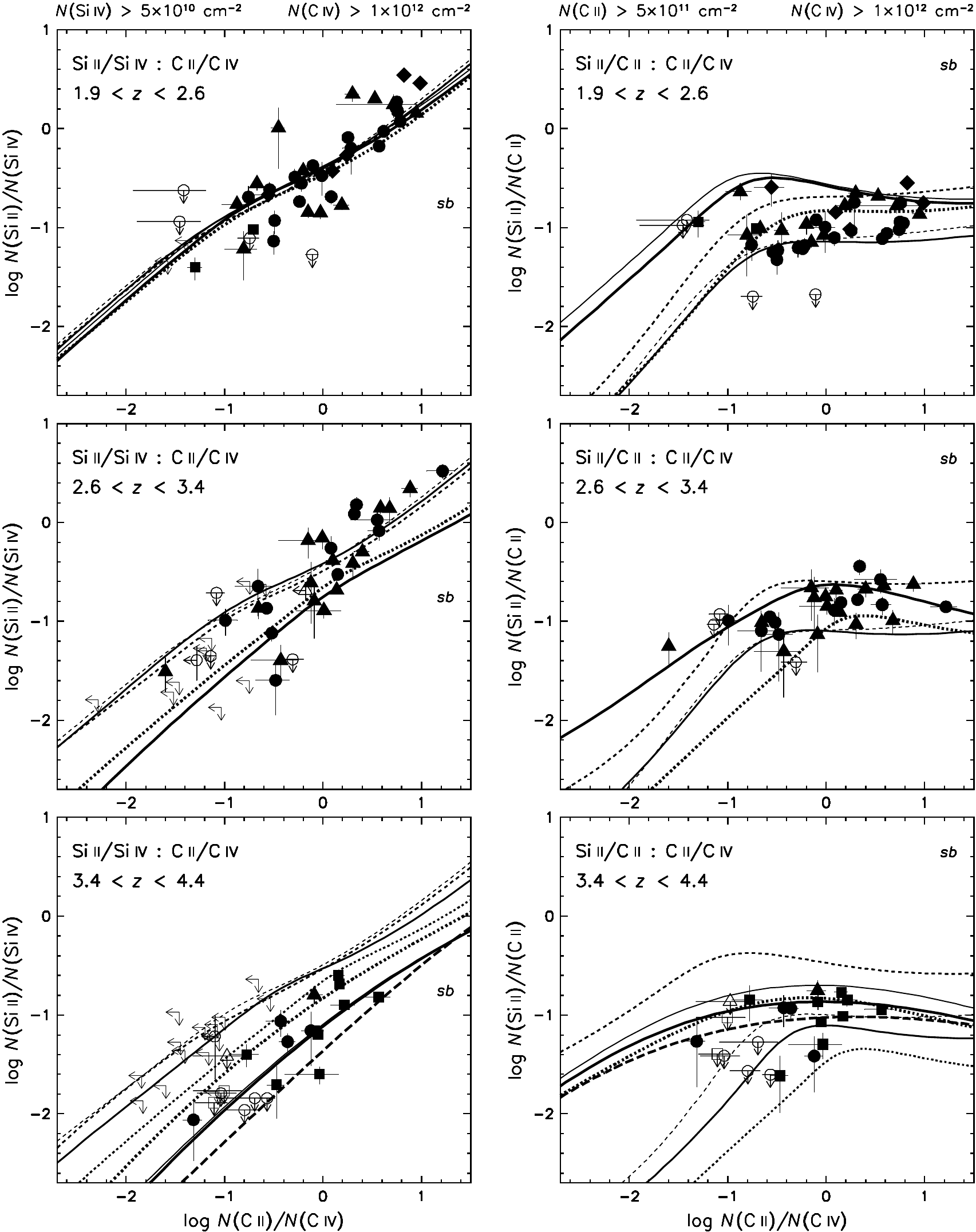}
\caption{\scriptsize Continued. Note the caption to
the equivalent part of Figure 17(b).} 
\end{figure*}

\clearpage
\begin{figure*}
\figurenum{\scriptsize 24}
\epsscale{1.92}
\plotone{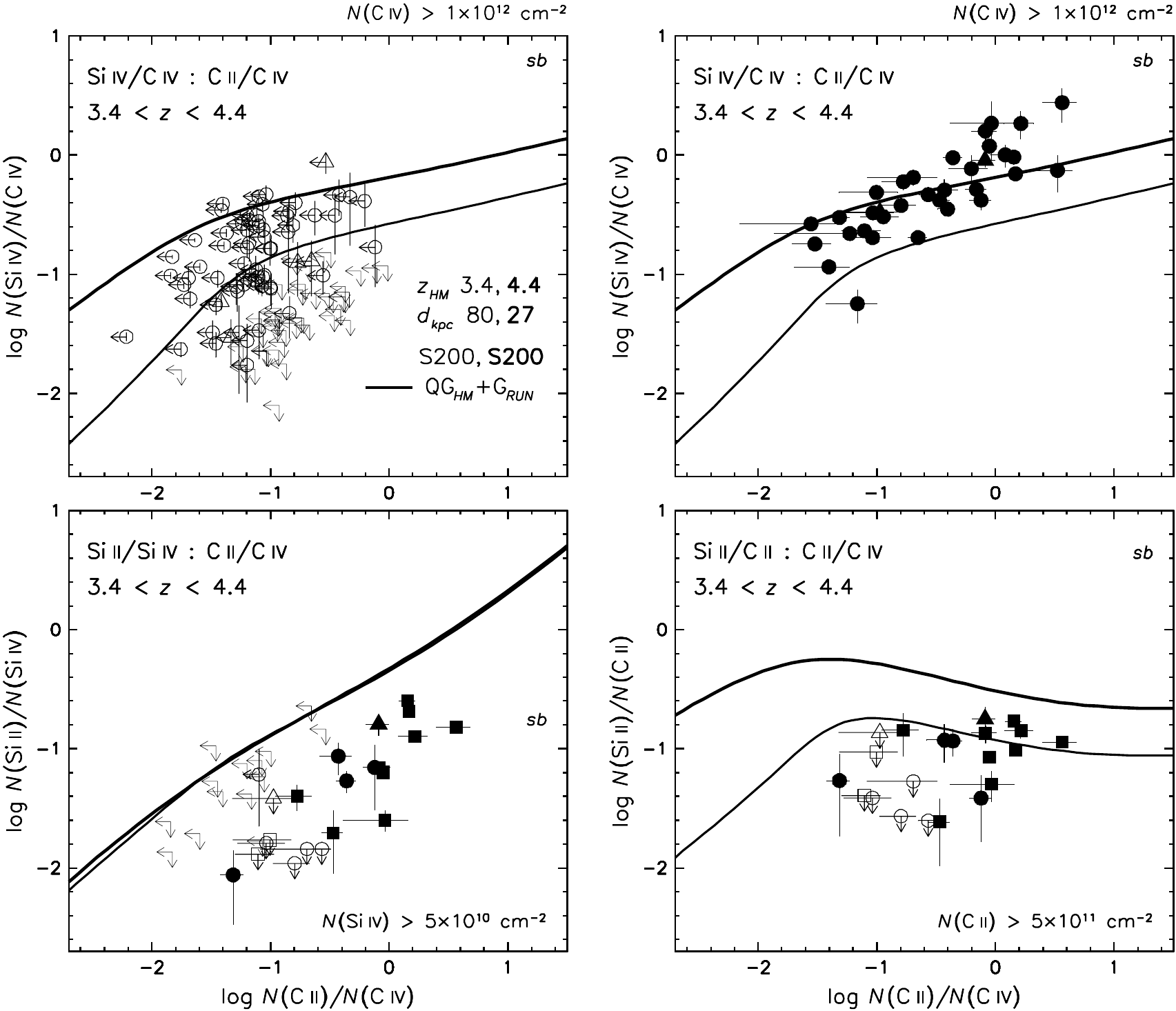}
\caption{\scriptsize \Cloudy\ model predictions of 
two-dimensional column density ratios for the fiducial absorbers exposed 
to metagalactic ionizing radiation at the bounding redshifts 
$z_{HM} =$ 3.4, 4.4 and local radiation received at the indicated 
distances from \emph{dynamically ejected} runaway stars (DE) escaping a 
spheroidal galactic structure S200, described in the text; the cosmic 
microwave background at these redshifts is included. The model curves are 
\emph{thin-thick} redshift-coded as indicated. The \emph{upper left panel} 
contains all \SiIV/\CIV~:~\CII/\CIV\ upper limit cases from the 
corresponding panel in Figure 23(b); the associated full detections are 
shown in the \emph{upper right panel}. In the \emph{lower panels} are the 
Figure 23(b) counterpart cases for \SiII/\SiIV~:~\CII/\CIV\ and 
\SiII/\CII~:~\CII/\CIV, here integrally including the upper limits 
(however, as explained  in Figure 17(b), in particular for the latter 
only \SiII\ upper limits can be present because \emph{detected} \CII\ is 
needed as the base for the ratio \SiII/\CII). Refer also to Figure 23(b) 
for \NV/\CIV~:~\CII/\CIV\ relating to the above and all panels for 
$1.9 < z < 2.6$ and $2.6 < z < 3.4$: the balance between full detections 
and indications from upper limits highlighted here for $3.4 < z < 4.4$ 
develop progressively over the whole observed redshift range.}
\end{figure*}


\begin{thebibliography}{}
\bibitem[Adelberger et al.(2003)]{ade03} Adelberger, K. L., Steidel, C., C., 
    Shapley, A., E., \& Pettini, M. 2003, \apj, 584, 45
\bibitem[Adelberger et al.(2005a)]{adb05} Adelberger, Steidel, C., Pettini, M.,
    et al. 2005a, \apj, 619, 713
\bibitem[Adelberger et al.(2005b)]{ade05} Adelberger, K. L., Shapley, A.E.,  
    Steidel, C., et al. 2005b, \apj, 629, 636
\bibitem[Agafonova et al.(2007)]{aga07} Agafonova, I. I., Levshakov, S. A., 
    Reimers, D., et al. 2007, \aap, 461, 893
\bibitem[Aguirre et al.(2002)]{agu02} Aguirre, A., Schaye, J., \& Theuns, T.
    2002, \apj, 576, 1
\bibitem[Aguirre et al.(2004)]{agu04} Aguirre, A., Schaye, J., Kim, T.-S., 
    et al. 2004, \apj, 602, 38 	
\bibitem[Aguirre et al.(2005)]{agu05} Aguirre, A., Schaye, J., Hernquist, L., 
    et al. 2005, \apj, 620, 13 	
\bibitem[Anosova(1986)]{ano86} Anosova, J. P. (1986), \apss, 124, 217
\bibitem[Aracil et al.(2004)]{ara04} Aracil, B., Petitjean, P., \& 
    Bergeron, J. 2004, \aap, 419, 819
\bibitem[Bajtlik, Duncan, \& Ostriker(1988)]{bdo89} Bajtlik, S., Duncan, R. C.,
    \& Ostriker, J. P. 1988, \apj, 327, 570
\bibitem[Barlow \& Sargent(1997)]{bas97} Barlow, T. A., \& Sargent, W. L. W. 
    1997, \aj, 113, 136 
\bibitem[Bechtold et al.(1987)]{bec87} Bechtold, J., Weymann, R. J., Lin, Z., 
    \& Malkan, M. A. 1987, \apj, 315, 180
\bibitem[Becker et al.(2011)]{bec11} Becker, G. D., Bolton, J. S., 
    Haehnelt, M. G., \& Sargent, W. L. W. 2011, \mnras, 410, 1096
\bibitem[Becker \& Bolton(2013)]{bab13} Becker, G. D., \& Bolton, J. S. 
    2013, \mnras, 436, 1023
\bibitem[Becker et al.(2013)]{bec13} Becker, G. D., Hewett, P. C., 
    Worseck, G., \& Prochaska, J., X. 2013, \mnras, 430, 2067
\bibitem[Benson et al.(2001)]{ben01} Benson, A. J., Frenk, C. S., Baugh, C. M., 
    Cole, S., \& Lacey, C. G. 2001, \mnras, 327, 1041
\bibitem[Bergeron \& Stasi\'nska(1986)]{bes86} Bergeron, J., \& Stasi\'nska, G. 
    1986, \aap, 169, 1
\bibitem[Bergeron \& Boiss\'e(1991)]{bab91} Bergeron, J., \& Boiss\'e, P. 1991, 
    \aap, 243, 344
\bibitem[Bergeron et al.(2002)]{ber02} Bergeron, J., .Aracil, B., 
    Petitjean, P., \& Pichon. C. 2002, \aap, 396, L11
\bibitem[Bianchi et al.(2001)]{bck01} Bianchi, S., Cristiani, S., 
    \& Kim, T.-S. 2001, \aap, 376, 1
\bibitem[Binette et al.(2005)]{bin05} Binette, L., Magris C. G., Krongold, Y., 
    et al. 2005, \apj, 631, 661
\bibitem[Blaauw(1961)]{bla61} Blaauw, A. 1961, Bull. Astron. Inst. Neth., 
    15, 265
\bibitem[Boksenberg(1997)]{bok97} Boksenberg, A. 1997, in Structure and 
    Evolution of the Intergalactic Medium from QSO Absorption Line Systems, 
    P. Petitjean, \& S. Charlot, Edition Frontieres, 85
\bibitem[Boksenberg \& Sargent(1978)]{bas78} Boksenberg, A., \& 
    Sargent, W. L. W. 1978, \apj, 220, 42
\bibitem[Boksenberg et al.(2001)]{bsr01} Boksenberg, A., Sargent, 
    W. L. W., \& Rauch, M. 2001, in The Birth of Galaxies, Proceedings of 
    the Xth Rencontres de Blois, June 28--July 4 1998, B. Guiderdoni, 
    F. R. Bouchet, T. X. Thu\^an, J. T. T. V\^an, The Gioi Publishers, 
    Vietnam, 429
\bibitem[Boksenberg et al.(2003)]{bsr03} Boksenberg, A., Sargent, 
    W. L. W., \& Rauch, M. 2003, astro-ph/0307557
\bibitem[Bolton et al.(2006)]{bol06} Bolton, J. S., Haehnelt, M. G., Viel, M.,
    \& Carswell, R. F. 2006, \mnras, 366, 1378
\bibitem[Bolton \& Haehnelt(2007)]{bah07} Bolton, J. S.; Haehnelt, M. G. 2007, 
    \mnras, 382, 325
\bibitem[Bond et al.(2001)]{bon01} Bond, N. A., Churchill, C. W., 
    Charlton, J. C., \& Vogt, S. S. 2001, \apj, 562, 641
\bibitem[Boutsia et al.(2011)]{bgg11} Boutsia, K., Grazian, A., Giallongo, E., 
    et al. 2011, \apj, 736 41
\bibitem[Bouwens et al.(2004)]{bou04} Bouwens, R. J., Illingworth, G. D.,
    Blakeslee, J. P., et al. 2004, \apjl, 611, L1 
\bibitem[Bouwens et al.(2009)]{bou09} Bouwens, R. J., Illingworth, G. D., 
    Franx, R.-R., et al. 2009, \apj, 705, 936
\bibitem[Bouwens et al.(2011)]{bou11} Bouwens, R. J., Illingworth, G. D., 
    Oesch, P. A., et al. 2011, \apj, 737, 90
\bibitem[Bouwens et al.(2012a)]{boa12} Bouwens, R. J., Illingworth, G. D., 
    Oesch, P. A., et al. 2012a, \apj, 752, L5
\bibitem[Bouwens et al.(2012b)]{bob12} Bouwens, R. J., Illingworth, G. D., 
    Oesch, P. A., et al. 2012b, \apj, 754, 83
\bibitem[Bruzual \& Charlot(2003)]{brc03} Bruzual, A. G., \& Charlot, S. 2003, 
    \mnras, 344, 1000
\bibitem[Calzetti et al.(1994)]{cal94} Calzetti, D., Kinney, A. L., \& 
     Storchi-Bergmann, T. 1994, \apj, 429, 582
\bibitem[Calzetti et al.(2000)]{cal00} Calzetti, D., Armus, L., Bohlin, R., 
     et al. 2000, \apj, 533, 682
\bibitem[Carswell et al.(2002)]{car02} Carswell, R. F., Webb, J. K., 
    Cooke, A.J., \& Irwin, M.J. 2002, Institute of Astronomy, University of 
Cambridge, http://www.ast.cam.ac.uk/\~{}rfc/vpfit.html
\bibitem[Cen et al.(1994)]{cen94} Cen, R., Miralda-Escud\'e, J., 
    Ostriker J. P., \& Rauch, M. 1994, \apjl, 437, L9
\bibitem[Cen \& Chisari(2011)]{cac11} Cen, R., \& Chisari, N. E. 2011, 
    \apj, 731, 11
\bibitem[Chaffee et al.(1986)]{cha86} Chaffee, F. H., Foltz, C. B., 
    Bechtold, J., \& Weymann, R. J. 1986, \apj, 301, 116
\bibitem[Charlton \& Churchill(1998)]{cc98} Charlton, J. C., \& 
    Churchill, C. W. 1998, \apj, 499, 181
\bibitem[Chen et al.(2007)]{che07}  Chen, H.-W., Prochaska, J. X., \& 
    Gnedin, N. Y. 2007, \apjl, 667, L125
\bibitem[Churchill et al.(2003)]{cvc03} Churchill, C. W., Vogt, S. S., 
    \& Charlton, J. C. 2003, \aj, 125, 98
\bibitem[Churchill et al.(2001)]{chu01} Churchill, C. W., Mellon, R. R., 
    Charlton, J. C., et al. 2001, \apj, 562, 641
\bibitem[Clarke \& Oey(2002)]{cao02} Clarke, C. \& Oey, M. S. 2002, 
    \mnras, 337, 1299
\bibitem[Conroy \& Kratter(2012)]{cak12} Conroy, C., \& Kratter, K. M. 2012, 
    \apj, 755, 123
\bibitem[Cooke et al.(1997)]{cec97} Cooke, A. J., Espey, B., 
    \& Carswell, R. F. 1997, \mnras, 284, 552
\bibitem[Cowie et al.(1995)]{cow95} Cowie, L. L., Songaila, A., Kim, T. -S., 
    \& Hu, E. M. 1995, \aj, 109, 1522
\bibitem[Cowie \& Songaila(1998)]{cas98} Cowie, L. L., \& Songaila, A. 1998,
    Nature, 394, 44
\bibitem[Cristiani et al.(1997)]{cri97} Cristiani, S., D'Odorico, S., 
    D'Odorico, V., et al. 1997, \mnras, 285, 209
\bibitem[Croft at al.(2002)]{cro02} Croft, R. A. C., Hernquist, L., 
    Springel, V., Westover, M., \& White, M. 2002, \apj, 580, 634 
\bibitem[Crowther(2000)]{cro00} Crowther, P. A. 2000, \aap, 356, 191
\bibitem[Dav\'e \& Tripp(2001)]{dat01} Dav\'e, R., \& Tripp, T. M. 2001, 
    \apj, 553, 528
\bibitem[Cruise(1993)]{cru93} Cruise, A.M. 1993, \mnras, 265, 881
\bibitem[Dinshaw \& Impey(1996)]{dai96} Dinshaw, N., \& Impey, C. D. 1996, 
    \apj, 458, 73
\bibitem[D'Odorico et al.(2010)]{dod10} D'Odorico, V., Calura, F., 
    Cristiani, S., \& Viel, M. 2010, \mnras, 401, 2715
\bibitem[D'Odorico et al.(2013)]{dod13} D'Odorico, V., Cupani., G.,
    Cristiani, S., et al. 2013, \mnras, 435, 1198
\bibitem[Donahue \& Shull(1987)]{das87} Donahue, M., \& Shull, J. M. 1987, 
    \apjl, 323, L13
\bibitem[Dove \& Shull(1994)]{das94} Dove, J. B., \& Shull, J. M. 1994, 
    \apj, 430, 2220
\bibitem[Dove et al.(2000)]{das00} Dove, J. B., Shull, J. M., \& Ferrara, A.
    2000, \apj, 531, 846
\bibitem[Draine \& Lee(1984)]{dal84} Draine, B. T., \& Lee, H. M. 1984, 
    \apj, 285, 89
\bibitem[Eldridge \& Stanway(2009)]{eas09} Eldridge, J. J., \& Stanway, E. R. 
    2009, \mnras, 400, 1019
\bibitem[Ellison et al.(2000)]{ell00} Ellison, S. L., Songaila, A., 
    Schaye, J., \& Pettini, M. 2000, \aj, 120, 1175
\bibitem[Erb et al.(2006)]{erb06} Erb, D. K., Shapley, A. E., Pettini, M., 
    et al. 2006, \apj, 644, 813
\bibitem[Erb et al.(2010)]{erb10} Erb, D. K., Pettini, M., Shapley, A. E., 
    et al. 2010, \apj, 719, 1168
\bibitem[Fathivavsari et al.(2013)]{fat13} Fathivavsari, H., Petitjean, P.,
    Ledoux, C., et al. 2013, \mnras,435, 1727
\bibitem[Faucher-Giguère et al.(2008)]{fau08} Faucher-Giguère, C.-A., 
    Prochaska, J. X., Lidz, A., Hernquist, L., \& Zaldarriaga, M. 2008, \apj, 
    681, 831
\bibitem[Fan et al.(2006)]{fan06} Fan, X., Strauss, M. A., Becker, R. H. et al. 
    2006, \aj, 132, 117
\bibitem[Ferland et al.(1998)]{fer98} Ferland, G.  J. et al. 1998, PASP, 110, 
    761
\bibitem[Ferland et al.(2011)]{fer11} Ferland, G. J. et al. 2011, Hazy, a 
    brief introduction to CLOUDY C10, Cloudy \& Associates, www.nublado.org  
\bibitem[Fernandez \& Shull(2011)]{fas11} Fernandez, E. R., \& Shull, J. M. 
    2011, \apj, 731, 20
\bibitem[Fern\'andez-Soto et al.(1996)]{fso96} Fern\'andez-Soto, A., 
    Lanzetta, K. M., Barcons, X., Carswell, R. F., Webb, J. K., \& Yahil, A. 
    1996, \apjl, 460, L85
\bibitem[Fern\'andez-Soto, Lanzetta, \& Chen(2003)]{fso03} Fern\'andez-Soto, 
    A., Lanzetta, K. M., \& Chen, H.-W. 2003, \mnras, 342, 1215
\bibitem[Fioc \& Rocca-Volmerange(1997)]{far97} Fioc, M., \& 
    Rocca-Volmerange, B. 1997, \aap, 326, 950 
\bibitem[Fujita et al.(2003)]{fuj03} Fujita, A.,  Martin C., Low, M. -M. M., 
    \& Abel, T. 2003, \apj, 599, 50 
\bibitem[Fox et al.(2007a)]{fox07a} Fox, A. J., Petitjean, P., Ledoux, C., 
    \& Srianand, R. 2007a, \aap, 465, 171 
\bibitem[Fox et al.(2007b)]{fox07b} Fox, A. J., Ledoux, C., Petitjean, P., 
    \& Srianand, R. 2007b, \aap, 473, 791 
\bibitem[Giallongo et al.(2002)]{gia02} Giallongo, E., Cristiani, S., 
    D'Odorico, S., \& Fontana, A. 2002, \apjl, 568, L9 
\bibitem[Gies \& Bolton(1986)]{gab86} Gies, D. R., \& Bolton, C. T. 1986, 
    \apjs, 61, 419
\bibitem[Giroux \& Shull(1997)]{gas97} Giroux, M. L., \& Shull, J. M. 1997, 
    \aj, 113, 1505 
\bibitem[Gnat \& Steinberg(2007)]{gna07} Gnat, O., \& Steinberg, A. 2007, 
    \apjs, 168, 213
\bibitem[Gnedin et al.(2008)]{gne08} Gnedin, N. Y., Kravtsov. A. V., \& 
    Chen, H.-W. 2008, \apj, 672, 765
\bibitem[Grevesse et al.(2010)]{gre10} Grevesse, N., Asplund, M., Sauval, A. J.,
    \& Scott, P. 2010, Astrophysics and Space Science, 328, 179
\bibitem[Haardt(2014)]{haa14} Haardt, F. 2014, private communications from 1998 
    to 2000 and 2010 to 2014
\bibitem[Haardt \& Madau(1996)]{ham96} Haardt, F., \& Madau, P. 1996, \apj, 
    461, 20
\bibitem[Haardt \& Madau(2012)]{ham12} Haardt, F., \& Madau, P. 2012, \apj, 
    746, 125
\bibitem[Haehnelt et al.(1998)]{hsm98} Haehnelt, M. G., 
    Steinmetz, M., \& Rauch, M., 1998, \apj, 495, 647
\bibitem[Haehnelt et al.(2001)]{hae01} Haehnelt, M. G., Madau, P., 
    Kudritzki, R., \& Haardt, F. 2001, \apjl, 549, L151 
\bibitem[Haiman \& Loeb(1997)]{hal97} Haiman, Z., \& Loeb, A. 1997, \apj, 483, 
    210, 2220
\bibitem[Hawkins et al.(2003)]{haw03} Hawkins, E. et al. 2003, \mnras, 346, 78 
\bibitem[Heap et al.(2000)]{hea00} Heap, S. R., Williger, G. M., Smette, A., 
    et al. 2000, \apj, 534, 69 
\bibitem[Heckman et al.(2001)]{hec01} Heckman, T. M., Sembach, K. R., 
    Meurer, G. R., et al. 2001, \apj, 558, 56
\bibitem[Heisler et al.(1989)]{hhw89} Heisler, J., Hogan, C. J., 
    \& White, S. D. M. 1989, \apj, 347, 52
\bibitem[Henry et al. (2000)]{hek00} Henry, R. B. C., Edmunds, M. G., 
    \& K\"oppen, J. 2000, \apj, 541, 660  
\bibitem[Henry \& Prochaska (2007)]{hap07} Henry, R. B. C., \& Prochaska, J. X. 
    2007, PASP, 119, 962
\bibitem[Hernquist et al.(1996)]{her96} Hernquist, L., Katz, N., 
    Weinberg, D.II., \& Miralda-Escud\'e, J. 1996, \apjl, 457, L51
\bibitem[Hopkins et al.(2007)]{hop07} Hopkins, P. F., Richards, G. T., 
    \& Hernquist, L. 2007, \apj, 654, 731
\bibitem[Hoogerwerf et al.(2001)]{hoo01} Hoogerwerf, R., de Bruijne, J. H. J.,
    de Zeeuw, P. T. 2001, \aap, 365, 49
\bibitem[Hurley et al.(2000)]{hur00} Hurley, J. R., Pols, O. R., Tout, C. A.
    2000, \mnras, 315, 543
\bibitem[Hurwitz, Jelinski, \& Dixon(1997)]{hjd97} Hurwitz, M., Jelinski, P., 
    \& Dixon, W. v. D. 1997, \apjl, 481, L31
\bibitem[Inoue et al.(2006)]{ino06} Inoue, A. K., Iwata, I., \&  
    Deharveng, J.-M. 2006, \mnras, 371, L1
\bibitem[Iwata et al.(2009)]{iwa09} Iwata, I., Inoue, A. K., Matsuda, Y., 
    et al. 2009, \apj, 692, 1287
\bibitem[Janknecht et al.(2006)]{jan06} Janknecht, E., Reimers, D., Lopez, S., 
    \& Tytler, D. 2006, \aap, 458, 427
\bibitem[Kaiser(1987)]{kai87} Kaiser, N. 1987, \mnras, 227, 1
\bibitem[Kawata \& Rauch(2007)]{kar07} Kawata, D.,  \& Rauch, M. 2007, \apj,
    663, 38
\bibitem[Kewley et al.(2001)]{kew01} Kewley, L. J., Dopita, M. A., 
    Sutherland, R. S., Heisler, C. A., \& Trevena, J. 2001, \apj, 556, 121
\bibitem[Kewley \& Kobulnicky (2007)]{kak07} Kewley, L., \& Kobulnicky, H. A. 
    2007, in Island Universes: Structure and Evolution of Disk Galaxies, ed. 
    R. S. de Jong (Dordrecht: Springer), 435
\bibitem[Kim et al.(2002a)]{kcd02} Kim, T.-S., Cristiani, S, \& 
    D'Odorico, S. 2002a, \aap, 383, 747
\bibitem[Kim et al.(2002b)]{kim02} Kim, T.-S., Carswell, R.F., Cristiani, S, 
    \& D'Odorico, S. 2002b, \mnras, 335, 555
\bibitem[Kimm \& Cen (2014)]{kac14} Kimm, T., \& Cen, R. 2014, 
    arXiv:1405.0552v
\bibitem[Kolatt et al.(1999)]{kol99} Kolatt, T. S., Bullock, J. S., 
    Somerville, R. S., et al. 1999, \apjl, 523, L109
\bibitem[Kollmeier et al.(2014)]{kol14} Kollmeier, J. A., Weinberg, D. H., 
    Oppenheimer, B. D., et al. 2014, \apjl, 789, L32
\bibitem[Komatsu et al.(2011)]{kom11} Komatsu, E., Smith, K.M., Dunkley, J.
    et al. 2011, \apjs, 192, 18
\bibitem[Kriss et al.(2001)]{kri01} Kriss, G. A., et al. 2001, Science, 293, 
    1112 
\bibitem[Kuhlen \& Faucher-Gigu\`ere(2012)]{kaf12} Kuhlen, M., \& 
    Faucher-Gigu\`ere, C.-A. 2012, \mnras, 423, 862
\bibitem[Lanzetta et al.(1991)]{lan91} Lanzetta, K. M., Wolfe, A.M., Turnshek,
    D. A., Lu, L., McMahon, R. G., \& Hazard, C. 1991, \apjs, 77, 1
\bibitem[Law et al.(2007)]{law07} Law, D. R., Steidel, C. C., Erb, D. K., 
    et al. 2007, \apj, 656, 1
\bibitem[Law et al.(2012)]{law12} Law, D. R., Steidel, C. C., Shapley, A. E., 
    et al. 2012, \apj, 745, 85
\bibitem[Le Brun et al.(1997)]{leb97} Le Brun, V., Bergeron, J., Boiss\'e, P., 
    \& Deharving, J. M. 1997, \aap, 321, 733
\bibitem[Ledoux et al.(1998)]{led98} Ledoux, C., Petitjean, P., Bergeron, J., 
    Wampler, E. J., \& Srianand, R. 1998, \aap, 337, 51
\bibitem[Leitherer et al.(1995)]{lei95} Leitherer, C., Ferguson, H. C., 
    Heckman, T. M., \& Lowenthal, J. D. 1995, \apjl, 454, L19
\bibitem[Leitherer et al.(1999)]{lei99} Leitherer, C., Schaerer, D., 
    Goldader, J. D., et al. 1999, \apjs, 123, 3
\bibitem[Leitherer et al.(2010)]{lei10} Leitherer, C., Ortiz Ot\'alvaro, P. A., 
    Bresolin, F., et al. 2010, \apjs, 189, 309  
\bibitem[Levesque et al.(2012)]{lev12} Levesque, E. M., Leitherer, C.,
    Ekstrom, S., Meynet, G., \& Schaerer, D. 2012, \apj, 751, 67
\bibitem[Loveday et al.(1995)]{lov95} Loveday, J., Maddox, S. J., 
    Efstathiou, G., \& Peterson, B. A. 1995, \apj, 442, 457
\bibitem[Madau \& Haardt(2009)]{mah09} Madau, P., \& Haardt, F. 2009, \apj, 
    693, L100
\bibitem[Madau et al.(1999)]{mhr99} Madau, P., Haardt, F., \& Rees, M. J.       
    1999. \apj, 514, 648
\bibitem[Makino \& Aarseth(1992)]{maa92} Makino, J., Aarseth, S.J. 1992, 
    \pasj, 44, 141.
\bibitem[Mannucci et al.(2009)]{man09} Mannucci, F., Cresci, G., 
    Maiolino, R., Marconi, A., et al. 2009, \mnras, 398, 1915
\bibitem[Marri \& White(2003)]{maw03} Marri S., \& White, S. D. M. 2003, 
    \mnras, 345, 561
\bibitem[Martin \& Rouleau(1990)]{mar90} Martin, P. G., \& Rouleau, F. 1990,
    in Extreme Ultraviolet Astronomy, ed. R. F. Malina \& S. Bowyer
    (Oxford: Pergamon), 341
\bibitem[Martin et al.(2010)]{mar10} Martin, C. L., Scannapieco, E., 
    Ellison, S. L., et al. 2010, \apj, 721, 174
\bibitem[McDonald \& Miralda-Escud\'e(1999)]{mam99} McDonald, P., \& 
    Miralda-Escud\'e, J. 1999, \apj, 519, 486
\bibitem[McDonald et al.(2000)]{mcd00} McDonald, P., Miralda-Escud\'e, J., 
    Rauch, M., et al. 2000, \apj, 543, 1
\bibitem[McDonald et al.(2001)]{mcd01} McDonald, P., Miralda-Escud\'e, J., 
    Rauch, M., et al. 2001, \apj, 562, 52
\bibitem[McQuinn et al.(2009)]{mcq09} McQuinn, M., Lidz, A., Zaldarriaga, M.,
    et al. 2009, \apj, 694, 842 
\bibitem[McWilliam(1997)]{mcw97} McWilliam, A. 1997, \araa, 35, 503
\bibitem[Meiksin(1994)]{mei94} Meiksin, A. 1994, \apj, 431, 109
\bibitem[Meiksin \& Madau(1993)]{mam93} Meiksin, A., \& Madau, P. 1993, \apj, 
    412, 34
\bibitem[M\'enard et al.(2011)]{men11} M\'enard, B., Wild, V., Nestor, D., 
    et al. 2011, \mnras, 417, 801
\bibitem[Misawa et al.(2002)]{mis02} Misawa, T., Tytler, D., Iye, M., 
    et al. 2002, \aj, 123, 1847
\bibitem[Morton(2003)]{mor03} Morton, D. 2003, \apjs, 149, 205 (erratum  
    \apjs, 151, 403 [2004])
\bibitem[Miyamoto \& Nagai(1975)]{min75} Miyamoto, M., \& Nagai, R. 1975, 
    \pasj, 27, 533
\bibitem[Nagamine et al.(2004)]{nsh04} Nagamine, K., Springel, V., 
    \& Hernquist, L. 2004, \mnras, 348, 421 
\bibitem[Navarro et al.(1997)]{nfw97} Navarro, J. F., Frenk, C. S., \& 
    White, S. D. M. 1997, \apj, 490, 493
\bibitem[Norberg et al.(2002)]{nor02} Norberg, P., et al. 2002, \mnras, 332, 
    827
\bibitem[Oke \& Gunn(1983)]{oag83} Oke, J. B., \& Gunn, J. E., 1983, \apj, 
    266, 713
\bibitem[Oesch et al.(2010)]{oes10} Oesch, P. A., Bouwens, R.J., 
    Carollo, C. M., et al. 2010, \apj, 709, L21
\bibitem[Ono et al.(2013)]{ono13} Ono, Y., Ouchi, M., Curtis-Lake, E., et al.
     2013, \apj, 777, 155 
\bibitem[Oppenheimer et al.(2009)]{opp09} Oppenheimer, B. D., Dav\'e, R., \& 
    Finlator, K. 2009, \mnras, 396, 729
\bibitem[Oppenheimer et al.(2012)]{opp12} Oppenheimer, B. D., Dav\'e, R., 
    Katz, N., Kollmeier, J., A., \& Weinberg, D. 2012 \mnras, 420, 829
\bibitem[Ouchi et al.(2009)]{ouc09} Ouchi, M., Mobasher, B., Shimasaku, K., 
    et al. 2009, \apj, 706, 1136O
\bibitem[Papovich et al.(2005)]{pap05} Papovich, C., Dickinson, M., 
    Giavalisco, M., et al. 2005, \apj, 631, 101
\bibitem[P\^aris et al.(2011)]{par11} P\^aris, I., Petitjean, P., 
    Rollinde, E., et al. 2011, \aap, 530, 50 
\bibitem[Pascarelle et al.(2001)]{par01} Pascarelle, S., M., Lanzetta, K. M.,  
    Chen, H.-W., Webb, J., K. 2001, \apj, 560, 101
\bibitem[Patnaik et al.(1992)]{pbw92} Patnaik, A. R., Browne, I. W., 
    Walsh, D., Chaffee, F. H., \& Foltz, C. B. 1992, \mnras, 259, 1P  
\bibitem[Pauldrach et al.(2001)]{pau01} Pauldrach, A. W. A., 
    Hoffmann, T. L., \& Lennon, M. 2001, \aap, 375, 161 
\bibitem[Pawlik, Schaye, \& van Scherpenzeel(2009)]{pss09} Pawlik, A. H., 
    Schaye, J., \& van Scherpenzeel, E. 2009, \mnras, 394, 1812
\bibitem[Pei(1992)]{pei92} Pei, Y. C. 1992, \apj, 395, 130
\bibitem[P\'erez-Gonz\'alez et al.(2008)]{per08} P\'erez-Gonz\'alez, P., 
    Rieke, G. H., Villar, V., et al. 2008, \apj, 675, 234 
\bibitem[Petitjean \& Bergeron(1990)]{pab90} Petitjean, P., \& Bergeron, J. 
    1990, \aap, 231, 309 
\bibitem[Petitjean \& Bergeron(1994)]{pab94} Petitjean, P., \& Bergeron, J. 
    1994, \aap, 283, 759 
\bibitem[Petitjean et al.(2008)]{pls08} Petitjean, P., Ledoux, C., \& 
    Srianand, R. 2008, \aap, 480, 349 
\bibitem[Pettini et al.(2000)]{pet00} Pettini, M., Steidel, C. C., 
    Adelberger, K. L., et al. 2000, \apj, 528, 96
\bibitem[Pettini et al.(2001)]{pet01} Pettini, M., Shapley, A. E., 
    Steidel, C. C., et al. 2001, \apj, 554, 981
\bibitem[Pettini et al.(2008)]{pet08} Pettini, M., Zych, B. J., Steidel, C. C.,
    \& Chaffee, F. H. 2008, \mnras, 385, 2011 
\bibitem[Pettini \& Cooke(2012)]{pet12} Pettini, M., \& Cooke, R. 2012, in XII 
    International Symposium on Nuclei in the Cosmos, August 5-12, 2012, Cairns, 
    Australia, http://pos.sissa.it/cgi-bin/reader/conf.cgi?confid=146, id.71  
\bibitem[Plummer (1911)]{plu11} Plummer, H. C. 1911, \mnras, 71, 460
\bibitem[Prochaska \& Wolfe(1998)]{paw98} Prochaska, J. X., \&  
    Wolfe, A. M. 1998, \apj, 507, 113
\bibitem[Poveda et al.(1967)]{pov67} Poveda A., Ruiz J., Allen C. 1967, 
    Boletin de los Observatorios Tonantzintla y Tacubaya, 4, 86
\bibitem[Prochaska et al.(2009)]{pro09} Prochaska, J. X., Worseck, G., \& 
    O'Meara, J. M. 2009, \apj, 705, L113
\bibitem[Quashnock \& Vanden Berk(1998)]{qav98} Quashnock, J. M. \& 
    Vanden Berk, D. E. 1998, \apj, 500, 28
\bibitem[Quashnock et al.(1996)]{qvy96} Quashnock, J. M., 
    Vanden Berk, D. E., \& York, D. G. 1996, \apj, 472, L69 
\bibitem[Rauch et al.(1992)]{rau92} Rauch, M., Carswell, R. F., Chaffee, F.H.,
    et.al. 1992, \apj, 390, 387 
\bibitem[Rauch et al.(1996)]{rau96} Rauch, M., Sargent, W. L. W., Womble, 
    D. S., \& Barlow, T. A. 1996, \apjl, 467, L5
\bibitem[Rauch et al.(1997)]{rhs97} Rauch, M., Haehnelt, M. G.,  
    \& Steinmetz, M. 1997, \apj, 481, 601
\bibitem[Rauch et al.(1999)]{rsb99} Rauch, M., Sargent, W. L. W.,  
    \& Barlow, T. A. 1999, \apj, 515, 500
\bibitem[Rauch et al.(2001)]{rsb01} Rauch, M., Sargent, W. L. W.,  
    \& Barlow, T. A. 2001, \apj, 554, 823
\bibitem[Razoumov \& Sommer-Larsen(2006)]{rsl06} Razoumov, A. O., \& 
    Sommer-Larsen, J. 2006, \apjl, 651, L89
\bibitem[Razoumov \& Sommer-Larsen(2007)]{rsl07} Razoumov, A. O., \& 
    Sommer-Larsen, J. 2007, \apj, 668, 674
\bibitem[Razoumov \& Sommer-Larsen(2007)]{rsl10} Razoumov, A. O., \& 
    Sommer-Larsen, J. 2010, \apj, 710, 1239
\bibitem[Reddy et al.(2006)]{red06} Reddy, N. A., Steidel, C. C., Charles C., 
    et al. 2006, \apj, 644, 792
\bibitem[Reddy et al.(2008)]{red08} Reddy, N. A., Steidel, C. C., Pettini, M., 
    et al. 2008, \apjs, 175, 48
\bibitem[Reddy \& Steidel(2009)]{ras09} Reddy, N. A., \& Steidel, C. C. 2009,
    \apj, 692, 778
\bibitem[Reddy et al.(2010)]{red10} Reddy, N. A., Erb, D. K., Pettini, M., 
    Steidel, C. C., \& Shapley, A. E. 2010, \apj, 712, 1070
\bibitem[Reimers et al.(1997)]{rei97} Reimers, D., Kohler, S., Wisotzki, L., 
    et al. 1997, \aap, 327, 890 
\bibitem[Ricotti \& Shull(2000)]{ras00} Ricotti, M., \& Shull, J. M. 2000, 
   \apj, 542, 548
\bibitem[Robertson et al.(2013)]{rob13} Robertson, B. E., Furlanetto, R.,
    Schneider, E., et al. 2013, \apj, 768, 71
\bibitem[Rudie et al.(2013)]{rud13} Rudie, G.C., Steidel, C. C., 
    Shapley, A. E., \& Pettini, M. 2013, \apj, 769, 146
\bibitem[Ryan et al.(1996)]{rnb96} Ryan, S. G., Norris, J. E., \& 
    Beers, T. C. 1996, \apj, 471, 254
\bibitem[Salpeter(1955)]{sal55} Salpeter, E. E. 1955, \apj, 121, 161
\bibitem[Sargent \& Steidel(1987)]{sas87} Sargent. W. L. W., \& Steidel, C. C. 
    1987, \apj, 322, 142
\bibitem[Sargent et al.(1988a)]{sbs88} Sargent, W. L. W., 
    Boksenberg, A., \& Steidel, C. C. 1988, \apjs, 68, 539    
\bibitem[Sargent et al.(1988b)]{ssb88} Sargent, W. L. W., 
    Steidel, C. C., \& Boksenberg, A. 1988, \apj, 334, 22    
\bibitem[Sargent et al.(1980)]{sar80} Sargent, W. L. W., Young, P. J., 
    Boksenberg, A., \& Tytler, D. 1980, \apjs, 42, 41    
\bibitem[Savage \& Sembach(1991)]{sas91} Savage, B. D., \& Sembach, K. R. 1991,
    \apj, 379, 245
\bibitem[Scannapieco et al.(2006)]{sca06} Scannapieco, E., Pichon, C., 
    Aracil, B., et al. 2006, \mnras, 365, 615 (erratum: 2006, \mnras, 366, 1118)
\bibitem[Schaye et al.(2000a)]{sch00} Schaye, J., Rauch, M., Sargent, W. L. W., 
    \& Kim, T.-S. 2000a, \apjl, 541, L1
\bibitem[Schaye et al.(2000b)]{str00} Schaye, J., Theuns, T., Rauch, M., 
    Efstathiou, G., \& Sargent, W. L. W. 2000b, \mnras, 318, 817 
\bibitem[Schaye et al.(2003)]{sch03} Schaye, J., Aguirre, A., Kim, T.-S., 
    Theuns, T., Rauch, M., \& Sargent, W. L. W., 2003 \apj, 596, 768
\bibitem[Scott et al.(2000)]{sco00} Scott, J., Bechtold, J., Dobrzycki, A.,
    \& Kulkarni, V. P. 2000, \apjs, 130, 67
\bibitem[Seibert et al.(2002)]{sei02} Seibert, M., Heckman, T., M., \& 
    Meurer, G. R. 2002, \apj, 124, 46
\bibitem[Shapiro \& Giroux(1987)]{sag87} Shapiro, P. R., \& Giroux, M. L. 1987,
    \apjl, 321, L107
\bibitem[Shapiro \& Moore(1976)]{sam76} Shapiro, P. R., \& Moore, R. T. 1976,
    \apj, 207, 460 
\bibitem[Shapley et al.(2003)]{sha03} Shapley, A. E., Steidel, C. C., 
    Pettini, M., Adelberger, K.L. 2003, \apj, 588, 65
\bibitem[Shapley et al.(2006)]{sha06} Shapley, A. E., Steidel, C. C., 
    Pettini, M., Adelberger, K.L., \& Erb, D.K. 2006, \apj, 651, 688
\bibitem[Shen et al.(2012)]{she12} Shen, S., Madau, P., Aguirre, A., 
    Guedes, J., Mayer, L., \& Wadsley, J. 2012, \apj, 760, 50
\bibitem[Shen et al.(2013)]{she13} Shen, S., Madau, P., Guedes, J., Mayer, L., 
    Prochaska, J. X., \& Wadsley, J. 2013, \apj, 765, 89
\bibitem[Shull et al.(2010)]{shu10} Shull, J. M., France K., Danforth, C. W., 
    et al. 2010, \apj. 722, 1312
\bibitem[Shull et al.(2014)]{shu14} Shull, J. M., Danforth, C. W., \& 
    Tilton, E. M. 2014, \apj (in press),
\bibitem[Siana et al.(2010)]{sia10} Siana, B., Teplitz, H. I., Ferguson, H. C., 
    et al. 2010, \apj, 723, 241
\bibitem[Simcoe et al.(2002)]{ssr02} Simcoe, R. A., Sargent, W. L. W.,
    \& Rauch, M. 2002, \apj, 578, 737
\bibitem[Simcoe et al.(2006)]{{sim06}} Simcoe, R. A., Sargent, W. L. W.,
    Rauch, M., \&  Becker, G. 2006, \apj, 637, 648
\bibitem[Sim\'on-D\'iaz \& Stasi\'nska(2008)]{sas08} Simon-Diaz, S., \& 
    Stasinska, G. 2008, \mnras, 389, 1009
\bibitem[Smette et al.(2002)]{sme02} Smette, A., Heap, S. R., Williger, G. M., 
    et al. 2002, \apj, 564, 542 
\bibitem[Smith et al.(2002)]{snc02} Smith, L. J., Norris, R. P. F., 
    \& Crowther, P. A. 2002, \mnras, 337, 1309
\bibitem[Smith et al.(2011)]{smi11} Smith, B. D., Hallam, E. J., Shull, J. M.,
    \& O'Shea, B. W. 2011, \apj, 731, 6
\bibitem[Sokasian et al.(2002)]{sok02} Sokasian, A., Abel, T., \& Hernquist, L.
    2002, \mnras, 332, 601
\bibitem[Songaila(1998)]{son98} Songaila, A. 1998, \aj, 115, 2184
\bibitem[Songaila(2001)]{son01} Songaila, A. 2001, \apjl, 561, L153 (erratum: 
    2002, \apjl, 568, L139)
\bibitem[Songaila(2005)]{son05} Songaila, A. 2005, \aj, 130, 1996
\bibitem[Songaila \& Cowie(1996)]{soc96} Songaila, A., \& Cowie, L. L. 1996, 
    \aj, 112, 335
\bibitem[Springel \& Hernquist(2002)]{sah02} Springel, V., \& Hernquist, L.
    2002, \mnras, 333, 649 
\bibitem[Stasi\'nska \& Schaerer(1997)]{sas97} Stasi\'nska, G., \& Schaerer, D. 
    1997, \aap, 322, 615
\bibitem[Steidel(1990)]{ste90} Steidel, C. C. 1990, \apjs, 72, 1 
\bibitem[Steidel \& Sargent(1989)]{sas89} Steidel, C. C., \& Sargent, W. L. W. 
    1989, \apjl, 343, L33
\bibitem[Steidel \& Sargent(1992)]{sas92} Steidel, C. C., \& Sargent, W. L. W. 
    1992, \apjs, 80, 1
\bibitem[Steidel et al.(1994)]{ste94} Steidel, C. C., 
    Dickinson, M., \& Persson, E. 1994, \apjl, 437, L75
\bibitem[Steidel et al.(1996)]{ste96} Steidel, C., Giavalisco, M., Pettini, M.,
    et al. 1996, \apjl, 462, L17
\bibitem[Steidel et al.(1999)]{ste99} Steidel, C. C., Adelberger, K. L., 
    Giavalisco, M., et al. 1999, \apj, 519, 1
\bibitem[Steidel et al.(2001)]{spa01} Steidel, C. C., Pettini, M., \&  
    Adelberger, K. L. 2001, \apj, 546, 665
\bibitem[Steidel et al.(2002)]{ste02} Steidel, C. C., Kollmeier, J. A., 
    Shapley, et al. 2002, \apj, 570, 526
\bibitem[Steidel et al.(2003)]{ste03} Steidel, C. C., Adelberger, K. L., 
    Shapley, A. E., et al. 2003, \apj, 592, 728
\bibitem[Steidel et al.(2010)]{ste10} Steidel, C. C., Erb, D. K., 
    Shapley, A. E., et al. 2010, \apj, 717, 289
\bibitem[Stone(1991)]{sto91} Stone. R. C. 1991, \aj, 102, 333
\bibitem[Sutherland \& Dopita(1993)]{sad93} Sutherland, R. S., \& Dopita, M. A. 
    1993, \apjs, 88, 253
\bibitem[Syphers \& Shull(2013)]{syp13} Syphers, D., \& Shull, J. M. 2013,
    \apj, 765, 119 
 \bibitem[Telfer et al.(2002)]{tel02} Telfer, R. C., Zheng, W., Kriss, G. A., 
    \& Davidsen, A. F. 2002, \apj, 565, 773
\bibitem[Tetzlaff et al.(2011)]{tet11} Tetzlaff, N., Neuh\"auser, R., \&
    Hohle, M. M. 2011, \mnras, 410, 190
\bibitem[Theuns et al.(2002)]{the02} Theuns, T., Schaye, J., Zaroubi, S., 
    et al. 2002, \apjl, 567, L103
\bibitem[Tripp, Lu, \& Savage(1995)]{tls95} Tripp, T. M., Lu, L., \& 
    Savage, B. D. 1995, \apjs, 102, 239
\bibitem[Tripp et al.(2002)]{tri02} Tripp, T. M., Jenkins, E. B., 
    Williger, G. M., et al. 2002, \apj, 575, 697 
\bibitem[Tytler et al.(1995)]{tyt95} Tytler, D., Fan, X. -M., Burles, S., 
    et al. 1995, in QSO Absorption Lines, G. Meylan, Berlin: Springer-Verlag, 
    289
\bibitem[Vanzella et al.(2010)]{van10} Vanzella, E., Giavalisco, M., 
    Inoue, A. K., et al. 2010, \apj, 725, 1011
\bibitem[V\'azques et al.(2007)]{vas07}  V\'azques, G. A., Leitherer, C.,
    Schaerer, D., et al. 2007, \apj, 663, 995
\bibitem[V\'azques \& Leitherer(2005)]{val05} V\'azques, G. A., 
    \& Leitherer, C. 2005, \apj, 621, 695
\bibitem[Verner et al.(1996)]{ver96} Verner D. A., Ferland, G. J., 
    Korista, K. T., \& Yakovlev, D. G. 1996, \apj, 465, 487 
\bibitem[Viel et al.(2002)]{vie02} Viel, M., Matarrese, S., Mo, H. J., et al. 
    2002, \mnras, 336, 685
\bibitem[Vogel \& Reimers(1993)]{var93} Vogel, S., \& Reimers, D. 1993, \aap, 
    274, L5
\bibitem[Vogt et al.(1994)]{vog94} Vogt, S. S., et al. 1994, S.P.I.E.E. 2198, 
    362
\bibitem[Wang \& Heckman(1996)]{wah96} Wang, B., \& Heckman, T.M. 1996 \apj,
     457, 645
\bibitem[Weinberg et al.(2004)]{wei04} Weinberg, D. H., Dav\'e, R., Katz, N.,
    Hernquist, L. 2004, \apj, 601, 1 
\bibitem[Weingartner \& Draine(2001)]{wad01} Weingartner, J. C., 
    \& Draine, B. T. 2001, \apj, 548, 296
\bibitem[Wild et al.(2008)]{wil08} Wild., V., Kauffmann, G., White, S. 2008,
    \mnras, 388, 227
\bibitem[Wise \& Cen(2009)]{wac09} Wise, J. H., \& Cen, R. 2009, \apj, 693, 984
\bibitem[Wolfe \& Prochaska(2000)]{wap00} Wolfe, A. M., \& Prochaska, J. X. 
    2000, \apj, 545, 603
\bibitem[Womble et al.(1996)]{wsl96} Womble, D. S., Sargent, W. L. W., 
    \& Lyons, R. S. 1996, in Cold Gas at High Redshift, M. Bremer,
    H. Rottgering, C. Carilli, \& P. van de Werf, Dordrecht: Kluwer, 249
\bibitem[Yajima et al.(2011)]{yaj11} Yajima, H., Choi, J.-H., \& Nagamine, K.
    2011, \mnras, 412, 411
\bibitem[Young et al.(1982)]{ysb82} Young, P., Sargent, W. L. W., 
    \& Boksenberg, A. 1982, \apjs, 48, 455
\bibitem[Zaldarriaga et al.(2001)]{zht01} Zaldarriaga, M., Hui, L.,
    \& Tegmark, M. 2001, \apj, 557, 519
\bibitem[Zehavi et al.(2002)]{zeh02} Zehavi, I., et al. 2002, \apj, 571, 172
\bibitem[Zehavi et al.(2004)]{zeh04} Zehavi, I., Weinberg, D. H., Zheng, Z., 
     et al. 2004, \apj, 608, 16
\bibitem[Zemp et al.(2011)]{zem11} Zemp, M., Gnedin, O. Y., Gnedin, N. Y., \&  
     Kravtsov, A. V. 2011, \apjs, 197, 30
\bibitem[Zhang et al.(1995)]{zan95} Zhang, Y., Anninos, P., 
    \& Norman, M. L. 1995, \apjl, 453, L57 
\end{thebibliography}
\end{document}